\documentclass[letterpaper,12pt]{report}

\usepackage{fullpage}
\usepackage{graphicx}
\usepackage{subfigure}
\usepackage[linesnumbered,norelsize]{algorithm2e}

\usepackage{cuthesis}																																													

\usepackage{makeidx}

\usepackage{morefloats}

\usepackage[numbers]{natbib}
\usepackage{bibentry}

\makeatletter\let\saved@bibitem\@bibitem\makeatother
\usepackage[linktocpage,pdfpagelabels]{hyperref}
\makeatletter\let\@bibitem\saved@bibitem\makeatother

\nobibliography*

\newcommand{\xf}[1]{Figure~\ref{#1}}
\newcommand{\xp}[1]{page~\pageref{#1}}
\newcommand{\xs}[1]{Section~\ref{#1}}
\newcommand{\xa}[1]{Appendix~\ref{#1}}
\newcommand{\xc}[1]{Chapter~\ref{#1}}

\newcommand{\xg}[1]{Algorithm~\ref{#1}}
\newcommand{\xl}[1]{Listing~\ref{#1}}

\newcommand{\etal}{\emph{et al}}

\newcommand{\C}{{C\index{C}}}
\newcommand{\cpp}{{C++\index{C++}}}
\newcommand{\csharp}{{C\#\index{C\#}}}

\newcommand{\java}{{Java\index{Java}}}
\newcommand{\python}{{Python\index{Python}}}

\newcommand{\mono}{{Mono\index{Mono}}}
\newcommand{\monoxna}{{MonoXNA\index{MonoXNA}\index{XNA!MonoXNA}}}

\newcommand{\todo}[0]
{
	{\Large \[TODO\]}
}

\newcommand{\file}[1]{\url{#1}\index{Files!#1}}
\newcommand{\tool}[1]{\texttt{#1}\index{Tools!#1}}

\newcommand{\api}[1]{\texttt{#1}\index{API!#1}}
\newcommand{\apipackage}[1]{\url{#1}\index{API!Packages!#1}\index{Packages!#1}}

\newcommand{\puredata}{PureData\index{PureData}\index{Tools!PureData}}

\newcommand{\opengl}[0]{OpenGL\index{OpenGL}\index{Libraries!OpenGL}\index{API!OpenGL}}
\newcommand{\glut}[0]{GLUT\index{GLUT}\index{Libraries!OpenGL!GLUT}\index{API!OpenGL!GLUT}}
\newcommand{\glui}[0]{GLUI\index{GLUI}\index{Libraries!OpenGL!GLUI}\index{API!OpenGL!GLUI}}
\newcommand{\libsdl}[0]{SDL\index{SDL}\index{Libraries!SDL}\index{API!SDL}}
\newcommand{\cugl}[0]{CUGL\index{CUGL}\index{Libraries!OpenGL!CUGL}\index{API!OpenGL!CUGL}}
\newcommand{\directx}[0]{Direct X\index{Direct X}\index{Libraries!Direct X}\index{API!Direct X}}
\newcommand{\xna}[0]{XNA\index{XNA}\index{Libraries!XNA}\index{API!XNA}}
\newcommand{\gpu}[0]{GPU\index{GPU}}
\newcommand{\glsl}[0]{GLSL\index{GLSL}\index{OpenGL!GLSL}}
\newcommand{\hlsl}[0]{HLSL\index{HLSL}}
\newcommand{\mel}[0]{MEL\index{MEL}\index{Maya!MEL}}

\newcommand{\macos}[1]{\index{Mac~OS~#1@{\sc{Mac~OS~#1}}}{\sc{Mac~OS~#1}}}
\newcommand{\linux}{\index{Linux@{\sc{Linux}}}{\sc{Linux}}}

\newcommand{\win}[1]{\index{Windows #1@{\sc{Windows #1}}}{\sc{Windows #1}}}

\newcommand{\lucidL}[1]{{$\mathit{Lucid}$}($L$) }

		{}

\def\myvert{\raise 2.27pt \hbox{\vrule depth 0pt height 8pt width 0.2mm}}
\def\myarrow{\hspace*{0.43mm}%
             \raise 2.29pt\hbox{\vrule depth 0pt height 8pt width 0.16mm}%
             \hspace*{-0.32mm}%
             $\longrightarrow$
             \ %
             }

\newcommand{\softbodysys}{Softbody Simulation System\index{Softbody Simulation System}}
\newcommand{\jellyfish}{jellyfish\index{jellyfish}\index{Softbody Simulation System!Jellyfish}}

\newcommand{\oglsf}{OGLSF\index{Frameworks!OGLSF}\index{OpenGL Slides Framework}}

\newcommand{\motek}{Motek\index{Motek}}
\newcommand{\caren}{CAREN\index{CAREN}}
\newcommand{\sensics}{Sensics\index{Sensics}}
\newcommand{\falcon}{Falcon\index{Falcon}}
\newcommand{\kinect}{Kinect\index{Kinect}}
\newcommand{\openkinect}{OpenKinect\index{Kinect!OpenKinect}}
\newcommand{\xbox}{Xbox\index{Xbox}}
\newcommand{\wii}{Wii\index{Wii}}

\newcommand{\ogre}{OGRE3D\index{OGRE3D}\index{Tools!OGRE3D}}
\newcommand{\maxmsp}{Max/MSP\index{Max/MSP}\index{Tools!Max/MSP}}
\newcommand{\jitter}{Jitter\index{Jitter}\index{Tools!Jitter}}
\newcommand{\maya}{Maya\index{Maya}\index{Tools!Maya}}
\newcommand{\blender}{Blender\index{Blender}\index{Tools!Blender}}
\newcommand{\vizard}{Vizard3D\index{Vizard3D}\index{Tools!Vizard3D}}
\newcommand{\ipi}{i$\pi$\index{iPi}\index{mocap!i$\pi$}}

\newcommand{\bassaudiolib}{BASS\index{BASS}\index{Libraries!BASS}\index{Audio!BASS}}

\newcommand{\slideimagewidth}{\columnwidth}

\newcommand
	{\worktitle}
	[2]
	{\emph{#1}\index{\emph{#1}}\index{#2!\emph{#1}}\index{Works!\emph{#1}}}

\newcommand
	{\filmtitle}
	[1]
	{\worktitle{#1}{Film}}

\newcommand
	{\docutitle}
	[1]
	{\worktitle{#1}{Documentary}}

\newcommand
	{\animtitle}
	[1]
	{\worktitle{#1}{Animation}}

\newcommand
	{\gametitle}
	[1]
	{\worktitle{#1}{Game}}

\newcommand
	{\playtitle}
	[1]
	{\worktitle{#1}{Performance}}
	
\newcommand{\simmonds}{Dr.~Simmonds}

\renewcommand{\etal}{\emph{et al.}}
\renewcommand{\macos}[1]{\index{Mac~OS~#1@{{Mac~OS~#1}}}{{Mac~OS~#1}}}
\renewcommand{\linux}{\index{Linux@{{Linux}}}{{Linux}}}
\renewcommand{\win}[1]{\index{Windows #1@{{Windows #1}}}{{Windows #1}}}

\newcommand
	{\myphdthesistitle}
	{Computer-Assisted Interactive Documentary and Performance Arts in Illimitable Space}

\degree{Doctor of Philosophy in Special Individualized Program/Computer Science}
\dept{SIP-Computer Science and Software Engineering}
\PhD
\titleOfPhDAuthor{}

\title{{\myphdthesistitle}}
\author{Miao Song}

\makeindex

\newif\ifshortpaper
\shortpapertrue

\newcommand
	{\longpaper}
	[1]
	{\ifshortpaper\relax\else#1\fi}

\begin{document}

\begin{abstract}%
This major component of the research described in this thesis is 3D computer graphics, specifically
the realistic physics-based softbody simulation and haptic responsive environments. 
Minor components include advanced human-computer interaction environments, 
non-linear documentary storytelling, and theatre performance.
The journey of this research has been unusual because it requires a researcher
with solid knowledge and background in multiple disciplines; who also has to be creative
and sensitive in order to combine the possible areas into a new research direction.
Thus, we summarize the innovative research work surrounding the topic as
``{\myphdthesistitle}''.
This work encompasses a lot of research performed in each of the disciplines.
It focuses on the advanced computer graphics
and emerges from experimental cinematic works and theatrical artistic practices.
Some development content and installations are completed
to prove and evaluate the described concepts and to be convincing.

More specifically, on one hand, the major research component includes
the advanced rendering in real-time of an interactive physics-based 
softbody object simulation and visualization with {\opengl}. 
Its immediate follow-up work in this thesis extends that system onto
an artistic interactive jellyfish simulation controlled with advanced
interaction devices, such as Falcon haptics.
Some more advanced rendering techniques have been applied, such as
stereoscopic effects, {\glsl}, LOD etc. in order to bring the CG objects to life and
to increase the level of realism and speed.

On the other hand, the installation work, \docutitle{Tangible Memories}
transforms the award-winning personal documentary film
\docutitle{I Still Remember}~\cite{song-still-remember-movie-bjiff2011}, 
into a non-linear interactive
audience-controlled piece using the new media technologies and devices. 
The audience and society have already appreciated the personal documentary 
\docutitle{I Still Remember}
with the social values resulting in portrayal of immigration,
divorce, re-unification with the family, and other memories.
Turning it into an interactive new media work makes it a much more profound and sensory
in-depth storytelling approach that can be educational as well as help the audience to feel the story
by interacting with it and making it available via many media sources.
Moreover, the audience's participations and feedback are themselves well-preserved in the new ``memory bubbles'',
so that the same documentary project could be eternal and ever evolving,
which may determine the concept of tomorrow's documentary film production.
Additionally, another audacious approach extended from \docutitle{Tangible Memories} installation is for theatrical practice
with the same set of technical tools.
Theater performers could use their body movements, gestures, and facial expressions
to achieve the perceptual and emotional digital effects in sound and images dynamically.

To summarize, the resulting work
involves not only artistic creativity, but solving or combining technological hurdles
in motion tracking, pattern recognition, force feedback control, etc., with the
available documentary footage on film, video, or images, and text via a variety of
devices (input and output, projection, stereoscopic viewing) and programming, and
installing all the needed interfaces such that it all works in real-time.
Thus, the contribution to the knowledge advancement is in solving these
interfacing problems and the real-time aspects of the interaction that have uses
in film industry, fashion industry, new age interactive theatre, computer games,
and web-based technologies and services for entertainment and education.
It also includes building up on this experience to integrate {\kinect}- and
haptic-based interaction, artistic scenery rendering, %(the jellyfish),
and other forms of control. 
This research work connects all the research disciplines,
seemingly disjoint fields of research, such as computer graphics, 
documentary film, interactive media, and theatre performance together.

\end{abstract}

\begin{acknowledgments}
I am deeply grateful to my supervisory committee
for their valuable advices and support 
as well in guiding this dissertation and providing all the necessary resources.
Choosing an incredible committee might be the best decision I have made.

The first person I have to thank is my principal supervisor, Dr.~Peter Grogono.
Whenever I approach him with different questions in various disciplines,
he could always give me an intelligent and insightful advice not only in Computer Science,
but also in theatre, film, music, and interactive media fields.
Later I discovered the reason why he is so knowledgeable and
has very artistic disposition is because he had
been working in electronic music,
theatre performance, and even film production early in his career.
Dr.~Grogono is wise, senseful, and allowed me
to be creative in this uncommon research
journey with the freedom and space I needed. Moreover, he gives me a lot of encouragement,
spiritual, and financial support. There is no doubt that I wouldn't be even on the right track in
the research work and
would have drowned from the disastrous family situation
without his continuing trust, patience, and tolerance for the last several years.

I am grateful to Professor Jason Lewis,
whom I knew the earliest among other professors.
He always gives me incredible support and very valuable advice. I remember when I was in my
undergraduate studies, we had quite friendly and pleasant conversations. From that time
I was dedicated to achieve a higher degree, a master's, and now a doctoral degree as well.

I would like to show my appreciation to Professor Marielle Nitoslawska
who agreed to be on my supervisory committee
when I applied to the SIP doctoral program.
It was thanks to her trust that I could achieve my destination
even though at that time when I proposed the research topic, ``interactive documentary'' really sounded a novel idea.
I got a lot of inspirations from her own research work, every meeting and conversation with her.

I am very lucky to have met Dr.~Maureen Simmonds three years ago and to have worked in her research lab.
It might have been a completely different research experience if I did not have the chance to collaborate with her.
Working with her not only broadened my view in the interdisciplinary research, 
but also was the first time that I could apply some academic knowledge to a real medical research case.

I am very grateful to Dr. Peter Rist,
who arrived at my supervisory committee
at the later stage of my research. He is so knowledgeable in Chinese cinema and animation and
gave me the opportunity to rediscover the beauty of Chinese culture and arts, which I had forgotten
about even after having grown up in China.

I was very honoured to be accepted to the SIP program and to have received the support from
Dr.~David Howes and Ms.~Darlene Dubiel.
Dr.~Howes has been encouraging me and providing with
all the possible solutions in front of me
when I encountered a dilemma and had to make a
hard decision in my interdisciplinary studies.

I believe that it is my fate to meet Dr.~Don Marinelli who would become the turning point in my career.
I thank him for inviting me to the ETC, Carnegie Mellon to see the ``world''.
The trip gave me tremendous experience not only with a new way of seeing things,
but also a chance to demonstrate my own research projects to peers.

Also, a token of appreciation for helpful advice
and comments go to professionals and collaborators on all the different multidisciplinary
projects I worked with giving me an enriching experience.
Thanks go out to Professors Jean-Claude Bustros,
Alison Loader, Robert Reid, Ruru Ding, Guojun Ma, Thomas Waugh, Alice Jim, and Tongdao Zhang.
Many thanks to the Montreal Herstory Theatre Performance Group artists---it was
a soul exchange experience to collaborate with all of them.

I would express my gratitude to Marco Luna who gave me very precious and valuable
advice in documentary filmmaking. Without his encouragement, I wouldn't be brave
enough to make the short documentary \docutitle{I Still Remember}.
Thanks to the curator, Ms.~Annie Briard who selected \docutitle{I Still Remember} at the HTMlles festival.
I also thank the BJIFF ``See the World through Films'' Competition Committee to have given me
the best short documentary award. I also would like to acknowledge filmmakers Mr.~Paul Carvalho
and Ms.~Beverly Shaffer to collaborate with me on the future fiction film. 
Many thanks to Joel Taylor, Mark Baehr, Momoko Allard, Phil Hawes from Hexagram and CDA
who gave me a lot of technical support.
Thanks for their tolerance when every time I had a technical emergency and 
asked their help at the last minute.

Thanks to Ms.~Catherine LeBel and the CSLP team. Without her trust and encouragement, 
I would not have been in the position to continue my work and studies.
Moreover, I need to thank my lawyer Me~Raphael Levy, my daughter's lawyer Me~Beatrice Clement,
and Judge Pierre B\'{e}liveau, to having saved a desperate woman's and a poor child's lives.
I am truly and deeply indebted to so many friends that 
helped and saved me and my daughter during our family divorce crisis, 
and supported and cared about us.
They are Xuemei Ye and her family, Yichuan Zhang and his family, Jin Cao and her family,
Li Ma and her family, Ying Chen and her family, Yi Zhao and her family,
Shulei Guo, Mi Zhou, Helen Wang, Debbie Davis, Sharon Nelson and others whom
I could not name them all.

I would like to express my appreciation to my family to support me
to bring this project to completion.
Thanks to my daughter Deschanel, for her inspiration, collaboration, artistic creativity,
and contributions to my life and research work.
She shared truthful feelings with me so that we together made \docutitle{I Still Remember}.
Thanks to my son, Timothy, who brings me a lot of happiness, hopes, and courage.
Many thanks to my brother Liu Song, for his emotional and financial support to me and my family.
I cannot thank enough and am forever indebted to my parents, Lin Song and Fang Liu, 
for their invaluable support to my life and my family.
It is because they have been helping me to bring up my two kids,
I would be able to stay in school late every evening and concentrate to my studies.
I need to thank to my husband, Serguei Mokhov, for never leaving me (the poor soul) 
behind no matter what happened for so many years.
Thanks to his love, clinging together in times of trouble, and to
his support and belief that I could eventually complete my studies.

In the end, I would like to acknowledge the SIP program, the Department of Computer Science and
Software Engineering, Faculty of Engineering and Computer Science,
Hexagram Concordia, and Graduate School at Concordia University
as well as FQRSC doctoral scholarship, CCSEP scholarship, McGill University, 
CSLP Concordia University, the Central Academy of Drama (China), China Scholarship Council, 
for various financial, space, hardware, and software resources
provided to make this work possible.
\end{acknowledgments}

\chapter{Introduction}
\label{chapt:introduction}

This chapter begins by stating the research question this thesis answers
in \xs{sect:research-question}.
Motivation for this research crossing multiple disciplines is
detailed in \xs{sect:ndisciplinary-research}.
\xs{sect:research-overview} reports on the creative and research achievements, the process,
and the thesis scope.
\xs{sect:organization} outlines the remaining content of this
thesis.

\section{Summary Statement of Research Questions}
\label{sect:research-question}

Theatre, Film, Television:
from the very ancient art form with thousands of years of history
to the recent main new media sources of entertainment in the last several decades,
whether played by a live performance or a pre-recorded program,
all these media have something in common: performer, audience, and space.
These media have played a pivotal role in the socialization, civilization,
education, and human connection.
Moreover, in each of their traditional contexts,
the performance is typically linear, the actor is scripted,
the audience is observationally passive, and the space is physically constrained.

After the appearance of the first electronic computers developed in the mid-20th century,
accompanied by the arrival of computer networking and the Internet, technology has
dramatically evolved. Consequently, new media products and devices, such as
digital theatre, video games, Internet TV, computer-graphics-based animation documentary film,
``smart'' mobile phones, virtual/augmented reality, stereoscopic movies, web based docudrama,
building projection, etc., gave audiences more participatory power to the control of artistic works,
thus creating more new challenges to the production team and artists.

Today, the new computer technology has already affected the well-established traditional art form,
especially in the aspects of actors' theatre performance, filmmaker's production decisions, and
the audience's participation, and even in the redefinition
of the nature and boundaries of ``space'' in the digital era.

Will virtual reality (VR), computer generated images (CG) and environment destroy
the ``holy'' aspect in theatre or, conversely, improve its representation?
Will the new digital technology subvert the definition of realism of documentary film arts?
What does the computation do to change the relation or difference
between theatrical and cinematic experiences?
What will the next generation interactive theatre and documentary film be like?
What unique experience could the live event bring to audience other than pre-recorded media
in consideration of HCI?
What have the theatre and documentary been? How could we redefine them today in the digital age?

The purpose of this research is to investigate an innovation on 
how multidisciplinary aspects, such as computer graphics,
interactive media, cinema montage, and theatre performance
could coherently form and blend in harmony.

\section{Motivation for Multi-, Inter-, and Transdisciplinary Research Interests}
\label{sect:ndisciplinary-research}
\index{research!multidisciplinary}
\index{research!interdisciplinary}
\index{research!transdisciplinary}

When I first heard about SIP, the Special Individualized Program at Concordia University, 
which allows students to pursue an interdisciplinary research degree, 
I was extremely happy that my artistic background I acquired in China 
would be useful at last!
Even though I had already completed another bachelor's and a master's degree in Computer Science at
Concordia University, it had not been an easy transition for me
to switch to Computer Science given my background in China
was in theatre performance, and following that I became a TV journalist having
graduated from Shandong University of Arts in 1997.

Thus, my research gradually and incrementally focuses on the mentioned aspects and
includes the following: enhanced softbody simulation and specialist training in virtual reality,
interactive media, documentary film production, and theatre arts as the applications of my research.
None of these were offered by a single specific Faculty or Department at the PhD
level making the SIP (Special Individualized Program) unique and appealing
to blend the multiple disciplines throughout my research.
I have worked on my proposed research areas and made some contributions
and achievements in documentary film making, theatre performance, and computer graphics. 
I am especially interested in how humans interact with computers and computer-generated
graphics not only in science and everyday life, but also in artistic creativity.

It has been already a challenge to work on interdisciplinary research that usually requires 
collaborations of experts and professionals from specified areas.
They have to understand the concepts underlying a discipline other than their own
and find a common language to communicate ideas.
A lot of courage is required to face up to the highest degree in interdisciplinary research,
which becomes more and more a buzzword these days,
and have one person with all the related skills and knowledge.
Additional difficulty arises when that person's English is not their first language. 
However, there are also some advantages, such as the innovative ideas 
could be more naturally, unitedly, and efficiently developed because
everything is deeply rooted within \emph{myself}.
Meanwhile, the SIP gave me a great opportunity to work with a team of highly
qualified professors in computer science, film studies and production, virtual reality, theatre
and digital media arts, who have strongly showed their enthusiasm
in supporting and guiding me throughout my endeavor.

Regardless of the actual contents of the research itself,
my whole research experience shaped up this work from two fronts:
I have been studying a technical degree after originally being trained as an artist.
Then as a result I could combine the technical knowledge back to the artistic creativity.
Thus, this work has become
very valuable to me and my life personally, I believe the society as well,
as a pioneer of future interdisciplinary researchers.
I am quite confident that standing in between
science and art, western and eastern, with hard work, passion, experience,
and knowledge definitely makes a great contribution to the rapidly
developing era.

To me, a \emph{multidisciplinary study} means that research could be done
in a mixture of disciplines, in each of which there exist their own methodologies.
These disciplines don't typically integrate or interface
with each other. Additionally, there are two other terms, {\em interdisciplinary}
and \emph{transdisciplinary} that
have also been increasingly used these years. The former combines two or more 
academic fields into one single discipline; the later crosses many disciplinary 
boundaries to create a holistic approach, as to quote:

\begin{quote}
\emph{``Multidisciplinarity draws on knowledge from different disciplines but,
stays within their boundaries. Interdisciplinarity analyzes, synthesizes and 
harmonizes links between disciplines into a coordinated and coherent whole. 
Transdisciplinarity integrates the natural, social and health sciences in 
a humanities context, and transcends their traditional
boundaries.''}~\cite{mul-inter-trans-disciplinary}
\end{quote}

My studies started from a multidisciplinary viewpoint
and have been expanded to interdisciplinary and transdisciplinary research methods.
First of all, the proposed research area of focus for the SIP doctoral research, ``{\myphdthesistitle}''
is leaning towards the softbody simulation (inherited from my master's research topic~\cite{msong-mcthesis-2007})
in haptic-enabled environment and specialist-training fused together. 
In that work I proposed
a variety of \emph{inexpensive} techniques for human-computer interaction
in various domains as well as advanced 3D modeling and animation in a rich
spectrum of application domains.
Thus, I explored options to come up with a set of tools, techniques, and
methodologies that use haptics- and other-interface-enabled devices
such that the toolbox/framework we develop in this work is affordable
and suitable with the least number of modifications to the respective
application domains.

Secondly, the thesis work also has promised a research and development on a
new method and approach on how technical invention effects artistic production,
such as documentary film making and theatre performance.
It demonstrates how important the new technology could help 
develop the contemporary studio art works and satisfy artist's wild imagination
compared to the traditional art forms.

\section{Research, Creation, and Development Process Overview}
\label{sect:research-overview}

I can still remember when the first time I proposed this interdisciplinary research,
the idea seemed very novel but uncertain.
After the past several years of research and study in Computer Graphics (CG), Virtual Reality (VR),
Interactive Digital Media, 
Theatre Performance,
and Documentary Film Production, today, I could conclude the research work was worthwhile
with some convincing evidence.
This is manifested through the publications, film screening, and media installations.

The SIP program required a diverse range of courses taken from different disciplines.
From the courses offered from Concordia University's Faculty of Engineering and Computer Science, such as
``Advanced Rendering and Animation'', ``Computer Graphics for Television Production'' 
(offered by my principle supervisor Dr.~Peter Grogono), and ``3D Graphics and Computer Animation''
(offered by Dr.~Nizar Bouguila).

I acquired deeper knowledge and understanding in theory of algorithms for advanced computer
graphics and animation, advanced rendering, and stereoscopic techniques.
The course co-offered by Concordia's same faculty and McGill University's Faculty of Medicine,
School Of Physical and Occupational Therapy, ``Computer Graphics in Virtual Realities''
(co-supervised by Drs.~Peter Grogono and Maureen Simmonds), gave me the precious opportunity to access to fancy motion
capture (MoCap) system, advanced virtual reality applications, augmented reality equipment, and to discover a closer relationship
among computer technology and its application to medical research.

The courses
offered by Concordia University Mel Hoppenheim School of Cinema,
``Expanded Documentary'' (given by Dr.~Marielle Nitoslawska) and
``PhD Film Studies Seminar in Film and Moving Image History'' (taught by Dr.~Thomas Waugh),
supplemented my needs in theoretical and practical applications to documentary film and 
their possible expansion resulted from the rapidly developing computer technology.
I started to become aware that traditional documentary form has been more and more affected
by computer-generated imagery and focused on how human beings' stories could transform from
the traditional linear storytelling to non-linear interactive approach.

I was excited to find out that theatre arts, which have always been a love knot to me,
as an art form, actually is extremely rich and powerful through ``Enchantment, Matter, and Topological Media''
(offered by Dr.~Sha Xinwei) at Concordia's Design and Computation Arts Department.  
Another new discovery will be some research of arts in Asia,
more explicitly, the film and animation arts in China.
Moreover, I was very impressed by Dr.~Peter Rist's knowledge of Asian cinema and Asian traditional theatre.
In his course, ``Film Studies in Chinese Cinema and Water Ink Animation Film'',
offered by Film Study Program at Concordia University, we reviewed famous Chinese feature films and 
international winning animated films and their animation techniques.
Furthermore, from September 2011 to July 2012, I have been awarded the CCSEP (Canada-China Scholar's Exchange Program) scholarship
and was studying ``Traditional Chinese Theatre and its History'' (by Dr.~Guojun Ma) at Theatre Literature Department and
``Theatre Directing, from Ideas to Performances'' (with Dr.~Ruru Ding) at Theatre Direction Department of
the Central Academy of Drama in Beijing, China.

It was a great pleasure to rediscover theatre from a brand new angle and gain more knowledge in Chinese traditional art form and the Contemporary arts
after many years computer science studies and related technical training.
Additionally, I have also audited other courses related to my research area, such as Dr.~Robert Reid's theatre class in the Theatre Department
and Dr.~Jean-Claude Bustros's ``Film Production'' course, and had very deep discussion with them about my proposed research.

Additionally, in the last three years, I also obtained various opportunities to explore and work on different
projects with professors from many research disciplines which lasted
from several weeks to many months.
From collaborations at close quarters with the professors,
I derived a lot of first-hand experience from ideas generated from scratch that led
into productions, the team cooperation, and their critical and creative approach
to the artistic and research work.
More importantly, I have also witnessed several excellent examples in the
interdisciplinary research, and have been fascinated about 
the outcome of the combinations.

The very early collaboration was with Dr.~Alice Ming Wai Jim, Concordia Art History Professor,
whose research focuses on media arts, spatial culture and contemporary Asian and Asian-Canadian art.
From the research work done in her project, \worktitle{Augmented Reality in Asian Contemporary Arts}{Project},
I realized that there have been some emerging Asian media artists already experiencing virtual and augmented reality
into their creative installations.
The short research assistance experience in Dr.~Jason Lewis's project \worktitle{CitySpeak}{Installation}~\cite{cityspeak}
and Dr.~Marielle Nitoslawska's
and Alison Reiko Loader's
\docutitle{The Grey Nuns Project} gave ideas about how the programming and computer graphics
could be applied to interactive media installation and film production projects.
In the \worktitle{3D Maya Stereoscopic Plugin}{Project}~\cite{stereo-plugin-interface,stereo3d} project I collaborated
with Professor Alison R. Loader; this project was the
first that tested my multidisciplinary background because it used the {\opengl} and {\mel} programming
skills to implement a stereoscopic tool for 3D {\maya}~\cite{maya} software interface in order
to help a 3D animation filmmaker to preview stereoscopic effects in realtime prior to final rendering.
The \worktitle{Interactive Cinema}{Project} project led by Dr.~Jean-Claude Bustros opened up my mind about
how traditional film production could be
benefited from interactive media, how the interactive cinema/film could be.
I have not only participated in the implementation work
as a programmer, but also got close look at the artistic design of their interactive cinema system.
The two years of working experience as a research assistant at the ``Virtual Reality in Pain, Mind and Movement Lab''~\cite{pain-performance-vr-pmmIII}
originally led by Dr.~Maureen Simmonds at McGill University, was
extremely important to me and my research as well.
From there, I could have access to a very modern MoCap system, sensor-enabled treadmill, 
HMD (Head Mounted Display) equipment and the software system, where one could implement applications
for medical research (and of course entertainment).
Moreover, the experience of being a teaching and research assistant and lab instructor
in Computer Graphics, Animation for Video Games, and various courses
was very instrumental to keep up my knowledge in computer graphics
and video games development up-to-date.

\subsection{Achievements and Contributions}

Throughout the past years I have been active in various areas of research in terms
of actual investigation, publications, lab research work, and the like. %teaching assistance work.
I have published or have accepted-for-publication several related conference papers and posters
on each or combined domains included in my proposed research, such as
real-time physical based animation, conceptual work on haptic devices and interactive cinema,
stereoscopic plug-in and its applications in medical and cinematic research.
There were also debate of positions when I presented my work
and research results at conferences and public presentations.
Moreover, I got my art works and documentary film presented
and showcased in different conferences and festivals respectively.
See \xs{sect:conclusion-contributions} for a detailed list.

In my primary research direction in computer graphics,
I have made some achievements in softbody simulation system framework and 
3D game development as well
~\cite{jellyfish-c3s2e-2012,%
vr-medical-research-docu,%
msong-mcthesis-book-2010,%
stereo-plugin-interface,%
softbody-framework-c3s2e08}.
Throughout the duration of this aspect the knowledge of how to adapt inexpensive haptic devices,
algorithms, techniques, and methodologies to new human computer interaction in order for it to be more accessible
to digital media artists, augmented reality researchers, and computer scientists alike to further
research in the area, and make it accessible to audiences at home and at school.

Since I have had more opportunities in interdisciplinary studies, including the mentioned
expanded documentary film studies, Asian cinema studies, the role of computer graphics
in TV and theatre productions, there also have been some achievements in theoretical research
and studio works in cinema/documentary film, interactive media~\cite{i-still-remember-opengl-remake-2011,%
song-still-remember-movie,%
presentation-hexagram-2010,%
water-ink-animation-film-2011,%
haptics-cinema-future-grapp09,%
role-cg-docu-film-presentation-2009}.

As a studio component of my research,
a very recent award is for my documentary film, \docutitle{I Still Remember},
included in the top ten video art work showcase~\cite{song-still-remember-movie} at HTMlles 2010 festival and,
moreover, it won the best short documentary film~\cite{song-still-remember-movie-bjiff2011}
at 1st Beijing International Film Festival (BJIFF).
One of the sub-directions of my SIP PhD research topic centered in Interactive Documentary.
The first {\opengl} prototype of an interactive media installation, \docutitle{Tangible Memories},
which is based on my short documentary film, has been implemented and demoed
at Concordia Humanities student conference and was also presented
at the ICEC~2011 conference in October~2011~\cite{i-still-remember-opengl-remake-2011}.

Additionally, I got involved with {\simmonds} and the previously mentioned supported course
and her McGill lab on virtual reality techniques as a research assistant and a lab
technician of the {\motek} equipment and a head-mounted display VR
systems in pain, movement, and VR graphics research.
This overall research experience covers, at various levels,
conducting a literature research, collection and summarization of the sensor
measurements and results,
and developing a testable research question and research proposals.
It also included development of basic proof-of-concept prototype walk-through game to demonstrate the idea
on human-computer interactivity, learn the API, as well as make a more real-life application for patient testing
and operating of the VR lab within the context of the research 
\cite{vr-medical-research-docu}.
In the end, I was able to apply back that experience directly to the
studies and research in computer graphics and interactive techniques
and documentary~\cite{vr-medical-research-docu}.

Overall, these years of working experience in multidisciplinary and 
interdisciplinary fields enables me to think critically and open-mindedly on how future cinema and 
theatre lead the direction of the new media. I am able not only
to be very creative in art production, but also to handle the associated technical difficulties.
I could see how mature now and how much I have progressed as a researcher 
compared to three years ago.
The knowledge earned from multidisciplinary research is all blended in together
in my thoughts.
Such knowledge acquired from making the installations, setups, software development and re-use,
and the mathematics
behind
is a contribution made
available as a guideline of how to make such setups,
the process, and select the appropriate tools and
techniques~\cite{haptics-cinema-future-grapp09}.
Such contributions are important in the age of digital media not only
for artists and small audiences, but also for educational purposes of children through interaction,
as it is widely known that children develop better their abilities when one interacts with them more.
Telling interactive stories, teaching decision making in various life situations are specifically
important applications of this research~\cite{haptics-cinema-future-grapp09}.

\subsection{Process}
\label{sect:creative-rnd-process}

My creative research, development, and production process is thematically
depicted and described in this section following one of the main themes
of ``memory bubbles''.
Conceptually, the five high-level bubbles in \xf{fig:overall-research} representing
each of the elements of the research corresponding to these themes.

\begin{figure*}[htpb]%
	\centering
	\includegraphics[width=\textwidth]{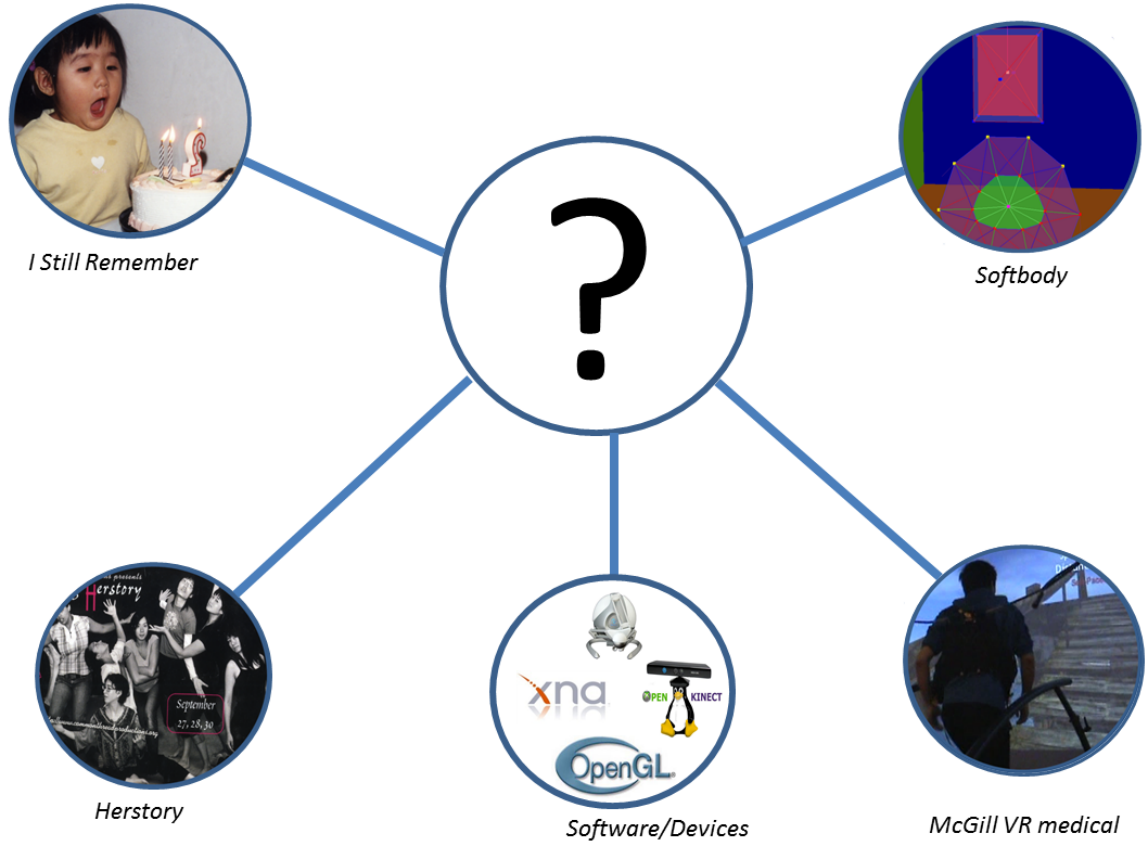}%
	\caption{Overall Major Research Points Converging}%
	\label{fig:overall-research}%
\end{figure*}

Bubble~1 in \xf{fig:overall-research} represents
the real-time softbody simulation.
This project is physical based simulation to render 1D, 2D, and 3D 2- and 3-layer
elastic objects and some applications of it~\cite{softbody-framework-c3s2e08}.
Bubble~2 represents the Virtual Reality Lab for pain, mind, and movement research
that was originally founded by Dr.~Simmonds from McGill University, Faculty of Medicine.
The lab contained a motion capture (MoCap) system for motion data acquisition,
a treadmill as a part of the interactive environment, and an arc screen for realistic
computer graphics projective display~\cite{pain-performance-vr-pmmIII}.
Bubble~3 represents affordable devices and software APIs used for application implementation.
The haptic device acts not only as an interaction tool with computer generated virtual environment, but also a
resource of responsive force and feedback of simulated CG objects back into the real.
It gives more realistic and advanced interactive perceptual stimulation between a human and a computer.
Other possible consumer devices for human-computer interaction
are the now well-known {\kinect}, {\wii} and {\xbox} controller and APIs are {\opengl} and {\xna}.
Bubble~4 represents \worktitle{Unraveling Her Story}{Theatre production}, which was a theatre
performance piece co-produced with a group of female artists including
myself in 2007~\cite{her-story-miao-2007}.
Six of us were also the script writers and directors
for our own story pieces, which were based on the actresses'
ancestors or own experience.
Bubble~5 represents the \docutitle{I Still Remember} documentary film
based on my personal story~\cite{song-still-remember-movie-bjiff2011}.
It is also a project about memories, collected between me and my daughter.
The story was told by the little girl about what she could remember from
when she was small.
The bubble in the middle is the
the research question (detailed in \xs{sect:research-question})
combining computer graphics, virtual reality, interactive media,
theatre performance, responsive environments, and documentary production. 

In subsequent figures, specifically in
\xf{fig:etc1}, \xf{fig:etc2}, \xf{fig:etc3}, \xf{fig:etc4}, 
\xf{fig:etc5}, \xf{fig:etc6}, and \xf{fig:etc7},
the process is refined and connections are
drawn between media, performance, and technologies.

\begin{figure*}[htpb]%
	\centering
	\includegraphics[width=\textwidth]{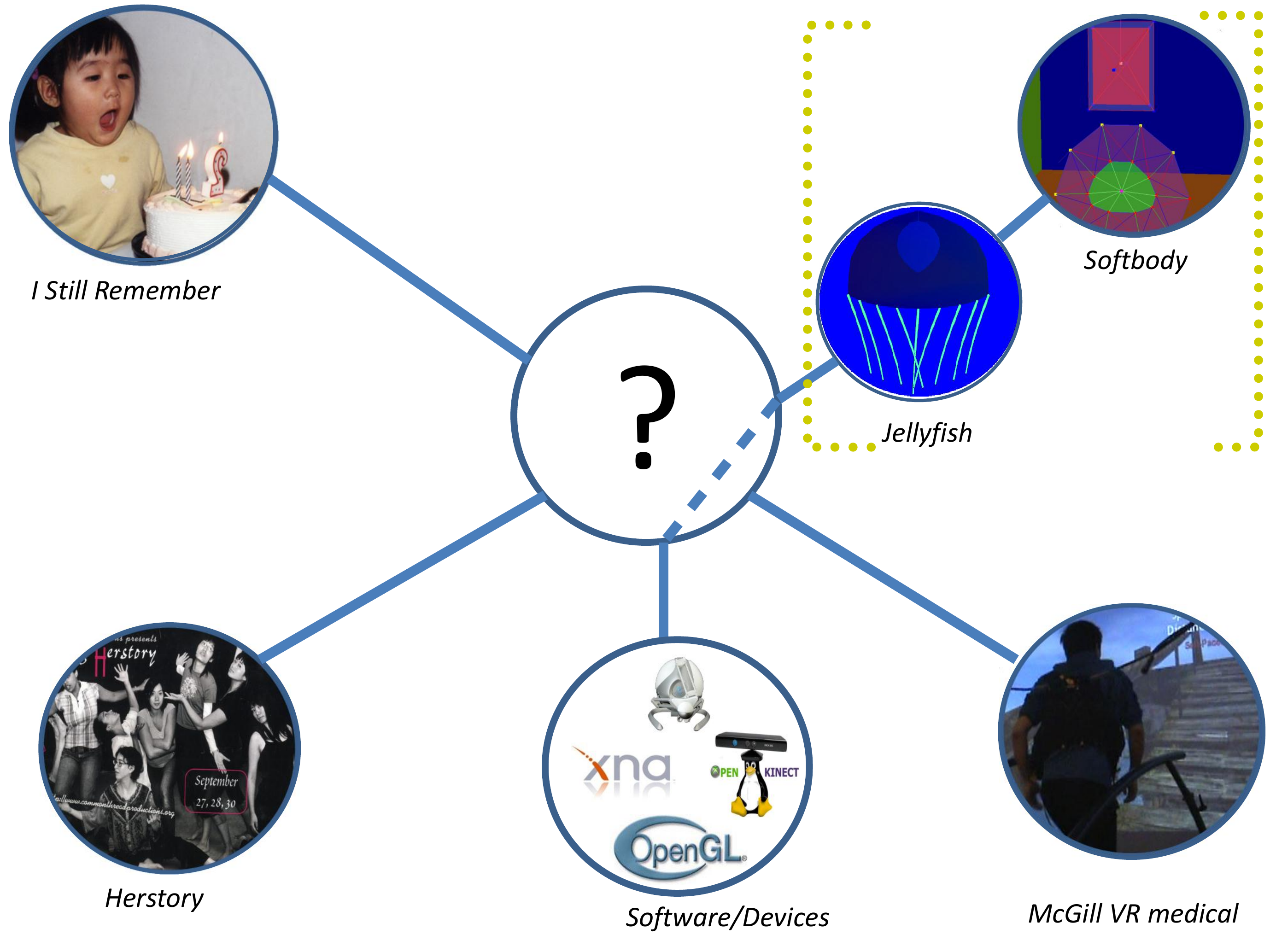}%
	\caption{Overall Major Research Points Converging Progression 1}%
	\label{fig:etc1}%
\end{figure*}

In \xf{fig:etc1} the earlier contribution of the {\softbodysys} is
extended and applied to the modeling and interactive physical based animation of
the {\jellyfish} with added {\falcon} haptic touch to it.

\begin{figure*}[htpb]%
	\centering
	\includegraphics[width=\textwidth]{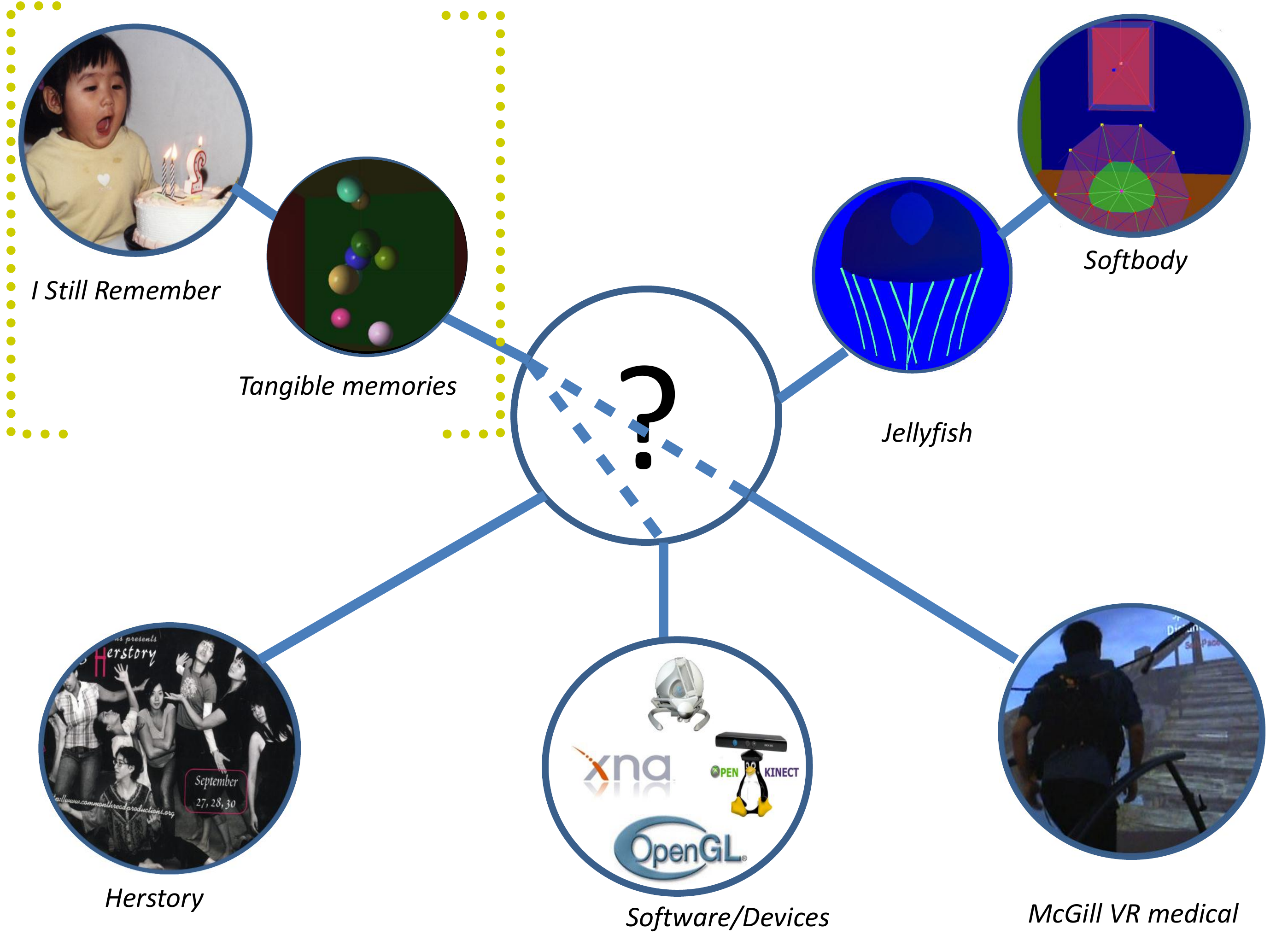}%
	\caption{Overall Major Research Points Converging Progression 2}%
	\label{fig:etc2}%
\end{figure*}

The concepts of VR interaction and computer graphics are expanded into
an interactive documentary depicted by the additional bubble in \xf{fig:etc2}.
In the process of fusing the linear documentary footage, CG
and VR environments, connections are made to make the new media
interactive documentary \docutitle{Tangible Memories} with the memory bubbles
(see \xs{sect:interactive-docu-intro-scope}) that we detail in this
thesis in \xc{chapt:interactive-docu}.

\begin{figure*}[htpb]%
	\centering
	\includegraphics[width=\textwidth]{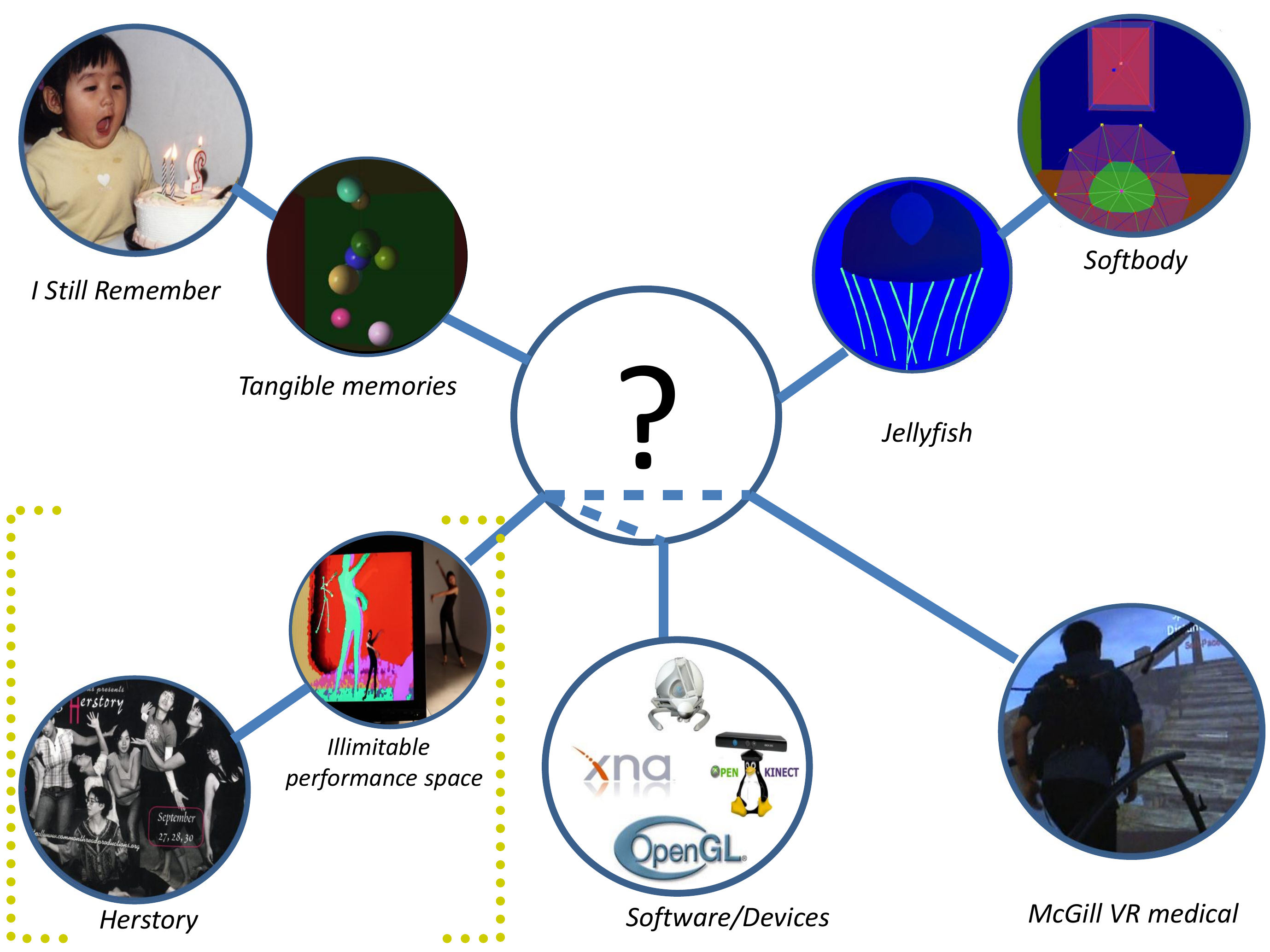}%
	\caption{Overall Major Research Points Converging Progression 3}%
	\label{fig:etc3}%
\end{figure*}

In \xf{fig:etc3}, the bubble of theatrical performance, \worktitle{Illimitable Performance Space}{Theatre production}
inspired by previous theatre production \worktitle{Herstory}{Theatre production}, connects to the same interactive computer graphics
and VR techniques. This innovation gives the audience not only unlimited imagination space,
but also actors real improvisation experience by interacting with computer generated environment.  

\begin{figure*}[htpb]%
	\centering
	\includegraphics[width=\textwidth]{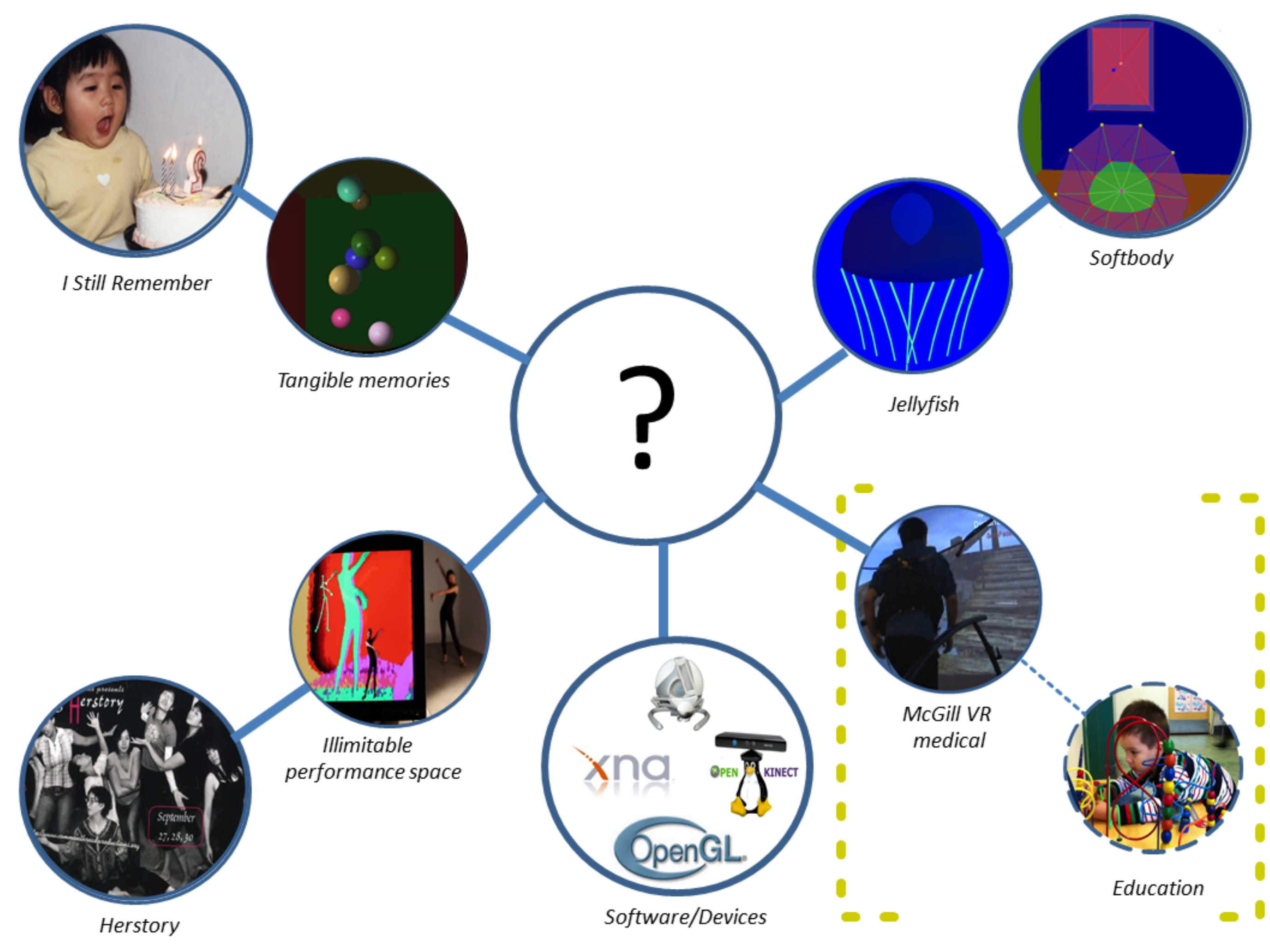}%
	\caption{Overall Major Research Points Converging Progression 4}%
	\label{fig:etc4}%
\end{figure*}

\xf{fig:etc4} indicates that the VR system and its applications could be
used not only for medical research purpose, but also for education
(in progress, outside the scope of this thesis, see \xs{sect:scope}).

In \xf{fig:etc5} the process expands further from the short linear
documentary \docutitle{I Still Remember} to a feature length fiction
film ``Snowflake'' (in progress, also outside the scope of this thesis).

\begin{figure*}[htpb]%
	\centering
	\includegraphics[width=\textwidth]{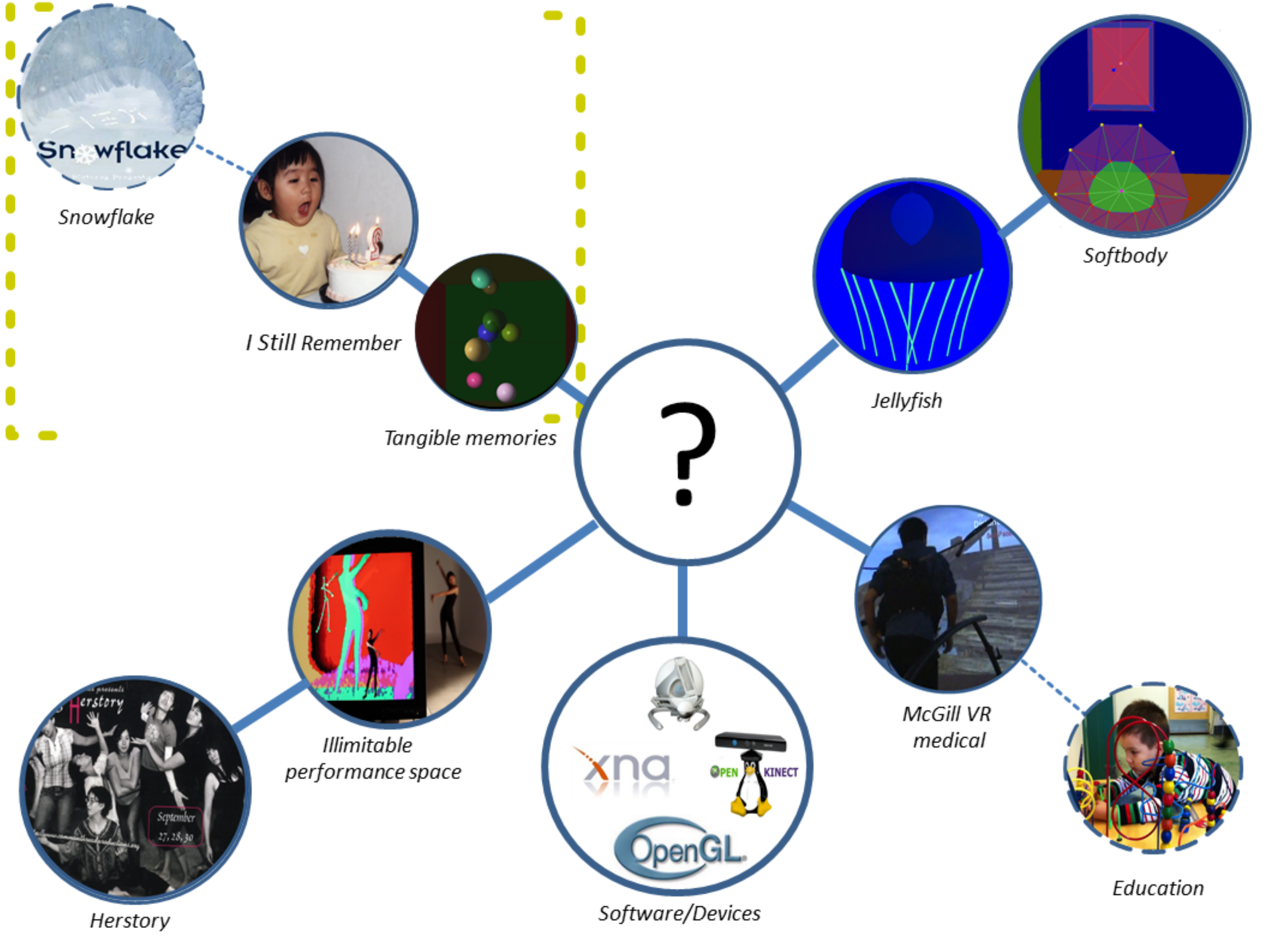}%
	\caption{Overall Major Research Points Converging Progression 5}%
	\label{fig:etc5}%
\end{figure*}

At the same time, in \xf{fig:etc6} the additional bubble for the
projected \docutitle{Zooming Through the Generations} interactive documentary
(which is outside of the scope of this thesis), connects to the
the same interactive computer graphics and VR techniques, theatrical
performance of real people portraying scenes footage for which can no
longer be otherwise recorded augmented with the audience participation.

\begin{figure*}[htpb]%
	\centering
	\includegraphics[width=\textwidth]{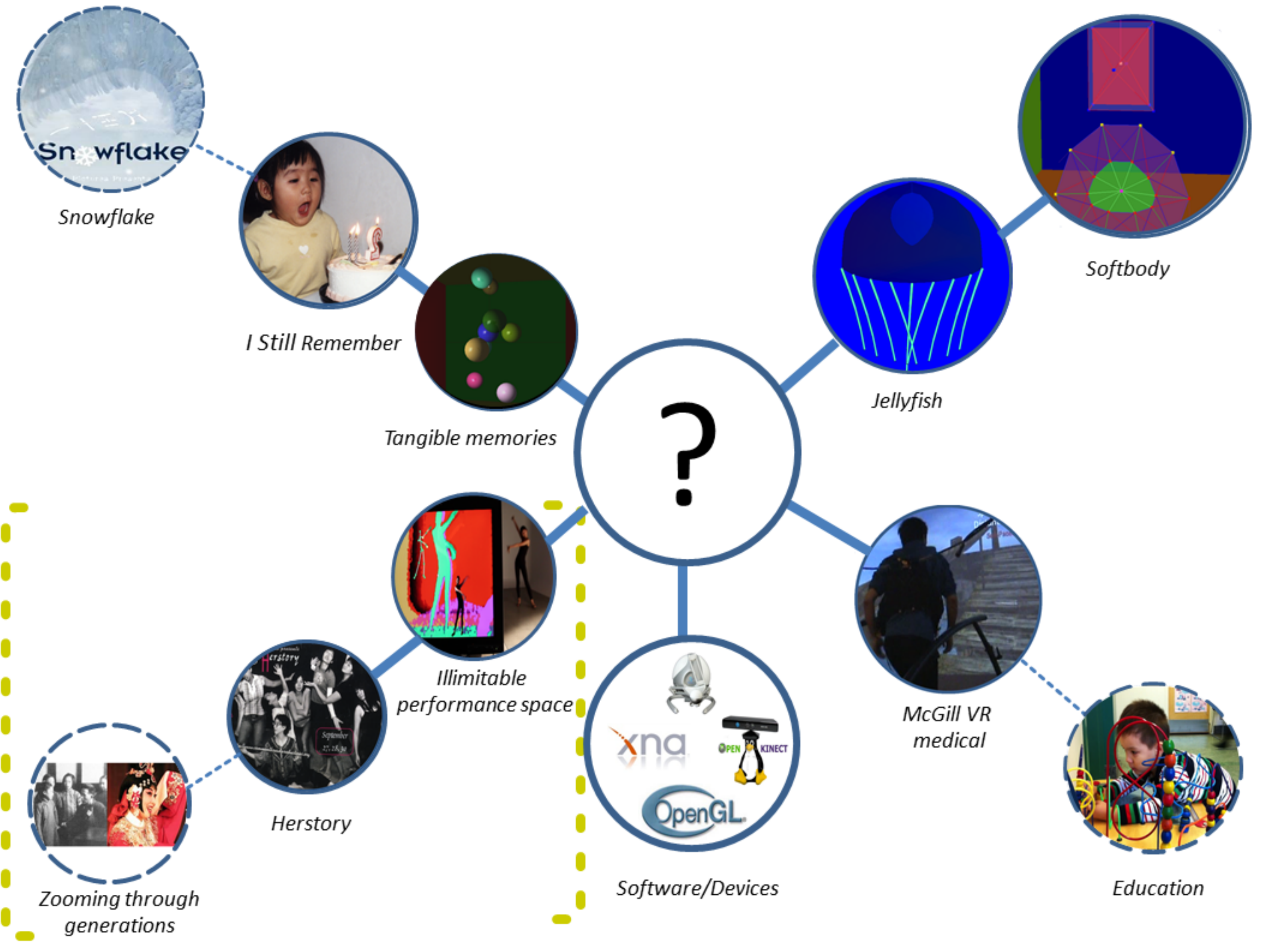}%
	\caption{Overall Major Research Points Converging Progression 6}%
	\label{fig:etc6}%
\end{figure*}

\xf{fig:etc7} is the future work in computer graphics research direction.
Physical based softbody/jellyfish simulation could be even applied
onto skeleton-based characters in order to make their animation more realistic.

\begin{figure*}[htpb]%
	\centering
	\includegraphics[width=\textwidth]{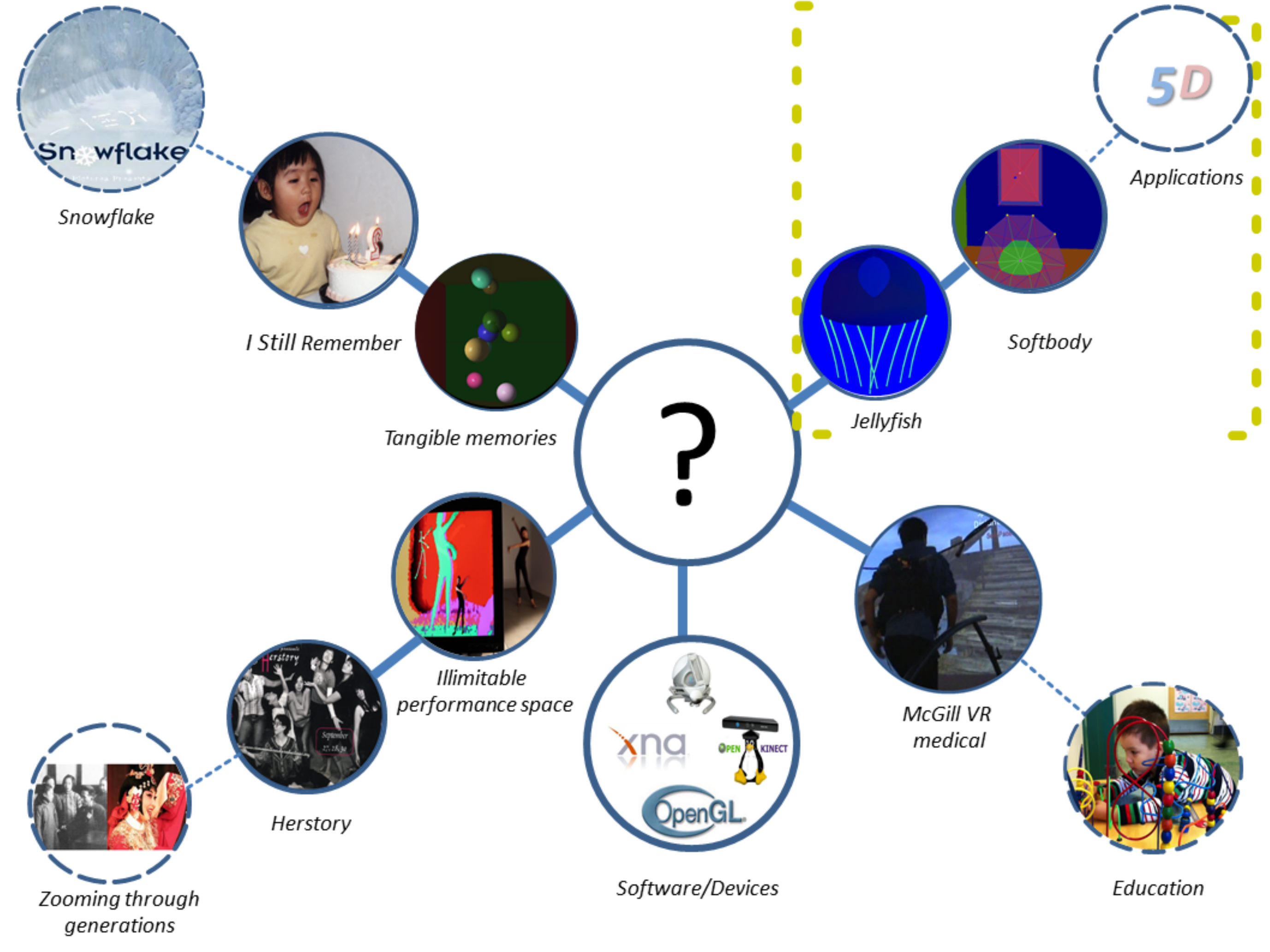}%
	\caption{Overall Major Research Points Converging Progression 7}%
	\label{fig:etc7}%
\end{figure*}

Thus the multidisciplinary nature of the process of connecting
CG, art, performance and production (\xf{fig:etc7}) pose some
interfacing challenges, which this thesis addresses in part
and lays ground for the ongoing and future work.

\subsection{Scope}
\label{sect:scope}

Here we declare limitations put on the research question based on the feasibility 
with the time constraints (how much of the proposed work is intended to
be done and how much is intended to be left for the future).

\subsubsection{Softbody Simulation and Jellyfish}

This part focuses on the analysis and review of the
relevant
advanced rendering and animation techniques in
computer graphics used for 3D games, some
of which we juxtapose with the physically based softbody
simulation framework we designed and implemented earlier.
We also discuss such implementation
results and the future directions. The techniques we touch upon
are various algorithms for animation and physical based
modeling as well as {\gpu} programming with {\opengl}.
We break down our contribution into two main parts:
the analysis and review of the applicability of the techniques to
softbody rendering and animation with the summary of
the results to date~\cite{adv-rendering-animation-softbody-c3s2e09}.

Then, we derive a set of requirements from the detailed case study
of the {\softbodysys} design and specification in an attempt
to come up with the requirements that all similar systems need to have and
to further generalize our findings to other physical-based simulation
systems involving real-time computer graphics~\cite{soen-spec-cg-simulation-systems}.

Moreover, we describe details of modeling and implementation of an
interactive {\jellyfish} simulation
using {\opengl} and the {\softbodysys} as a result.
We outline our process, setup, planned exhibition installation, and the difficulties encountered
and the ways we solve some and propose to solve some other difficulties
in this work as a possible guideline for others doing similar installation
with some educational value.
We review a number of related work items considered in
the design and implementation of this work as well as future
plans and considerations~\cite{jellyfish-c3s2e-2012}.

\subsubsection{Interactive Documentary}
\label{sect:interactive-docu-intro-scope}

It is or at least has been difficult in general to make
documentary interactive.
The interactivity in documentaries converges from several streams of
interactive media, such as interactive TV, the Internet, computer graphics,
interactive cinema, computer games, and associated realities (virtual,
augmented, etc.) and their associated interaction
devices.
In our research we look into how common interactive TV and film
as well as computer graphics techniques may help with the
interactive documentary production and what it means
to interact with a documentary.
We also review some related literature and existing documentary
projects and the logic and techniques behind them. Furthermore, by comparing
some documentary projects which are about similar topics, with linear and nonlinear storytelling
approach, we present our position on future documentary production, and the importance
of interactivity of documentary film making.

Making the \docutitle{I Still Remember} documentary's floating memory bubbles
interactive with audience's participation first used a cross-platform
{\opengl} (aimed at later enhancements with haptics).
We describe a simple process of making a passive documentary interactive using available
tools and preserving the aesthetic and emotional appeal.
Moreover, careful comparison of the linear conventional film and 
its nonlinear narrative version with audiences' body movement involvement
may give answers to some of the artists who are still hesitant to adapt their projects
to the dramatically developed new technology~\cite{i-still-remember-opengl-remake-2011}.

The interactive (non-linear) documentary story-telling approach explored
here is somewhat different from what was known by interactive documentary
as a ``web documentary'' of database-based compiled/submitted footage
played in the order desired by the audience or merged into games
\cite{wiki:web-docu,interactive-docu-golden-age-2009,korsakow-system,highrise-window-docu}.
Instead, we render the documentary footage in a graphics environment inside
bubbles, like ``memory bubbles'', based on a personal story (including the
concept of the ``memory bubbles'' themselves) which can be called out by
their color via voice or a key/click.

We rendered the documentary in two instances, one is based on
{\opengl}~\cite{i-still-remember-opengl-remake-2011} and the newly
redone one is based on {\xna} and {\kinect}.

The {\opengl} version features the
AVI documentary footage, photographs, or tidgets, and the bubbles and
{\glui} and mouse-based interaction.

The {\xna}+{\kinect} version has more advanced bubble effects,
speech recognition, and {\kinect} interaction on top of the keyboard-based
version. This later version has various perspectives being rendered for
projection onto various surfaces in the performative blackbox environment.
Two of the ``fancy'' soap bubbles can be hand-controlled. The other bubbles
are split into media-containing ones (that can be called out by their color)
and soap bubbles.
The content is our own production including the short documentary film.
The bubbles float in 3D space and collide with each other.

\subsubsection{Illimitable Performative Space Installation}

Illimitable performance space installation, while bounded in a particular environment,
such as a theatrical black box, removes the limits of participation of local and remote
audience. In the first prototype installation it involves real-time animation with
audience participation as well as effects of fancy bubble animation along with
background live feed video projected on the floor (from the ceiling). With {\kinect}
we work with color and depth streams for fascinating motion tracking effects of
the main performer---a dancer wearing traditional Chinese dance long sleeves.
The bubbles move along with the dancer movements to different directions. Other
cameras and possibly {\kinect} sensors observe the audience who are also projected
on different displays, walls, etc. The dancer and audience video is preserved from one session
into a new bubble and another and is ``memorized'' and documented and played back
along with the subsequent performances.
This way each performance is contributed to and enables multiple dancers projections
at the same time. A similar dance or audience participation can be submitted
as well via the Internet and played back on the Internet creating a global show.

\section{Organization}
\label{sect:organization}

This chapter starts with the introduction and overview of my cross disciplinary %research 
background, the overall multidisciplinary research overview, and proposed research questions
in \xs{sect:research-overview}.
\xc{chapt:background} describes all the necessary background and related work.
Specifically, we briefly review
the related work and literature on various topics
of interactive cinema, documentary, and our position about
it as well as new media and interaction research in TV, cinema,
and documentary in general.
\xc{chapt:softbody-objects} details the methodology of real-time softbody object
simulation applied to a jellyfish modeling and animation.
\xc{chapt:interactive-docu} presents an interactive documentary approach based
on the application of some very common computer graphics tools and techniques 
and the personal linear \docutitle{I Still Remember} documentary \cite{song-still-remember-movie-bjiff2011}
told non-linearly inside ``memory bubbles'' as \docutitle{Tangible Memories}
with advanced user interaction features.
\xc{chapt:illimitable-space} introduces the design and implementation of
an illimitable performative space and installation.
\xc{chapt:conclusion}
concludes
with the contributions, advantages, and limitations of our approach and
an outline of the near and longer term future work.
\longpaper
{
\xc{chapt:discussion} discusses various aspects of this work from a positional
point of view.
}
\xa{appdx:user-manuals} details some operational instructions and user manuals
on the developed systems.

\paragraph*{Fair use of imagery.}

The author Song declares the use of screen captures and the related imagery
produced by others
in this academic work are for the illustrative reasons under the fair
use rationale (similar to Wikipedia's~\cite{wikipedia}) where such work may still be
under copyright and the author does not make any claims of ownership
or authorship of the said images. The screen captures and the like
are attributed in the text and are necessary to illustrate the points
presented in this academic research work.
A certain number of images from the cited documentary films are already
in the public domain or have compatible documentation licensing.

\chapter{Background and Related Work}
\label{chapt:background}

In this chapter, some historical aspects of the topics related
to the present research are reviewed to give deeper context
into the foundations of this dissertation's work.
These related works involve
computer animation and advanced rendering (\xs{sect:cg-background}), interactive
media and documentary film production (\xs{sect:interactive-media-docu-background}),
and performance arts and theatre production (\xs{sect:theatre-background}).
The related technology used in or considered for this research is described
in \xs{sect:tech-background}.

More specifically, the multidisciplinary background described in this chapter
is of relevance and the source of inspiration for the chapters that
follow:

\begin{itemize}
	\item 
many of the the computer graphics, animation, and rendering techniques
described further are employed primarily in \xc{chapt:softbody-objects}
(\emph{Toward Realtime Jellyfish Simulation in Computer Graphics})
as well as throughout the other main contribution chapters (\xc{chapt:interactive-docu}
and \xc{chapt:illimitable-space})

	\item
the interactive media and documentary film production background serves
as a source of inspiration and analysis of the novelty needed for non-linear
storytelling and advanced interaction to produce the prototype of the
interactive installation documentary \docutitle{Tangible Memories}
detailed in \xc{chapt:interactive-docu} (\emph{Tangible Memories in Interactive Documentary})

	\item
the performance arts and theatre production provides the theoretical and
practical foundation to the discovery and design of the illimitable performative
space concept using the new technology for traditional performance art forms
detailed subsequently in \xc{chapt:illimitable-space} (\emph{Illimitable Space in Responsive Theatre})

	\item
the technological background describes the affordable technologies today
that aid with the prototyping for and experimenting with the above main three
aspects and these technologies are used throughout one more of mentioned
chapters

	\item
the topics discussed in this chapter continue to be relevant not only in the
current contributions summarized in \xc{chapt:conclusion}, but form basis for
the vast open-ended future work possibilities summarized there as well to
complete the missing pieces and serve as a guide for direction or source of
inspiration and improvement over for the purposes of the modern evolution
of the present art forms
\end{itemize}

\section{Computer Animation and Advanced Rendering}
\label{sect:cg-background}
\label{sect:bg-animation-rendering}

Graphics are drawings and visual representations which can
be traced back from 40,000 to 10,000 BC or even
earlier~\cite{OSU-CG-history,animation-asia-pacific-2001}\footnote{\url{http://en.wikipedia.org/wiki/Graphics}}.
``Animation is a graphic representation of drawings to show
movement within those drawings.''\footnote{\url{http://en.wikipedia.org/wiki/History_of_animation}}
Computer graphics and animation
are the drawings and sequences of drawings generated by computer,
which has had a profound impact 
on television, movies and the video game industry. It has significant value for 
medical research and media arts achievement as well 
with relative long history since 1960s~\cite{cg-with-opengl-4ed}.

In this section we review only
some of the most prominent computer graphics techniques for advanced rendering and animation
in 3D games and beyond instead of traditional graphics and animation.
The analysis is based on 
the various literature, tutorials, and practice reviews,
some of which are~\cite{realtimerendering2002,advancedrealtimerendering2006,cs5243,mudur-comp7661,rost2004,losasso2004}
as well as other works cited throughout the text~\cite{adv-rendering-animation-softbody-c3s2e09}.

\subsection{Traditional and Asian Animation Influence on Computer Animation}
\label{sect:bg-animation}
\label{sect:bg-traditiona-asian-animation-to-cg}

We briefly review traditional animation\longpaper{ history},
some of the early animated film in both Western (e.g., \xf{fig:early-animated-feature-films})
and Asian (e.g., \xf{fig:early-asian-color-animations}) countries and their traditional techniques.
Moreover, several computer animated films have also been
included in the discussion, particularly, having Asian artists and  
their contributions assessed.

\begin{figure*}[ht]
\hrule\vskip4pt
\begin{center}
	\subfigure[\textit{Snow White} (1937) Poster]
	{\includegraphics[width=.45\columnwidth]{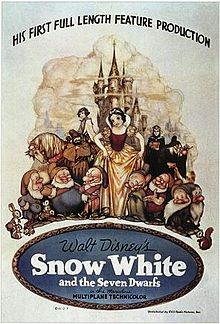}
	 \label{fig:snow-white}}%
	\hspace{.1in}
	\subfigure[\textit{Princess Iron Fan} (1941) Poster]
	{\includegraphics[width=.45\columnwidth]{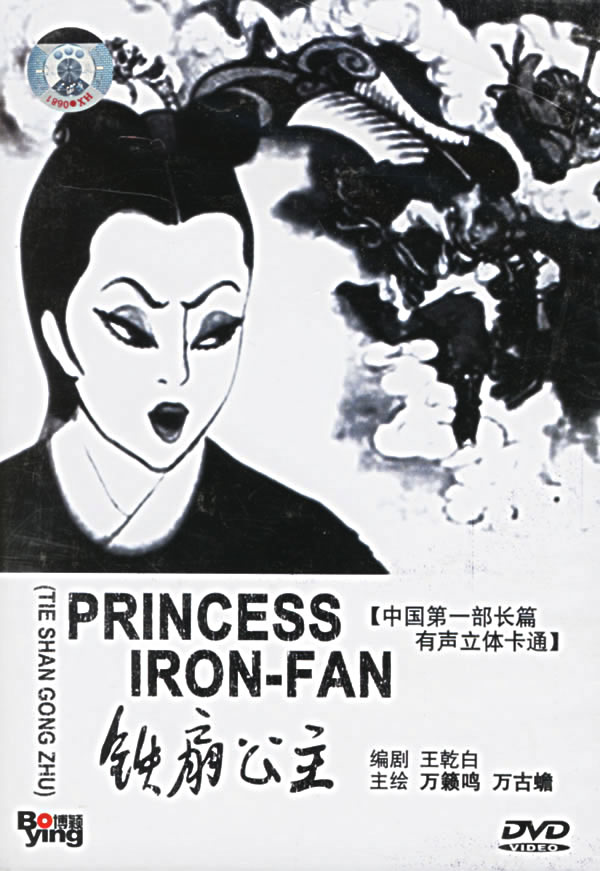}
	 \label{fig:princess-iron-fan}}
\caption{Early Animated Feature Films}
\label{fig:early-animated-feature-films}
\end{center}
\hrule\vskip4pt
\end{figure*}

Animation has a long history. 
The first feature animation film was \animtitle{Snow White} (\xf{fig:snow-white}) made in the US in 1937.
And the second one was \animtitle{Princess Iron Fan} (\xf{fig:princess-iron-fan}), made in 1941 in China.
Chinese animation played a very important role in Asian animation, especially
influenced Japanese animation film. It was interesting to know that the
first Japanese color animation feature film, \animtitle{The Tale of the White Serpent},
was even based on the Chinese folk story.

\begin{figure*}[ht]
\hrule\vskip4pt
\begin{center}
	\subfigure[\textit{Uproar In Heaven} (1961,1964) Poster]
	{\includegraphics[width=.45\columnwidth]{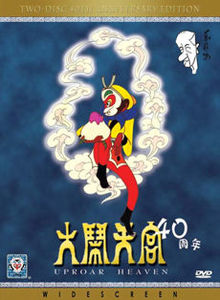}
	 \label{fig:havoc-in-heaven}}%
	\hspace{.1in}
	\subfigure[\textit{Astro Boy} (1963) Poster]
	{\includegraphics[width=.4\columnwidth]{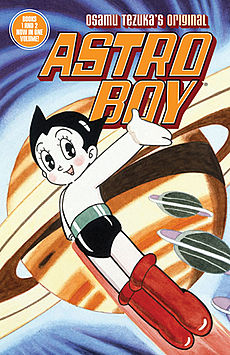}
	 \label{fig:astro-boy}}
\caption{Early Asian Color Animations}
\label{fig:early-asian-color-animations}
\end{center}
\hrule\vskip4pt
\end{figure*}

Traditional 2D animation was drawn by hand, such as Chinese water-ink animation.
Other animation art forms use puppets, paper cutting, and folding-paper.
Some excellent animation works in the 20th century were 
\animtitle{Uproar In Heaven} (\xf{fig:havoc-in-heaven}, in China, 1960s), Asian first feature animation in color, based on Chinese
the earliest chapters of the classic story \worktitle{Journey to the West}{Story};
\animtitle{Astro Boy} (\xf{fig:astro-boy}), Japanese manga series broadcast on TV program (in Japan, 1963);
\animtitle{Hong Gildong}, the earliest Korean animation movie in which the main character 
was from an old Korean novel (in Korean, 1967);
Hong Kong's first puppet animation film of theatrical animation, 
\animtitle{Prince of the Big Tree} (in Hong Kong, 1948);
and Malaysian animated cartoons, \animtitle{Hikayat Sang Kancil} (in Malaysian, 1978)
was produced for television~\cite{animation-asia-pacific-2001}.

After CG technology appeared and since CG animation techniques stemmed originally from the
traditional animation, many cartoonists turned to animators in order to
accommodate themselves in the new digital era producing first CG-animated films (e.g., \xf{fig:early-cg-feature-animations}).
After the first CGI feature-length animation film, \animtitle{The Toy Story}~\cite{wiki:toy-story} (\xf{fig:toy-story}, in the US, 1995), was created, 
ten years later, first Asian 3D-CG film, \animtitle{Thru the Moebius Strip} (\xf{fig:moebius-promo}) was made (in China, 2005),
(whose cost was about 20 million in US Dollars), and has finally been released.

\begin{figure*}[ht]
\hrule\vskip4pt
\begin{center}
	\subfigure[\textit{Top Story} (1995) Poster]
	{\includegraphics[width=.4\columnwidth]{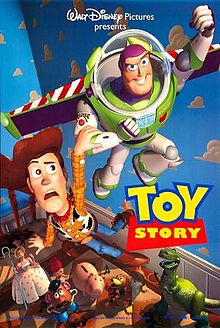}
	 \label{fig:toy-story}}%
	\hspace{.1in}
	\subfigure[\textit{Thru the Moebius Strip} (2005) Poster]
	{\includegraphics[width=.45\columnwidth]{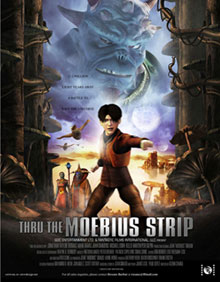}
	 \label{fig:moebius-promo}}
\caption{Early CG Feature Animation Film}
\label{fig:early-cg-feature-animations}
\end{center}
\hrule\vskip4pt
\end{figure*}

In general, there are 2D and 3D CG animation types.
Looking at the Chinese market as an example (e.g., \xf{fig:cg-application-China}), there are two categories of animation:
conventional animation (usually well-financed projects by government
or production companies) and ``webtoons'' (mostly produced by individuals
who post them on the web).
Chinese TV series and animation film box office winner, 
\animtitle{Pleasant Goat and Big Big Wolf-The Super Snail Adventure}
(in China, 2009, \xf{fig:pleasant-goat-big-wolf}) was criticized however by the experts
because it aimed at audience aged from 6 to 8.
In contrast to the Western CGI film usually attracts a wider range of
audience and could portray very serious topics in films.

Similarly to the roles of CG in North America, in Asia, 
CG animation has been widely used for theatre performance, 
film, advertisement, special effects in post production, 
education, medical training, military rehabilitation,
and video games.

Some examples includes
an augmented reality installation, \worktitle{Citizens Comfort}{Installation} (in Singapore, 2008)~\cite{citizens-comfort};
motorized installation, \worktitle{Cloud}{Installation} (Tai Wan Interactive Technology Laboratory)~\cite{motorized-installation};
mixed reality installation, \worktitle{The Swimming Fish}{Installation} (Beijing, Tsinghua University)~\cite{a-swimming-fish}
(where an interactive virtual reality installation translates the visitor's hand gestures to 
generate a swimming fish as if he or she reenacts the magical touch of the painter
Ma Liang\index{Ma Liang}---a Chinese legend). Moreover, some new developing CG companies, such as \worktitle{E-GO Computer Graphics}~\cite{e-go-cg},
focus on high-end digital animation in film, television, theater, and large scale outdoor projections (e.g., \xf{fig:e-go-cg}).

\begin{figure*}[ht]
\hrule\vskip4pt
\begin{center}
	\subfigure[\textit{Pleasant Goat and Big Big Wolf-The Super Snail Adventure} (2007) Poster]
	{\includegraphics[width=.3\columnwidth]{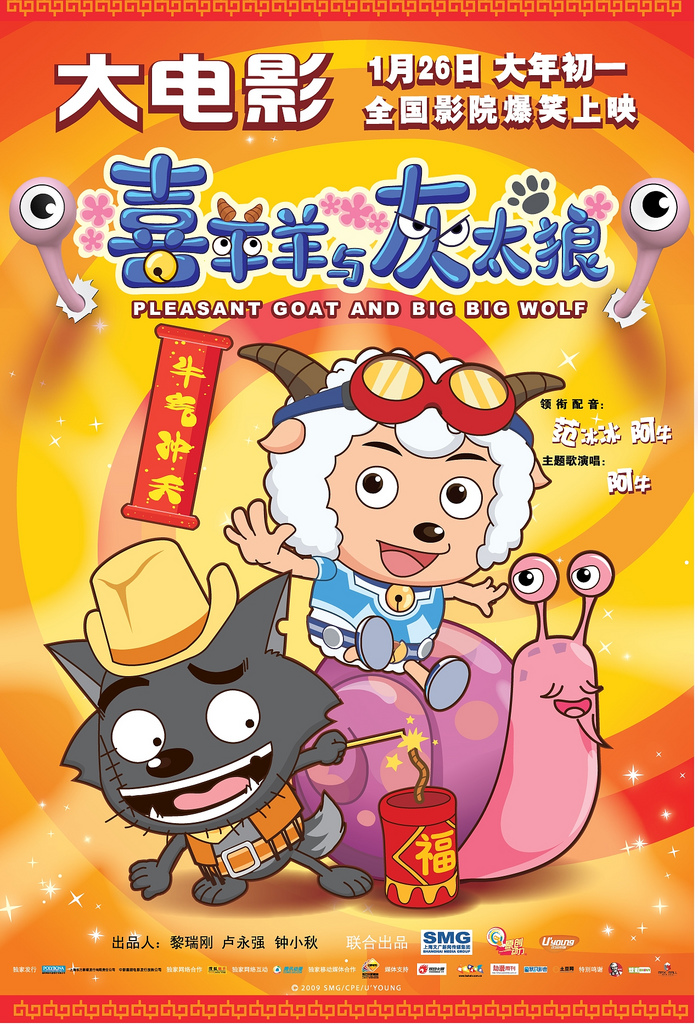}
	 \label{fig:pleasant-goat-big-wolf}}%
	\hspace{.1in}
	\subfigure[\textit{Tsinghua University 100 Anniversary Celebration Building Projection} (2011)]
	{\includegraphics[width=.6\columnwidth]{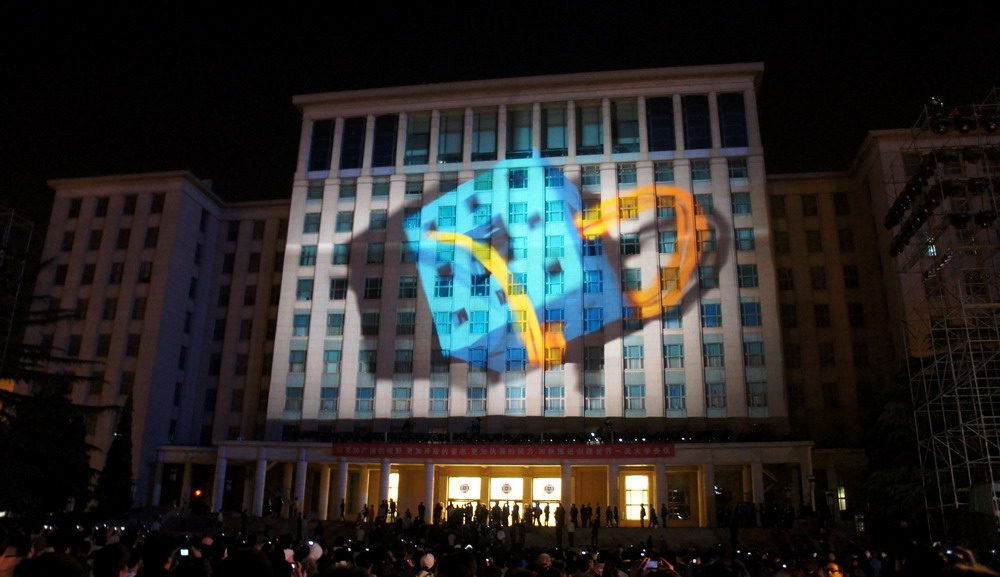}
	 \label{fig:e-go-cg}}
\caption{CG Applications in China}
\label{fig:cg-application-China}
\end{center}
\hrule\vskip4pt
\end{figure*}

The burst of CG and innovation in animation in Asia can be witnessed by
creation of the major branch of the SIGGRAPH conferences---SIGGRAPH Asia
in 2008 and continuing to today that went through Hong Kong, Singapore, Japan, Korea
to additionally accommodate ever-growing dissemination of works in computer
graphics that include not only technical papers, but also art papers,
exhibitions, animation festivals, apps, and the like, featuring numerous artist
from Asia and abroad
\cite{%
siggraph-asia-2010-proceedings,%
siggraph-asia-2010-cg-anim-fest,%
siggraph-asia-2010-art-tech-showcase,%
siggraph-asia-2009-art-gallery-emerge-tech,%
siggraph-asia-2009-cg-anim-fest,%
siggraph-asia-2008-cg-anim-fest,%
siggraph-asia-2008-art-gallery-emerge-tech,%
siggraph-asia-2008-proceedings}.
There has also been a boom in video games (mostly from Japan (Sony) on PlayStation platforms)
and other vendors.

\subsection{Computer Animation}
\label{sect:bg-computer-animation}

Computer animation is a sub-field of computer graphics, which uses both
2D and 3D computer graphics~\cite{kerlow-3d-anim-2000} to create moving images.
The currently accepted rate of computer generated animation
is 30 frames-per-second (FPS) or 24 FPS for film.
The animation can be divided into two main categories:
offline and real-time.
Both off-line and real-time methods apply acceleration
methods, algorithms, and approximations in order to meet
the efficiency requirements as much as possible while
trying to retain some realism/believability elements
(e.g., \xf{fig:offline-realtime-animation-example-screenshots}).

\begin{figure*}[ht]
\hrule\vskip4pt
\begin{center}
	\subfigure[\textit{Offline Animation from Maya Software}~\cite{tradigital-maya-2011}]
	{\includegraphics[width=.7\columnwidth]{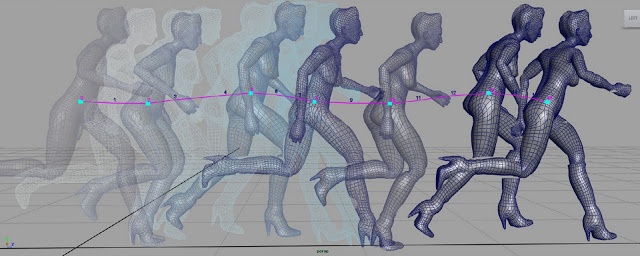}
	 \label{fig:maya-offline-animation}}%
	\hspace{.1in}
	\subfigure[\textit{Real-time Animation from Angry Bird Game}]
	{\includegraphics[width=.6\columnwidth]{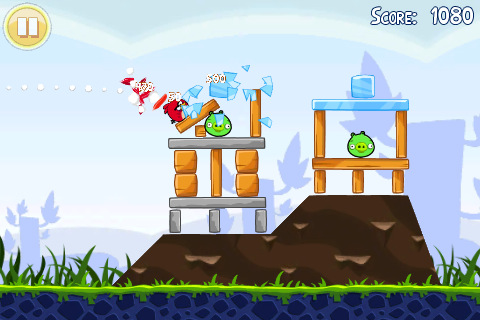}
	 \label{fig:angry-birds-realtime-animation}}
\caption{Offline and Real-time Animation Example Screenshots}
\label{fig:offline-realtime-animation-example-screenshots}
\end{center}
\hrule\vskip4pt
\end{figure*}

\subsubsection{Offline Animation}
\index{animation!offline}

Offline animation (also called ``precomputed animation''), constructs a list of fixed animated scenes without
responding to users' actions in real time. The animator or programmer has exact control of
the animation. The change of the scene database and scene viewing
happen independently. For example, offline animation generates a film 
and plays it back later. It is used mainly to create high-quality 
lifelike scenes for film industry.

The performance of the rendering is only the second priority compared to its realism. 
Most special effects in today's movies and entire animation films are created by computer graphics.
In order to achieve high-quality effects, the graphics hardware has become more programmable.
Some good examples of offline animations are
\animtitle{Final Fantasy} (\xf{fig:offline-final-fantasy}),
\animtitle{Toy Story}~\cite{wiki:toy-story},
\animtitle{Finding Nemo},
\animtitle{Cars} Pixar movies.

Some of the techniques for modeling and animation of the offline scenes are tool-based modeling,
data-driven modeling, procedural modeling, behavioral animation,
and dynamic physical based modeling, which we discuss next.

\paragraph*{Tool-Based Modeling.}
\index{modeling!tool-based}

We have a variety of techniques to model the animated scene, such as
using software modeling tools to directly specify 3D models and their transformations.
One of the traditional offline animation techniques is \emph{keyframing} (e.g., \xf{fig:maya-offline-animation}), which
depends on an artist to generate key frames and the in-between frames are drawn
automatically by a computer. Linear interpolation creates the in-between frames
at equal intervals along straight lines, whereas nonlinear interpolation
can create equally spaced in-between frames along curved paths.
Another example for tool-based modeling is that the artist directly controls
how a given skeleton specified by joint angles moves and the
computer solves for the angles that achieve the motion.
In general, tool-based modeling is tedious and inflexible.

\paragraph*{Data-Driven Modeling.}
\index{modeling!data-driven}

This technique measures the real world, and then uses that data 
to synthesize or alter models.
\emph{Motion capture} is an example of data-driven modeling,
which captures the motions of the actor and aligns motion
data with a CG character. Other examples include laser scanning
and eye tracking techniques. Moreover, the recently developed
{\wii}, {\kinect}, and haptic devices produce a lot of sensor
data that can be used in real-time data-driven modeling.

\paragraph*{Procedural Modeling.}
\index{modeling!procedural}

During procedural modeling,
one writes programs to automatically generate models and their transformations,
for example, to apply in special effects, such as explosions, water,
and flocking behavior by specifying simple physical rules of
motion~\cite{wiki:procedural-modeling}.

\paragraph*{Behavioral Animation.}
\index{animation!behavioral}

This technique provides realism to computer-generated animation,
which is defined by describing an actor's behavior or goal~\cite{beh2005}.
It gives the characters autonomous intelligent behavior and
is related to AI. For example, an actor's behavior defines
how the actor interacts with other actors and the environment.
It is particularly useful for crowd animation.

\paragraph*{Dynamic Physical Based Modeling.}
\index{modeling!physical based}
\index{animation!physical based}

Such modeling offers a concise, but rich and flexible
way of defining the behavior of animated objects and characters, 
based on the physics laws to determine or guide their motion. 
Particle systems are common examples here, which represent objects
such as fire, smoke, etc., as a collection of individual particles~\cite{TR83,pb98,aw99}.
Moreover, dynamic physical based modeling also applies to the rigid body animation,
deformable objects animation~\cite{deformable-obj-anim-reduced-control-2009,vega-deformable-lib},
cloth simulation, and fluid simulation.

\subsubsection{Real-Time Animation and Simulation}
\index{animation!real-time}
\index{simulation!real-time}
\index{real-time!animation}
\index{real-time!simulation}

Unlike offline animation, real-time animation %and simulation
allows changes of the scene database and scene viewing to happen at the same time. 
Representative real-time animation techniques are
\emph{machinima}\index{machinima}%
\footnote{\url{http://en.wikipedia.org/wiki/Machinima}}$^{,}$%
\footnote{\url{http://en.wikipedia.org/wiki/Machinima:_Virtual_Filmmaking}}%
~\cite{machinima-2000,machinima-art-practice-virtual-filmmaking,machinima-understanding-essays},
digital puppetry, and motion capture,
with which users could create a performance in real-time and capture it in a virtual world.
Real-time simulation is mostly physical based, usually imitating some real physical phenomena,
such as fluid, deformable objects, depth of field simulation, and others, with which users
can also interact in the virtual scenes in real time by altering the simulation data.
Real-time animation and simulation appear together in particular in interactive applications,
where it is important in those disciplines, such as video games (e.g., \xf{fig:angry-birds-realtime-animation}), flight, car, or surgery simulators.
The goal for real-time animation and simulation is to maximize realism at the minimum cost. 
Real-time animation and simulation represent a very wide range of research topics, and
we only discuss a few of the methods that are relevant to this dissertation work.
They are character animation, motion capture, softbody simulation,
depth of field simulation, and collision detection.

\paragraph*{Character Animation.} 
\index{animation!character}

Compared to traditionally animated characters, which would either be hand-drawn
or in 2D images, sprites, the demand for more realistic real-time character animation
has increased considerably in the recent past due to the advances 
in computer hardware and software.
At the time, many of the off-line animation techniques, which only were listed as
possible future developments after several years, like skeletal animation,
deformation of 3D skin meshes, are now established methods for real-time
3D character animation. Some of the techniques are used in character animation
are hierarchic articulated objects, animation blending, and skeletal animation.

In hierarchic articulated objects, different body parts are 
separately stored in a hierarchy joined at pivot points.
The transformation is recursively traversed from the root down to 
its children within the hierarchic data structure. 
Animation data can be generated using inverse kinematics (IK)
and applied to 3D model in real-time.
Animation blending uses more than one object, which 
represent same character showing in different poses.
The objects, which contain the same number of vertices 
are blended or just switched in order to achieve the animation effects. 
The original Quake game~\cite{wiki:quake-video-game}
used this method, such as vertex blending for character animation~\cite{anderson:ppsn2002:pp689}. 
Skeletal animation was designed to simplify the animation process for dealing
with articulated objects in order to provide more realism of characters.
It also uses a hierarchic structure of joints as skeleton, which drives 
a rigid or soft skin mesh.
The skeleton is animated, and the animation applied to the skin.    

\paragraph*{Motion Capture.}
\label{sect:bg-motion-capture}

Motion capture records the movement of objects and human beings and then maps
the collected data to computer generated objects and characters.
This technique is often used in film animation and computer games.
The MoCap systems and software were quite expensive in recent decades. They typically
place markers or sensors on the skin or clothing near actors' joints or even on the face
for facial animation.
However, these days, some inexpensive MoCap hardware and software systems for entertainment
purposes emerged, such as {\kinect} (see \xs{sect:bg-kinect}) and {\ipi} have been developed,
which are quite powerful, accurate, and easy to use~\cite{ipisoft}.

\paragraph*{Softbody Simulation.}
\label{sect:bg-softbody-simulation}
\index{simulation!softbody}

Softbody simulation also known as deformable object simulation,
uses physical based simulation to mimic non-rigid bodies, such as
human and animal soft body parts and tissue, and other non-living
soft objects, such as cloths, gels, liquids, and gas~\cite{wiki:soft-body-dynamics}.
It has been increasingly used for character animation, computer games,
and surgical training due to the improvement of the quality
and efficiency of the new generation of computer hardware.
However, there are still a lot of aspects that can be explored in the real-time
softbody deformation simulation visualization area because it has emerged
as a challenge in computer graphics, and, therefore, was not
fully exhausted yet~\cite{softbody-framework-c3s2e08,adv-rendering-animation-softbody-c3s2e09}.
We proposed in the past the {\softbodysys} framework to be released
open-source~\cite{softbody-framework-c3s2e08,adv-rendering-animation-softbody-c3s2e09}.
Additionally, a large profile deformable object library called
\emph{Vega}\index{Libraries!Vega}~\cite{vega-deformable-lib} has
also been released in the open-source and exhibited at SIGGRAPH~2012
showing continued interested in and usefulness of the subject.
We explore further on this topic in \xc{chapt:softbody-jellyfish}.

\paragraph*{Collision Detection.}
\label{sect:bg-collision-detection}
\index{collision detection}

Collision Detection (CD)~\cite{MJ88} is a very efficient way to increase the level of realism 
when animated objects collide on the scene (otherwise they would pass through each other)~\cite{sweep-and-prune-cd-ming-lin-phd}.
It is practically impossible to make realistic games,
movie production tools (e.g., \animtitle{Toy Story}~\cite{wiki:toy-story,mudur-comp7661})
without collision detection, because we would get ``quantum effects'' all the time~\cite{realtimerendering2002}.
In general, there are two major parts in most CD algorithms:
\emph{collision detection}\index{collision detection} itself, which is usually a geometric intersection problem; 
\emph{collision response}\index{collision response}~\cite{CH971}, which is the actual (approximated)
physics of determining the unknown forces (or impulses) of the collision~\cite{mudur-comp7661,MH02}.

The following are the
major algorithmic techniques to realize CD surveyed in some detail in~\cite{adv-rendering-animation-softbody-c3s2e09}:
\emph{ray tracing}\index{ray tracing}, which is relatively simple to implement,
not very accurate, relatively fast, and sufficient for most cases; and
\emph{bounding volume hierarchies}\index{bounding volume hierarchies}
(BVHs),
which are more complicated,
somewhat more accurate, and slower. 
Using BVHs one can compute exact results; 
\emph{efficient CD for several hundred objects}, which are
dedicated solutions~\cite{realtimerendering2002}.
Other optimization structures such as Octree, BSP (binary separating planes),
OBB tree (oriented bounding boxes, which is a popular instance of BVHs),
KD tree, K-dop tree, uniform grid, and hashing also exist and were explored
and exploited to various degrees by CG practitioners~\cite{mudur-comp7661,realtimerendering2002}.

CD is a very important part of games or any type of physical-based simulation
involving impact situations.
There are several different algorithms and frameworks to choose from,
which were
mentioned
earlier, so they should be selected
based on the situation being implemented. One way to approach
such an implementation, is to provide a collision detection
framework to allow a common API for the algorithms and
possibly run-time parameters adjustment of them, e.g.~\cite{SZ02}. In this work we
do not attempt to make such a general framework, but merely
expressing the idea in a proof-of-concept (PoC) implementation~\cite{adv-rendering-animation-softbody-c3s2e09}.

\subsection{Advanced Rendering}
\label{sect:bg-advanced-rendering}
\index{rendering}

Computer graphics rendering is a sampling and filtering process~\cite{wikirendering}.
Rendering research and development has been largely motivated by finding
efficient algorithms to simulate the most realistic scenes.
In a nutshell, it is a process of generating an image from a model by any 
specialized hardware, such as a graphics card. The model, which is a 
description of three-dimensional objects in a strictly defined language or 
data structures, typically contains geometry, viewpoint, texture, lighting, and
shading information.

Rendering, originally taking place entirely in the CPU,
has been used in architecture, video games, film or TV special effects, and design
visualization~\cite{wikirendering}. The advantages of rendering everything
with CPU is that it is not restricted to the specific capabilities of graphics hardware
providing a form of programmable ``card-independence''.
The obvious shortcomings for the CPU-based rendering are not taking the advantage
of the throughput and speed the specific graphics hardware is optimized
to provide and share the processing load~\cite{realtimerendering2002}.

Rendering rates of display monitors can accommodate refresh rates of up
to 100 frames per second (FPS). However, beyond 70--80 frames, the
human visual system cannot perceive the changes in moving images.
Movement is perceived by the human visual system when images
are refreshed at 16 frames per second or higher (called \emph{persistence of vision}).
Smooth motion cannot be perceived at much lower speeds. For
animation, one typically plays the images at 30 frames per second.

\emph{Real-time rendering and animation}\index{rendering!real-time} (see \xs{sect:bg-rendering-realtime})
usually denotes scene rendering computations at 15 frames per second or higher.
For 3D animation, it is possible to pre-compute some images (not necessarily in
real-time) and play back at the desired frame-rates~\cite{realtimerendering2002}.
\emph{Offline rendering}\index{rendering!offline} (see \xs{sect:bg-offline-rendering})
is more used to create realistic images and scenes for movies;
real-time rendering is frequently used to interactively render a scene,
such as in 3D games~\cite{3dgames-watt-policarpo-01,AlanFabio03,advancedrealtimerendering2006}.
From the rendering aspect, \emph{\textbf{the movie and game, the only difference is in the
number of samples and the quality of filtering}}~\cite{realtimerendering2002} \emph{and} the level
of \textbf{interaction}\index{interaction}.

\longpaper
{
Some of the classical and advanced rendering techniques for offline and realtime processing
including the following features~\cite{wikirendering}\footnote{\url{http://en.wikipedia.org/wiki/Rendering_(computer_graphics)}}:

\begin{itemize}
\item Depth of field: objects appear blurry or out of focus when too far in front of or behind the object in focus
\item Snell's of law of reflection, Fresnel Reflections/Refractions (water and light accuracy)
\item Beer Lambert Law (water absorption)
\item Shading: how the color and brightness of a surface varies with lighting
\item Fogging/participating medium: how light dims when passing through non-clear atmosphere or air
\item Soft shadows: varying darkness caused by partially obscured light sources
\item Shadows: the effect of obstructing light
\item Reflection: mirror-like or highly glossy reflection
\item Transparency, transparency or opacity: sharp transmission of light through solid objects
\item Translucency: highly scattered transmission of light through solid objects
\item Refraction: bending of light associated with transparency
\item Diffraction: bending, spreading and interference of light passing by an object or aperture that disrupts the ray
\item Area lights and soft shadows
\item Environment map
\item Texture mapping: a method of applying detail to surfaces
\item Bump mapping: a method of simulating small-scale bumpiness on surfaces, procedural textures
\item Photon mapping, high resolution caustics
\item Anti aliasing, super-sampling
\item Axis aligned bounding boxes (AABB) and their hierarchies
\item Plucker coordinates
\item Non-photorealistic rendering: rendering of scenes in an artistic style, intended to look like a painting or drawing
\item Motion blur: objects appear blurry due to high-speed motion, or the motion of the camera
\item Monte-Carlo path tracing
\item
	Indirect illumination -- surfaces illuminated by light reflected off other surfaces,
	rather than directly from a light source (also known as global illumination)~\cite{jannekontkanen2007}
\end{itemize}
}

Some examples of the advanced and classical rendering include shading and rending characters'
skin~\cite{skinhairshrek},
hair~\cite{finalfantasy} (e.g., \xf{fig:offline-final-fantasy}),
jellyfish~\cite{rendering-jellyfish-2004},
and a whole documentary~\cite{ryan-npar-2004}.

\begin{figure*}[ht]
\hrule\vskip4pt
\begin{center}
	\subfigure[\textit{Offline Rendering from Final Fantasy Film}]
	{\includegraphics[width=.6\columnwidth]{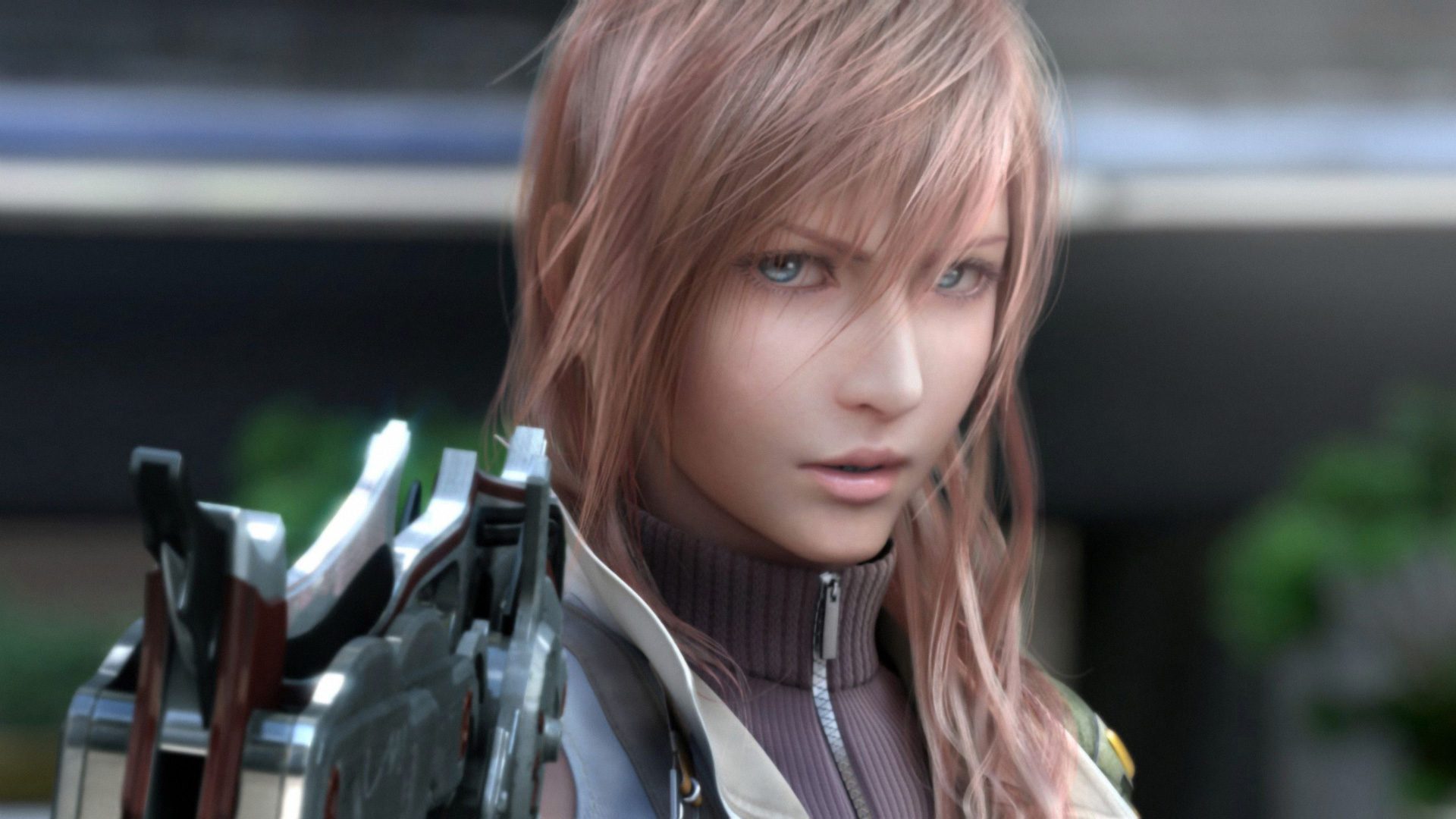}
	 \label{fig:offline-final-fantasy}}%
	\hspace{.1in}
	\subfigure[\textit{Real-time Rendering from Final Fantasy Video Game}]
	{\includegraphics[width=.6\columnwidth]{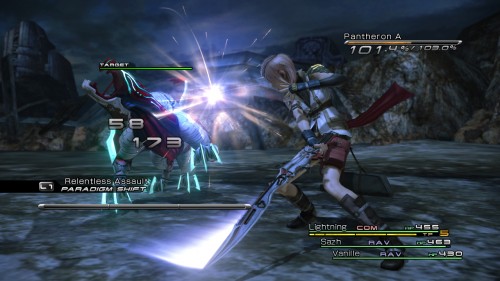}
	 \label{fig:realtime-final-fantasy}}
\caption{Final Fantasy Example Screenshots}
\label{fig:final-fantasy-example-screenshots}
\end{center}
\hrule\vskip4pt
\end{figure*}

\longpaper{IGNORED: \xf{fig:final-fantasy-example-screenshots}}

\subsubsection{Offline Rendering}
\label{sect:bg-offline-rendering}

To render a static scene is to determine the color to be assigned to every pixel of an image
depicting the scene (usually 3D) as viewed by the viewer~\cite{wikirendering,jannekontkanen2007}.
\emph{Image-based Rendering}\index{rendering!image-based} (IBR), which has images as the
primary data used for rendering, most commonly is photorealistic depiction for providing a
realistic viewing experience. 
Most IBR is done in offline rendering, which can tolerate one frame of animation 
taking even hours to render~\cite{wikirendering,jannekontkanen2007}. This is acceptable
in film production where real-time response is not needed.

\subsubsection{Realtime Rendering}
\label{sect:bg-rendering-realtime}

In practice, real-time applications often omit rendering correct soft shadows,
depth of field, small-scale surface detail, and realistic surface materials in order to
achieve the speed efficiency for user interaction~\cite{moller2002-RTR,advancedrealtimerendering2006}.
The performance of real-time rendering is extremely important because it has to allow
user interact with the computer-generated graphics in real-time.
Real-time rendering tries to achieve at least three goals: it allows more frames per second, 
generates higher resolution images, and more realistic objects.

Unlike the offline \emph{image-based rendering} (which in some cases
could be used for illustration or artistic purposes);
it is known as \emph{Non-photorealistic Rendering}\index{rendering!non-photorealistic } (NPR). 
NPR, also called \emph{stylistic rendering}, has a wide range of goals,
such as creating an image similar to technical illustrations and only the relevant
details would be displayed~\cite{realtimerendering2002,advancedrealtimerendering2006}.
NPR has always been related to computer games, in which each frame of animation is rendered in real-time (e.g., \xf{fig:realtime-final-fantasy}).
In order to achieve $\approx25$ FPS, we could only compromise the quality of images somewhat
to render dynamic images (change in scenes), such as
view position and view angle changes, object movements,
object deformations (in shape or other properties),
and changes in environment, such as lighting~\cite{advancedrealtimerendering2006,mudur-comp7661}.

\subsubsection{General Tools and Techniques}
\label{sect:bg-rendering-techniques}

\longpaper{{\todo}}

\paragraph{Ray Tracing.}
\label{sect:bg-ray-tracing}

Ray tracing produces a very high degree of photorealism by tracing the path of light.
The derived algorithms from ray tracing are beam tracing, cone tracing, path 
tracing, and metropolis light transport~\cite{advancedrealtimerendering2006,mudur-comp7661}.
\longpaper{Picture?, e.g. from \cite{uray-gpu-distributed-techrep}}

Ray tracing was traditionally an offline rendering technique until
relatively recently when with the {\gpu} support real-time and distributed
rendering became possible including using the
CUDA\index{CUDA}\index{Framework!CUDA}\index{Libraries!CUDA}\index{GPU!CUDA}
framework\footnote{\url{http://en.wikipedia.org/wiki/CUDA}}$^{,}$\footnote{\url{https://developer.nvidia.com/cuda-toolkit}},
URay~\cite{uray-gpu-distributed-techrep} and the like.

The technique is particularly good as in one algorithm it covers
shadows, reflections, and translucency at the same time. The classical
ray tracing does not do soft shadows but is adjustable with modifications
to make them possible in combination with other techniques~\cite{advancedrealtimerendering2006,mudur-comp7661}.

\paragraph{Global Illumination.}
\label{sect:bg-global-illumination}

Global illumination has been used more in offline rendering and it generates
the higher level of realism based on ray tracing, radiosity, and other techniques.
It uses a group of algorithms in 3D computer graphics
to demonstrates how realistic lights are reflected by surfaces in order
to increase the overall perception of and realism~\cite{wikiglobalillumination,jannekontkanen2007}.
The direct illumination refers to the light directly coming from a light source; indirect illumination
describes the reflected light rays from other surfaces. Some examples of global illumination
are reflection, refraction, color bleeding, and hard and soft shadows. The algorithms used in global
illumination are diffuse inter-reflection, radiosity, ambient occlusion,
photon mapping, and image based lighting~\cite{wikiglobalillumination,jannekontkanen2007}.

\longpaper
{
\begin{enumerate}
\item Object representations 
\item Object-ray intersections 
\item Shading models 
\item Acceleration techniques 
\item 2D and 3D texture mapping 
\item Anti-aliasing and filtering 
\item shadows and recursive reflections
\end{enumerate}

reflection and refraction, scattering, and chromatic aberration

{\todo}
} % \longpaper

\paragraph{Level of Detail (LOD).}
\label{sect:bg-lod-techniques}

The basic idea of LOD is it use simpler versions of an object when it
makes less and less of a contribution to the rendered image.
In computer graphics, accounting for level of detail involves
decreasing the complexity of a 3D object representation as it moves
away from the viewer or according other metrics such as object
importance, eye-space speed or position~\cite{wikilod,hlod-2000}. Different
levels of detail are used at different distances from the viewer. There is not
much visual difference, but rendering can be significantly more efficient. We could use area of projection
of bounding volumes (BVs) to select appropriate LOD~\cite{hlod-2000}.
When the object is far away, we replace it with a quad of some color;
when the object is really far away, we do not render it (called \emph{detail culling})~\cite{wikilod,hlod-2000}.

Although most of the time LOD is applied to geometry detail only,
the basic concept can be generalized. Recently, LOD techniques included
also shader management to keep control of pixel complexity~\cite{pplod-shader-2003,gpu-geometry-lod-2010}.
A form of level of detail management has been applied to textures for years,
under the name of mipmapping~\cite{fast-terrain-mipmapping-2000},
also providing higher rendering quality~\cite{advancedrealtimerendering2006}.

The LOD selection is to make a choice for which object to render or to blend. 
A metric called the benefit function, is evaluated for the current viewpoint 
and the location of the object. The metric may be based on the distance from 
the viewpoint to the object, or the projected area of the bounding volume of 
the object~\cite{wikilod,fast-terrain-mipmapping-2000}.

\paragraph*{Subdivision.}
\label{sect:bg-subdiv-techniques}

Subdivision surfaces are powerful in defining smooth, continuous, crackless 
surfaces~\cite{realtimerendering2002,wikilod}.
They also provide the infinite LODs. The staring mesh is called the \emph{control mesh}.
The first phase called the \emph{refinement phase}, which 
creates new vertices and reconnects to create new, smaller triangles. The 
second phase is called the \emph{smoothing phase}, which computes new
positions for some or all vertices in the mesh.

DLOD is the simplest type of LOD algorithm, which is based on subdividing the
space in a finite amount of regions, each with a certain level of detail~\cite{wikilod}. The
result is discrete amount of detail levels. There's no way to support a
smooth transition between LOD levels at this level. Moreover, DLOD could
provide various models to represent the same object~\cite{wikilod}.

\longpaper
{

To associate the different LODs of an object with different ranges. The most 
detailed LOD has a range from zero to user 
defined $r_0$. The next range is $r_1$ to $r_2$.  
} % \longpaper

\paragraph{{\gpu}.}
\label{sect:bg-gpu}

Hardware-based rendering on the graphics card's graphics processing unit {\gpu} adds a significant
computational power to the CPU-based rendering by offloading a good amount
of computation and rending in parallel to the CPU and multiple computational
units on the card.
The {\gpu}s in most graphics pipelines primarily work with the geometry
(vertices) and the pixel fragments after rasterization. The shader programs that
provide access to both of these such points in the pipeline have
standards defined for them~\cite{vertex_program,fragment_program}.
The programs one writes are the shaders that run on the {\gpu}
and can do more than traditional graphics API provide faster.
The shading programming languages used for the GPU programming
typically are either cross-vendor
assembly~\cite{vertex_program,fragment_program}\footnote{\url{http://en.wikipedia.org/wiki/ARB\_(GPU\_assembly\_language)}}
or higher-level shading languages, such as
{\glsl}~\cite{rost2004,opengl-glsl-quick-guide} \longpaper{(see \xs{sect:opengl-shading-language})}
or {\hlsl}~\cite{advancedrealtimerendering2006}\footnote{\url{http://en.wikipedia.org/wiki/HLSL}}.

The concrete sections that describe the use of the {\gpu} shading (using the shading languages mentioned)
provide some more implementation details in 
\xs{sect:softbody-jellyfish-api-glsl},
\xs{sect:softbody-jellyfish-gpu},
\xs{sect:impl-xna},
\xs{sect:interactive-docu-modeling-xna}, and
\xs{sect:illimitable-api-xna}.

\longpaper{{\todo}}

\paragraph{Stereoscopic Rendering.}
\label{sect:bg-stereo}

The basic stereoscopic rendering has to do with rendering the scene twice
at slightly different angles and recombining the two images back into a single image such
as it is seen as a real three-dimensional scene through red-blue glasses.
This is achieved by rendering the scene offset by two stereo synthetic cameras.
The distance
between these stereo cameras, called the \emph{interaxial}\index{interaxial}, is adjustable 
and the cameras may be set to remain parallel as in \xf{fig:Parallel} or 
to \emph{toe in} (angling towards a center of
interest) as shown in \xf{fig:ToeIn}.
Supplemental attributes
need to be added to
deliver predictable stereoscopic effects and to 
facilitate stereoscopic post-production. 
A greater understanding of stereoscopy and the notion of
parallax (illustrated in \xf{fig:parallax})
key to building such camera rigs~\cite{arloader-thesis-08,stereo-plugin-interface}.

\begin{figure*}[ht]
\hrule\vskip4pt
\begin{center}
	\subfigure[Parallel Stereo Viewing]
	{\includegraphics[width=.49\columnwidth]{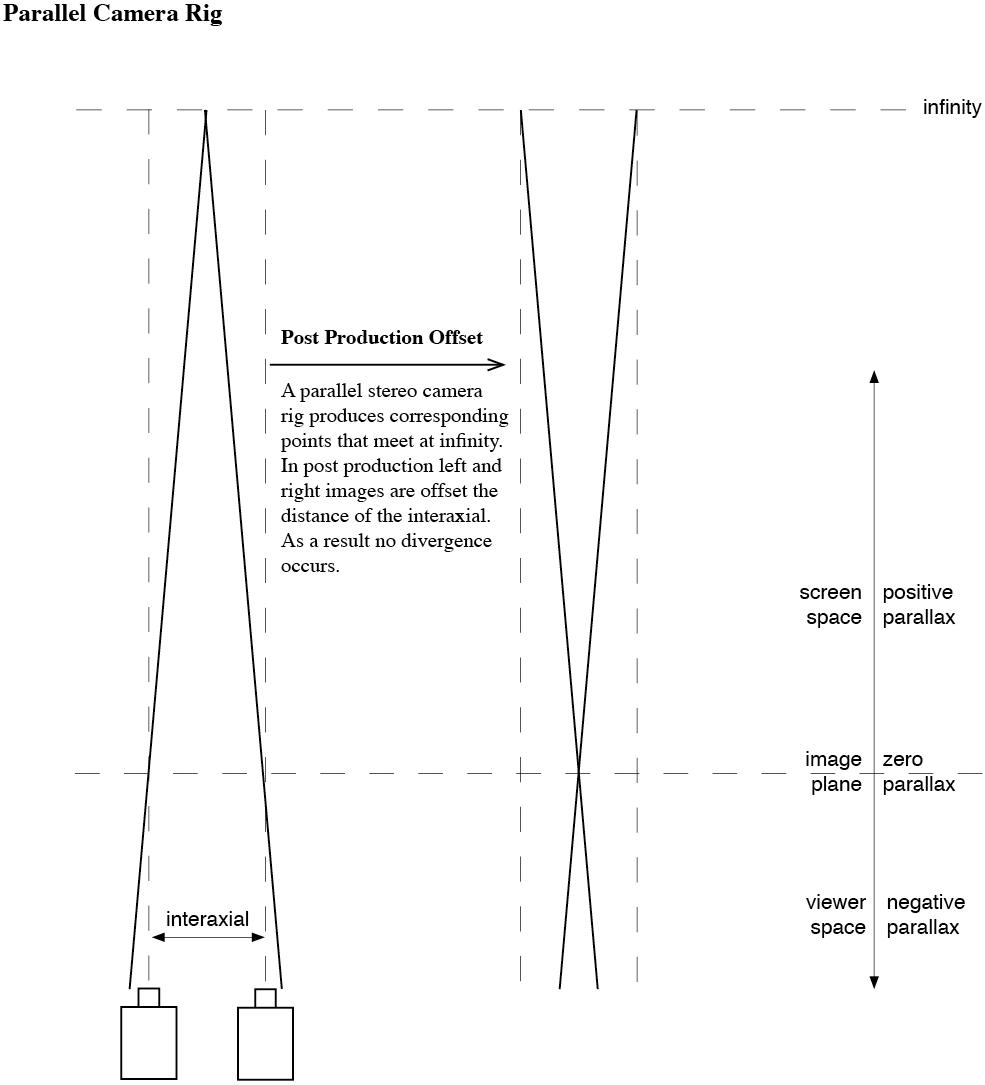}
	 \label{fig:Parallel}}%
	\subfigure[Toed-In Stereo Viewing]
	{\includegraphics[width=.49\columnwidth]{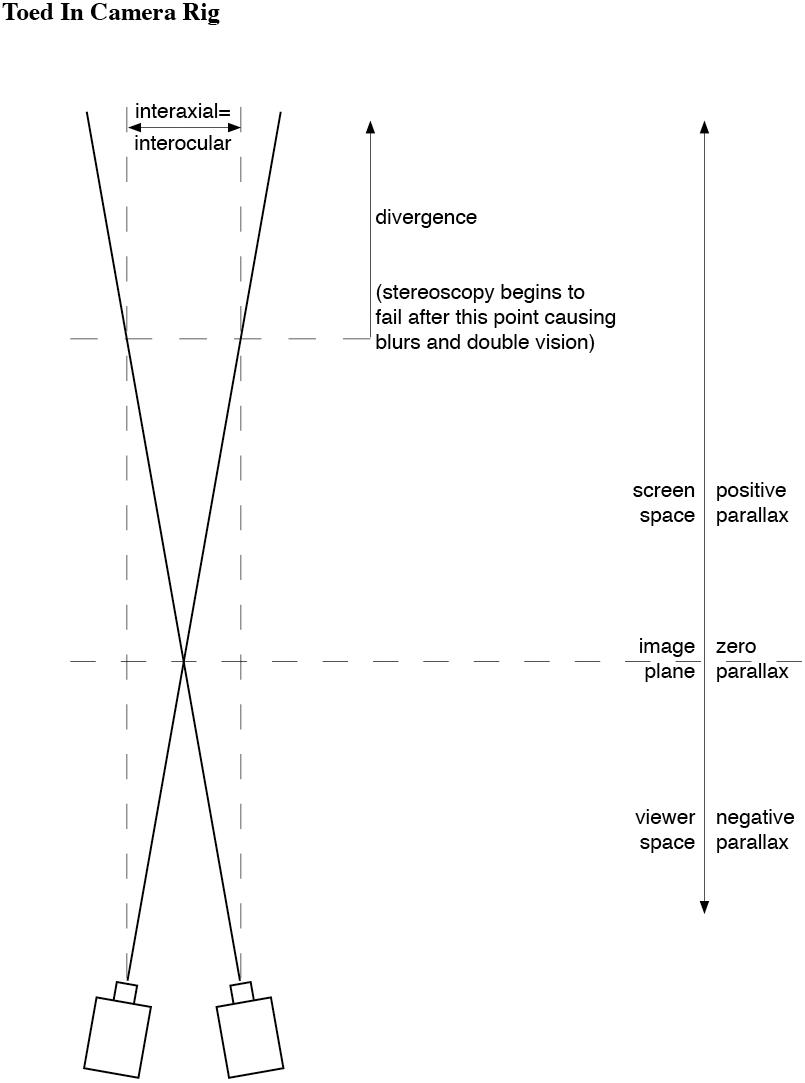}
	 \label{fig:ToeIn}}
\caption{Parallel vs. Toed-In Stereo Viewing~\cite{arloader-thesis-08}}
\label{fig:stereo-viewing-parallel-toedin}
\end{center}
\hrule\vskip4pt
\end{figure*}

\longpaper{IGNORED: \xf{fig:stereo-viewing-parallel-toedin}}

\begin{figure}[htb!]
	\centering
	\includegraphics[width=.7\columnwidth]{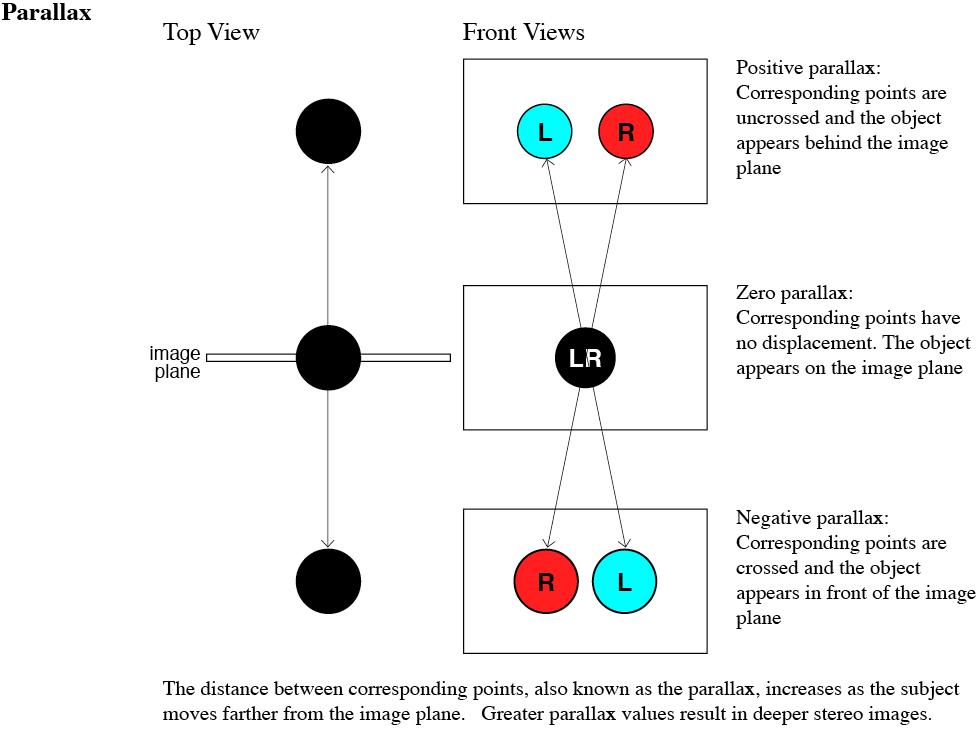}
	\caption{The Concept of Parallax~\cite{arloader-thesis-08}}
	\label{fig:parallax}
\end{figure}

\longpaper
{ 
In many 3D modeling and visualization tools, it is generally
difficult to create well-composed stereoscopic animation because
stereo previews cannot be seen in real-time. Unable to work interactively 
in stereo, modelers and animators must work
with flat previews and subsequently perform software renders and 
compositing to preview the stereoscopic effect.
We propose an interactive plug-in-based framework and a case-study user interface
that uses it for the {\maya} 3D animation
software to remedy this problem. Our work is {\opengl}-based and
open-source and we have designed it for extension to tools other than Maya. 
We present the first working iteration of the real-time stereoscopic
multi-view visualization interface and its challenges. Despite starting {\maya}~2009
includes built-in stereoscopic support, our plug-in works for {\maya}~2008 and {\maya}~8
that did not support it natively as well as our plug-in does not require the
use of the graphics cards that support stereo buffers. Due to being based on {\opengl}--an
open standard and open-source, and the generality of the architecture, the framework can be
included in any {\opengl} application that wishes to do stereoscopy as either a plug-in
or a library. It can also serve for educational purposes.
}

\paragraph*{Film Production.}
\label{sect:bg-stereo-film-production}

Stereoscopy is increasingly used
in the animated stereo film production, e.g., in the films like \animtitle{Folding}~\cite{loader-folding-movie} and
\animtitle{Journey To The Center Of The Earth}~\cite{jce-stereo-07} 
and the ones that inspired the work on the stereoscopic plug-in at the time~\cite{stereo-plugin-interface}
and many others.
\xs{sect:interactive-media-docu-background} describes some of this as well.

\paragraph*{Stereoscopy in VR and Medical Research.}
\label{sect:bg-stereo-vr-medical}

Simulation of stereoscopic softbody objects in haptic environments~\cite{softbody-framework-c3s2e08,%
adv-rendering-animation-softbody-c3s2e09,%
haptics-cinema-future-grapp09%
} can help surgeons to train for real an virtual (tele) surgeries on virtual responsive subjects.
Stereoscopic virtual reality (VR) environments for medical and rehabilitation research are
other applications where it has potential for use as a cheaper alternative to expensive
head-mounted display units. For more details in VR in medical research please refer
to \xs{sect:vr-and-medical-research}.

\subsection{Additional Related Work}
\label{sect:bg-related-work-cg}

We review a number of related work items considered in
the design and implementation of this work as well as future
plans and considerations. This subsection primarily covers
the related work on the modeling, animation, and rendering
of a {\jellyfish} (\xs{sect:related-work-cg-jellyfish}) and
{\opengl} slides in \xs{sect:bg-oglsf}.
The rest of the related work is cited and described throughout
the rest of the section and where appropriate throughout the
thesis.

\subsubsection{Jellyfish Modeling, Rendering, and Animation}
\label{sect:related-work-cg-jellyfish}

The inspiration and aesthetics related work along with modeling, rendering, and
animation of the {\jellyfish} project can be found
in~\cite{rendering-jellyfish-2004,dancing-jellyfish-2006,jellyfish-underwater-scene}.
An advanced interactive jellyfish game implemented in the Unity engine
is presented in~\cite{blush-jelly-fish-game}. The main issue with all these
works, which look realistic and nice is that most of the animation is precomputed
offline/with keyframes and is modeled in tools such as {\blender} or {\maya}~\cite{blender,maya}
and then played in a scene or a game. We'd like to make a similar type of
animation but in real-time based on physics~\cite{bio-mechs-jellyfish-swimming-2002,synthetic-engineered-jellyfish}
with the aim at a real-time
tangible interaction or manipulation of the jellyfish in a game or an interactive
installation with haptics and motion tracking~\cite{jellyfish-c3s2e-2012}.

\subsubsection{OpenGL Slides Framework ({\oglsf})}
\label{sect:bg-oglsf}
\longpaper
{
The major pieces of the related work that contribute
to this works, are two frameworks alongside with their
implementation, put together with other items,
to eventually form a teaching module for computer graphics.
Here we extrapolate from our previous poster presentation~\cite{softbody-opengl-slides}
on this topic with more details on the actual design and 
implementation of the physical-based softbody
animation techniques in an OpenGL power-point-like
presentation tool with the demonstrated results. We describe the two CG systems
in this section in some detail for the unaware reader.
\cite{softbody-teaching-opengl-slides}
} % \longpapers

{\oglsf} gives ability to make {\opengl} ``slides'', navigate between them
using various controls, and allow for common bulleted textual widgets --- the \emph{tidgets}\index{tidgets}.
It also allows to override the control handling from the main
idle loop down to each individual (current) slide (a scene).
All slides together compose a concrete instance of \api{Presentation},
which is a collection of slides
that uses the Builder pattern to sequence the slides. Each
slide is a derivative of the generic \api{Slide} class and represents
a scene with the default keyboard controls for the tidgets
and navigation.
It is understood
that the tidgets can be enabled and disabled to allow the
main animation to run unobstructed~\cite{softbody-teaching-opengl-slides}.
Each scene on the slide is modeled using traditional
procedural modeling techniques~\cite{wiki:procedural-modeling} and is set as a developer
or artist desires. It can include models and rendering
of any primitives, complex scenes, texturing, lighting,
{\gpu}-based shading, and others, as needed and is fit
by the presenter~\cite{softbody-teaching-opengl-slides}.
The main program delegates its handling of the {\glut} callback
controls for keyboard, mouse, and idle all the way down to the
presentation object that handles it and passes it down
to each current slide~\cite{softbody-teaching-opengl-slides}.
Our {\softbodysys} was integrated into such a collection of {\opengl} slides
for educational and demonstration purposes.

\section{Documentary Film and Interactive Media}
\label{sect:interactive-media-docu-background}

We briefly review the topic on the role of computer graphics in the production
of documentaries, 
which was quite often ignored in favor of other topics~\cite{role-cg-docu-film-prod-2009}.
Typically, except for some rare occasions (the number of which is now
currently growing in this emerging trend), documentary producers
and computer scientists and or digital artists who do computer graphics
are relatively far apart in their domains and infrequently intercommunicate
to have a joint production; yet it happens, and perhaps more so
in the present and the future.

We attempt to classify the documentaries on the amount and techniques of computer
graphics used for documentaries. We come up with the initial
categories such as ``plain'' (no graphics), ``mixed'' (or ``hybrid''), ``all-out''---nearly 100\%
of the documentary consisting of computer-generated imagery. 
Computer graphics can be used to enhance the scenery, fill in the gaps in the
missing story-line pieces, or animate between scenes. It can incorporate
stereoscopic effects for higher viewer impression as well as interactivity
aspects. It can also be used simply in old archived image and film restoration.

We analyze the impact of the computer graphics on the present-day
documentaries in the two major sub-classes and project the future of the documentary films and TV production,
as well as potential cognitive pattern-recognition-based
interactivity with the documentary scenery might look like in the not-very-distant
future as the professor's responsive hologram in the motion picture \docutitle{I, Robot}~\cite{movie-irobot}.
Of course, the visual effects and animation is only one aspect
or role of the vast computer graphics topic.
Obviously, it counts for more than traditional animation
for documentary film production.

\subsection{Related Work}

Computer graphics (CG) today is nearly everywhere, including documentaries. First,
a very comprehensive review of CG itself is in the
online set of lecture notes~\cite{OSU-CG-history} describing the history and evolution of computer graphics
hardware, software, algorithms, and techniques and their use for animation, including
all kinds of CG standards, prolific conferences, artists, production companies,
etc.~(see also \xs{sect:cg-background}). CG is a key enabling point of new media that
influences the modern day productions~\cite{role-cg-docu-film-prod-2009}.

Perspective and cultural overview of new media forms from theoretical and philosophical
standpoints, and in particularly the interaction aspects
of the new media are described in \cite{new-screen-media-2002}.
Manovich's work~\cite{lang-new-media-2007} was deemed to be arguably as the
``most systematic and rigorous theory of new media'' along with its
reliance on the old media from historical point of view. This work covers
specific themes of interest including the fusion of cinematography
and computers as well as the notion of virtual reality slowly
merging with the real~\cite{lang-new-media-2007}.

Computer graphics has been widely used in physics, biology, video game industry,
education, and military matters. It has also been used in advertising,
music video, motion picture, and television such as international network promos
for CBC, e.g. shown in \xf{fig:cbc} produced in 1985,
and in the feature film \filmtitle{Jurassic Park} (1993) shown in
\xf{fig:jurassic_kitchen_raptor}~\cite{OSU-CG-history,wiki:jurassic-parks}.
In the past, computer graphics seemed only to be associated with big budget productions;
therefore, productions of the most of documentary films done in low budget
had rarely applied these new techniques in their work.
However, in recent years, more and more documentary films started to introduce
computer generated 2D and 3D images and CG elements in a partially or fully animated
film. Some other documentary films have used stereoscopic techniques, 
such as \docutitle{Hubble} and \docutitle{Born to be Wild}
have been very successful in IMAX 3D theatres.

No matter whether a hand-drawn animated documentary or computer-generated animated documentary,
they have certain aspects in common---they are ``broadcast'' to the audience.
There is no doubt that animation is taking on a new hybrid-like style for 
documentary film. However, some questions arise. Can animation bring more realism to the
documentary films? Is it necessary for the artists to apply this non-traditional
medium to documentary film production? What is the motivation of
the artists? How will animation further impact in the documentary film
production process in the future~\cite{chris-robinson-2004}?

\begin{figure*}[ht]
\hrule\vskip4pt
\begin{center}
	\subfigure[Canadian Broadcasting Promo]
	{\includegraphics[height=2.0in]{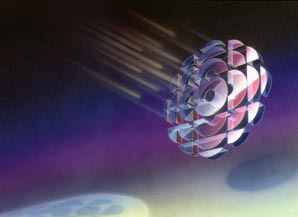}
	 \label{fig:cbc}}
	\hspace{.3in}
	\subfigure[Jurassic Park's Raptor]
	{\includegraphics[height=2.0in]{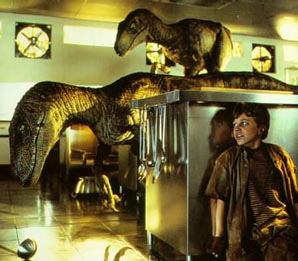}
	 \label{fig:jurassic_kitchen_raptor}}
\caption{Some Example Graphics for Production}
\label{fig:cg-production-examples}
\end{center}
\hrule\vskip4pt
\end{figure*}

\longpaper{IGNORED: \xf{fig:cg-production-examples}}

Moreover, the interactivity between
human and computer generated images is a particular special characteristic related to computer
technology. Thus, there is ``interactive documentary'' --- a brand newly emerged genre,
which has various forms, such as 3D and web-based, where the audience can
participate in one way or another to as what is happening in the documentary.

\subsection{Traditional Animated Documentary}
\label{sect:trad-anim-docu}

The animated documentary is a genre of film,
which combines the genres of animation and documentary.
Non-fiction documentary animation, which deals with 
non-fiction material can utilize documentary audio interviews, or it can be 
an interpretation and re-creation of factual events~\cite{sheila-sofian-2005}.
Some audiences have reacted negatively because they found out they had the 
feeling of ``forced empathy'' and the animated documentary was ``propaganda''.
They more associate the cartoons to children and could not
accept this to documentary which expresses truth.
However, animated documentary is more obvious and transparent for audience to
distinct the reconstructed scenes and bring them the extraordinary emotion
than live-action. Thus, to me, animated documentary films have more truth-value.

\subsubsection{Animation, Realism, and the Truth-Value}

``Truth-value'' and ``film realism'' are two often opposing topics debated in academic discussion
since 1990s~\cite{life-reproduced-drawings-2005}.
Some filmmakers try to enhance the realism of a film because some of the scenes in it are not ``real''.
Some other filmmakers say that there is no perfect truth in documentary film
because as soon as the film has been edited, some of the truth
has been hidden and is a subject to the opinion of the filmmaker.
For some hybrid documentary films, some scenes have to be reconstructed.
Do the reconstructed scenes still have the truth value?

Compared to traditional non-animated documentary film, the scene
in an animated documentary does not ``exist'' at all. It is neither a
live-action camera footage nor a photograph, but it is created by artists based on the
real world objects or based completely on their imagination.
How much truth-value will remain in an animated documentary film after reconstruction?
Does it exist there at all?
We always talk about how to make a film more realistic and increase its realism
in order to make the ``faked'' scene look more believable. Obviously, for such a film,
the truth-value is not comparable with the one that contains real footage and photographs.

As shown, the biggest challenge for most of the documentary filmmakers is to capture truth.
Moreover, it is a difficult task for them to make a film when there is a shortage of footage.
There are various solutions to fill in the gaps, such as scene reconstruction,
photographs, texts according to historical events, and the needs from artists and the makers.
Some artists decide to use the photos to represent the past, such as
American filmmaker Ken Burns who uses the ``photographing of live-action material''~\cite{wiki:ken-burns}.
Some use either traditional animation or computer graphics, or both to reconstruct the scenes because
they feel it is more vivid and can make their scenes closer to the truth compared to photos.
Some explain the necessity to use animation in documentary to show the footage
that one can not capture in the real life, such as imagination, dream, hallucination, and memory.
A typical example such as in documentary \docutitle{American Teen}
(2008)~\cite{paramount:american-teen,wiki:american-teen}, animation was
used for ``teen dream'' and ``teen imagination''.

\begin{quote}\textsl{``Realism is not what animation is best at, instead, freer invention,
fantasy, and exaggerating reality are its privilege. The realer-than-real environment
follows the concept of \emph{hyper realism}, which offers a completely
artificial environment as a representation of the real.''}~\cite{understanding-animation-1998}
\end{quote}

\subsubsection{Traditional Animation in Documentaries}

The use of animation in documentary production is not new.
The traditional animated documentary can be traced back all the way
to 1918 by Winsor McCay in his 12-minute-long film
\docutitle{The Sinking of the Lusitania} (1918)~\cite{sink-lusitanian-1918},
the first animated documentary.
It used animation to portray the 1915 ``sinking of RMS Lusitania after it was struck
by two torpedoes fired from a German U-boat''~\cite{wiki:sinking-lusitania}.
Quite obviously, there was no live-action footage recorded when the event occurred.
The animation used in this silent documentary film for things such as the underwater
fish swimming and the explosion and sinking of Lusitania, some of which shown
in \xf{fig:sinking-lusitania}. The explaining texts were throughout
the film between the animated scenes. This is very traditional story
telling in silent film. The only difference is the animated scene replaced
the live-action footage.

\begin{figure*}[ht]
\hrule\vskip4pt
\begin{center}
	\subfigure[Example 1]
	{\includegraphics[height=2.0in]{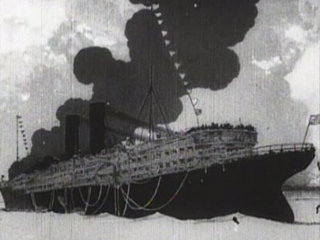}
	 \label{fig:sinking-lusitania1}}
	\hspace{.3in}
	\subfigure[Example 2]
	{\includegraphics[height=2.0in]{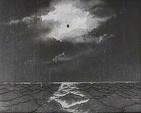}
	 \label{fig:sinking-lusitania2}}
\caption{\docutitle{The Sinking of the Lusitania} (1918) Example Screenshots}
\label{fig:sinking-lusitania}
\end{center}
\hrule\vskip4pt
\end{figure*}

\longpaper{IGNORED: \xf{fig:sinking-lusitania1}, \xf{fig:sinking-lusitania2}}

Animation has been used in other educational and social guidance documentary
films, such as \docutitle{The Einstein Theory of Relativity} (1923). The purpose
to use animation in earlier times is because of the need to be able to
illustrate abstract concepts in mainly live-action examples of these
genres~\cite{truth-toons-documentary-1997,walking-life-animation-truth}.
In traditional animation, inbetweening, cell animation~\cite{TJ84}, and rotoscoping~\cite{rotoscoping-2006}
have been later introduced by Disney in order to be efficient when working with many single-frame images.
Disney's film exaggerate reality in order to create a greater impression of realism
is often called ``ultra-realism''.
However, these old techniques required artists to illustrate thousands
of pictures to be filmed when close to the end of the production process.

\subsection{Computer Animated Documentary}
\label{sect:comp-anim-docu}

With the appearance and rapid development of computer hardware and software,
traditional animation techniques mentioned above,
have been slowly replaced by computer generated graphics and animation.
Since \docutitle{Walking with Dinosaurs} (1999),
a six-part documentary series produced by the BBC
in 1999~\cite{walking-with-dinosaurs-1999}, the style of a traditional documentary
was simulated. More and more documentary films these days include visual
effects and CGI (Computer Generated Imagery) as a norm.
Comparing to \docutitle{Walking with Dinosaurs},
still the most expensive documentary series per minute ever made,
production costs have come down in recent years with the rapid development of
computer hardware and software.
Thus, even for documentary films with a low budget,
directors could consider introducing CG animations into their film nowadays.
This new genre of documentary films usually reconstructs the historical
and informational footage, which is not available.
Moreover, computer generated graphics and special effects
can illustrate very serious and heavy topics with a humorous connotation
in order to attract the audience's attention.

\begin{quote}
\textsl{``If you have several visual effects shots that work together to 
tell a story, using them separately earlier on in the documentary 
can increase their impact. ... Earlier in the documentary, 
each shot will work well in sections of the film detailing the different 
aspects of the threat, and then when the entire animation is brought together 
for the climax of the documentary, 
the effect will be even greater.''}~\cite{3d-animation-motion-cgi-documentary}.
\end{quote}

Computer animation has a lot of similarities when compared to the traditional
animation. However, it is more advanced than the traditional one and more and more
adopted by the current film industry because of its efficiency, accuracy, lower cost, and more
advanced rendering effects, which can enhance the realism of a film more than the traditional animation could.
Even though computer graphics has been playing a significant role in formulating
new aesthetic grounds for both fiction and non fiction films, will this kind of film
abandon the most important concept, the ``truth-value'', which is precious in documentary film?
How this new type of documentary film will affect the older audience and new generation audience?
How much could they believe CGI documentaries?
Through the discussion about this and other topics, we will reach to a
conclusion in that regard.

\subsubsection{CGI and Realism of Documentary}
\label{sect:cgi-and-realism}

According to the recent work by Landesman~\cite{ohad-landesman-2008},
the documentary film overall in the past years has been experiencing
a notion of formal change from traditional ``observation and omniscient narration''
in terms of being less strict in the need to objectively portray
the material. More and more the documentary film has been embracing
the paradigm switch to performance rather than recording of an observation,
some subjective rather than objective aspects and even fiction; equally
no longer requiring to be as certain and complete as possible
in the argumentation in the film instead of just showing dry
factual knowledge~\cite{ohad-landesman-2008}. CGI techniques
are here to help with the emerging trends of documentary film concepts
and production Landesman described.

More and more documentary films use computer animation and special effects in their production.
As mentioned earlier, some spectators found out that animated documentary is more obvious and transparent for audience
to distinguish the reconstructed scenes and bring them the extraordinary emotion than the live-action.
Moreover, it could sometimes reach an audience that might not watch live-action documentary
and is easier to incorporate difficult, delicate, and controversial
topics~\cite{interview-doc-anim-directors,vr-medical-research-docu}.
There is no doubt that animation is taking on a new hybrid-like style for documentary film~\cite{vr-medical-research-docu}.
However, 

\begin{itemize}
\item
Can animation bring more realism to the documentary films?
\item
Is it necessary for the artists to apply this non-traditional medium to documentary film production?
\item
What is the inspiration that lead artists apply this new layer to documentary film?
\item
How does the audience respond?
\end{itemize}

Some of the notions and beginning of such documentary style
appear in different documentary film early and later-day works
\cite{%
truth-in-pictures-2005,%
interview-doc-anim-directors,%
levin-15-docu-interviews,%
nichols-docu-theory-practice,%
nichols-ideology-image,%
nichols-intro-to-docu,%
history-myth-narrative-docu,%
nichols-representing-reality,%
rogosin-interpreting-reality,%
new-challenges-docu-1988,%
artist-in-docu,%
wahlberg-docu-time,%
white-narrativity-reality%
} as conveniently compiled by Lee and Nitoslawska~\cite{expanded-docu-bib}.

Rowley talks about the quest
of traditional animated film (hand-drawn style) for a particular kind
of realism in order to succeed in feature film making, such as
``visual realism'', ``aural realism'', ``realism of motion'',
``narrative and character realism'', and ``social realism''~\cite{life-reproduced-drawings-2005}.
We borrow some of these terms for completely different perspective of analysis
in ``computer animated documentary''.
We not only explain their concepts, but also look at computer graphics
role on how enhance theses realism, and which realism types are applicable
for our study.

\paragraph*{Visual Realism.} It evaluates the extent to which the animated
environment and characters are understood by the audience compared to
the ones from the actual physical world~\cite{life-reproduced-drawings-2005}.
Dimensionality and the level of detail (LOD) are
two main aspects of visual realism. Dimensionality refers to the extent that an
illusion of depth is created, whereas LOD describes the extent to which the background
depicts complex particularities of the environment.
3D modeling software, e.g. {\maya}~\cite{maya} or {\blender}~\cite{blender}, can not only provide
the advance modeling tools for shaping the objects, but also supply
the advanced rendering techniques for artists to build the very realistic environment
with vivid textures and lighting. Moreover, even an individual without any drawing skills,
can use most of the current 3D modeling software to model the objects,
characters, and landscapes.
The visual realism type is the most applicable from the CG point of view in our study as it
depicting complex particularities of the environment
where it impacts and impresses the audience most.

\paragraph*{Realism of Motion.} It contrasts
the extent to the characters moves
and motion in artificial environment and physical world and laws of physics.
Traditional animation relies on persistence of vision and refers to a series
motion illusion resulting from the display of static images in rapid-shown
succession~\cite{life-reproduced-drawings-2005}. Artists have to use not only their drawing skills and intuition,
but also possess some knowledge of physics to make the objects behave as if
they are in the real world or close to it. The motion of the virtual objects
will not convince audiences if no natural laws of physics are applied~\cite{life-reproduced-drawings-2005}.
Moreover, drawing the virtual objects moving from one frame to another frame
is an inefficient way without functionality provided by software. However,
one of the computer techniques, motion capture is very efficient and accurate
to describe virtual objects motion. It attaches sensors on actors bodies and
records the data for their movements and apply these data to a computer
generated characters. This technique increases the realism of
motion dramatically.
Additionally, some physics engines combined with computer graphics rendering techniques, e.g., softbody
simulation~\cite{msong-mcthesis-2007,softbody-framework-c3s2e08,adv-rendering-animation-softbody-c3s2e09},
contribute a lot in this realism type by tweaking physical simulation parameters
of the laws, such as gravity, material properties, inertia, one can produce
interesting visual motion outcomes to make a point in a film.

\paragraph*{Narrative and Character Realism.} This aspect attempts
to make audience believe the fictitious events and characters of the animated
film actually exist. For example, in order to portray the animated character vividly,
artists use the squash-and-stretch method exaggerated for soft parts
of the character in traditional animation techniques~\cite{life-reproduced-drawings-2005}.
However, it is a very time consuming procedure.
Today's computer graphics software often provides a group of functionalities and a library,
such as hair, skin, clothes animation, skeleton animation in order to simplify
artists' work and achieve more realistic results.

\paragraph*{Social and Psychological Realism.} Social Realism makes audience believe
that the event is taking place in the fictitious animated world is as complex and diverse
as the real world. This concept applies both on traditional animation and
computer animation. In order to achieve social realism, artists not only rely on other
visual, character, motion realism, but also count on the writing
of the documentary production.
Psychological Realism was the first time brought up by Chris Landreth,
the director of animated short documentary \docutitle{Ryan}~\cite{ryan-orion-2005}.
It does not consider the physical
based motion as the priority, nor use some techniques, such as rotoscoping
and motion capture. Instead, he uses CGI animation in his work with added
elements, such as an original, personal, and hand animated
three-dimensional world.
Using psychological realism technique puts surrealist styling into the animated documentary,
so that peoples' psychological traumas are represented by twisted, surreal characters and their deformable faces. 
Films intend to use psychological realism to show the emotions of the characters in a way never seen before.

\begin{quote}
{\it ``There is no pre-existing reality, no pro-filmic event captured in its occurrence,
an animated film exists only when it is projected.
With no any existence in the world of actuality, the animated film like the partially
dramatized documentary, rely on a kind of artistic re-enactment, depending, in part,
on imaginative rendering as a compensation for the camera's non-presence at the
event.''}~\cite{truth-toons-documentary-1997}
\end{quote}

\subsubsection{Types of Visual Effects and Animation in Documentaries}
\label{sect:visual-effects-types}

Christian Darkin categorized the types of computer graphics and the corresponding techniques
used in documentary film as following in several categories~\cite{3d-animation-motion-cgi-documentary} 
that we recite below to complement our study.

\paragraph*{Explanatory or Explanation Graphics.} According to Darkin,
using \emph{explanation (explanatory) graphics} is a quite acceptable means to provide explanatory notes, ideas,
as well as information when the available footage
cannot portray it sufficiently well. The CG-animated explanatory
supplements can be very exciting and creative; 2D or 3D; text, cartoon,
moving characters---it is
up to the director, animator, artists, and their creativity,
resources available, and the corresponding needs of what to
portray. Such as CG type can arguably be fitting with any
documentary style~\cite{3d-animation-motion-cgi-documentary}.

\paragraph*{Animation for Color Shots.} This is a general CG mechanism
that is applicable and relevant to many types of documentary
film production, especially if at some point there is more of the
narration material than footage.
Darkin also gives good examples of such color shots, such as
a 20-second 3D animation where a virtual camera rushes
through a bloodstream of a patient and blood cells and others
``fly'' past can easily be used in many medical-related documentaries
and be ``on topic'' and not boring. Similarly, in the crime-related
documentary, one can animate a ``fly-through'' a CG-generated
building where the crime happened while narration is running and
prior to when the real footage begins~\cite{3d-animation-motion-cgi-documentary}.

\paragraph*{Visual Effects Reconstructions.} As opposed to the explanation graphics,
visual effects constructions are required in the absence of footage for the most prominent and necessary
events to portray in the documentary film that could not have been possibly
shot, physically inaccessible (the scale of cellular biology or the Universe),
or too far in the past, e.g. the assault on Baghdad,
dinosaurs, an assassination, the Big Bang, or if needed to show the
fat molecules getting from a burger to one's thighs, the 3D animated
CG reconstruction can greatly help to portray such
events in a documentary film~\cite{3d-animation-motion-cgi-documentary}.

\paragraph*{Text and Title Animation.}
This type of graphics is commonly employed to bring up documentary titles,
scene announcements, some short on-screen paragraphs
or questions raised by the maker, and subtitles.
An introduction to the film, its topics, and ideas are
good candidates for this type of animation in order to
set the tone the documentary film is to proceed with. Some
of the Motion Graphics techniques mentioned earlier can
be very well applied to text~\cite{cityspeak} as well here as the CG
techniques are typically common across the board, except
the text animation is optional and can be just
a still~\cite{3d-animation-motion-cgi-documentary}.

\subsubsection{CG Role in Documentary Production}
\label{sect:cg-role-docu}

We will have a closer look at some of the following documentary films
followed by an analysis of the impact of the computer graphics
featuring computer graphics techniques to a various degree that
feature CG-based animation:

\paragraph*{\docutitle{Super Size Me} (2004)}
is a personal experimental documentary with some computer animation sequence.
This is a small budget documentary film,
produced by an independent filmmaker Morgan Spurlock. In this film, he did a test on himself that
he had to only eat McDonald's three meals a day for 30 days~\cite{super-size-me}.
The film is very fast-paced and
full of sense of humor. It is a ``hybrid'' animated documentary, which contains partially
animated graphics and scenes according to our classification of the animated documentary film
stated earlier.
It combines computer generated graphs, charts, and animation, with the tune in humor,
which is the best way to get an audience interested in topics like that.

The simple graphics captured from the film and shown in \xf{fig:super-size-me-1},
is 100 times more visually vivid for audience to understand the information about
how fast food would cause a 10 year old to a 14 year old girl to gain so much weight.
Another example of animation used in this film shown
in \xf{fig:super-size-me-2} exaggerates how McDonald's make the chicken nuggets from
``unusually large breast chicken''.

\begin{figure*}[ht]
\hrule\vskip4pt
\begin{center}
	\subfigure[Example 1]
	{\includegraphics[height=2.0in]{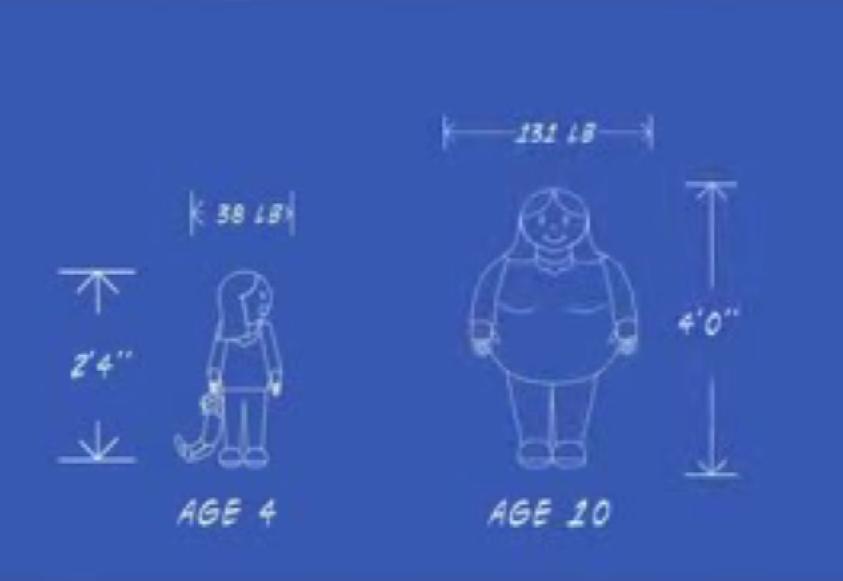}
	 \label{fig:super-size-me-1}}
	\hspace{.1in}
	\subfigure[Example 2]
	{\includegraphics[height=1.9in]{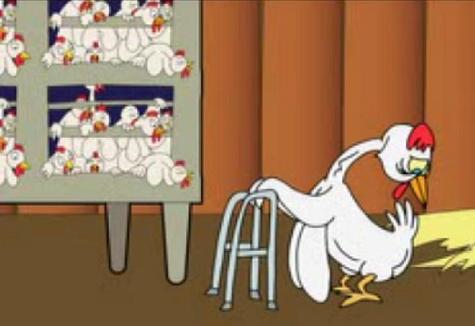}
	 \label{fig:super-size-me-2}}
\caption{\docutitle{Super Size Me} (2004) Example Screenshots}
\label{fig:super-size-me-screenshots}
\end{center}
\hrule\vskip4pt
\end{figure*}

\longpaper{IGNORED: \xf{fig:super-size-me-screenshots}}

When Morgan Spurlock was asked of the fact he used ``a lot of video games, a lot of computer graphics
and cell animation'' and if it is the way to make his points
``more accessible to the mainstream'', he replied,

\begin{quote}\textsl{``Definitely. Absolutely. One of my beliefs as a filmmaker is that if you can make somebody
laugh, you can make them listen. With laughter, you can get somebody's guard down,
you can open them up to listening to you.
They don't feel like they're being preached to or talked down to.
I think it helps, it makes really hard to understand information
a little more accessible and palatable. And at the end of the day,
it makes a movie a little more fun. It doesn't feel so heavy handed.''}
\end{quote}

\paragraph*{\docutitle{Little Voices} (2003)} is a hybrid documentary and a computer animation film.
The film's director, Jairo Eduardo Carrillo,
made a number of interviews of displaced children in Colombia's capital, Bogota, during the
Colombian Civil War~\cite{little-voices}. The core theme of the film runs through the real stories
told by the real children in their own voices, but the stories themselves were
illustrated initially by the children's drawings and paintings of the scenes
they were describing. Then Carrillo took those 2D drawings of characters, scenery,
etc., made by the children and turned them into the animated 3D CG models.
A combination of the children's art, computer animation, virtual and augmented reality
techniques~\cite{cost-refs-ar-gi08,arpracticalguide2008,7thingsar,augmented-reality-2010}
together with the children's voices create an impressive environment
and an art piece for the audience that not only engaging, but also preserving
the charm, integrity, and energy of the original art work of the children in their
3D animated counterparts~\cite{little-voices}. Following the creation of this
hybrid documentary film, Carrillo further created an educational game for displaced
children, with the same title \gametitle{Little Voices}
while he was staying during his residency at the Banff New Media Institute (BNMI).
In \xf{fig:little-voices-screenshots} are examples of a drawing,
then a 3D-redone scene, and the photograph of two children participants.

\begin{figure*}[ht]
\hrule\vskip4pt
\begin{center}
	\subfigure[Example 1]
	{\includegraphics[height=1.3in]{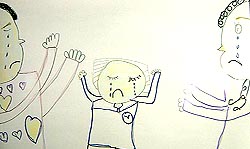}
	 \label{fig:littlevoices-1}}
	\hspace{.03in}%
	\subfigure[Example 2]
	{%\includegraphics[width=1.7in]{images/littlevoices1.jpg}
	 \includegraphics[height=1.3in]{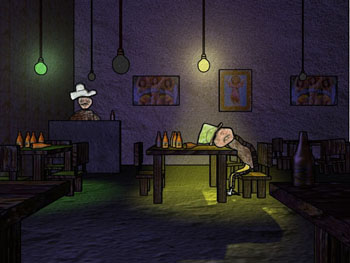}
	 \label{fig:littlevoices-2}}
	\hspace{.03in}%
	\subfigure[Example 3]
	{%\includegraphics[width=1.5in]{images/littlevoices2.jpg}
	 \includegraphics[height=1.3in]{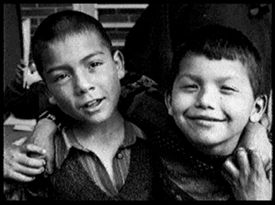}
	 \label{fig:littlevoices-3}}
\caption{\docutitle{Little Voices} (2003) Example Screenshots}
\label{fig:little-voices-screenshots}
\end{center}
\hrule\vskip4pt
\end{figure*}

\longpaper{IGNORED: \xf{fig:littlevoices-1}, \xf{fig:littlevoices-2}, \xf{fig:littlevoices-3}}

\begin{quote}
\textsl{``There is a certain poignancy in the Little Voices project, that can be found in the
intimate self-portraits drawn by the young people of Bogota recontextualized
into a large-scale animated world that is both compelling and deeply disturbing,''}
says BNMI director Susan Kennard. \textsl{``Carrillo draws our attention to this
juxtaposition through a representation of place that is both real and unreal.''}~\cite{speaking-through-little-voices}
\end{quote}

\paragraph*{\docutitle{Born Under Fire} (2008)} is another computer animated
documentary by Carrillo that follows the similar style as the
\docutitle{Little Voices}, but this time, it is based on the
interviews with and drawings by the new generation of children who
were 8 to 13 years old at the time of the interview, and have grown up
in midst of violence and chaos in Colombia~\cite{born-under-fire}.

\paragraph*{\docutitle{Chicago 10} (2007)} is a unique and unconventional documentary
uses motion-capture animation to portray the ``Chicago Conspiracy Trial''.
This is known as a very good and commended example where
the visual effects reconstruction type of graphics used for animating
the missing footage of the court room scenes of the Chicago Conspiracy Trial proceedings
all the way back in 1968 where the animation is nicely blended with some available
footage archives back from 1968 in order to accent the development
of the story and its emotion more sharply~\cite{chicago-10}.
Some example screenshots from the animation are in \xf{fig:chicago-10-screenshots}.

\begin{figure*}[ht]
\hrule\vskip4pt
\begin{center}
	\subfigure[Example 1]
	{\includegraphics[height=2.0in]{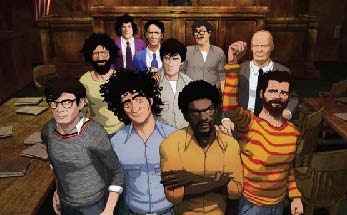}
	 \label{fig:chicago10-1}}
	\hspace{.3in}
	\subfigure[Example 2]
	{\includegraphics[height=2.0in]{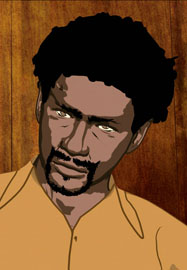}
	 \label{fig:chicago10-2}}
\caption{\docutitle{Chicago 10} (2007) Example Screenshots}
\label{fig:chicago-10-screenshots}
\end{center}
\hrule\vskip4pt
\end{figure*}

\longpaper{IGNORED: \xf{fig:chicago10-1}, \xf{fig:chicago10-2}}

\paragraph*{\docutitle{Ryan} (2005)} won an Oscar at the 77th Annual Academy Awards for Best Animated Short Film.
It is a prominent example of an ``all-out'' graphics documentary.
In this documentary, based on a period of life of a Canadian animator Ryan Larkin,
the audience perceives the voice of Ryan and the surrounding people, but
3D CGI characters visualizing Ryan and the others appear a bit strange,
twisted, see-through, sometimes broken and disembodied, which are humorous
or disturbing at times~\cite{ryan-interview-2005,ryan-orion-2005} (as, e.g.,
shown in \xf{fig:ryan-screenshots}).
While the rendering of the CGI scenes in \docutitle{Ryan} is non-photorealistic,
the 3D characters and the virtual environment are very detailed and make an
impression of being very realistic despite the fact that this documentary was created
not with the use of rotoscoping or motion capture techniques presented earlier,
but rather by using a 3D modeling software~\cite{maya} with some extra modifications
and plug-ins to enable the non-photorealistic rendering and mixed perspective and
non-linear projection~\cite{ryan-npar-2004}. Scholars classify this as an
autobiographical documentary, but non-traditional, because the whole film,
even the interviews were turned 100\% into 3D computer graphics scenes
instead of being filmed by a live-action camera.

\begin{figure*}[ht]
\hrule\vskip4pt
\begin{center}
	\subfigure[Example 1]
	{\includegraphics[height=2.0in]{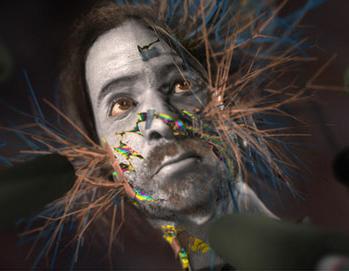}
	 \label{fig:ryan}}
	\hspace{.3in}
	\subfigure[Example 2]
	{\includegraphics[height=2.0in]{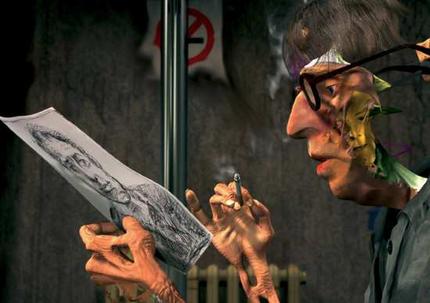}
	 \label{fig:ryan2}}
\caption{\docutitle{Ryan} (2005) Example Screenshots}
\label{fig:ryan-screenshots}
\end{center}
\hrule\vskip4pt
\end{figure*}

\longpaper{IGNORED: \xf{fig:ryan}, \xf{fig:ryan2}}

Director of this film, Chris Landreth,
says that after he learned Ryan Larkin's story:
\begin{quote}
\textsl{``There's a lot wrapped up in that, as far as a great story to tell, and I wanted to tell that in a
way that was as powerful as possible. Making an animation out of
a documentary was the best way, in my opinion, to do that.''}~\cite{ryan-interview-2005}
\end{quote}

\paragraph*{\docutitle{Waltz with Bashir} (2008)}~\cite{ide-waltz-with-bashir} is a CGI-animated
documentary film, which was advertised as being the first
feature-length CG animated documentary except the short part in the ending
that was featuring the real documented results of the Sabra and Shatila
found in an archived footage from the news at the time.
The film uses the animation to portray the memory of Ari Folman (the director of this film),
in the first Lebanon War twenty years after the war.

In this documentary Israeli director Ari Folman attempts to reconstruct
his missing memories from his time as a soldier in the 1982 Lebanon War by using
the animation.
First, each his drawing was sliced into hundreds of pieces that were moved
in relation to one another in order to create the movement. Then, the film was
preliminary shot in a sound studio as a 90-minute video, which was then transferred
to a storyboard. From there about 2300 original illustrations were drawn
with respect to the the storyboard~\cite{israeli-filmmakers-2008}.
All those illustrations, which together
eventually formed the actual film scenes, were composed using the aforementioned Flash animation,
classic animation, and other 3D technologies~\cite{israeli-filmmakers-2008}%
---in reality a combination of Adobe Flash scenes and
classic animation techniques.
Folman by using the freedoms that animation provides to take the
file into scenes that could not have been possible to shoot in the
traditional way. It also impacts the audience's preconceptions
of cartoons always belonging to the realm of narrative filmmaking
and attempting to emphasize where the line between the
fiction and reality lies~\cite{ide-waltz-with-bashir,sony-waltz-with-bashir}.
Overall, it took four years for the director to complete the film,
which is ``a combination of Flash animation, classic animation, and 3D''~\cite{sony-waltz-with-bashir}.
Two example screenshots from the documentary are in \xf{fig:waltz-bashir-screenshots}.

\begin{figure*}[ht]
\hrule\vskip4pt
\begin{center}
	\subfigure[Example 1]
	{\includegraphics[height=1.9in]{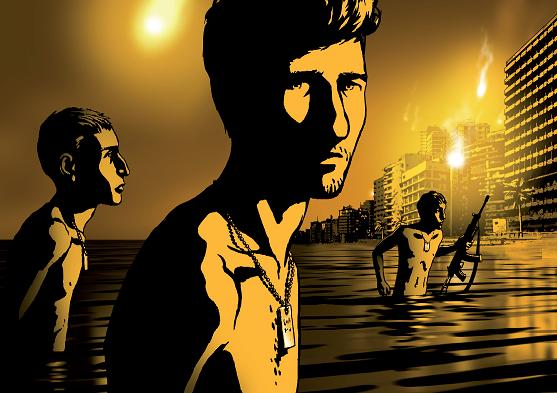}
	 \label{fig:waltz}}
	\hspace{.3in}
	\subfigure[Example 2]
	{\includegraphics[height=1.9in]{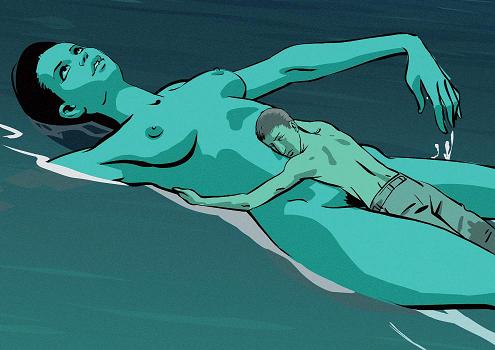}
	 \label{fig:waltz2}}
\caption{\docutitle{Waltz with Bashir} (2008) Example Screenshots}
\label{fig:waltz-bashir-screenshots}
\end{center}
\hrule\vskip4pt
\end{figure*}

\longpaper{IGNORED: \xf{fig:waltz}, \xf{fig:waltz2}}

This opens up the tools and techniques used in the production process to complement
the missing footage. Despite the documentary being mostly animated, the Folman's
story was told undistracted and still impacting the audience. On the team the
animator was Yoni Goodman, who produced the majority of the types of styles of
vivid and stunning animation, sometimes hand-drawn, he did not obscure the main
point of the story~\cite{drawing-from-memory-2009}.

\begin{quote}{\it ``For a few years, I had
the basic idea for the film in my mind but I was not happy at all to do it in real life video. How would that
have looked like? A middle aged man being interviewed against a black background, telling stories that
happened 25 years ago, without any archival footage to support them. That would have been SO BORING!
Then I figured out it could be done only in animation with fantastic drawings. War is so surreal, and memory
is so tricky that I thought I'd better go all along the memory journey with the help of very fine illustrators.
The animation, unique with its dark hues representing the overall feel of the film, uses a unique style
invented by Yoni Goodman at the Bridgit Folman Film Gang studio in Israel.''}~\cite{sony-waltz-with-bashir}
\end{quote}

Some other talented documentary filmmakers and artists
who creatively applied computer technology, theatrical elements
into documentary film making, such as Robert Legage and his work,
\docutitle{The Image Mill} (see, e.g., \xf{fig:image-mill-screenshots});
Dennis Del Favero with his \worktitle{iCinema}{Installation} installation.

\begin{figure*}[ht]
\hrule\vskip4pt
\begin{center}
	\subfigure[Example 1]
	{\includegraphics[height=1.5in]{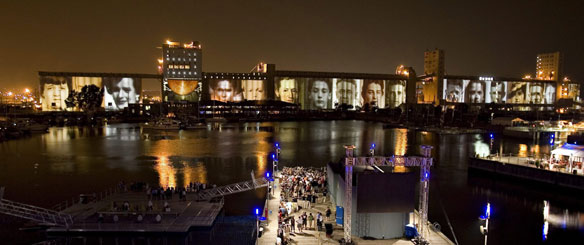}
	 \label{fig:image-mill1}}
	\hspace{.3in}
	\subfigure[Example 2]
	{\includegraphics[height=1.9in]{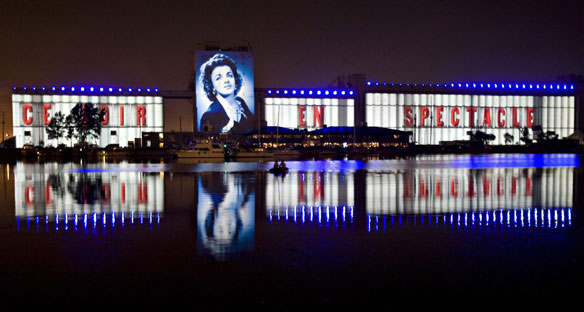}
	 \label{fig:image-mill2}}
\caption{\docutitle{The Image Mill} (2008) Example Screenshots}
\label{fig:image-mill-screenshots}
\end{center}
\hrule\vskip4pt
\end{figure*}

\longpaper{IGNORED: \xf{fig:image-mill1}, \xf{fig:image-mill2}}

\subsection{Emergence of Interactive Documentaries}
\label{sect:bg-interactive-docu}

Everything has been affected by arising new technologies, including newspaper, 
surveys, applications, communications, and so on,
change from paper format to mobile or computer web-based.
The same has been happening to the interactive media, TV, and documentary production.
As we explained in \xs{sect:trad-anim-docu} and \xs{sect:comp-anim-docu},
there are several different types of animation and computer graphics (CG) techniques used in
the traditional and modern documentary production, such as
computer-generated 2D and 3D images (CGI), computer animation.

Moreover, there are even ``manipulation''
of the data generated by the computer, which is also called computer-human
interaction~\cite{improving-hc-dialogue,wiki:computer-graphics,wiki:human-computer-interaction}.
Interactivity is another important element of the emerging documentary film production,
in which other broadcast media cannot compete with.
The interactivity of the new genre of documentary film enables
the audience to make the decision on what will
be going on in the film. They could participate on what should
be included in the documentary film and in which order.
Today, mass media can be easily used for documentary production
with video clips shot by virtually anyone, uploaded to Youtube
or similar services.
Some documentary-specific online projects allow
user control and interactive documentary making from either
pre-compiled scenes or video clips or user-generated video content
on a particular topic (e.g., weddings, city streets, etc.).

\subsubsection{Interactive TV}
\label{sect:interactive-tv}

Interactive TV was arguably the first to introduce the interaction aspects
into viewers' experience (outside of computers themselves), followed by
the interactivity being introduced into cinema and specifically documentaries.
The United Kingdom is particularly worthy of detailed study because 
it is the most well-developed interactive television market in the 
world~\cite{interactivetelevisionproduction2003}.
A good classical example of interactive TV was when before Christmas 2000, Sky One announced the
launch of the UKs first interactive TV program,
\worktitle{Harrods Christmas Special}{TV program}~\cite{harrods-shopping-documentary-2000},
which was in itself an entertainment for audience and additionally offered them interactive
shopping experience. It was the pioneering TV of shopping service, which
arguably made the broadcasting history in the world. The viewers shop
from the world's most famous store while they watch the program
of how Harrods has celebrated Christmas from its beginning
in 1849 to the present day with their TV hand set.
As~\cite{KCDS2004} put it, ``the recent advances in the STB (Set-Top Box) technology have introduced real-time video 
capturing and rich multimedia at consumers' homes.'' 
Digital STBs, like TiVo store television content, while the user controls 
the television flow with an on-screen user interface~\cite{personalized-television-2003,KCDS2004}.
Moreover, television content can be augmented with rich computer generated content, 
like animated characters and Internet information sources. 
Consumers are starting to have the need for a multimedia experience that 
seamlessly integrated diverse sources of information and entertainment 
content~\cite{press-teleputer-1990,telecomputer1992}. 
Even before the emergence and widespread adoption of the Internet and the Web, 
researchers were suggesting an integrated computer-television product~\cite{social-tv-and-ui}. 
However, immature technology, not enough consumer demand, 
and the success of Web have been postponing the convergence 
between the computer and the television~\cite{personalized-television-2003,KCDS2004}.
Interactive TV is a hot research topic and a subject of some conferences~\cite{social-tv-and-ui},
such as \emph{Euro ITV} and others where the interactive TV also naturally invades
the Internet is quite common in people's households and Set-Top Boxes.
The interactive TV is relevant as one medium of interaction with interactive
TV documentary productions as opposed to web-based or installation-based
productions.

\subsubsection{Interactive Cinema}
\label{sect:interactive-cinema}

\emph{Interactive Cinema} and \emph{Interactive Movie} are not very new terms,
which produce new storytelling experiences for audience by 
combining traditional linear storytelling with interactive digital media.
They can be considered as a generalization of interactive TV.
Interactive cinema is the evolutionary approach to traditional movie,
which gives the audience an active role in the showing movies.
Compared to traditional movie, the interactive cinema allows viewers to interrupt the movie from time to time,
and to choose among different possibilities of how the story goes on.
\docutitle{Kinoautomat} (1967) was the world's first interactive movie~\cite{kinoautomat-1967}
shown at Expo~'67 in Montreal.
The audience can change the plot by voting, by pushing buttons at nine points 
during the film the action stops. A moderator appears on stage to ask the audience 
to choose between two scenes; following an audience vote, the chosen scene is played.
Another example of the interactive movie is \filmtitle{I'm Your Man} (1992), 
which allows audience to decide which direction the plot should move forward. 
The film has a story-decision-tree where a choice must be made from a list of options
and then eventually it reaches one of the leaf concluding points.
A complete research overview on the evolution of cinema in general,
modern digital art (to 2003), and specifically including the
interaction aspects are well described in the book \worktitle{Future Cinema}{Book}~\cite{future-cinema-2003}.

In recent years, with the overbearing tide of video games,
interactive cinema is normally taken to be blended with 3D computer games,
which gives viewer a strong amount of control in the characters' decisions,
typically of the adventure or quest type (e.g., reincarnation of \gametitle{The Lord of the Rings: The Return of the King}
with the scenes of Peter Jackson's 2003 movie with the rolled into the role-playing game).
A good example of such a technique is \filmtitle{Policenauts} written and directed by Hideo Kojima.
In this work, a cinematic adventure game with a hard science fiction storyline, 
allows viewers or players to interact with the game through a point-and-click interface.
Video game based interactive cinema, evolving from the mathematical models and procedural programming, 
becomes alive when some of its parameters are controlled by a viewer. 
These works generally focus on continuous story playout environments, story authoring systems,
and scenarios for interaction~\cite{dab00}.
The real-time challenges of the aspect are
the most interesting to solve in a less expensive way than traditional approaches.

\subsubsection{Interactive Documentary}
\label{sect:interactive-docu}

Accompanied with the development of digital media and computer-based technologies,
new forms of documentary are challenging the traditional ones everyday.
Except the CGI documentary we discussed in previous section, beside interactive cinema,
interactive documentary is another new media type
directly related to interactive computer graphics~\cite{ea03},
which had been widely used in video game programming. Only since the 21st century,
interactive computer graphics has been slowly introduced to documentary production.
The related projects are mostly web-based to enable user to not only view but also
participate in the making of documentary film. Some pioneers of
``Interactive Online Documentary''~\cite{interactive-online-documentary} such as
Australian Film Commission (AFC)~\cite{afc} and Australian Broadcast Corporation (ABC)~\cite{abc}
collaborated with local Australian filmmakers and digital media artists to work on
these new documentary projects.
Interactive documentary, as a brand new field in new
digital media arts, is the revolutionary approach to traditional
documentary, which gives the audience an active role in the showing documentary.
The notion of interactive documentary has roots in traditional documentary story
telling and narrative, and web-based computer-based techniques with user-interaction,
interactive TV, and computer games.
Some research work has been done in this new field:

\begin{itemize}
\item Perspective and cultural overview of new media forms from theoretical and philosophical
standpoints~\cite{media-manifestos-1994,virtual-life-film-2007}, and in particularly the interaction aspects
of the new media, are described in Rieser and Zapp's book~\cite{new-screen-media-2002}.

\item Bill Nichols is the only documentary theorist who uses documentary
mode as a conceptual scheme to distinguish various styles
of documentary film. The interactive mode in documentary theory appears in the introduction
of the \emph{four modes} in documentary theory, such as ex-positional, observational,
\emph{interactive}, and self-reflexive, as detailed in his work~\cite{voice-of-docu}.

\item Barfield discusses what interactive documentaries could be back in 2003 and the
problem of narrative vs. interaction and interactive story telling
with 4 main structures~\cite{interactive-docus-2003}.

\item Tarrant published a very relevant description of technical and 
non-technical sides of making an interactive video archive
by a brother of a person with the degrading Usher syndrome (who can't hear or see any longer)
who in turn made home-shot family audio/video footage spanning across 20
years~\cite{planet-usher-interactive-movie-2004}.

\item Another relatively recent article details the production process of the documentary 
\docutitle{A Golden Age} in England as an interactive configurable documentary (\emph{interactive narrative})
and the use of the ``ShapeShifting Media'' technology with the technical implementation
details and the Narrative Structure Language (NSL)~\cite{interactive-docu-golden-age-2009}.

\item Then there is also a similar shift in education and drama portrayal.
A recent book~\cite{drama-edu-digital-2009} describes drama teaching using computer games with the intent to
make a memorable learning experience in a simulated environment including the
documentation of the research and practice of the approach~\cite{docu-videogame-2011}.

\end{itemize}

\paragraph{Systems and Tools.}

Current existing interactive documentary production system and technology:

\paragraph*{Diamond Road Online (DRO).} DRO is an experimental interactive
documentary system with user interface and recommendation systems to present
the documentary stories (in a keyword indexed database) of diamond trade allowing
semantic links between clips to make a continuous story off those
clips~\cite{diamond-road-user-docu-2008}.

\paragraph*{Interactive Drama Engine (IDE).} In the journal 
article~\cite{interactive-rt-3d-drama} the authors implement
an IDE based on theoretical foundations of narratives and drama 
as well as practicality and interactivity
of 3D first-person fiction/adventure/etc. games where participants can deeply affect the
storyline unlike in traditional games nor documentaries.

\paragraph*{Korsakow.} An open-source software system, Korsakow~\cite{korsakow-system}
in {\java} is used to build one own's documentaries in tree-like structure.
Florian Thalhofer, the Korsakow inventor says,
``linear storytelling, a special behavior, 
the story has been told same way every time...''

\paragraph{Prominent Examples.}

More generally, interactive web-based documentaries are a rapidly emerging medium. 
Now, we have a more detailed review of the artists' new approach 
to documentary filmmaking and the associated
technologies found in their works here:

\paragraph*{\docutitle{The Unexplained} (1996)} produced by the company
FlogTower could arguably be considered as the first interactive documentary
because in its manual of \docutitle{The Unexplained}, it says,
``FlagTower has named this concept
the Interactive Documentary, a title which reflects the televisual appeal
of our style of production''.

\paragraph*{\docutitle{Jerome B. Wiesner} (2004)}
The project \docutitle{Jerome B. Wiesner (JBW) (1915--1994): A Random Walk through the 20th Century},
developed in 1996 by Golrianna Davenport's MIT Media Lab on interactive cinema, is part of the
``evolving documentary'' genre~\cite{mit-2004}.
This ``hyper-portrait'' introduces the audience to a remarkable man whose life centered on science,
government, education and issues of cultural humanism~\cite{mit-2004}. In this hyper portrait (which runs on
the World Wide Web), audience is invited to explore the 20th century through
an extensible collection of stories about and recollections by the central figure.
Audience who knew JBW are also invited to share a memorable story
with the growing society of audience~\cite{mit-2004}.

\paragraph*{\docutitle{Traveling in Zagori} (2004)}
Another similar project developed by the MIT Media Lab is \docutitle{Traveling in Zagori}.
Through visual and textual snapshots of the landscape, architecture and people,
the audience is invited to construct a story about a local legend while
discovering aspects of Zagori's extraordinary history and legacy~\cite{mit-2004}.

\paragraph*{\docutitle{Man With a Movie Camera: The Global Remake} (2007)}
is the best example to represent
the concept of interactivity of documentary~\cite{man-with-a-movie-camera-2007}.
It is inspired by Vertov's \docutitle{Man With A Movie Camera} made in 1929.
The original documentary film records the progression of one full day synthesizing
footage shot in Moscow, Riga, and Kiev. It begins with titles that declare it
``an experiment in the cinematic communication of visible events without
the aid of intertitles, without the aid of a scenario,
without the aid of theater.''~\cite{perry-bard-interview}
Vertov's footage was shot in three different cities,
the industrial landscape of the 20's.
What images translate the world today? The current project is a participatory
video shot by people around the world who are invited to record images
interpreting the original script of Vertov's film and upload them to the website.
Anyone can upload footage and contribute as part of a worldwide montage.
The artist Perry Bard says,

\begin{quote}
\textsl{``Vertov's 1929 film is a great point of departure
for the Internet because it has so many dimensions from the documentary to the
performative to the effects along with its use of an archive which translates
to a database and it's a film within a film...
going global was obvious and the rhythm is very contemporary, there's no shot in
the film longer than twenty seconds. It seemed like a perfect vehicle for
global input and in keeping with Vertov's intentions as a filmmaker.''}~\cite{perry-bard-interview}
\end{quote}

\paragraph*{\docutitle{Out My Window} (2010)}
is the most recent item and outstanding example of a web-based and database-driven interactive documentary 
by the National Film Board of Canada's project \docutitle{Out My Window} designed
for digital storytelling~\cite{bits-blogger-digital-storytelling}.
This project explores our urban planet through through highrise windows.
Later, this $360^\circ$ web-based project is transformed into the \worktitle{Highrise StorySpace}{Installation}
installation, which extends the original documentary 
into the larger than life-size physical environment. 
The installation takes the audience on a journey into the center of a
``spatialized cinematic experience''. In \xf{fig:out-of-my-window} are
two example illustrations from this work.

\begin{figure*}[ht]
\hrule\vskip4pt
\begin{center}
	\subfigure[Example 1]
	{\includegraphics[height=2.0in]{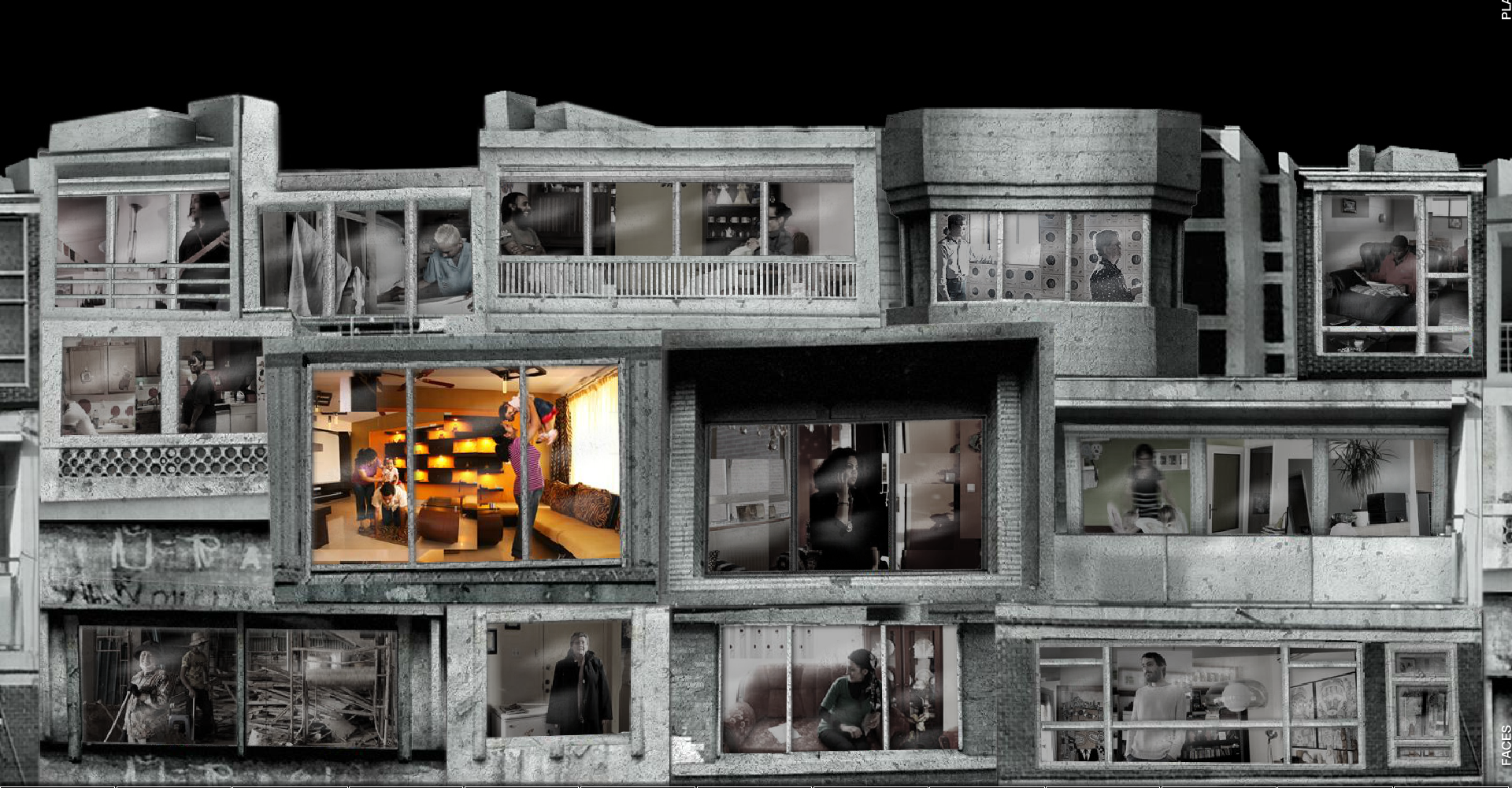}
	 \label{fig:out-of-window-1}}
	\hspace{.3in}
	\subfigure[Example 2]
	{\includegraphics[height=2.0in]{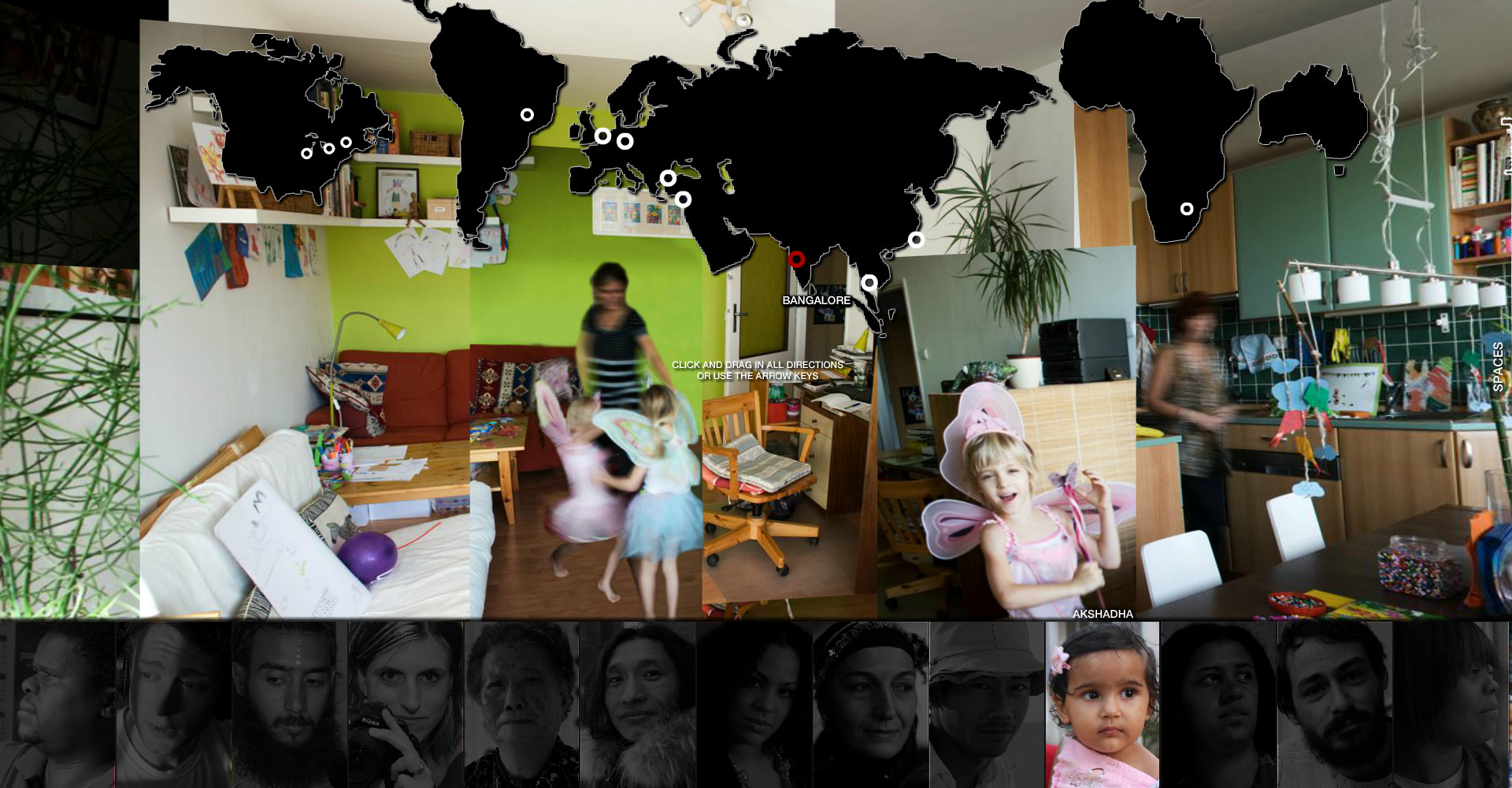}
	 \label{fig:out-of-window-2}}
\caption{\docutitle{Out My Window} (2010) Example Screenshots}
\label{fig:out-of-my-window}
\end{center}
\hrule\vskip4pt
\end{figure*}

\longpaper{IGNORED: \xf{fig:out-of-window-1}, \xf{fig:out-of-window-2}}

\paragraph*{\docutitle{Ceci N'est Pas Embres} (2012)} is a Korsakow film,
created by the ARC team (Adventures in Research/Creation), co-directed by Matt Soar.
This project is a database-driven, diary-style to tell a story about a family
from Quebec moving to southwestern France.
ARC is a research group co-directed by Professors Monika Kin Gagnon and Matt Soar
at Concordia University, Department of Communication 
Studies.
The audience could interact the system which include personal testimony,
animated scenes, and soundtracks~\cite{bits-blogger-digital-storytelling}.

\paragraph*{\docutitle{Sea Monsters} and \docutitle{U2} (2008).}
Another 3D aspect---stereoscopy---that comes with the new
stereoscopic movies documenting potential life of animals
millenia ago, or the animals now in the see or simply
documenting a concert in 3D, such as done by IMAX for
their animations and stereo-enhanced motion pictures
such as \docutitle{Sea Monsters} (2008) and \docutitle{U2} (2008) concert
(and the previously mentioned \docutitle{Hubble}). 
The audience in this case, while swimmingly passive and watching,
interacts with the show through the stereoscopic illusion
their eyes receive as a part of the show enhancing the
perception and impression of the show. One can argue
it is one-way interactive, but nonetheless should not
be neglected. Needless to say, the stereoscopic shows
can also be made interactive, and even more so in the
future with the techniques presented throughout this and other
chapters.

\section{Performance Arts and Theatre Production}
\label{sect:theatre-background}

Theatre is not only an art form to give the actors
an opportunity to present a live performance in front
of a passive audience within a specific space, but also a structure
of such space itself.

The ancient Greek drama as the earliest history in theatre,
could be traced back to 550--220~BC\footnote{\url{http://en.wikipedia.org/wiki/Theatre_of_ancient_Greece}}.
There were some scenic techniques that had been used in Greek theatre
as early as the fourth century BC, such as, e.g., a \emph{mechane} was used to
lift flying actors from or onto the stage; 
trap doors or openings in the ground to bring actors onto the stage;
\emph{Pinakes}, pictures hang or built on the stage to create scenery;
a \emph{Mask} was also a significant element in ancient Greek theatre.

Western theatre, which is derived from ancient Greek drama and 
developed and expanded under Roman theatre,
has been a very important art form 
not only for entertainment, but also had a significant impact on their countries's
cultures. 

Asian theatre also has a long and complex history, 
for instance, Chinese Shang theatre 
has more than 2500 years history. 
There were some traditional techniques used in Asian theatre.
Masks, which have also been widely used in
Asian theatre performance, especially in today's Japanese Noh performance, 
are still one of those very important performance elements~\cite{japanese-theatre-nihon}.
Other technology used in conventional theatre stagecraft such as puppetry, sound, 
theatrical fog are also still quite popular in modern theatre production.
Today Chinese theatre
is not only limited to traditional Chinese opera,
but also refers to modern Chinese drama.
For example, in Beijing Opera (one of the traditional Chinese opera), 
facial makeup and costumes are extremely important to
distinguish historical characters, their roles and social ranks in the scene.  

Theatre focuses on live performers enacting in front
of the audience within the same space. However, it also involves other contributions
from a playwright, costume designer, makeup artist, or set and props designer--all
introducing some kind of ``technology'', albeit often primitive, into a theatre setting.

Subsequently, we review the theatre background from the classic point of view in \xs{sect:classic-theatre},
Asian theatrical influence in \xs{sect:asian-theatre}, and the current digital theatre further in
\xs{sect:theatre-new-digital}.

\subsection{Classic}
\label{sect:classic-theatre}
\label{sect:bg-classic-theatre}

In the $20^{th}$ century, there have been different voices about 
the relationship between theatre and conventional non-computer technology:
how to use technology in performances,
and how much technology should be applied in theatre productions.

Some important literature and theories include such as those of
Artaud's \emph{Theatre of Cruelty}~\cite{theatre-its-double-1970}, 
Grotowski's \emph{Poor Theatre}~\cite{poor-theatre-1968}, 
Brook's \emph{Holy Theatre}~\cite{empty-space-1968}, 
and Bertolt Brecht's \emph{Epic Theatre}~\cite{brecht-on-theatre-1977}.
And some of these theatre practitioners were against the use of technology
in theatre production, most notably Grotowski 
(who is considered the father of contemporary theatre).
He declared in his book \worktitle{Towards a Poor Theatre}{Book}~\cite{poor-theatre-1968} that theatre 
should not, because it could not, compete against the overwhelming spectacle of film 
and should instead focus on the very root of the act of theatre: 
actors co-creating the event of theatre with its spectators.
I support his position, because to me, the performance of actors,
and the moments of their communication to audience are the most
precious elements in a theatre performance.
However, at that time he could not have imagined what we have today: computers, the Internet, cellphones, e-newspaper,
iDevices, and other different types of new media technology, which all have a potentiality to
influence theatre performance noticeably. If technology could only enhance actors' performance
and broaden audience imagination,
rather than overwhelming audience with unrelated digital stuff, why cling to those theories?

On the other side, an artist such as 
Artaud, was very interested in movements, sounds and 
encouraged that theatre needs to 
``recapture from cinema, music-hall, the circus and life itself, 
those things that always belong to it'' and indicated that
``our sensibility has reached the point where we surely need theatre 
that wakes us up heart and nerves''~\cite{theatre-its-double-1970}.
His theatre of cruelty whose sentiments could 
still easily apply in today's cyber theatre space. 
Artaud's theory to me, seems holy, an awesome power that is both a creator and destroyer.
Whereas Grotowski's reference to theatre generally applies to the dedication of the actor,
in giving himself as a gift over performance which is transcendent in a much more
human-sized way. Moreover, he insisted that there was no point in trying to compete with film
but that theatre should rather convert back to its roots,
``If it [the stage] cannot be richer than the cinema, 
then let it be poor.''~\cite{richards-physical-actions-95}

Peter Brook's recent research on theatre says,
``Holy Theatre for short, but it could be called The Theatre of the Invisible-Made-Visible: 
the notion that the stage is a place where the invisible can appear has a deep hold on our
thoughts.''~\cite{empty-space-1968}

The German artist Brecht,
the most significant artist in Germany (Bavaria at the time), was very interested in 
Chinese and Japanese traditional theatre.
His epic theatre theory, in particular has been influenced by the Chinese theatre, 
also employed technology to theatre production.
Brecht expected the audience to be aware that they were always watching a play and 
always be rational so that could provoke their self-reflection, 
``It is most important that one of the main features of the ordinary theatre 
should be excluded from [epic theatre]: 
the engendering of illusion''~\cite{brecht-on-theatre-1977}.
Moreover, he applied some practices of using bright lighting, loud sounds,
images and texts to interrupt audience in order to remind
them that ``the play is a representation of reality and not reality
itself.''~\cite{brecht-on-theatre-1977}

At the beginning, one may get the impression that they raised up different voices as
to whether technology and theatre performance could coexist;
however, to me, their positions are not in conflict at all.
The acting and performance are always the most important element
in theatre production; the role of technology is to widen, enrich, and
enhance the experience of live performance instead of obstruct,
replace, or destroy the art of performance.

\subsection{Asian Influence}
\label{sect:asian-theatre}
\label{sect:bg-asian-theatre}

There exist several theatrical forms in Asia, such as the earliest Sanskrit drama ($8^{th}$~Century BCE) in India; 
Beijing Opera in China; traditional Kabuki puppet theatre and Noh drama in Japan. 
The story \playtitle{A Treatise on Theatre} is the most complete work and evidence of Sanskrit drama. 
It addresses ``acting, dance, music, dramatic construction, architecture, costuming, make-up, 
props, the organization of companies, the audience, competitions, 
and offers a mythological account of the origin of theatre''~\cite{richmond-farley-1995}.
There are four main types of traditional theater in Japan: \emph{noh}, \emph{kyogen}, \emph{kabuki}, and \emph{bunraku}.  
Each of these forms of theater performance is very distinct and unique from the others.
In Noh theatre (e.g., \xf{fig:noh-stage}), most of the characters in these plays are concealed by masks, 
and men play both the male and female roles; however, in Kyogen theater, which is performed between
Noh's acts intermission, actors do not wear masks. Bunraku theater uses puppets. 
The puppets about three to four feet tall are controlled by puppeteers who dress completely in black.

\begin{figure}[hptb!]%
	\centering
	\includegraphics[width=.7\columnwidth]{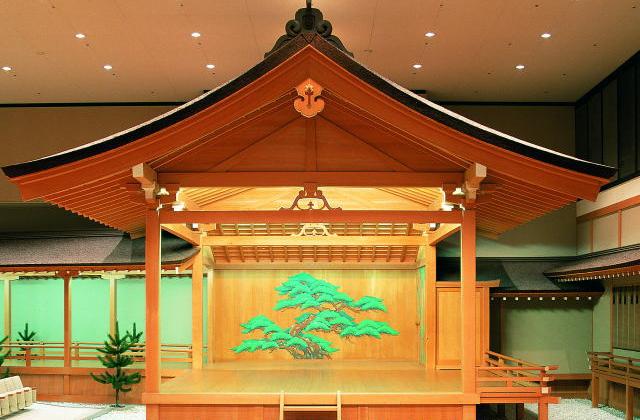}%
	\caption{Noh Drama Stage}%
	\label{fig:noh-stage}%
\end{figure}

Shang theatre as early as 1500 BC could be considered the earliest Chinese theatre, which
often involved music, clowning and acrobatic displays.
During the Han Dynasty, shadow puppetry first emerged as a recognized form of theatre in China.
The rods used to control puppets were attached perpendicularly to the puppets' heads so that
puppeteers were not seen by the audience when the shadow was created.
In the Tang Dynasty, Emperor Xuanzong formed an acting school, the Pear Garden,
to produce a form of drama that was primarily musical.
In the Sung and the Yuan Dynasty, there were many popular plays involving acrobatics and music
with a four- or five-act structure.
Yuan drama spread across China and diversified into numerous regional forms,
the best known of which is Beijing Opera (e.g., \xf{fig:san-cha-kou}), which is still popular today.
It combines music, vocal performance, mime, dance and acrobatics into one entertaining show.

\begin{figure*}[htp!]
	\centering
	\includegraphics[height=3.0in]{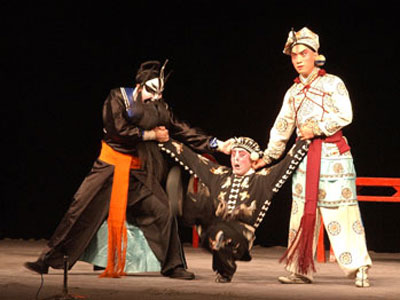}
	\caption{Beijing Opera \playtitle{San Cha Kou} Example}
	\label{fig:san-cha-kou}
\end{figure*}

\paragraph*{Facial Makeup.}
Facial makeup has obtained the reputation as ``painting of the heart and soul'',
which enable the audience to get a glimpse of the characters' inner world through
their symbolic facial makeup. It utilizes the color of red, purple, black, white, etc.,
with each color representing a unique character stereotype:
red symbolizes utter devotion and loyalty; 
black represents faithfulness and integrity; white implies craft.

\paragraph*{Costumes.}

Costumes could distinguish the rank of the character being played, for example,
Emperors are in yellow robes, and high-ranking characters wear brilliant colors.
The gowns for high-level ranking female characters usually have water sleeves, and
the characters of no rank wear simple clothing without embroidery~\cite{halson-elizabeth-1966}.
Actors use the long flowing sleeves to facilitate emotive gestures (about hundreds of gesticulations),
such as sadness and shyness are expressed by one hand pulling another water sleeve to cover the face; 
raising and putting up two persons' water sleeves to embrace each other.

Modern Chinese Drama started to develop in 1907 in Shanghai.
Unlike traditional Chinese opera, it realistically reflected the changes
in the lives of Chinese before and after the founding of New China.
A number of Chinese playwrights have realistically portrayed the lives of
common folks greatly affected by Western playwrights such as Shakespeare, Chekhov, and Moliere.
The most memorable Chinese plays 
include \playtitle{Teahouse} (\xf{fig:drama-teahouse}), \playtitle{The Thunderstorm}, and \playtitle{The Family}.
Younger generations of playwrights have tried to develop a more modern style.
The most innovative theatre director, Meng Jinghui, who employs a lot of theatrical techniques,
such as electronic music, acoustics, lighting, and novel stage settings in his plays
in addition to the new ideas in order to impress the audience both mentally and visually.
However, Modern Chinese drama is still quite ignored in the West in 
favor of traditional theatre.

\begin{figure*}[htp!]
	\centering
	\includegraphics[height=3.0in]{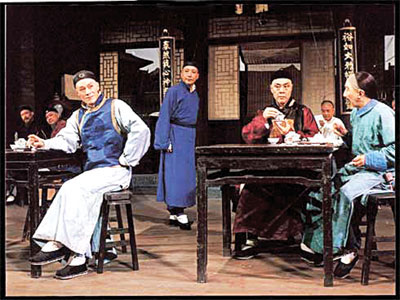}
	\caption{Modern Chinese Drama \docutitle{Teahouse} Example Screenshot}
	\label{fig:drama-teahouse}
\end{figure*}

\subsection{Current Theatre and its Technology in Digital Era}
\label{sect:theatre-new-digital}
\label{sect:bg-theatre-new-digital}

\emph{Digital theatre}, \emph{cyberspace theatre},
\emph{multimedia performance}, \emph{interactive theatre}, \emph{virtual theatre},
and more possible terms associated with today's theatre,
have all appeared based on the new technology and digital media innovation.
We are in the new digital era, performers, audiences, and
space---are they still all inter-related as conventional theatre proposed?
Is it necessarily that actors and audience in today's live performance
are within the same physical space?
What if performers are not real human beings?
What is the new definition of THEATRE now or for future?

In Laurel's very early work in 1991, \worktitle{Computers as Theatre}{Book}~\cite{computers-as-theatre-93},
she applied her knowledge of theatre to Computer-User interface design.
She says, ``In many ways, the role of the graphic designer in human-computer 
interaction is parallel to the role of the theatrical scene designer.''
Both create representations of objects and environments that provide a context for action.
She thinks of the computer, not as a tool, but as a medium.
Moreover, she examines Aristotle's six elements of theatre,
``action, character, thought, language, melody or pattern,
and spectacle or enactment'' in his \worktitle{Poetics}{Book}~\cite{poetics-aristotle} versus that of HCI.
She proposes the reader various experiments addressing touch, smell and taste and
notes the possible site-specific interactive plays and performance art model.

Another very extreme artist in this field is an Australian performance artist, known as \emph{Stelarc},
whose research is in \emph{Cyborg Theatre}. Cyborg theatre uses cybernetics as a method, and
establishes the relationship between a human being and technology~\cite{virtual-theatre-intro-2004}.
Stelarc's work, \playtitle{Stomach Sculpture} uses nanotechnology,
``to insert an art work into the body.''
As described by Stelarc, the viewer not only would observe inside their body,
but are also an ``artist'' himself.
His intention is to use cyber-systems to break the barrier of skin in order to
extend his body's performance.
He is both an artist and his body, even his organs, is an art work at the same time.
Stelarc believes, ``New technologies tend to generate new perceptions and paradigms of the world, 
and in turn, allow us to take further steps.''~\cite{extended-body-stelarc}

There is a significant amount of
innovative multimedia artists, performers eager to explore
the possibilities that multimedia could bring to theatre production.
There are conventional techniques for theatre productions,
such as stage lighting, sound effects, set design, costumes,
and special effects, such as fog and explosions, still have been
widely used for most of the theatre production companies.
Likewise, digital technology helps to revolutionize the design
of these theatrical techniques in order to make 
the theatre the true sense of LIVE performance.

Some of the multimedia artists who apply digital technology to theatre productions,
such as Chris Salter with his real-time audio, image, gesture in
responsive space~\cite{entangled-2010}, Natasha Tsakos's live 3D
animated show~\cite{natasha-tsakos},
Marc Hollogne's cinema-theatre, in which one cannot decide 
between a movie and a play~\cite{marc-hollogne},
Silhouettes Dance Group's shadow performance combined with photographs~\cite{silhouettes-dance-group},
and a Taiwanese artist Stan Lai,
who applied many creative staging innovations, such as bringing a live camera
on stage and not only present the scene of the stage, but also broadcast
another dimensional image directly from various cameras.
We catch a glimpse of their projects and exploited technological advances:

\begin{itemize}

\item
Salter's artistic creativity focuses on ``dynamic and temporal processes
over static objects and representations''. For instance, in his work, \playtitle{SCHWELLE II},
a live dance theatre performance, a solo dancer
experienced a traumatic transformation from death to rebirth.
During the live performance period, the dancer wears several
wireless acceleration sensors, the input of which dynamically affected the audio
and visual performance output based on the sensor data obtained from the performer.

\item
Natasha Tsakos' \playtitle{Up Wake} is a live performance with only one actor
and a bare stage lit with perfectly synchronized computer graphics imagery,
digital sound effects to demonstrate a cartoon character's dream and
wake throughout his day. The projection techniques of CG
animation and graphics have been widely
used in her work, which is a very important solution for environment/space
transformation and time vicissitude~\cite{natasha-tsakos}.
In \xf{fig:up-wake-army} is the
example illustration from her work.

\begin{figure*}[htp!]
	\centering
	\includegraphics[height=3.0in]{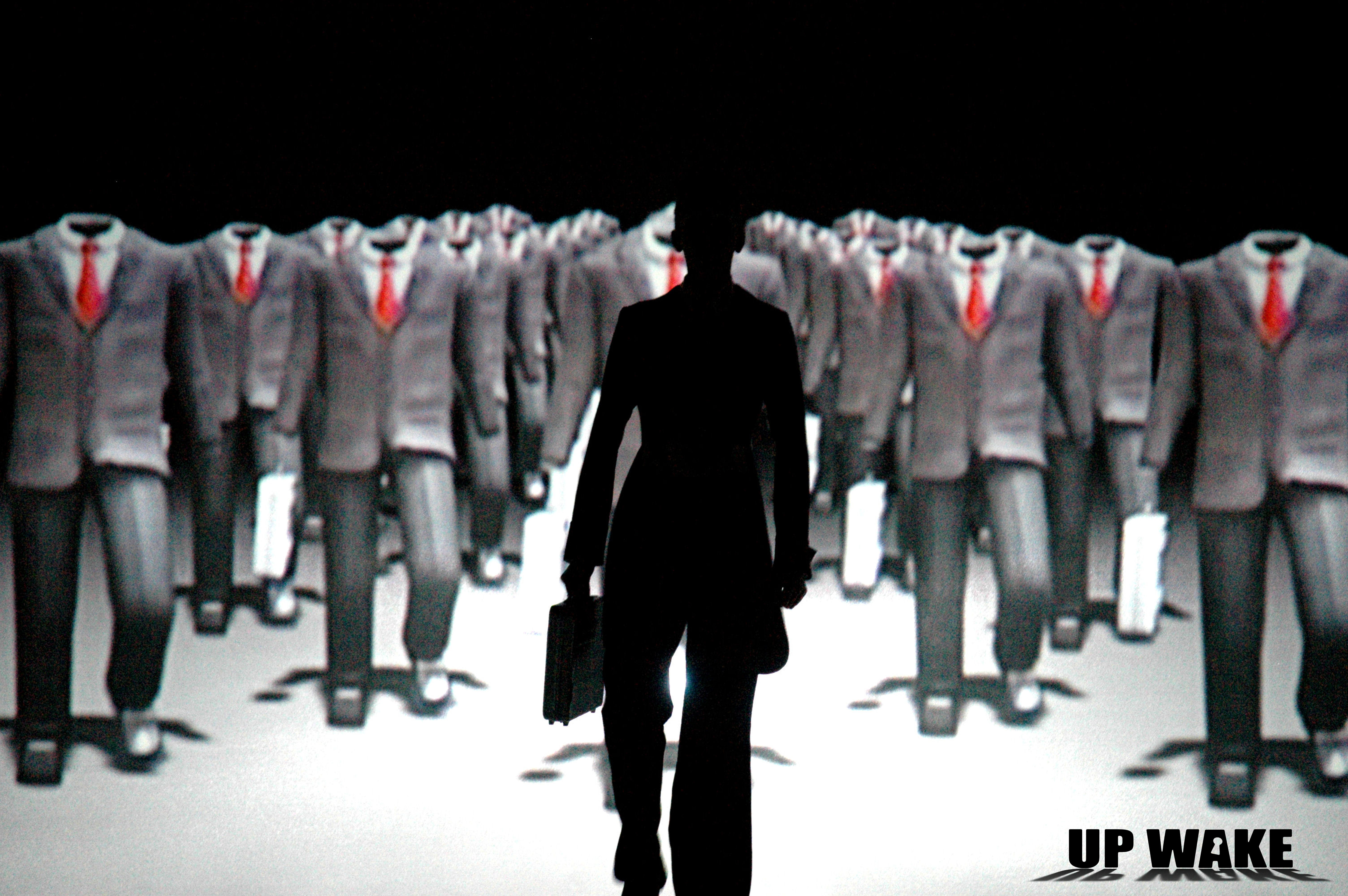}
	\caption{\docutitle{Natasha Tsakos Performance} Example Screenshot}
	\label{fig:up-wake-army}
\end{figure*}

\item
Most Marc Hollogne's works include a cinema screen on the theatre
stage, projected with filmed sequences. Everything happened on the stage,
the live actors' performance and broadcasting stories on the screen,
have to be extremely well-synchronized as if the story continually happens within the same
physical spacetime. The setup requires a great deal of precision on dialogue,
gestures, and motions, for example, when the moment actors walk 
in and and out from the side of the screen to behind while
the scenarios about the actors are continuing smoothly on the screen,
one could not tell what is real in the film and what is not real on the stage.
In \xf{fig:marc-hollogne} are
two example illustrations from his work.

\begin{figure*}[ht]
\hrule\vskip4pt
\begin{center}
	\subfigure[Example 1]
	{\includegraphics[height=2.3in]{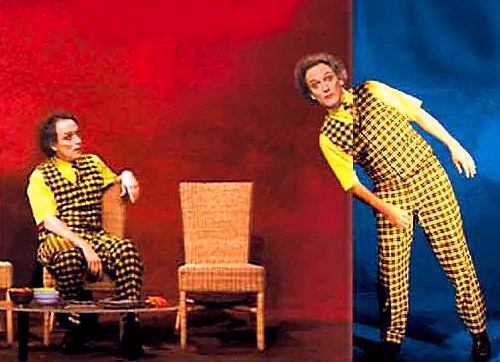}
	 \label{fig:marc-hollogne-1}}
	\hspace{.3in}
	\subfigure[Example 2]
	{\includegraphics[height=2.0in]{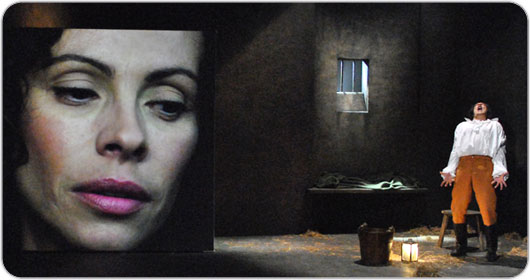}
	 \label{fig:marc-hollogne-2}}
\caption{\docutitle{Marc Hollogne Performance} Example Screenshots}
\label{fig:marc-hollogne}
\end{center}
\hrule\vskip4pt
\end{figure*}

\longpaper{IGNORED: \xf{fig:marc-hollogne-1}, \xf{fig:marc-hollogne-2}}

\item
Denis Marleau, a director, production designer and creator of stage installations, 
is a major figure in Quebec theatre. His work is distinguished by innovative use 
of audio and video technology. A unique example is the play showed in 2002,
\playtitle{Les Aveugles}, in which twelve mannequin faces were animated by prerecorded
video without any actor on stage. In about summer of 2011, 
Denis Marleau and St\'{e}phanie Jasmin collaborated with
Jean Paul Gaultier's fashion show exhibition in Montreal Fine Arts Museum.
They placed 30 mannequins throughout the galleries and animated these faces
through an ingenious projection system with prerecorded videos~\cite{denis-stephanie-mmfa}.
In \xf{fig:denis-marleau} are
two example illustrations from his work.

\begin{figure*}[ht]
\hrule\vskip4pt
\begin{center}
	\subfigure[Example 1]
	{\includegraphics[height=2.2in]{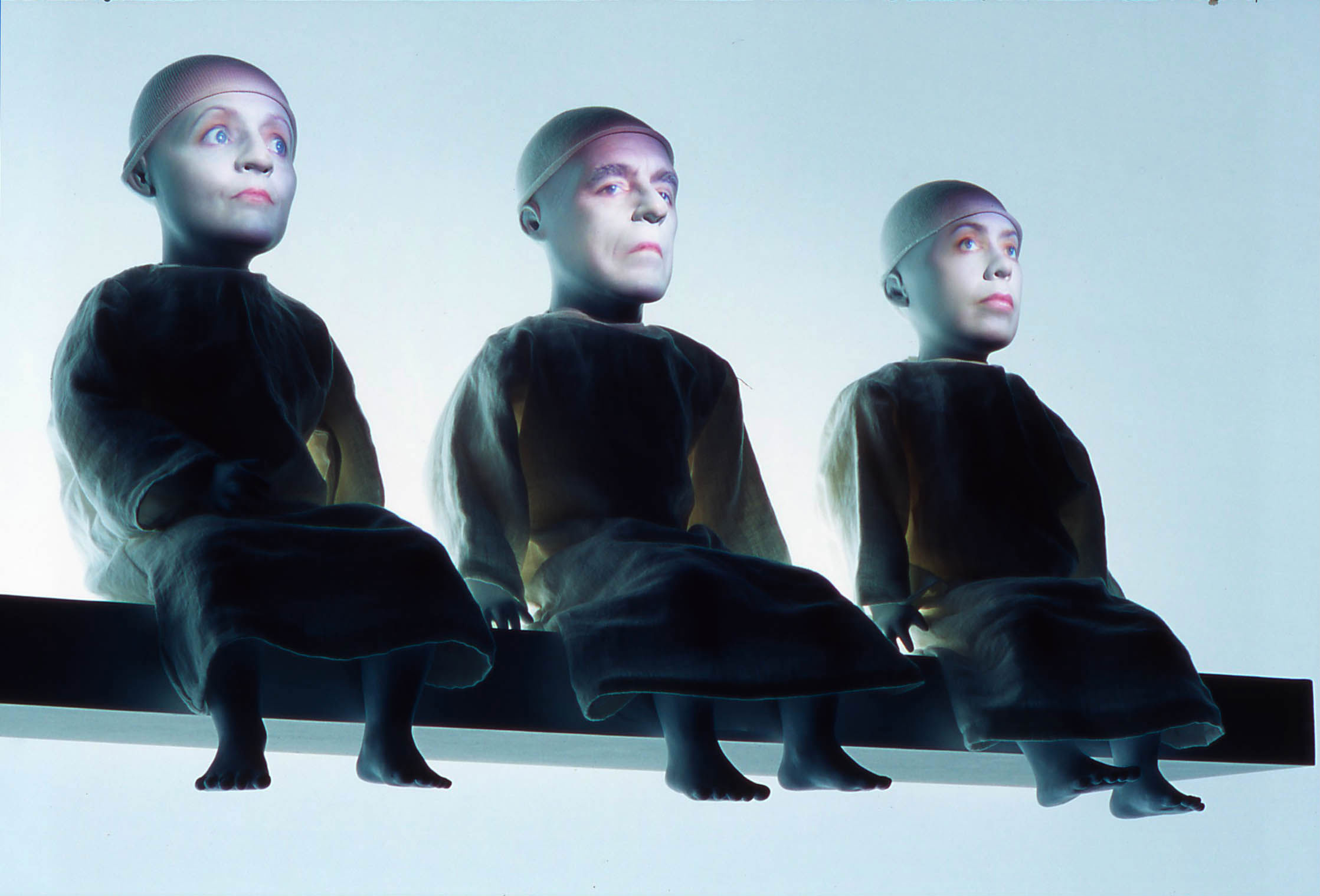}
	 \label{fig:denis-marleau-1}}
	\hspace{.3in}
	\subfigure[Example 2]
	{\includegraphics[height=2.0in]{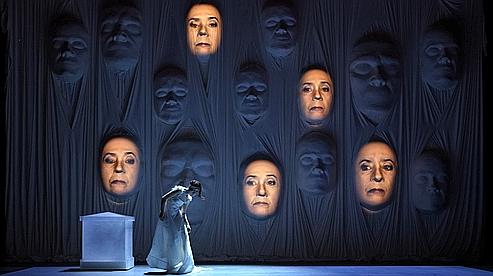}
	 \label{fig:denis-marleau-2}}
\caption{\docutitle{Denis Marleau Performance} Example Screenshots}
\label{fig:denis-marleau}
\end{center}
\hrule\vskip4pt
\end{figure*}

\longpaper{IGNORED: \xf{fig:denis-marleau-1}, \xf{fig:denis-marleau-2}}

\item
The inspiring performance from The Silhouettes dance group makes the shape of the pictures 
with dancers' bodies and then projects the photographs mapped along to appear on the stage.
This very traditional technique which has been widely used in shadow theatre 
as early as the 18th century, now is innovatively applied to The Silhouettes's performance
combined with new digital technology.
In the performance \playtitle{Believe}, a number of figures were created on the stage by
actors' shadows: White House, a ballet dancer, swimmer, and an astronaut on the Moon;
following that, the real photographs of the figures projected on the canvas blending with the shadows.
In \xf{fig:silhouettes-dance} are
two example illustrations from this dance group's performance.

\begin{figure*}[ht]
\hrule\vskip4pt
\begin{center}
	\subfigure[Example 1]
	{\includegraphics[height=2.2in]{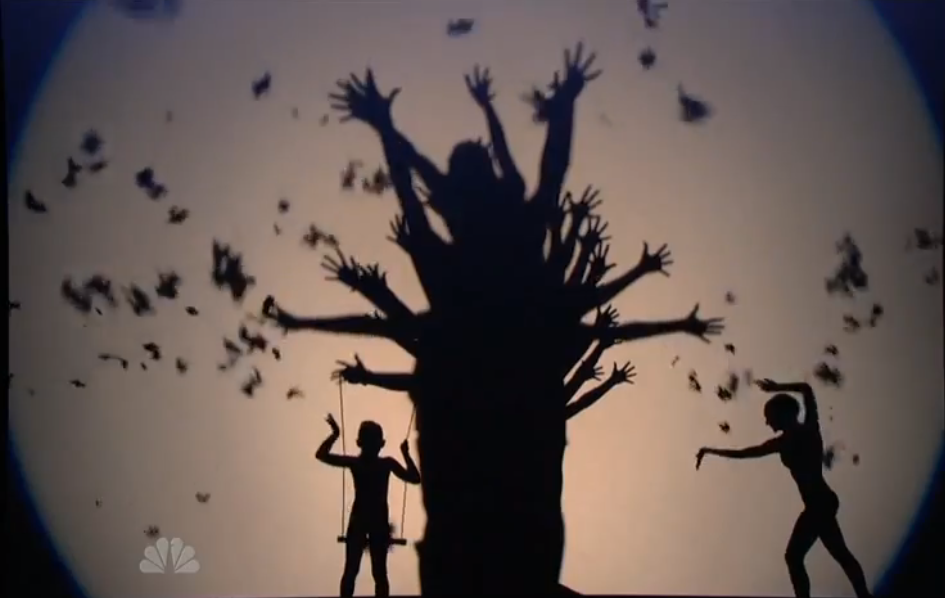}
	 \label{fig:silhouettes-dance-group-1}}
	\hspace{.3in}
	\subfigure[Example 2]
	{\includegraphics[height=2.0in]{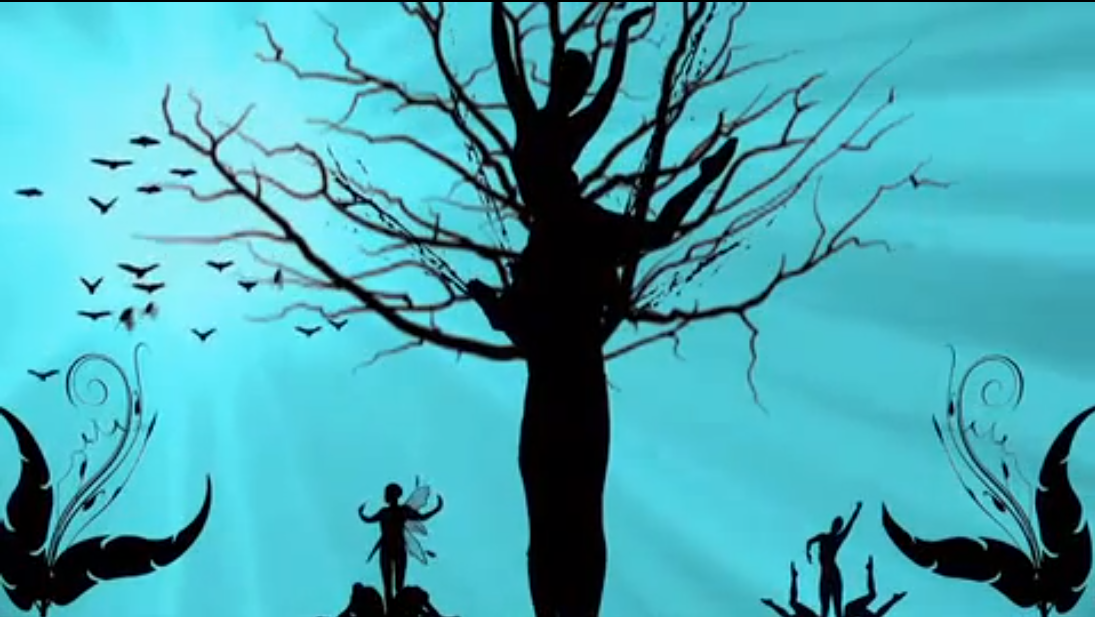}
	 \label{fig:silhouettes-dance-group-2}}
\caption{\docutitle{The Silhouettes Dance Group Performance} Example Screenshots}
\label{fig:silhouettes-dance}
\end{center}
\hrule\vskip4pt
\end{figure*}

\longpaper{IGNORED: \xf{fig:silhouettes-dance-group-1}, \xf{fig:silhouettes-dance-group-2}}

\item
Stan Lai is the most celebrated Chinese language playwright and director for not only
his memorable works, but also creativity of new genres and staging innovations.
In his recent work, \playtitle{Watch TV with Me}, a seven-act drama incorporates several stories 
involving both television and everyday people in mainland China since the 1980s.
Lai used the most advanced sound and lighting technology to create a time and space change with
development of television. The montage effect, the interactive theatrical performances between 
live performance and virtual characters on the same stage, and the multimedia concept
leave audience with the burdens of memory, history, longing, love and appreciation of the power of theater itself.

\end{itemize}

To summarize,
the technology itself has no vitality without artists' creativity, 
without the connection to the audience, without the context of society and humanity.
Recent computer technology developments have improved the theatre performance, have
extended theatre performance and expanded it to more dimensions.
It has closed in on the gap among physical world, human body, and digital world.
I would conclude that ``Poor Theatre'' is never actually poor, 
it is the richest
medium.

\section{Technology}
\label{sect:tech-background}

In this section
we review the relevant technological aspects pertinent to our work in
terms of visualization interaction terminology, hardware, and software.
We cover the recent interaction hardware in \xs{sect:bg-devices},
viewing and projection aspects in \xs{sect:bg-viewing-projection},
the related application programming interfaces (APIs) and libraries
in \xs{sect:bg-api}, followed by the credits in \xs{sect:bg-code-samples} to the open and shared
source communities for providing programming examples to individual
features, components, or effects that we got inspired from, relied on, re-used,
improved and built up like supporting puzzle pieces to the development
of the prototype installation work in the chapters that follow. 

\subsection{Devices}
\label{sect:bg-devices}

Advanced interaction devices emerged recently and became accessible
and affordable to many increasing the landscape of human-computer interaction
in a variety of disciplines. The devices have to do either with some kind of
motion, gesture, posture, etc., tracking or haptic force feedback
allowing for the sense of touch of the virtual in the real~\cite{wiki:haptic-technology}.
Any such tracking and force feedback garnered a number of
research contributions such as~\cite{scene-pose-tracking-2007,gw2007-proceedings,advances-in-haptics-2010}
and many others. We subsequently briefly review two such devices that we use
in our research and creation: Microsoft {\kinect} (see \xs{sect:bg-kinect}) and
Novint {\falcon} (see \xs{sect:bg-falcon-haptics}).

\subsubsection{Microsoft {\kinect}}
\label{sect:bg-kinect}

Microsoft {\kinect} (see \xf{fig:kinect}) is a modern and affordable powerful sensor device
with multiple camera and audio inputs that allow exploring new ways
of media interaction accessible to a large audience base and provides
for nearly limitless ways of creativity for artists, a truly formidable
device. It goes beyond devices like {\wii} in such that the motion tracking does
not require the users holding other devices by providing robust
skeletal tracking with visible and depth data; it also can act as an audio source.

Combined with its API and SDK, besides its common use for motion tracking in gaming,
it was used for stereoscopy, security tracking, and fashion projection on people,
(see the Wikipedia page~\cite{wiki:kinect}).

Since the introduction of the device by Microsoft for XBOX360, a number of
APIs and drivers came into existence because Microsoft at the time hasn't released
its official SDK yet, being in Beta testing and only for Microsoft OS platforms.
(Microsoft have released their own 1.0 in February 2012 along with a somewhat more expensive
{\kinect} for Windows with the developers beta version from last summer and now it is in 1.5.x series.)
Open source versions from {\openkinect} and OpenNI by PrimeSense were also available earlier
to make the device accessible from other operating systems. For example,
{\openkinect}'s open-source library/API/driver are available for \win{}, \linux{}, and \macos{X}
(uses {\glut}, etc.)~\cite{kinect-openkinect}.

\begin{figure}[hptb!]%
	\centering
	\includegraphics[width=.5\columnwidth]{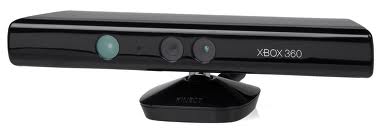}%
	\caption{Kinect Device}%
	\label{fig:kinect}%
\end{figure}

The most recent Microsoft SDK supports primarily {\cpp} and {\csharp} (and VB) and provides skeletal
tracking and viewing~\cite{kinect-ms-sdk} and many other
recent code samples and the toolking from \cite{kinect-ms-toolkit} to deal with
depth, audio, color data streams.
In \xs{sect:impl-kinect} we elaborated on the actual API used in a part
of this work from \cite{kinect-ms-sdk} and primarily in {\csharp}.
The SDK provides Kinect Studio for depth debugging~\cite{kinect-ms-studio}.
Additionally, their documentation provides Human Interface Guidelines
and design principles for skeleton tracking, depth processing, speechs
processing and the like~\cite{kinect-ms-hci-guidelines}.

\subsubsection{Novint {\falcon}}
\label{sect:bg-falcon-haptics}

The inexpensive haptic {\falcon} device (see \xf{fig:falcon}) is used in this research as another
means of tangible interaction. It has means of integration with {\opengl}/{\glut}
and others. Its main goals include~\cite{falcon-www,falcon-manual}:

\begin{itemize}
	\item Give the user a sense of touch by applying forces, vibrations, or motions to the user
	\item Used for medical research, art design and video games
	\item Novint's {\falcon}, the first inexpensive 3D touch device
	\item Three-dimensional degree of  force feedback allow the haptic simulation of objects, 
	textures, recoil, momentum, and the physical presence of objects in games
\end{itemize}

\begin{figure}[hptb!]%
	\centering
	\includegraphics[width=.3\columnwidth]{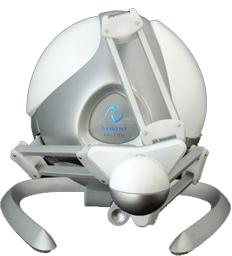}%
	\caption{Falcon Haptic Device~\cite{falcon-www}}%
	\label{fig:falcon}%
\end{figure}

{\falcon} comes with the SDK, drivers, examples, and integration with various Windows platforms
via their new community tool F-Gen and its SDK~\cite{falcon-sdk,novint-f-gen,novint-f-gen-sdk}.

\longpaper
{
\subsubsection{Mobile (iDevices)}

{\todo}
} % \longpaper

\subsection{Viewing and Projection}
\label{sect:bg-viewing-projection}

Viewing and projection techniques have to do with various technologies
capabilities that they provide to the users for eventual rendering and 
projection for performance, interaction, or other purposes.

\subsubsection{Depth}
\label{sect:bg-viewing-projection-depth}

The notion of depth is important in a number of aspects.
It is used in stereo viewing and image generation, as well
as 3D gesture/motion/posture recognition via either two
cameras observing the same target within interaxial distance
and/or with the infrared mapping. Devices such as {\kinect}
provide \emph{depth frames} alongside the color image readings
under a specified resolution and consisting of 16-bit values of
the depth sensor readings within its field of view. These frames
can later be used for green screening, skeleton tracking, and other applications.
Since depth frames have no natural color assigned to them, to be
visualized, the depth pixels (\emph{dexels}\footnote{\url{http://en.wikipedia.org/wiki/Dexel}}) have to be mapped somehow to
the visual domain pixels first. A depth mask can help, e.g., with the
green screen effect.

\subsubsection{Green Screen}
\label{sect:bg-viewing-projection-green-screen}

The notion of green screen\footnote{\url{http://en.wikipedia.org/wiki/Green_screen}}
more formally known as chroma key screen, is the idea of having a color screen
present behind a host or a performer, a color that is unlikely to be present on the
host or a performer, can be replaced during production with static or moving imagery
as if it were a true background originally when viewed or projected. The notion
of depth allows us with the devices such as {\kinect} to implement such a green
screen like effect dynamically in real-time~\cite{kinect-ms-sdk} and use it in
a real-time installation and performance. Here the depth frames would serve
in part as the ``green screen''.

\subsubsection{Skeleton Tracking}
\label{sect:bg-viewing-projection-skeleton-tracking}

Not strictly viewing and projection but in the context of {\kinect} it helps
with green screen abstraction, depth, and motion tracking making in the viewing of 
avatars, skeletons, or green-screened video feed possible in real-time.
Skeletons can also be visualized, of course. Tracking skeleton and joints allows
for interaction of the virtual character as well as filter out non-player pixel
data to help with the reduction of the depth/color data/mask processing.
{\kinect} allows about 2 skeletons but 6 depth-frame ``players'' to be tracked
that way.

\longpaper{{\todo}}

\subsubsection{Stereoscopic}
\label{sect:bg-viewing-projection-stereo}

Stereoscopic viewing and project has to do here either with traditional anaglyph (red-blue)
glasses or a $180^\circ$ projection screen (see \xs{sect:bg-vr-and-medical-research}) as
opposed to a 3D graphics rendered on a 2D monitor screen or a traditional projection.

\longpaper{{\todo}}

\subsection{APIs}
\label{sect:bg-api}

We briefly review various APIs used in this work.

\subsubsection{{\opengl}}
\label{sect:bg-opengl}

{\opengl}\footnote{\url{http://en.wikipedia.org/wiki/OpenGL}}\cite{opengl,opengl-redbook-ed7}
is an industry standard API and a library for cross-platform graphical processing, rendering,
animation, {\gpu} programming, visualization, game development, and the like (starting with version 1.0 in 1992
and the most recent is 4.3 in 2012, including support for OpenGL ES 3.0 for mobile devices running
iOS and Android). It is used on desktops
and mobile devices by various vendors and has bindings to many programming languages.
It implements a programmable graphical processing pipeline enabling various CG algorithms
and techniques. The pipeline works to transform initial 3D space virtual world objects
(their geometry, lighting and texture information, etc.) to rasterized pixels rendered into the final
image on the computer screen. There are a number of books, tutorials, rendering engines,
and publications that deal with {\opengl}~\cite{opengl,%
opengl-redbook-ed7,opengl-redbook-ed8,%
realtimerendering2002,%
cg-with-opengl-4ed,%
cg-thru-opengl-theory-experiments-2010,%
openglExtensions,%
opengl-codecolony}.
There are a number of common extensions to {\opengl} that work together with such
as {\glut}~\cite{opengl-glut-examples}, {\glui}~\cite{gluiManual}, and others.

\longpaper
{
\paragraph*{OpenGL Pipeline Stages.}

We briefly review the OpenGL pipeline stages as a precursor
to discussion of the GPU and shaders, which are popular
and important technologies and techniques in the advanced
rendering and animation as they extend programmable portions
of the OpenGL stages. The high-level overview of
the stages is shown in \xf{fig:opengl-pipeline}.

\begin{figure*}[htp!]
	\centering
	\includegraphics[width=\textwidth]{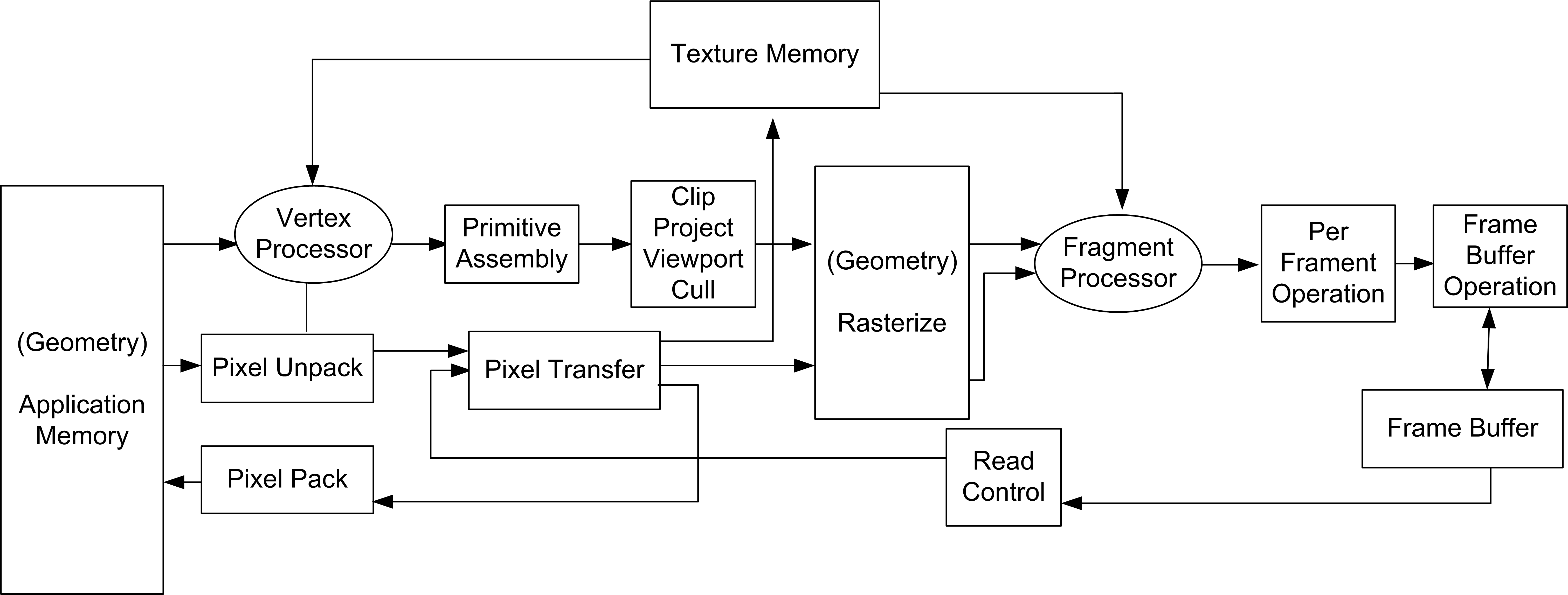}
	\caption{OpenGL 2.0 Logical Diagram}
	\label{fig:opengl-logical-diagram}
	\label{fig:opengl-pipeline}
\end{figure*}

\noindent
There are two major ways of getting better performance: \emph{pipelining} and \emph{parallelization}.
Pipelining refers to the evolution of graphics hardware which has started reverse the pipeline, 
for example rasterizer was put into the hardware first, and then the geometry stage.
Moreover, optimization of the pipeline is also very important. 

The OpenGL pipeline is the state machine ``engine'' that creates images from the geometric 3D scenes.
There are three conceptual stages of the pipeline:
application (executed on the CPU), geometry, and rasterizer~\cite{realtimerendering2002}.
The programmer ``sends'' down geometric primitives to be rendered through the pipeline using
the appropriate OpenGL API calls at the right time within the state machine.

\paragraph*{Application Stage.}

The application stage is executed on the CPU, which means that the programmer
decides what happens, such as collision detection, speed-up optimization
techniques, and animation.
The most important task is to send rendering primitives
(e.g. triangles) to the graphics hardware.

\paragraph*{Geometry Stage.}

The geometry stage, which does the per-vertex operations,
is to do ``geometrical'' operations on the input data (e.g. triangles).
It allows to move objects (matrix multiplication), move the camera (again matrix multiplication),
compute lighting at vertices of triangles, project onto screen (3D to 2D),
apply clipping (to avoid processing triangles outside the screen),
and map to the containing window.

At the geometry stage one processes the projection aspect as a necessary step in
conversion between 3D to a flat screen of two dimensions.
Two major ways of geometry projection are:

\begin{enumerate}
\item
Orthogonal, which useful only in a very few applications.

\item
Perspective, which is most often used to achieve better realism
and believability by mimicking how humans perceive the world, i.e., objects'
apparent size decreases with distance.
\end{enumerate}

We first get a $x,y$ square after the projection.
Then we clip the primitives to that square.
We then perform screen mapping, scale and translate the square such 
that it ends up in a rendering window.
These ``screen space coordinates'' together
with the $z$ (depth) dimension are sent to the rasterizer stage
for further processing.

One way of optimizing performance of the
geometry stage is by using triangle strips
and meshes.

\paragraph*{Rasterizer Stage.}

The rasterizer stage, which does per-pixel operations, takes the output
from the geometry stage and turns it into the actual visible pixels on
the screen.
It also applies textures and various other per-pixel operations.
The pixel visibility is also resolved here: it sorts the
primitives in the $z$-direction in order to decide, which
pixel to keep.
} % \longpaper

\subsubsection{Graphics Processing Unit (GPU) and Shaders}
\label{sect:bg-gpu-shaders}

Since around the year 2000, CG has placed more emphasis on real-time photo-realism.
Gaming and 3D cinema are the driving forces to such development of 3D acceleration hardware performance.
GPUs are seen as co-processors of the main CPU
and general-purpose computing applications are being developed
to use GPUs (so-called \emph{General Purpose GPU Computing}---GPGPU).
With the increasing power of mainstream programmable GPUs made the SGI type of graphics less pertinent.
NVIDIA and ATI are presently dominating the consumer graphics cards with GPUs even in commodity hardware.
They are also major contributors to the Vertex and Fragment
extension programs' specifications alongside NVIDIA and
Microsoft~\cite{vertex_program,fragment_program,openglExtensions}.
With the GPUs a lot of new features have been appearing very fast
such as programmable pipelines, floating-point support,
and hardware occlusion support~\cite{3dgames-watt-policarpo-01,AlanFabio03,fly3d,mudur-comp7661}.
Especially, in 2002 the video game industry throughput
surpassed the film industry.

\longpaper
{

\paragraph*{Advantages.}

\begin{enumerate}
\item Make the inexpensive power of the GPU available to developers as a sort of computational co-processor. The example can be conventional computational science and in-game physics simulation.
\item The wide availability of the GPUs has generated an explosion of innovative algorithms suitable for rendering complex virtual worlds at interactive rates~\cite{cs5243}.
\item Parallel linear algebra programming and computation are extremely easy and fast.
\item Easier to use with high-level shading languages like {\glsl}.
\end{enumerate}

\paragraph*{Disadvantages.}

\begin{enumerate}
\item For the most part GPUs designed for and driven by video games.
\item Difficult to use in the graphics card assembly language.
\item Programming model is a bit unusual.
\item Programming paradigms tied specifically to CG.
\item Programming environment is tightly constrained
(underlying architectures are inherently parallel, but not the programming models)
\item Rapidly evolving (even in basic feature set grows fast).
\item Largely secret by the manufacturers.
\item Can't simply ``port'' a CPU-bound code (GPUs are fast because they are specialized),
have to rewrite the program for GPU from scratch.
\item Poorly suited to sequential, ``pointer-chasing'' code.
\end{enumerate}

\paragraph*{Programmable Shading.}

Programmable shading has become a hot topic in the few recent years.
Vertex shaders and fragment (aka pixel) shaders make it become possible
to have a lot more control and much more possibilities for
the programmer~\cite{mudur-comp7661}.

When comparing the OpenGL pipeline and GPU's graphics pipeline solutions
we notice vertex processor does all transforms and lighting computation
and the fragment shader decides the final illumination of a pixel in
fragments.
Additionally, the pipe widths vary: intra-GPU pipes are usually wider
than the CPU-to-GPU pipes.
OpenGL does not natively allow program the vertex and fragment shaders,
but with the programmable GPU we now can program them achieving
greater flexibility.
} % \longpaper

\longpaper
{
\paragraph*{OpenGL Shading Language.}
\label{sect:opengl-shading-language}

This is a very brief review of the OpenGL Shading Language (GLSL)
primarily based on the book by Randi J. Rost of the same title~\cite{rost2004}, also now known as the ``Orange Book'',
as well as a report and slides and online tutorial
materials~\cite{glsl-clockworkcoders,opengl-codecolony,glsl-lighthouse3d,glsl-typhoonlabs,opengl-glsl-quick-guide}
by various vendors and contributors.
Rost works for 3D Labs, Fort Collins, Colorado and is a part of
OpenGL 2.0 driving force. He was
previously employed by the Hewlett Packard's graphics software lab as well as
Kubota Graphics Corp. Rost did various workshops, presentations,
tutorials in numerous computer graphics-related conferences, such as
SIGGRAPH, Eurographics, and Game Dev. SIGGRAPH 2004 was the one
when GLSL and the ``Orange Book''~\cite{rost2004} were introduced.
OpenGL Shading Language was approved as OpenGL Architecture Review Board (ARB) extensions in
June 2003 and is a part of the OpenGL 2.0+ core.
Major reason for GLSL to exist--with the invent of GPUs the goal is allow
easier programmability in some areas up to now static in OpenGL--vertex
and fragment processing, which traditionally required
the GPU programmers to write the shaders in the low-level GPU assembly code~\cite{mokhovOpenGL}.
} % \longpaper

\longpaper
{
Vertex and fragment processing stages in the pipelines are responsible
for a variety of tasks.
Vertex processing involves computation of things like transformation and
lighting at each vertex at the logical geometry level.
Fragment processing has to do with per pixel data structures created by
rasterization of the graphics primitives. There are usually several
fragments per pixel.
OpenGL has those areas implemented as fixed functionality that
cannot change. GLSL augments the OpenGL
pipeline and makes these two stages programmable such that if a GPU or even
multiple CPUs are available, the programmers can make their graphics applications
much more flexible and achieve some realistic real-time effects in various areas.

GLSL's code is meant to be executed on the OpenGL
programmable processors and is referred to as a
\emph{shader}. Shaders can be written in any shading
language, including the graphics card assembly,
RenderMan, Cg, and others.
OpenGL defines two programmable processors;
hence, there are two types of shaders: vertex and fragment.

\paragraph*{The Need for Shaders.}

If a graphical application works with vanilla OpenGL,
and all its standard features satisfy
a given application, then there is no really any need for
shaders and nothing has to change. However, for
some applications OpenGL's features do not allow the flexibility
that today's graphics hardware provides access to
with the real-time interaction capabilities.
Traditional OpenGL pipeline has serious limitations (static) or does not allow
features like area lights, light calculation per vertex rather than per
fragment and a number of other aspects.
This is where the need for shaders comes into play.
The below become possible:

\begin{itemize}
\item
Increasingly realistic materials (metals, stone, wood, paints, etc.) and
lighting effects (area lights, soft shadows),
\item
Natural phenomena (fire, smoke, water, clouds),
\item
Non-photorealistic materials (painterly effects, pen-and-ink
drawings, simulations of illustration techniques),
\item
New uses of the texture memory (textures can also store
normals, gloss values, polynomial coefficients),
\item
Fewer texture accesses (procedural creation vs. texture maps),
\item
Image processing (convolution, unsharp masking, complex
blending),
\item
Animation effects (keyframe interpolation, particle systems,
procedurally defined motion),
\item
Programmable antialiasing methods.
\end{itemize}

GLSL addresses this by providing API hooks to the pipeline that can be
used to efficiently utilize the underlying graphics hardware in any
way an application programmer wants. It adds support for more complex rendering techniques in
hardware that OpenGL was not designed for.

\paragraph*{OpenGL Programmable Processors.}

GLSL provides access to only two of them so far:
Vertex Processor and Fragment Processor.
OpenGL Pipeline with
Programmable Processors is presented as an extract
from the SIGGRAPH 2004 introduction to OpenGL Shading
Language slide in \xf{fig:opengl-logical-diagram}.

\paragraph*{Vertex Processor.}

Vertex Processor, among other things, is usually responsible for the following:

\begin{itemize}
\item
Vertex transformation
\item
Normal transformation and normalization
\item
Texture coordinate generation and transformation
\item
Lighting
\item
Color material application
\end{itemize}

Vertex Processor I/O figure is present in \xf{fig:vertex-processor-overview}.

\begin{figure}[htp!]
	\centering
	\includegraphics[width=.8\textwidth]{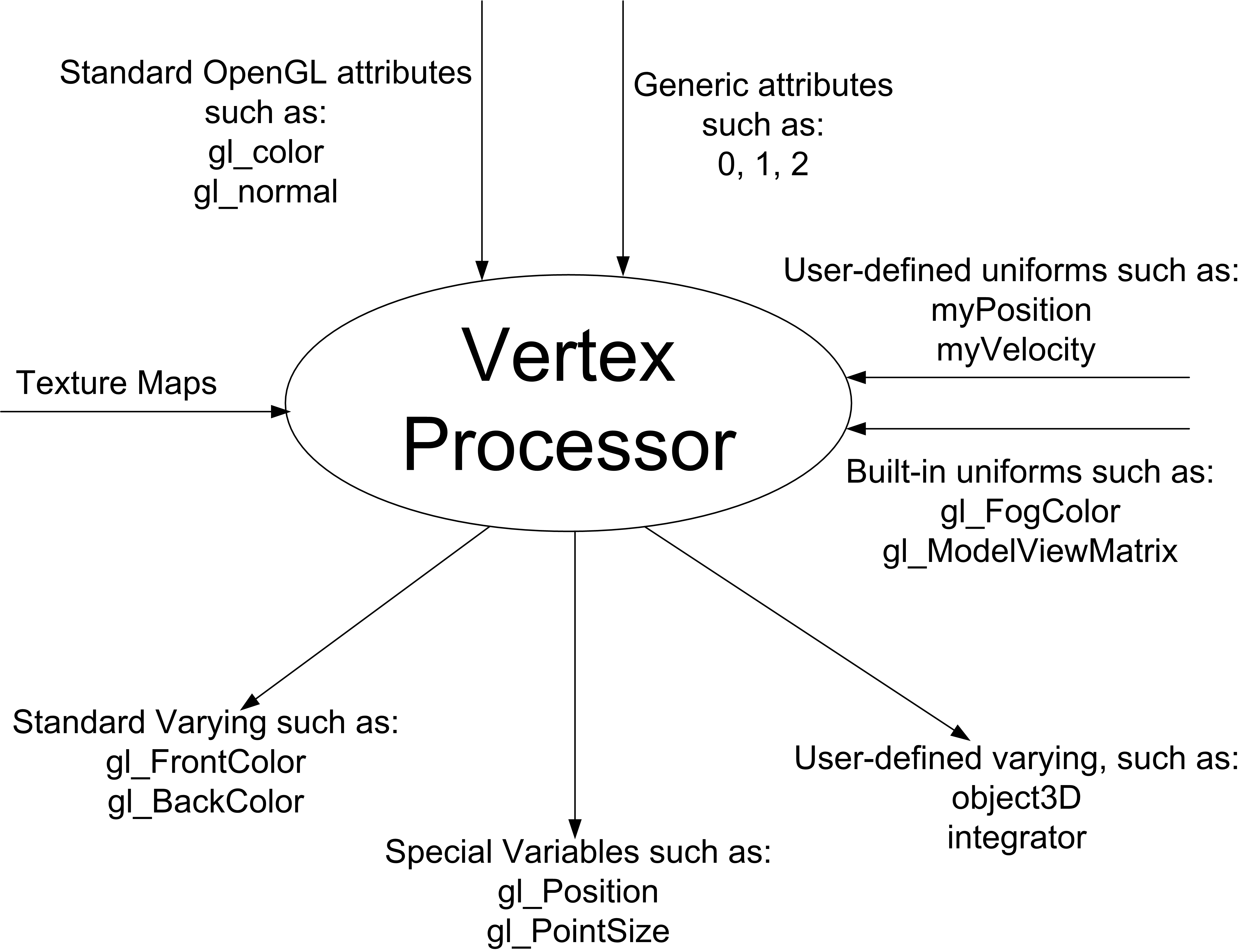}
	\caption{Vertex Processor Overview}
	\label{fig:vertex-processor-overview}
\end{figure}

\paragraph*{Fragment Processor.}

The fragment processor is primarily responsible for:

\begin{itemize}
\item
Operations on interpolated values
\item
Texture access and application (pixel zoom, convolution, color matrix,
scale and bias)
\item
Fog
\item
Color sum
\item
Operations on rasterized points, lines, pixel
rectangles and bitmaps
\end{itemize}

Advantages of a fragment processor typically include:

\begin{itemize}
\item
Access texture memory arbitrary number of times an
process the values read in arbitrary number of ways.
\item
Ray-casting can be implemented in FP.
\item
Custom filtering operations.
\end{itemize}

Fragment Processor I/O figure is in~\xf{fig:fragment-processor-overview}.

\begin{figure}[htp!]
	\centering
	\includegraphics[width=.8\textwidth]{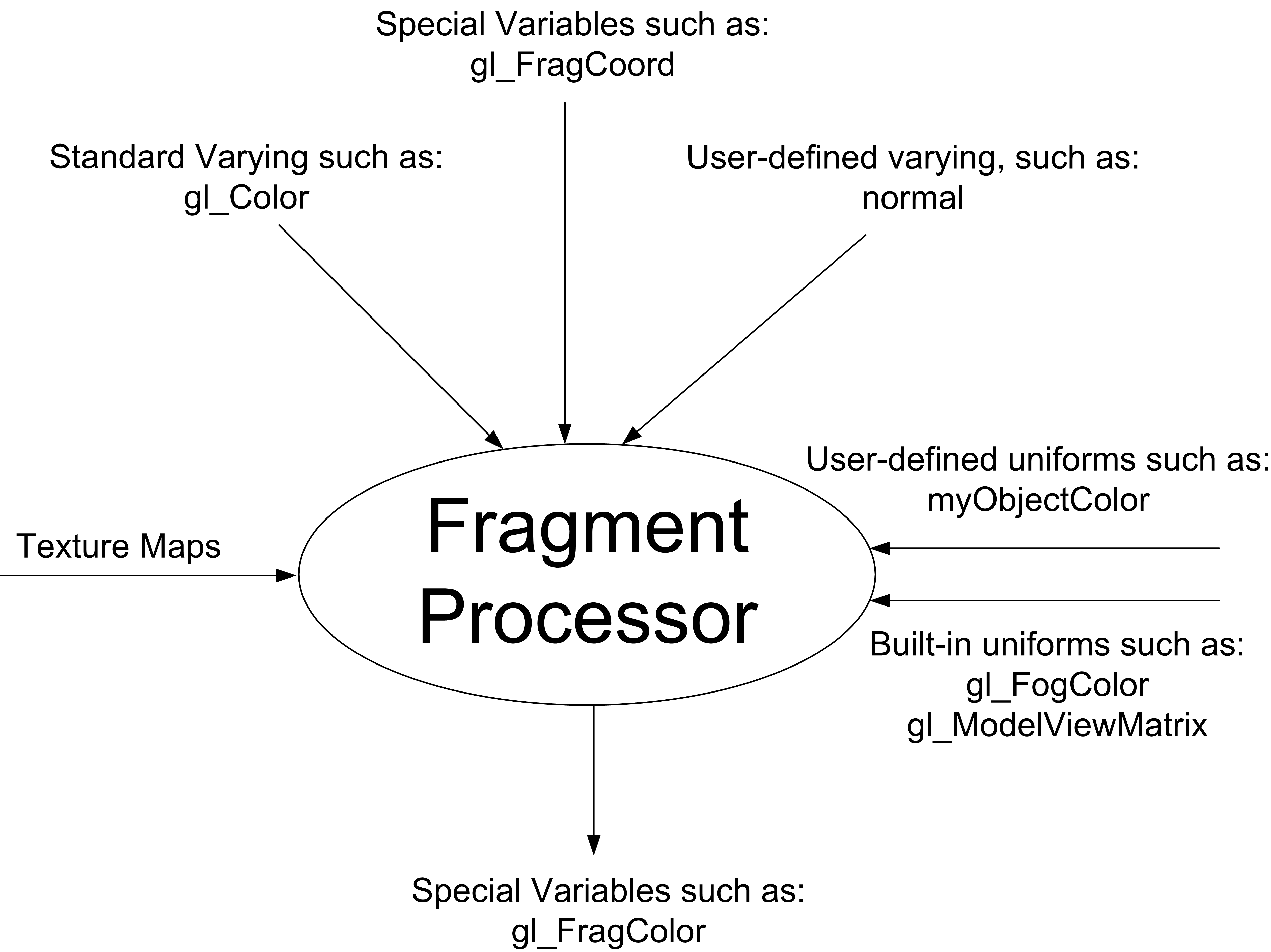}
	\caption{Fragment Processor Overview}
	\label{fig:fragment-processor-overview}
\end{figure}

\paragraph*{Limitations.}

Both or one of the processors may be responsible for some of the listed features,
but in OpenGL the most of them remain as a
fixed functionality that cannot be changed directly by the
shaders (only indirectly). The functionality involving
multiple vertices and Vertex Processor is not replaced,
such as
perspective divide,
primitive assembly,
clipping,
viewport mapping,
backface culling,
two-sided lighting selection,
some polygon operations,
selection of flat or smooth shading,
depth range.

For the Fragment Processor that does not replace fixed functionality of OpenGL
for the features like
pixel ownership test,
scissor,
stipple,
alpha test,
depth test,
alpha blending,
logical operations,
dithering, and
plane masking.

\paragraph*{Parallelism.}

It is possible to have more than one Vertex
and Fragment processors to process the data
in parallel on multiple hardware resources available,
so GLSL design planned for the
technology that becomes increasingly parallel.
The hardware resources may include multiple CPUs coupled
with one or more GPUs and replicated logic within
the GPUs themselves.

\paragraph*{Origins of OpenGL Shading Language.}

OpenGL Shading Language is a high-level procedural language that:

\begin{itemize}
\item
has roots in C,
\item
provides function overloading and variable declaration
as needed from C++,
\item
provides reach set of types (includes vector and matrix
types) and stricter types than that of C/C++,
\item
has standard loops, subroutine calls, conditionals that
follow C/C++,
\item
as features similar to RenderMan and others,
\item
uses type qualifiers to manage I/O (between application and the processors) instead of read/write.
\end{itemize}

\paragraph*{Language Design Considerations.}

\begin{itemize}
\item
Define a language that works well with OpenGL.

\item
Expose the flexibility of the today's hardware (GPUs).

\item
Provide hardware-independence (high-level).

\item
Performance is important (compiler).

\item
Define a language that is easy to use.

\item
Define a language that will stand the test of time (success
of C and RenderMan, ARB approval).

\item
Don't preclude higher levels of parallel processing.

\item
Ease of implementation (no pointers, there is value-return, strict types, hence compiler optimizations are much easier
than of C).
\end{itemize}

\paragraph*{Key Benefits.}

\begin{itemize}
\item
Tight integration with OpenGL.
\item
Runtime compilation.
\item
No reliance on cross-vendor assembly language.
\item
Lots of opportunities for compiler optimizations on wide range of hardware.
\item
Open, cross-platform standard.
\item
One HLL for all graphics processing (vertex and fragments).
\item
Modular programming support.
\item
No additional libraries or executables -- the compiler of GLSL is provided by OpenGL.
\end{itemize}

\paragraph*{Summary.}

Very brief summary of language features is presented below:

\begin{itemize}
\item
Vectors and matrices (of floats, ints, and bools) are included as basic types.
\item
Type qualifiers \texttt{attribute}, \texttt{uniform}, and \texttt{varying} are added to describe variables to manage shader I/O.
	\begin{itemize}
	\item
	\texttt{attribute} -- frequently changing values from an application to the vertex shader
	\item
	\texttt{varying} -- output from vertex shader, input to fragment shader,
	\item
	\texttt{uniform} -- application specifies infrequently changing values to both shader types.
	\end{itemize}
\item
Type sampler added to access textures.
\item
Access to OpenGL state through built-in variables.
\end{itemize}

\paragraph*{Summary}

The OpenGL Shading Language is generally a very programmer-friendly way to implement
shaders in the {\opengl} environment. Due to its ``high-levelness''
it's mostly platform-independent yet provides efficient access
to the underlying hardware (a subject to a good compiler backend,
of course).
Please refer to 3DLabs' website for GLSL demo, documentation,
and shaders source code~\cite{3dlabs1,3dlabs2} as well as
various GLSL
tutorials~\cite{glsl-clockworkcoders,opengl-codecolony,glsl-lighthouse3d,glsl-typhoonlabs,opengl-glsl-quick-guide}.
} % \longpaper

\subsubsection{{\xna} and {\hlsl}}
\label{sect:bg-xna-hlsl}

{\xna}~\footnote{\url{http://en.wikipedia.org/wiki/Microsoft_XNA}} is a Microsoft framework and toolset originally designed for XBOX and
simplified game development. Later it was generalized to Windows and Windows Phone
with the latest version being
4.0~\cite{xna-novice-pro-2009,xna-studio-express-dvd-2007,reed-learning-xna-3-2009,nitschke-xna-game-2007,carter-xna-2009}.
It runs on top of a lower-level API of {\directx}
(a Microsoft proprietary equivalent of {\opengl})
and uses primarily {\csharp} bindings to all the related libraries
for the manged code requirements by the XBOX and Phone that
are easier to program with {\csharp} than {\cpp}.
We adopt {\xna} because a large number of examples with easier programmability
and rapid prototyping for originally came out for {\kinect} in {\csharp}.
{\xna} also has a lot of internal support for media content built-in
and shader support via Microsoft's equivalent of {\glsl} described earlier---{\hlsl}---the 
High-Level Shader Language~\footnote{\url{http://en.wikipedia.org/wiki/HLSL}} that provides us with various GPU programming
effects we use in the installation.

\longpaper
{
Fancy bubble \cite{fancy-bubble-sample-xna3} adapted from XNA3 to XNA4 and
used \api{RenderTargetCube} to render on the bubble various effects.
{\todo}
}

\subsubsection{{\kinect} SDK}
\label{sect:bg-api-kinect-sdk}

{\kinect} SDK provides {\csharp} (and since version 1.5, {\cpp} and VB) bindings
to its API~\cite{kinect-ms-sdk,kinect-ms-toolkit} nestled under the
\apipackage{Microsoft.Kinect}~\cite{microsoft-kinect-namespace},
which provides all the classes, device status, access to color, depth,
skeleton, and audio streams from the {\kinect} as well as access to its
tilting motor that {\xna} applications can then process and use for
advanced mode of interaction and artistic performance.

\longpaper{{\todo}}

\subsubsection{BASS Audio Library}
\label{sect:bg-api-bass}

{\bassaudiolib} for audio processing for visualization is a fairly popular
tool~\cite{bass-audio-library-dot-net,bass-audio-library} with bindings
to several languages and frameworks, including a .NET version for {\csharp}
and {\xna} with the visualization--\api{TalkShowHostXNA}~\cite{talkshowhost-xna}.
It loads an audio file and uses signal processing techniques to extract
spectral characteristics of a tune, such as beats and their frequencies
so the animation can run to the beat of the music.

\longpaper
{
\subsubsection{{\maxmsp}/{\jitter}}

\cite{maxmsp,jitter}

{\todo}
} % \longpaper

\subsection{Code Samples}
\label{sect:bg-code-samples}

The majority of the prototyping work in this thesis relies on the great
examples posted online of single or multiple effects or the use of APIs
under permissive or open-source licenses. The author Song expresses her
gratitude to the creators of the examples and providing their source code 
to learn from and re-use enabling rapid prototyping of the artistic work
presented in this thesis. Specifically, we credit the following works
that inspired, or got re-used and enhanced during this thesis, mostly
in chronological order. These works are also cited inline where appropriate.
You could find them in the reference here~\cite{nehe-lesson-35-avi,
opengl-glut-examples,xna-performance-sample,kinect-fundamentals-skeleton-tracking,
kinect-ms-sdk,kinect-toolbox,kinect-gesture-service,kinect-gestures-with-sdk,kinect-ms-shape-game,
fancy-bubble-sample-xna3,talkshowhost-xna,falcon-manual}.

\longpaper
{
\scriptsize
\begin{itemize}
	\item 
\bibentry{nehe-lesson-35-avi}

	\item 
\bibentry{opengl-glut-examples}

	\item 
\bibentry{xna-performance-sample}

	\item 
\bibentry{kinect-fundamentals-skeleton-tracking}

	\item 
\bibentry{kinect-ms-sdk}

	\item 
\bibentry{kinect-toolbox}

	\item 
\bibentry{kinect-gesture-service}

	\item 
\bibentry{kinect-gestures-with-sdk}

	\item 
\bibentry{kinect-ms-shape-game}

	\item 
\bibentry{fancy-bubble-sample-xna3}

	\item
{\xna} \api{VideoPlayback} 4.0 sample

	\item 
BASS.NET visualization example: \api{TalkShowHostXNA}
\bibentry{talkshowhost-xna}

	\item
Novint Falcon OpenGL code sample,
\bibentry{falcon-manual}
\end{itemize}
\normalsize
} % \longpaper

\section{Virtual and Augmented Reality for Medical Research and Beyond}
\label{sect:bg-vr-ar-medical-research}

This section reviews an example of the
combined multidisciplinary research experience
and collaboration~\cite{pain-performance-vr-pmmIII,stereo-plugin-interface}
that is of relevance and has given rise to some
ideas and inspiration in the follow up work.
It is a brief research experience report as to work with, experiment, and
develop computer graphics environments for a real-world VR
application.

My first encounter with VR was when doing a research survey job in mixed reality, 
which refers to the merging of real and virtual worlds to produce new
environments~\cite{context-controlled-flow-in-ar,cost-refs-ar-gi08,arpracticalguide2008,augmented-reality-2010}.
Then, I got the opportunity to work
as an operator of the Virtual Reality {\caren}/{\motek} equipment at the
\emph{Pain, Mind and Movement} research laboratories at the
Constance-Lethbridge Rehabilitation Center, directed by Simmonds
during
2008---2010, and to
use virtual
tools to better
understand and manage pain and movement difficulties.
The research
work for students included assisting them with Virtual Reality Research.
The major equipment included a virtual reality suite interfaced with a
self-paced instrumented treadmill on top of a motion platform with a
motion capture system.
It was inspiring to see {\simmonds}
design experiments and use the state of the art 
VR equipment in innovative research to disentangle the mental, physical, 
and social components of pain and its impact~\cite{pain-performance-vr-pmmIII}.
{\simmonds}'s clinical research greatly benefited the clinical community in the 
rehabilitation center and elsewhere.

The domain of the research in Computer Graphics is of direct relevance to, and
plays a very important role in, virtual and augmented
reality~\cite{augmented-reality-2010} research, allowing users to interact with the
computer-generated virtual environment at elevated sensory levels.
Associated gaming, motion capture, stereoscopic effects etc., used in the VR lab
are also the very related and relevant subjects in Computer Graphics and the
new interactive media.

\subsection{Overview}

We have done a \emph{Pain and Performance in Virtual Reality Environments: 
A Pilot Feasibility Study}~\cite{pain-performance-vr-pmmIII} alongside
with the prototype development that included
the {\caren} VR system,
the split-belt instrumented self-paced treadmill,
the large panoramic screen being used in rehabilitation research to test
how experimental pain threshold differed across VR environments with the
overall purpose of decreasing pain and improving movement.

Of relevance to this work we
extrapolate from some of our experience in
and propose further possibilities of 3D (stereo and non-stereo) computer graphics techniques
on perception of pain and rehabilitation in a virtual reality (VR) system setting and transposing
this into a preliminary discussion of going beyond the medical research into the realm of
interactive documentary production.
\longpaper
{
We research and report
on the equipment, techniques'
their usability, their applications, and the corresponding tools for VR development, such as that of treadmill
or a head-mounted display (HMD), as well as
the related
physical based softbody simulation and even interactive documentary
production aspects.
} % \longpaper

\subsection{Related Work}
\label{sect:bg-vr-cg-cinema-related-work}

We subsequently outline the research and
research
work done in various degrees of relevance to the virtual
reality, computer graphics, medical and cinematic documentary
research.\longpaper{ including
on elastic softbody
simulation systems.
In \xs{sect:vr-and-medical-research}
we relate the work to the medical research.
In \xs{sect:vr-and-cinema} we describe
a brief relationship with
the documentary production.
} % \longpaper
A list of the core sources or works considered includes the below with the rest
cited~\cite{what-is-real-in-vr,vr-hmds,sensics,merged-worlds}.

\paragraph*{VR and Medical Research.}
\label{sect:vr-and-medical-research}
\label{sect:bg-vr-and-medical-research}

In this body of work,
researchers did scientific and clinical studies with an expensive
head-mounted display for stereoscopic effects to demonstrate
the impact of virtual reality environments on pain reduction; more specifically
that manipulation of optic flow speed in
such a display was consistently and continuously influencing the
speed and locomotion of individuals and that it can improve walking.
They then interfaced a VR environment with a walking treadmill, which
further expanded to link the
perception of lower body chronic pain while walking to psychological
perception rather than physical condition and suggested the VR
trials for pain reduction do work and are recommended.
Currently such VR applications are used
by {\simmonds} and her colleagues to reduce the perception of pain and
improve active movement~\cite{%
stereo-display-speed-locomotion-2006,%
integrated-pain-mind-movement-2008,%
treadmill-vr-2009,%
vr-evidence-for-pain-reduction-2009,%
visual-flow-treadmill-walking-2007,%
optimizing-rehab-treadmil-vr-07,%
vr-effects-pain-post-stroke-2008,%
vr-effects-pain-threshold-2008,%
treadmill-vr-vs-overground-2009,%
vr-effectivness-pain-reduction-2009}.

The VR-treadmill interface used at the time,
while offered a $180^\circ$ projection screen, was not
stereoscopic~\cite{vr-medical-research-docu,stereo-plugin-interface,vr-poster-ecsga10}.
Stereoscopic virtual reality (VR) environments for medical and rehabilitation research
are some of the applications where our own \tool{stereo3d} plug-in~\cite{stereo-plugin-interface}
could be used.
It has a potential for such a use as a cheaper alternative to the expensive
head-mounted display units.
It is feasible that provision of a more affordable stereoscopic
solution would be effective and have clinical and research utility by requiring overall less resources
(hardware, software, etc.) to setup and use. This would certainly increase its
availability to many in-clinical and research
environments~\cite{vr-medical-research-docu,stereo-plugin-interface,vr-poster-ecsga10}.

\longpaper
{
We, therefore, would have liked to
investigate if we could help in that research by providing a more
affordable and cheaper stereoscopic solution and test its effectiveness
by requiring overall less resources (hardware, software, etc.)
to setup and use thereby increasing its availability to
masses~\cite{vr-medical-research-docu,stereo-plugin-interface,vr-poster-ecsga10}.
} % \longpaper

\begin{figure*}[ht]
\hrule\vskip4pt
\begin{center}
	\subfigure[Patient and VR System's Boat Demo]
	{\includegraphics[width=.49\columnwidth]{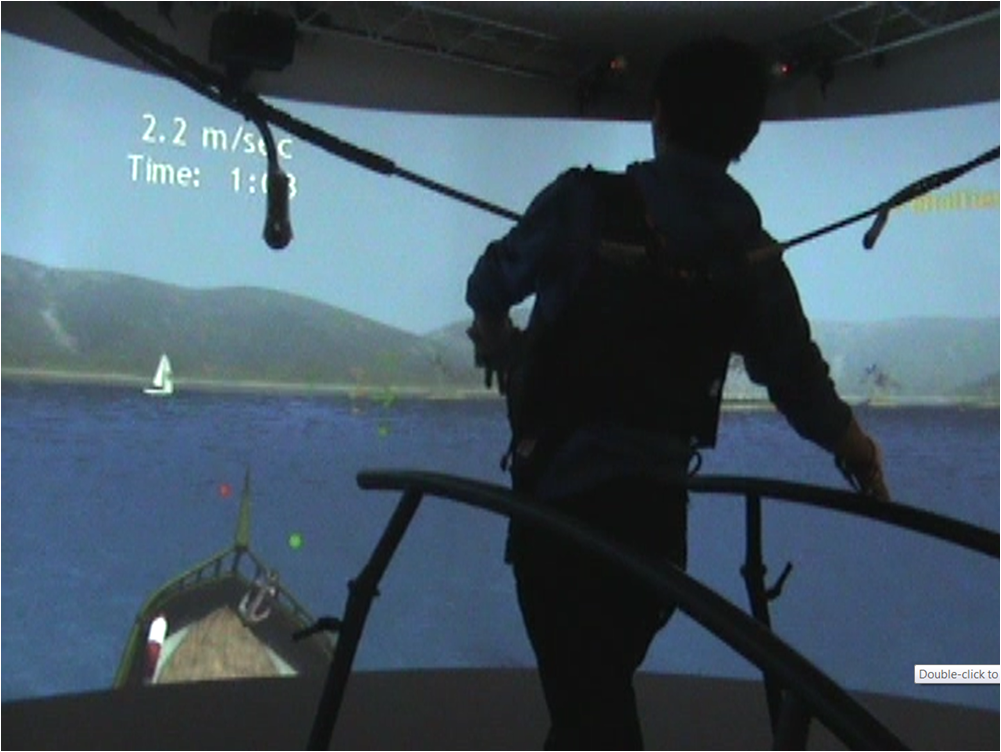}
	 \label{fig:song-liu-boat}}%
	\subfigure[Patient and VR System's Tropic Pathway Demo]
	{\includegraphics[width=.49\columnwidth]{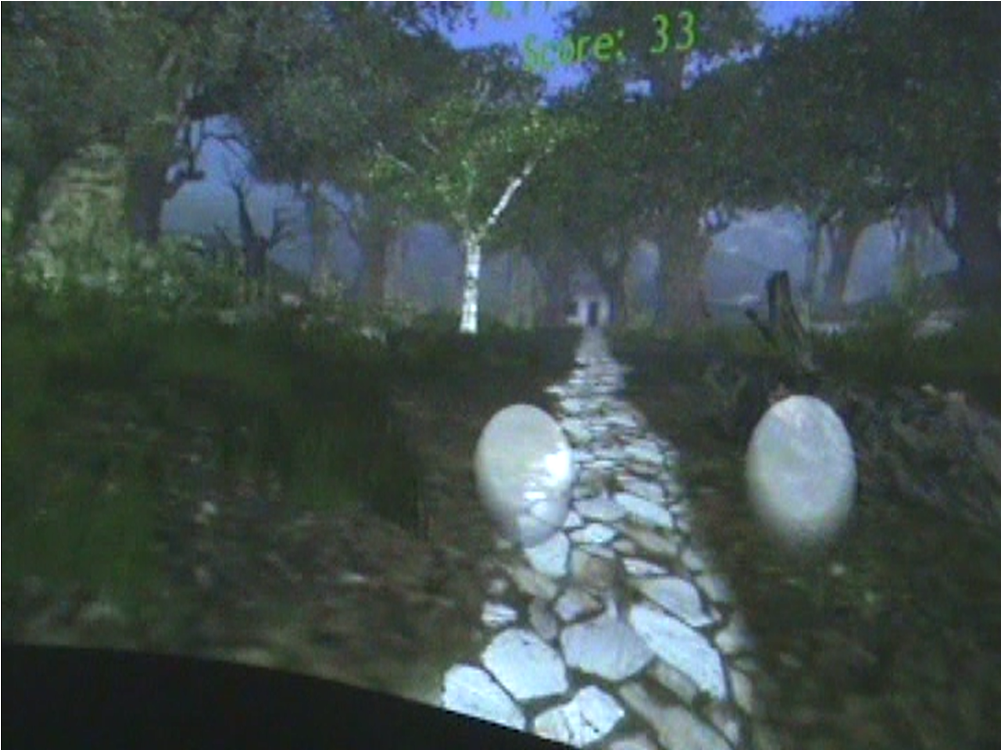}
	 \label{fig:song-liu-tropic-pathway}}
	\subfigure[Patient and VR System's Endless Road Demo]
	{\includegraphics[width=.49\columnwidth]{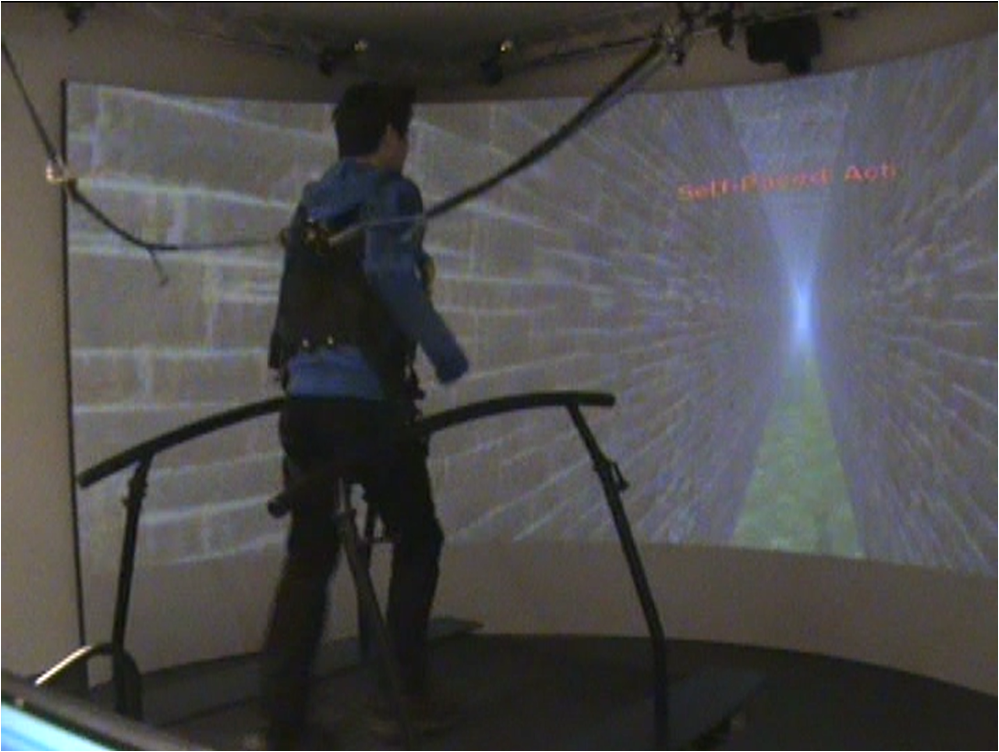}
	 \label{fig:song-liu-walls}}%
	\subfigure[Patient and VR System's Bridge Demo]
	{\includegraphics[width=.49\columnwidth]{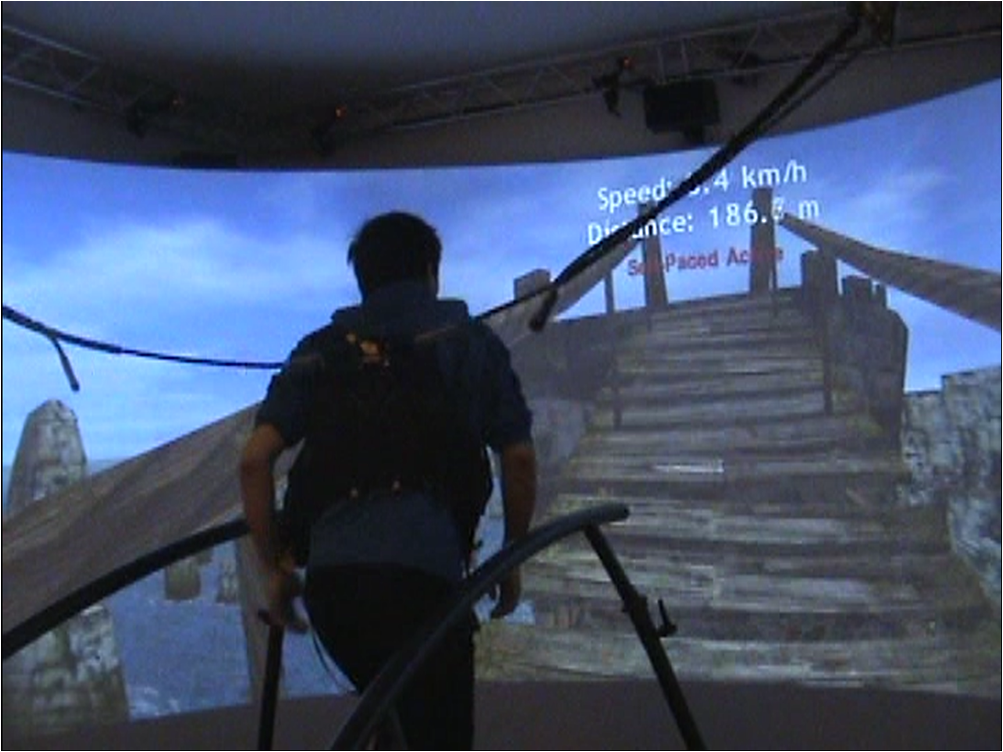}
	 \label{fig:song-liu-bridge}}
\caption{Various {\caren} Demo Examples}
\label{fig:song-liu-vr}
\end{center}
\hrule\vskip4pt
\end{figure*}

\longpaper{IGNORED: \xf{fig:song-liu-vr}}

\paragraph*{Implementation.}
\label{sect:bg-vr-implementation}

We successfully implement a VR virtual walk, \worktitle{Endless Road}{Simulation}, including
a successful implementation of the self-paced functionality with other
applications~\cite{vr-medical-research-docu}.

\longpaper
{

We began learning and exploring {\vizard} for HMD and started a Python-based
softbody simulation {\python}~\cite{core-python-prog-2007} plug-in in order to render it in the HMD.
\cite{vr-medical-research-docu}

\paragraph*{Tools and Equipment.}
\label{sect:tools}

At the beginning some main activities in the VR lab were training and
maintenance and demonstration of the system as
well as troubleshooting problems with it and
do development for it.
\cite{vr-medical-research-docu}

Motek

The majority of the equipment was provided by {\motek}, such
as the $180\circ$ projectors, motion markers, cameras, and the
treadmill all acting in unison and driven primarily by
the {\caren} system (see \xs{sect:caren}). We had technical
training from {\motek} with certification for lab technicians
to be able to operate, calibrate, and troubleshoot the lab
systems to some degree.
One of the problems was that for any patient or a healthy
person to get onto the treadmill required a relatively high
step up to climb. We ordered a small stairs set to remedy
the problem, but it posed another problem for the system
we did not anticipate----its metal handles were reflective
enough to act as pseudo markers that the system's cameras
erroneously picked up; upon this discovery of course we
covered the handles and were monitoring the patience do
not wear shiny reflective objects.
We also had to adjust level 2 camera's hight and calibrate
the entire system several times.
\cite{vr-medical-research-docu}
} % \longpaper

\paragraph*{{\caren} and Applications.}
\label{sect:caren}
\label{sect:bg-caren}

{\caren} is the core system behind the hardware and software
setup from {\motek} used in this research. It gathers readings from the cameras,
renders 3D graphics, has an API to build VR applications against,
and so on.
Our experience involves managing and operating the system,
its licensing, logging, wireless setup, maintenance of the manual and other resources,
training of, collaboration with and demonstration to various people and organizations
and more importantly patient examination with some initial pilot results~\cite{pain-performance-vr-pmmIII}.
\longpaper
{
This experience builds up with the initial and intermediate training from {\motek}
and follow up support related communication.
This our expertise covers versions 3.4.2 through 3.5.3 of {\caren}.
} %\longpaper
We made a simple application and test wireless controller with {\caren}
were troubleshooting bugs in several applications, such as Maze, random target;
testing self-paced modules, and its integration with {\vizard}~\cite{vizard}.
We also acquired Tracking Tools 2.0 beta 4 to be able to have the handle
on the tracking data as well as configure VR computer with multiple monitors
for optimized utilization and viewing and troubleshooting.
{\caren} software itself is based on the open-source 3D rendering engine
{\ogre}~\cite{ogre} and is written in {\cpp}~\cite{vr-medical-research-docu}.

\label{sect:vr-applications}

There are several applications that came with {\caren} for the {\motek} VR
system for the medical research and one developed by us. Example screenshots
of a patient navigating a virtual boat by leaning on the treadmill is in
\xf{fig:song-liu-boat}, walking a tropical pathway ``fighting off'' incoming
flying objects in \xf{fig:song-liu-tropic-pathway}, our own ``Endless Road'' in
\xf{fig:song-liu-walls}, and a high bridge walking is in \xf{fig:song-liu-bridge}.
Most applications require the patient on the treadmill
walking and leaning as well as arm motions with markers attached to perform
the research and capture the data~\cite{vr-medical-research-docu}.

Applications used as a demo, education activities for a University class,
the possible project in Virtual hospital,
and the patient testing. A pilot study was done to collect a baseline
for the healthy individuals prior to moving to the patients with
chronic pain~\cite{pain-performance-vr-pmmIII,vr-medical-research-docu}.

\longpaper
{

Maya, XSI, 3D Max

We have installed and configured various 3D modeling tools to
be able to import/export models between the tools and {\caren}.
We acquired {\maya}~\cite{maya} plug-ins in order to export 3D objects
to {\caren} or XSI software that {\caren} supported.
Then we tried the XSI~\cite{xsi} exporter with 3D Max~\cite{3dsmax}
file format. Part of this experience (e.g. the 3D model of the road) was used to develop the
a simple application ``Endless Road'' with force plate, texts, material properties, and expression module;
and further being investigated for other applications and uses.
The building process of such applications is critical to capture
and understand with systems so complex as {\caren} as to develop
custom applications for our needs in-house.
The supposedly newer version of {\caren} 3.6.3 would support 3D Max
and Maya directly.
}

\longpaper
{

Sensics, Vizard, and HMD

{\motek}'s setup is not the only piece of VR equipment and
software that comprise the lab. We also interacted with
{\sensics} for their head-mounted display (HMD) and
the accompanying software tools, SDK, API, etc.
We have received the training from {\sensics} of how
to operate the display and how project its view to either
the HMD (5 little embedded displays) or its equivalent
on computer monitors. As a result we were configuring
and debugging the HMD setup for pain and movement
research as well where the patients are fully immersed
into the virtual experience unlike in the treadmill.
Some of our own development effort and education took
place with the {\vizard} \cite{vizard} software and its plug-in system
in the Python programming language \cite{core-python-prog-2007}.
We acquired and configured higher-end graphics-oriented
two new laptops with the Maya~\cite{maya}, {\vizard}~\cite{vizard} software
for this development.
\cite{vr-medical-research-docu}

} % \longpaper

\subsection{Summary}
\label{sect:bg-vr-conclusion}

While covering a wide spectrum of activities, tools, and techniques,
by working at the {\simmonds}'s VR lab for pain and movement research,
expanded documentary production, VR and graphics techniques for
simulation, animation, modeling, and stereoscopy gave an unique
insight on the immediate future projects to undertake and to advance
the research%
~\cite{vr-medical-research-docu}.

In some relevance to the conducted research there are recent works%
~\cite{pain-performance-vr-pmmIII,stereo-plugin-interface,stereo3d,softbody-teaching-opengl-slides}
that we plan on investigating further and find a possibility of collaboration%
~\cite{vr-medical-research-docu}.

Finally, a similarly set up VR lab resources can also be used not only for medical
research, but also for documentary film and art extending the usefulness
and utilization of the expensive equipment and software in other
dimensions. Further, in \xs{sect:vr-and-cinema} we describe some
such plans of the audience immersion and interaction with and feedback from the
documentary film and computer graphics art~\cite{vr-medical-research-docu}.

\chapter{Toward Realtime Jellyfish Simulation in Computer Graphics}
\label{chapt:softbody-objects}
\label{chapt:softbody-jellyfish}

My research in Computer Graphics (CG) concentrates around physical based softbody simulation
because it is the continuation of the research and development originated in my master's studies~\cite{msong-mcthesis-2007}.
This work has also resulted in offshoot projects/applications that use the softbody simulation
library, specifically modeling and animation of a synthetic {\jellyfish} in 2D and 3D space.

The {\softbodysys} project itself has been further redesigned and developed from my master's
studies~\cite{msong-mcthesis-2007} in three primary sub-directions:
(a) optimizing its software simulation framework,
(b) generalizing its design and requirements into similar
types of systems to further the extensibility and usability,
(c) and then augmenting the design
to employ advanced rendering technology and algorithms, integrate the softbody
objects with haptic devices, and make use of it for more artistic applications, such as the
real-time softbody interactive {\jellyfish} prototype (the culminating case study of this chapter).

\section{Overview}
\label{sect:softbody-jellyfish-overview}

The {\softbodysys} and its framework, just like requirements engineering, are a moving
target, so the contribution we come up with here simply cannot cover all of CG-visualized
interactive simulation systems' needs, but we provide a basic foundation instead
with the goal of building upon it in our future work as well as that of other
researchers in the same field~\cite{soen-spec-cg-simulation-systems}.

This is also an academic project that had not been rigorously specified
and designed prior to its initial realization due to its relatively small scope
at the beginning. As the scope of the requirements grew over time with
subsequent iterations of the framework, the need to ``back-port'' the
requirements that arose and to plan for future extensions as well~\cite{soen-spec-cg-simulation-systems}.

We first begin by
iteratively synthesizing the guidelines for the functional and
non-functional requirements and design for interactive computer graphics physical based
simulation systems through a detailed case study of the original
{\softbodysys}~\cite{msong-mcthesis-2007,softbody-framework-c3s2e08,softbody-lod-glui-cisse08,soen-spec-cg-simulation-systems}.
We then attempt to generalize our findings in order to compile
a set of common requirement guidelines for the software architects,
so that they already have a reference list of artifacts to refer to when
specifying and designing similar new computer graphics
systems~\cite{soen-spec-cg-simulation-systems}.

We further review the {\softbodysys}, the framework it was
engineered into, and how such a framework evolved over time
until the present day with a constant influx of new requirements that had to
be back-ported into the system and the framework, sometimes requiring 
non-trivial redesign and re-implementation~\cite{soen-spec-cg-simulation-systems}.

Furthermore,
to validate the look and feel, physical properties and
realism in physical-based elastic softbody simulation
visualization requires a comprehensive interface to allow
``tweaking'' various simulation parameters at run-time
while the simulation is running, instead of editing, re-compiling and
re-starting the simulation program's source code
every time a single parameter is changed. The typical
values in our simulation that change are various forces
applied to the body's particles, such as four different types of spring
forces with elasticity, damping, and gas pressure
and even user-interaction with the object by dragging it
in a direction with a mouse as well as collision
response forces, spring stiffness, and so on. Since the
simulation is real-time, another version of level-of-detail (LOD)~\cite{wikilod,skinhairshrek} adjustments
includes the number of particles, springs, subdivisions,
at the geometry level. At the simulation level the variations
include the complexity and sensitivity of the physical-based
simulation algorithms and the time step that they can bear:
the finer the granularity of the algorithm, the more
computation time is required, but higher accuracy of the
simulation at the time step is achieved.
The problem here is how to visually study these properties,
validate them conveniently through either expert-mode
or less-than-expert-mode user interface included with
the simulation program, in real-time~\cite{softbody-lod-glui-cisse08}.

Thus, we additionally
propose an iteration of the {\glui}-based interface~\cite{gluiManual}
to our real-time softbody simulation visualization in
{\opengl}~\cite{ea03,opengl,opengl-redbook-ed7} that allows ``tweaking'' of the simulation
parameters. We introduce its visual design, as well as some details of the software design
and the mapping between the {\glui} components and the internal state of
the simulation system we are working with. We propose the current
interface in its current iteration of the design
and improvement~\cite{softbody-lod-glui-cisse08}.

This work expands even further
with the
integration of the OpenGL Slides Framework ({\oglsf}~\cite{softbody-teaching-opengl-slides},
see \xs{sect:bg-oglsf}) to make presentations with real-time animated graphics
where each slide is a scene with tidgets--and physical based
animation of elastic two-, three-layer softbody objects.
\longpaper
{
The whole project is very interactive,
and serves dual purpose--delivering the teaching material in a classroom-like setting
with real running animated examples as well as releasing the source code
to the students to show how the actual working things are made~\cite{softbody-teaching-opengl-slides}.
}

Finally, we forge the softbody objects into a {\jellyfish} character, which is
interactive and preserves all the properties of the softbody objects, but
is augmented with breathing and self-swimming animation. It inherits all the
advanced rendering and animation techniques, the UI developed for the {\softbodysys}
objects when it was redesigned, generalized, and modularized further as a
library and an API in itself.

\section{Organization}
\label{sect:softbody-jellyfish-organization}

All the related CG animation and rendering background and related work have
been described in \xs{sect:cg-background}.
More specifically,
in \xs{sect:bg-oglsf} we discuss the background and the related
work
and
the properties of the {\oglsf}~\cite{softbody-teaching-opengl-slides}.
Additionally, we have reviewed some of the previous related work done in {\jellyfish} modeling
and animation earlier in \xs{sect:related-work-cg-jellyfish}.

We then describe a brief methodology and layout in \xs{sect:softbody-jellyfish-methodology}.
What follows further in this chapter, is the description of the requirements and design
iterations of the system followed by the summary of derived
requirements and guidelines~\cite{soen-spec-cg-simulation-systems},
all the related design and implementation aspects of the
{\softbodysys} in \xs{sect:softbody-jellyfish-design-impl}.
Even more specifically here at the case study at hand, we describe our own
modeling approach as a transition from a static generated model to
a dynamic physical based model of the {\jellyfish} in 2D and 3D
along with the modeling of the tentacles and the surrounding
environment in \xs{sect:softbody-jellyfish-modeling}. The animation and implementation aspects are discussed
in \xs{sect:softbody-jellyfish-animation-interaction} and \xs{sect:softbody-jellyfish-design-impl}.

Most the sections are illustrated with the actual resulting screenshots
from the {\opengl} softbody system presentation slides referenced where
appropriate~\cite{softbody-teaching-opengl-slides}.

Then we conclude describing our achievement, the limitation of the approach~\cite{jellyfish-c3s2e-2012},
and highlight possible future directions in the overall summary
in \xs{sect:softbody-jellyfish-summary}.

\section{Methodology}
\label{sect:softbody-jellyfish-methodology}

This section discusses the methodology overview suitable for the exploration of
the proposed system with the step-wise process conceptualized in a general manner
on the focus to achieve individual sub-goals including the steps that have already been
done with the overall goal in mind (see \xs{sect:softbody-jellyfish-goal}).

\subsection{Goal}
\label{sect:softbody-jellyfish-goal}

We describe the overall design goals and describe a subset
of them realized in this work with the plans to have a complete
installation exhibited in various galleries after the thesis
completion and polishing the installation work.

First of all, in the general case
the intention is to optimize the current framework of the
{\softbodysys}~\cite{msong-mcthesis-2007,msong-mcthesis-book-2010}
and make it more usable and adaptable
to other human-computer interface libraries, bound to APIs for
various devices. This goal is largely met as shown
throughout \xs{sect:softbody-jellyfish-design-impl}.

Then we plan for the framework to be able to work with
a joystick-like or a sensor glove-like haptic devices to interact
and deform a softbody object in a 3D {\opengl} environment
that provides force feedback to the user restricting their movements based on the elasticity
and shape of the modeled virtual objects %(e.g. the softbody tissue of an organ)
acting as ``inverse sensors''. We succeed partially with this goal
with the Novint {\falcon} haptic device.

The other goals are
faster processing and response times
of the simulation by applying common ``speed-up techniques'',
especially to the subdivided 3D object with the higher LOD
algorithms (e.g., RK4's finer computations are more CPU-intensive than Euler's).
Thus, we in general need a framework and algorithm implementations for
faster real-time rendering,
intersection testing and collision detection,
and ray tracing and global illumination (see \xs{sect:bg-advanced-rendering}).
Another goal
is to have more realism in the material properties,
shadows, lighting effects, and so on.
Thus, the list below is a short list of techniques
that can help us with our goals. We did get to
try and implement some of them to some degree
or provide a ground work to enable them in the future.
The starred entries (*)
have been
at least
partially realized or streamlined the framework to
allow for easier extensions:

\begin{enumerate}
\item Object representations (GPU programming)
\item (*) Vertex and fragment shading (GPU programming)
\item (*) LOD
\item (*) Collision detection
\item Game engine architecture
\item Shadows of the softbody objects
\item Real-time animation of softbody parts attached to an existing skeleton
\end{enumerate}

\subsection{Proof of Concept}
\label{sect:softbody-jellyfish-poc}

The proof-of-concept covers many aspects present throughout the design
and implementation process in \xs{sect:softbody-jellyfish-design-impl}
in a number of screenshots taken and referred to in the chapter.
The current project on the softbody framework is to model a 3D character,
such as the {\jellyfish}~\cite{jellyfish-c3s2e-2012,jellyfish-grand-2011} (see \xs{sect:softbody-jellyfish-in-jellyfish}).
The audience is able to interact with the 3D creature in real-time.
After an associated haptic or MoCap system calibration, audience could drive the {\jellyfish}
by moving their body and perform various actions.
This application also could be used for digital theatre performance
with dynamic (input-sensitive) sound and real-time simulated computer graphics, 
such as the little mermaid dances with various sea creatures.
This would represent a more extensive version than the earlier collaboration prototype
(see \xs{sect:softbody-jellyfish-kinect})~\cite{fortin-interactive-fluid-flow-mcsthesis,jellyfish-grand-2011}
and follows
the conceptual design outlined in \xf{fig:ConceptualDesign-softbody-2}.

\section{Design and Implementation}
\label{sect:softbody-jellyfish-design-impl}

As previously mentioned,
the design- and implementation-related work centers around deformable softbody objects
simulation via physical based methods~\cite{deformable-obj-anim-reduced-control-2009}
applied to a certain type of softbody objects using the corresponding
simulation framework~\cite{softbody-framework-c3s2e08}. Its core is
{\opengl}~\cite{opengl,opengl-redbook-ed7} along with the {\cugl}~\cite{cugl} library
providing a set of convenient extensions to {\opengl}~\cite{jellyfish-c3s2e-2012}.
\longpaper
{
We further integrate with the OpenGL Slide Presentation
System~\cite{softbody-teaching-opengl-slides} and
the Stereoscopic 3D Framework~\cite{stereo3d}.
}
Thus we detail the conceptual design (\xs{sect:softbody-jellyfish-conceptual-design})
and implementation of this work with the results (\xs{sect:softbody-jellyfish-impl})
throughout this section.

\subsection{Conceptual Design}
\label{sect:softbody-jellyfish-conceptual-design}

The conceptual design centers around the operation of the {\softbodysys}.
The two-layered elastic object consists of inner and outer elastic mass-spring surfaces and compressible 
internal pressure.
The three-layered elastic object adds a center particle inside the inner layer.
The density of the inner layer can be set differently from the density of the outer layer; 
the motion of the inner layer can be opposed to the motion of the outer layer. These special features, which 
cannot be achieved by a single layered object, result in improved imitation of a soft body, such as tissue's 
liquid non-uniform deformation.
The inertial behavior of the elastic object is well illustrated in environments 
with gravity and collisions with walls, ceiling, and floor. The collision detection is defined by elastic collision 
penalty method and the motion of the object is guided by integrating Ordinary Differential Equations.
Users can interact with the modeled objects, deform them, and observe the response to their action in real-time
and we provide an extensible framework and its implementation for comparative studies of different 
physical-based modeling and integration algorithm implementations~\cite{msong-mcthesis-book-2010}.
The artistic rendering of such objects is the specific application of {\jellyfish}.

The conceptual component design of the system in \xf{fig:ConceptualDesign-softbody-1} is based on
the derived software methodology we detail further, where re-usable components for the modeling,
rendering, animation, and interaction are defined. The overall system pipeline follows the
process outlined in the conceptual design in \xf{fig:ConceptualDesign-softbody-2}, where the
users (participants) take on the center stage to interact with the system providing some input,
which then proceeds to the {\softbodysys} to do the required input processing, simulation,
and graphical processing prior to providing sensory feedback back to the users.
Regardless the actual modeling, all components follow the same operational simulation sequence
illustrated in \xf{fig:softbody-sequence-diagram}~\cite{msong-mcthesis-book-2010}.

\begin{figure*}[htpb]%
	\centering
	\includegraphics[width=\textwidth]{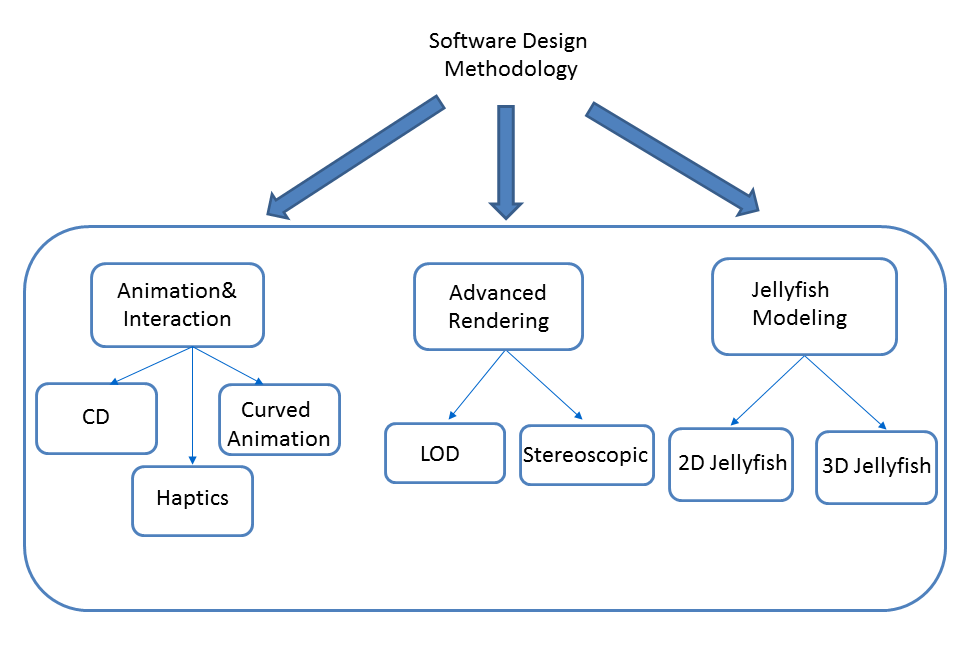}%
	\caption{Conceptual Design of a Softbody Simulation Installation Component Breakdown}%
	\label{fig:ConceptualDesign-softbody-1}%
\end{figure*}

\begin{figure*}[htpb]%
	\centering
	\includegraphics[width=\textwidth]{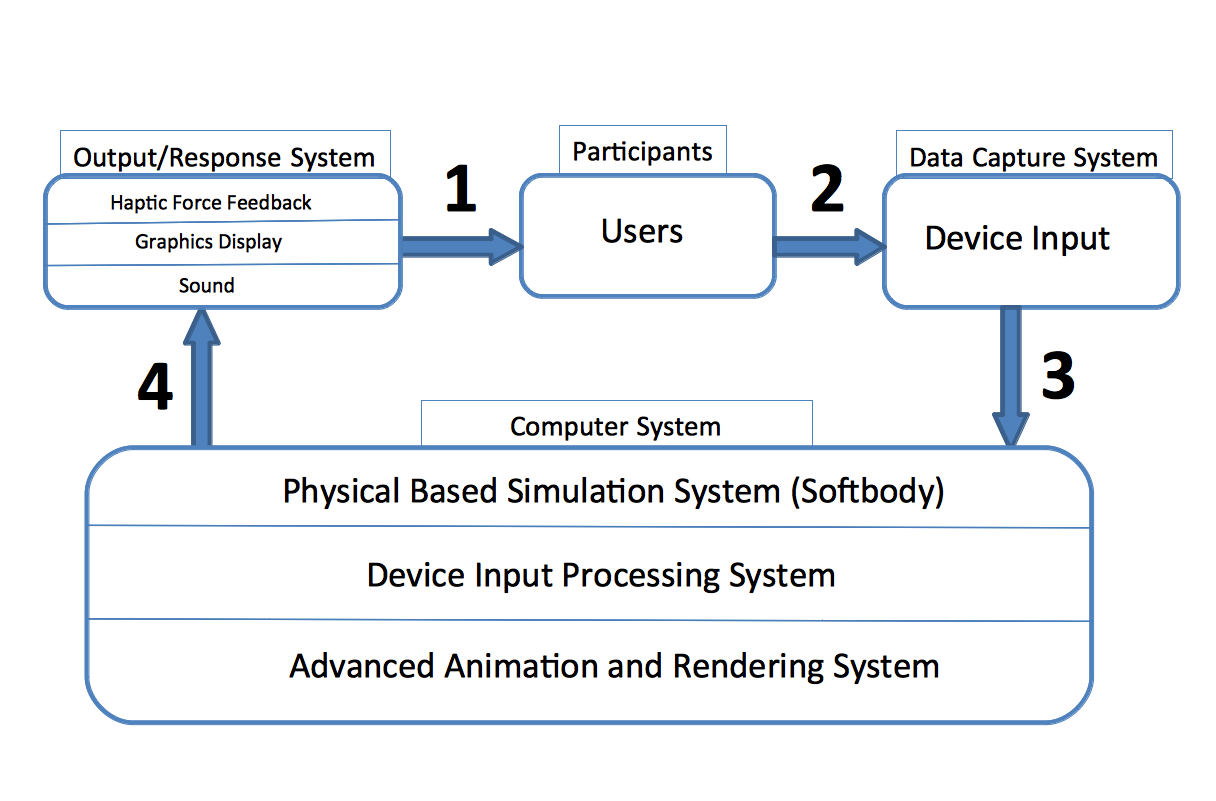}%
	\caption{Conceptual Design of a Softbody Simulation Installation}%
	\label{fig:ConceptualDesign-softbody-2}%
\end{figure*}

\begin{figure}[htpb]%
	\includegraphics[width=\columnwidth]{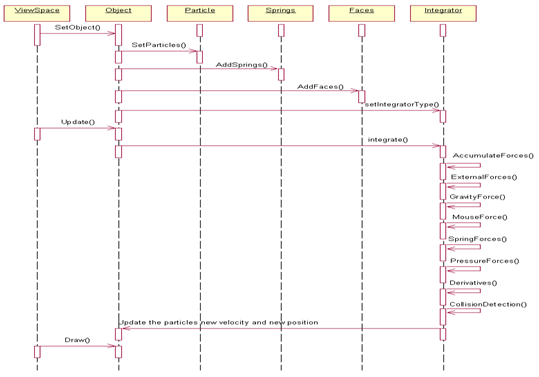}%
	\caption{Softbody System Simulation Sequence Diagram}%
	\label{fig:softbody-sequence-diagram}%
\end{figure}

\subsection{Implementation}
\label{sect:softbody-jellyfish-impl}

The implementation covers the use of a few {\opengl}, {\glut}, {\glui}, {\cugl} and {\glsl} API calls
(\xs{sect:softbody-jellyfish-api}) to support this installation and
its interaction mechanism. The founding core of the {\jellyfish}
installation is the {\softbodysys} and its underlying framework (\xs{sect:softbody-jellyfish-framework}).
The framework and system underwent significant (re)design updates to make
it more re-usable and extensible to support various applications, a case in point
here being the {\jellyfish}, as well as others, and the ways of interacting, shading,
control, collision detection, etc., that made the implementation aspects more streamlined
and simply easier from the time of its original realization~\cite{msong-mcthesis-2007}.
Thus in this contribution we detail the process of such transformation of the system
along with the generalization of its functional and non-functional requirements
along the way in \xs{sect:softbody-jellyfish-framework}. We then detail the
modeling details of various aspects in \xs{sect:softbody-jellyfish-modeling},
such as the introduction of the center particle ``layer'', Feynman algorithm,
{\glui} LOD controls, drag-based animation, curve-based animation, and finally the {\jellyfish} itself
constructed (\xs{sect:softbody-jellyfish-in-jellyfish}) as a real-time softbody object
and the screenshots of the PoC's runs in various rendering modes and effects. We then
review the animation, interactions, lighting, and projection issues.

\subsubsection{APIs}
\label{sect:softbody-jellyfish-api}

We briefly review the APIs used in this installation from
{\opengl}, {\glut}, and {\glui} in \xs{sect:softbody-jellyfish-api-opengl},
{\glsl} in \xs{sect:softbody-jellyfish-api-glsl},
{\cugl} in \xs{sect:softbody-jellyfish-api-cugl}, and
{\falcon} SDK API in \xs{sect:softbody-jellyfish-api-falcon}.

\paragraph{{\opengl}, {\glut}, and {\glui}.}
\label{sect:softbody-jellyfish-api-opengl}

The detailed overview on {\opengl} and related libraries is in \xs{sect:bg-opengl}.
We use a number of {\glui}-based functions~\cite{gluiManual} in our
\api{InitializeGUI()} to connect {\glut} callbacks (commonly present in many
{\opengl} applications, cf. \xs{sect:impl-opengl-glut}) to the {\glui} elements
via \api{GLUI\_Master.set\_glut*()}.
We then compose the UI itself via rendition of various widgets via
\api{GLUI\_StaticText},
\api{GLUI\_Separator},
\api{GLUI\_Rollout},
\api{GLUI\_Spinner},
\api{GLUI\_Checkbox},
\api{GLUI\_Panel},%\\
\api{GLUI\_RadioGroup},
\api{GLUI\_Button},
and others.

\longpaper{{\todo}}

\paragraph{Shaders and {\glsl}.}
\label{sect:softbody-jellyfish-api-glsl}

For the detailed background overview please see \xs{sect:bg-gpu-shaders}. The {\gpu} shaders
framework implementation is described in \xs{sect:softbody-jellyfish-gpu}.

The general principle is having to load and compile a source code
shader file and dynamically link it from within {\opengl}. Prior to
that the graphics card is queried to see which extensions it supports
via \api{glGetString(GL\_EXTENSIONS)}. Then we look for the
extensions called \api{GL\_ARB\_vertex\_program}~\cite{vertex_program}
and \api{GL\_ARB\_fragment\_program} \cite{fragment_program}. Once detected
we load the shader code into the buffer, compile, and link it by
registering functions to do so~\cite{rost2004,3dlabs1,3dlabs2}
to setup functions such as
\api{glGenProgramsARB},
\api{glDeleteProgramsARB},
\api{glBindProgramARB},
\api{glProgramStringARB}, and\\
\api{glProgramEnvParameter4fARB}.

The below sequence of operations required in {\opengl} 1.4 and 1.5 to install
a shader for {\glsl}, for example, as shown in~\xg{algo:arb-shader-extension-installation}.

\begin{algorithm}[hptb]
\hrule\vskip4pt
Create an empty shader: \api{glCreateShaderObjectARB()}\;
\If {successful} {
  Provide source code: \api{glShaderSourceARB()}\;
  \If {successful} {
    Compile: \api{glCompileShaderARB()}\;
    \If {successful} {
      Create object code: \api{glCreateProgramObjectARB()}\;
      \If {successful} {
        Attach the shader object to the program: \api{glAttachObjectARB()}\;
        \If {successful} {
          Link all the shader objects into a program: \api{glLinkProgramARB()}\;
          \If {successful} {
            Install the executable program as a part of current OpenGL's state: \api{glUseProgramObjectARB()}\;
          }
        }
      } 
    }
  }
}
\vskip4pt\hrule
\caption{Installation of a Shader ARB Extension}
\label{algo:arb-shader-extension-installation}
\end{algorithm}

\longpaper{{\todo}}

\paragraph{{\cugl}.}
\label{sect:softbody-jellyfish-api-cugl}

The synthetic camera object in this work is implemented using the API of \api{cugl::Quaternion}s, which relies on
the Concordia University Graphics Library ({\cugl})'s API~\cite{cugl}.
The $(x,y,z)$ axes are drawn at the scene's origin using the API \api{cugl::axes()}
call
primarily for debugging purposes.

\paragraph{Novint {\falcon}.}
\label{sect:softbody-jellyfish-api-falcon}

Please refer to the detailed overview \xs{sect:bg-falcon-haptics}. Here
we detail some specific API details that are in use as inferred from
the Novint manuals~\cite{falcon-manual,falcon-sdk}.

We use the application's \api{HapticsClass} sampled from the
{\falcon} SDK~\cite{falcon-sdk} to contains all the synchronization
and computation methods while deferring the actual softbody simulation
to our {\softbodysys}'s objects by retaining a reference on the object
in question.

The continuous Novint servo callback function from it called \api{ContactCB()} keeps polling
(\api{HDL\_SERVOOP\_CONTINUE}) the {\falcon} for its buttons' status and position,
and then calls our own computation method \api{softbodyContact()} that interacts with the softbody
object ``touch'' and ``response'' processing and putting a mass pressure
on to the device's handle perceived by the interacting user.

We also have an on-demand (\api{HDL\_SERVOOP\_EXIT}) synchronization callback
function \api{GetStateCB()} that uses non-blocking data synchronization
with {\falcon} instead of continuous tracking.

The data in both cases constitute handle button states, position $(x,y,z)$
of the sensor's handle in its local coordinates, and feedback forces
$(f_{x},f_{y},f_{z})$ transfer from and to the device. \api{vecMultMatrix()}
helps us to translate between the local device and world coordinates.
The corresponding Novint's SDK's calls \api{hdlToolPosition()}, \api{hdlToolButton()},
\api{hdlSetToolForce()} help us with that communication.

\subsubsection{Softbody Simulation Software Framework}
\label{sect:softbody-jellyfish-api-framework}
\label{sect:softbody-jellyfish-framework}

A two-layer physical based softbody deformation framework and the
implementing system were designed and developed throughout a
series of works~\cite{msong-mcthesis-2007,softbody-framework-c3s2e08,softbody-lod-glui-cisse08}
and simulating in using {\opengl}~\cite{opengl} in real-time.

The original {\softbodysys} evolved in a number of
ways that were not originally specified or planned for in the requirements or
were under-specified as ``nice-to-have'' high-level system
descriptions. A lot of integration work was planned or ongoing
with other frameworks and systems that also alter the requirements.
In this section we list the evolution that the system
has undergone or is currently undergoing through that ideally would need to be really
anticipated from the start (stated as requirements) and designed for subsequently.
Nonetheless, some earlier requirements and subsequent design
decisions have helped to ease up certain tasks and refine the
process through a number of smaller iterations.

The system we are working with was first conceived and implemented in my thesis studies%
~\cite{msong-mcthesis-2007,softbody-framework-c3s2e08},
originally did not have much user-interactive interface to LOD and
other details except the mouse drag and some function keys.
For every parameter change, the simulation had to be edited and
recompiled. The simulation program and its framework is written
in {\cpp} for 1D-, 2D-, and 3D- real-time physically-based two-layer elastic
softbody simulation system.
Originally, there was no provision to alter the simulation state at
run-time or startup via configuration parameters or GUI
control, so we made a contribution in improving the framework
by adding the notion of state, LOD hierarchy, and its link to the
{\glui}~\cite{softbody-lod-glui-cisse08}.

As our experiments continued we made the use of the softbody objects in other
projects to further validate the softbody's framework usability
on the design and the source code level to apply the studied
techniques in the softbody itself and then to use that in
another project.

The projects wishing to use the softbody objects have to be
able to set the include directory to where the softbody header
files are. The minimum required to \api{include} is
\api{ObjectXD} class where $X$ is 1, 2, 3, or 2-in-3. Then the project
has to make an instance of the desired dimensionality classes and
insert the corresponding drawing and
frame step update handling into the
appropriate callbacks of {\opengl}, such as \api{Display()} and \api{Idle()}.

Such projects have two choices---either add the code of the
{\softbodysys} to their directory structure and build system or
use it as a library to link against. Currently, we
provide only a static library compilation \file{libsoftbody.lib} under
\win{7}, using the Visual Studio 2010 platform and project
files. In the same solution set we use the said \file{libsoftbody.lib}
in our Bezier-curve-based animation where a softbody object is dragged
along the curve~\cite{adv-rendering-animation-softbody-c3s2e09}.

\paragraph{Early Design Decisions.}
\label{sect:softbody-jellyfish-early-design}

Some early design decisions, in the reverse order,
have put some requirements in place that allow
the system to sustain the change and extension
to some degree. As shown in the previous section
the model was broken down into objects of various
dimensionality and that can have some code and
functionality reused through inheritance.
The same applies to the hierarchy of integration
algorithms and the underlying data
structures~\cite{msong-mcthesis-2007,softbody-framework-c3s2e08,msong-mcthesis-book-2010}.
GUI at the time has been planned for the prototype system, but it was not
specified of how it will look like, which widgets
it would have and where it is to be placed within
the simulation window, and their respective functionality~\cite{soen-spec-cg-simulation-systems}.

\paragraph{Evolution.}
\label{sect:softbody-jellyfish-evolution}

Further evolution of {\softbodysys} is outlined below in a set of milestones.
The evolution began from its first version presented
in~\cite{msong-mcthesis-2007}.

\begin{itemize}

\item
The emphasized aspects of the level-of-detail (LOD)
``knobs-and-switches'' have been realized and more
formally addressed~\cite{soen-spec-cg-simulation-systems}. The hierarchy of the LOD has
also been specified, from primitive geometry
to the higher-level algorithms~\cite{softbody-lod-glui-cisse08}
(see \xs{sect:softbody-jellyfish-lod}).

\item
Then, the initial GUI based on {\glui}~\cite{gluiManual} has
been built to allow the run-time management of
the complexity and number of the LOD parameters~\cite{softbody-lod-glui-cisse08},
and this is where the first actual GUI design has
first appeared and materialized~\cite{soen-spec-cg-simulation-systems}
(see \xs{sect:softbody-jellyfish-framework-ui}).

\item
The architecture produced allowed dynamic integration
algorithm selection at run-time, e.g., Euler vs.\ midpoint
vs.\ Runge-Kutta 4 (RK4) vs.\ Feynman. It has necessitated
a plug-in like architecture to allow any number of such
algorithms to be available for the real-time simulation and
rendering as a part of the physical based integration~\cite{soen-spec-cg-simulation-systems}.

In particular, the Feynman integrator has been developed
(see \xs{sect:softbody-jellyfish-feynman}) significantly
later in the time frame from the other three. The architecture required
only minor adjustments to accommodate the new integrator, thus
validating the approach~\cite{soen-spec-cg-simulation-systems}.

\item
The next set of requirements for the redesign and additional
design came from the need to integrate the vertex
and fragment shader processing within the system---written in either
shader assembly or {\glsl} languages, so the framework of the system
had to be extended to allow parametrized generic loading,
compilation, linking, and enabling/disabling of the shaders
at run-time~\cite{soen-spec-cg-simulation-systems} (see \xs{sect:softbody-jellyfish-shader}).

\item
One of the items of the work involving the {\softbodysys}
is enhancing its interactivity with reactive and responsive controls
that include haptic devices to provide force feedback. Applications
based this can be used beyond gaming, e.g., for specialist training (medical),
or even interactive
cinema~\cite{haptics-cinema-future-grapp09,role-cg-docu-film-prod-2009,soen-spec-cg-simulation-systems}.
(See \xs{sect:softbody-jellyfish-api-cugl} and \xs{sect:softbody-jellyfish-haptics}).

\item
The original {\softbodysys} has implemented only a single
algorithm~\cite{msong-mcthesis-2007} for collision detection (CD), the penalty
method; however, now there is a requirement to be
able to use other collision detection algorithms,
so the implementation and design were updated to
reflect this new requirement to allow a collection
of CD algorithms, just like the collection of
integration algorithms there is, with a proper
API, abstraction, and so on~\cite{soen-spec-cg-simulation-systems}.
\longpaper{(See \xs{sect:softbody-jellyfish-cd}.)}

\item
Some library requirements were not explicitly specified
either because they were assumed ``obvious'' but their lack results the requirement
specification being incomplete when we widen the scope of applicability
of the system. The libraries include {\opengl}~\cite{opengl}
along with the drivers for {\glsl} support, {\glut}, {\glui}~\cite{gluiManual}
or even less obvious {\cugl}~\cite{cugl}. This include the HDL~\index{Libraries!HDL}
library for Novint haptic devices, such as {\falcon}~\cite{falcon-sdk}.
Other plans include the {\libsdl}~\cite{libsdl},
{\directx} and HD support, which were not planned for. For the latter
we can build up on the experience from {\ogre}~\cite{ogre} that
has solved that issue~\cite{soen-spec-cg-simulation-systems}.

\item
The system compiles and runs on the Microsoft \win{7} platform,
but with growth it is natural to accommodate other platforms,
such as \macos{X}, \linux{} and mobile such as iOS and Android.
The {\softbodysys} is written in {\cpp}, and the requirement is
to be source-code portable. It is not a very major effort
to make the system portable as {\opengl} itself is portable,
but there are build system efforts and code changes to make
it work reliably~\cite{soen-spec-cg-simulation-systems}. Some of the core softbody applications,
including {\jellyfish} have already been ported to \macos{X}.

\item
Other related projects
include
the integration of stereoscopic rendering of the softbody
objects as well as teaching some computer graphics and simulation
aspects and techniques and integration into the corresponding
OpenGL presentation slides framework~\cite{soen-spec-cg-simulation-systems}.

\item
Subsequent requirements
cover interoperability with different
game and rendering engines and systems (potentially distributed),
extended simulation systems, and the use of the system as
a library~\cite{soen-spec-cg-simulation-systems}.

\end{itemize}

All these past, new, or recent requirements put constraints on and
necessitate making re-design decisions of the system and its components since they were not
adequately planned for in the first place. (Not to mention
the testing aspect and quality assurance, which we leave
for another endeavor project~\cite{soen-spec-cg-simulation-systems}.)

\begin{algorithm}[htpb!]
\begin{small}
\hrule\vskip4pt

\api{ViewSpace} initializes the virtual world and provides the user an 
interactive environment in the form of an interface to allow the user to drag 
the objects, or choose the parameters. For example, the user can choose the 
object type, one-dimensional, two-dimensional, or three-dimensional. The user 
can choose the integrator type, Euler, Midpoint, Feynman, or Runge Kutta~4. 
The user can set up the springs' stiffness, damping variable, and the pressure\;

\api{SetObject()} creates an elastic object based on the interface variable 
set from Step 1\;

\api{SetParticles()} sets up the particles' position and their other initial 
properties, such as mass and velocity\;

\api{AddSprings()} connects particles with springs according to their index\;

\tcp{This step is omitted if the object is one-dimensional}
\api{AddFaces()} connects the springs with faces based on proper index\;

\api{SetIntegratorType()} tells the Controller, which integrator users select
through the interface\;

\api{Update()} updates the integrator's time step\;
\Begin
{
	\api{Integrate()} contains two major functions, \api{AccumulateForces()} and \api{Derivatives()}.
	It is based on all the object geometric information modeled and all the forces information accumulated,
	to integrate over the time step to get new object position and orientation\;

	\api{AccumulateForces()} sums up the forces accumulated on each particle\;
	\Begin
	{
		\api{GravityForce()} accumulates gravity force based on the particles' masses\;

		\api{MouseForce()} is the external force from the interface when user interacts with the object.
		It will be added or subtracted from the particles depending on the force's direction\;

		\api{SpringForce()} accumulates internal forces of the particles connected by springs\;

		\tcp{For one-dimensional object, this state is omitted}
		\api{PressureForce()} accumulates the internal pressure acted on the particles\;

		\api{CollisionForce()} checks if the object is out of boundaries after the
		integration state. If the new position is outside of the boundary,
		then it will be corrected and reset on the edge of the boundary.
		Moreover, the new collision force will be added to the object\;
	}

	\api{Derivatives()} does the real derivative computation of acceleration and velocity
	in order to get new velocity and position of elastic objects based
	on the integrator type defined by users\;
}

\api{Draw()} displays the object with new position, velocity, and deformed shape\;

\hrule\vskip4pt
\end{small}
\caption{General Softbody Simulation Algorithm}
\label{algo:softbody-simulation-algo}
\end{algorithm}

\paragraph{Requirements Synthesis Methodology.}

From the discussions in the previous sections and
the reverse engineering of the design and implementation
to some extent we synthesize the previously unstated
(or sometimes rather implicitly stated) requirements, both
functional and non-functional. This list is rather
complementary and is not replacing existing typical
software engineering requirements for information
systems~\cite{standards98ieee,swebok-2004}. The list also contains sometimes a checklist of guidelines rather
than just the requirements~\cite{soen-spec-cg-simulation-systems}.

\vfill

\paragraph*{Functional Requirements.}
\label{sect:softbody-jellyfish-func-req}

\begin{itemize}

\item
Most graphics systems adopt a version of the MVC architecture,
so we need to plan for user I/O, multiple displays, and the data
structure modeling.

\item
User I/O GUI and other input devices
should plan for an LOD GUI as well
as potentially multiple algorithms of
handling it. This is also prevalent among
typical PC games through their game settings
allowing them to run tuned on a wide range
of hardware.

\item
Implement statistics gathering for
various real-time performance metrics
not only for simulation, but also for
rendering.

\item
Plan to allow for multiple algorithms
selection at run-time for comparative studies
on any aspect of visual realism to run-time
performance and memory usage.

\item
Plan for support and the
corresponding GUI for the vertex
and fragment shaders written in either
cross-vendor GPU assembly, {\glsl}, and {\hlsl}.

\item
Plan for lighting and texture-mapping techniques.
This is tricky in particular with some softbody objects, so the
data structures should be planned ahead of time to support
normals, and so on.

\item
Provide ability for softbody-like objects
being able to be ``attached'' to ``hardbody'' (rigid body)
objects, e.g. to simulate muscles on a skeleton
and provide points of attachment of one to the other.

\item
Allow for modeling and alteration of the Archimedean-based graphs
and different types of them than just an octahedron~\cite{archimedean-graphs-wmsci08}
and just the number of iterations as a run-time LOD parameter.

\item
Interactivity through haptic devices~\cite{wiki:haptic-technology} with the softbody feedback
(e.g., for surgeon training or interactive cinema~\cite{haptics-cinema-future-grapp09}). Additional
interactivity through sensor devices like {\kinect}.

\item
Allow the state dump and reload functionality in order to display each
particle and spring state (all the force contributions, velocity, and the position)
at any given point in time in a text or XML file for further import into a relational
database or an Excel spreadsheet for plotting and number analysis, perhaps by external
tools. Reloading would enable to reproduce a simulation from some
point in time, a kind of a replay, if some interesting properties
or problems are found. This is useful for debugging as well.

\item
Plan for multiple rendering backends, potentially distributed or multithreaded
such as {\ogre}~\cite{ogre}, URay\index{Tools!URay}~\cite{uray-gpu-distributed-techrep},
and others.

\item
Allow for stereoscopic effects.

\end{itemize}

\paragraph*{Non-Functional Requirements.}
\label{sect:softbody-jellyfish-non-func-reqs}

\begin{itemize}

\item
\emph{Usability} may be of importance to researchers to test
different physical based phenomena at real-time~\cite{soen-spec-cg-simulation-systems}.

\item
\emph{Portability} of the source code (at minimum)
and plan for deployment and building under different
platforms and build systems, e.g., \linux{} with Makefiles~\cite{gmake} and autoconf,
or \macos{X} with XCode or also Makefiles, and mobile
device in order to cover larger user and researcher bases~\cite{soen-spec-cg-simulation-systems}.

\item
\emph{API hooks} should be always provided for plug-in architectures
for all algorithms where there could be more than one instance,
such as integration, subdivision, collision detection.
This will simplify the integration effort with other
projects and other programming languages~\cite{soen-spec-cg-simulation-systems}.

\item
Libraries, frameworks, APIs, such as {\opengl}, {\glsl},
{\glut}, {\glui}, {\cugl}, {\directx}~\cite{soen-spec-cg-simulation-systems}.

\item
Export and generation of the system as a library itself,
or a collection of libraries and APIs for use in other
applications~\cite{soen-spec-cg-simulation-systems}.

\item
Constrain the exported API and globals in the system's
own namespace to avoid clashes with external applications
during linking~\cite{soen-spec-cg-simulation-systems}.

\item
Plan for the academic value of teaching and learning computer
graphics~\cite{cgems} and physical based simulations, by
structuring the code, comments, and documentation per
consistent naming and coding conventions and the API.
This is especially valuable for open-source and academic
projects~\cite{soen-spec-cg-simulation-systems}.

\item
Allow for extension of the main algorithm, e.g., of \xg{algo:softbody-simulation-algo},
by subclassed applications so it is less rigid~\cite{soen-spec-cg-simulation-systems}.

\end{itemize}

\subsubsection{Augmented Softbody Modeling}
\label{sect:softbody-jellyfish-modeling}

\paragraph{Center Particle -- the Third Layer.}
\label{sect:softbody-jellyfish-framework-center-point-curve}

The core of the model design in the softbody system is that it has two interconnected
layers of tissue using the spring-mass system and is tested
against various integration algorithms (e.g., Euler, midpoint,
and RK4) as well as a hierarchy of various level-of-detail (LOD)
parameters as summarized earlier. The simulation is built upon one to three
softbody objects of 1D, 2D, or 3D which are simulated and interacted
within a limited ``boxed'' environment. To make the
softbody simulation as a tissue for example on an animated game
character's skeleton or any other type of animation, the
softbody needs to be ``attached'' somehow to the rigid body %``hardbody''
that it is moving along with. The softbody attachment aspect
in the softbody simulation framework was missing, thus we
contribute a first version of attachment of softbody objects
to hard surfaces for ``ride-along'' animation to achieve
higher realism.

We present an enhanced real-time two-layer softbody shape simulation system implemented
using {\cpp} and {\opengl} with a third, single center point layer for attachment of
the softbody objects onto hard surfaces and drag to simulate e.g. soft skin
tissues. We present the user interaction and the corresponding {\glui} interface
to the softbody objects for comparative purposes. We also add the Feynman
algorithm implementation to complement the original earlier framework
for comparative studies with classical Euler, mid-point, and RK4 algorithms.
Alongside our softbody attachment contribution
we add and discuss the Feynman algorithm implementation and
its properties as compared to the existing implementations
of Euler, midpoint, and RK4.
We also compare the generation of the modeling the 3D objects
spheres by subdivision or sphere mapping when using the attachment
points and the Feynman algorithm.
We attempt to make the simple extension that does not violate
framework constraints as a proof-of-concept and animate it
along a Bezier curve.

\paragraph{Curve-based Trajectory.}
\label{sect:softbody-jellyfish-bezier-curve}

Moving a softbody along a Bezier curve is the first simplest example
of advanced animation using the softbody object outside its
own simulation application. We were able to successfully
link the softbody features and use them through their
API exported as header files of the available softbody 
objects of various dimensions~\cite{adv-rendering-animation-softbody-c3s2e09}.

The animation is real-time and while the softbodies
are being dragged along the curve, they perform their
designated default physical based dynamic deformation.
We simulate the mouse drag along the Bezier curve
by the attachment center point and the \api{Drag} object.
The controls are detailed in \xa{appdx:curve-anim-controls}.
These key handling is implemented in the callback function \api{KeyboardKey()}
and \api{FunctionKey()} in the \file{controllerCallbacks.cpp} file.

The system also has mouse-based and {\glui}~\cite{gluiManual} controls.
The user can adjust the curve by adding more control
points.
The control points can be created either through the ``Add New Point'' button in the
GUI from a list of preset points or by middle button (or left and right together of only
two buttons) mouse click within the rendering view (see \api{Mouse()}
in \file{controllerCallbacks.cpp}).
The control points are properly connected by a dashed line in the order they are referenced.
The implementation of it is located in the function \api{void GUI::guiContorollerCallback()}
in the \file{gui.cpp}
and \file{globals.cpp} files.
An example screenshot is in \xf{fig:general-screenshot}.	

\begin{figure}[htp!]
	\centering
	\includegraphics[width=\textwidth]{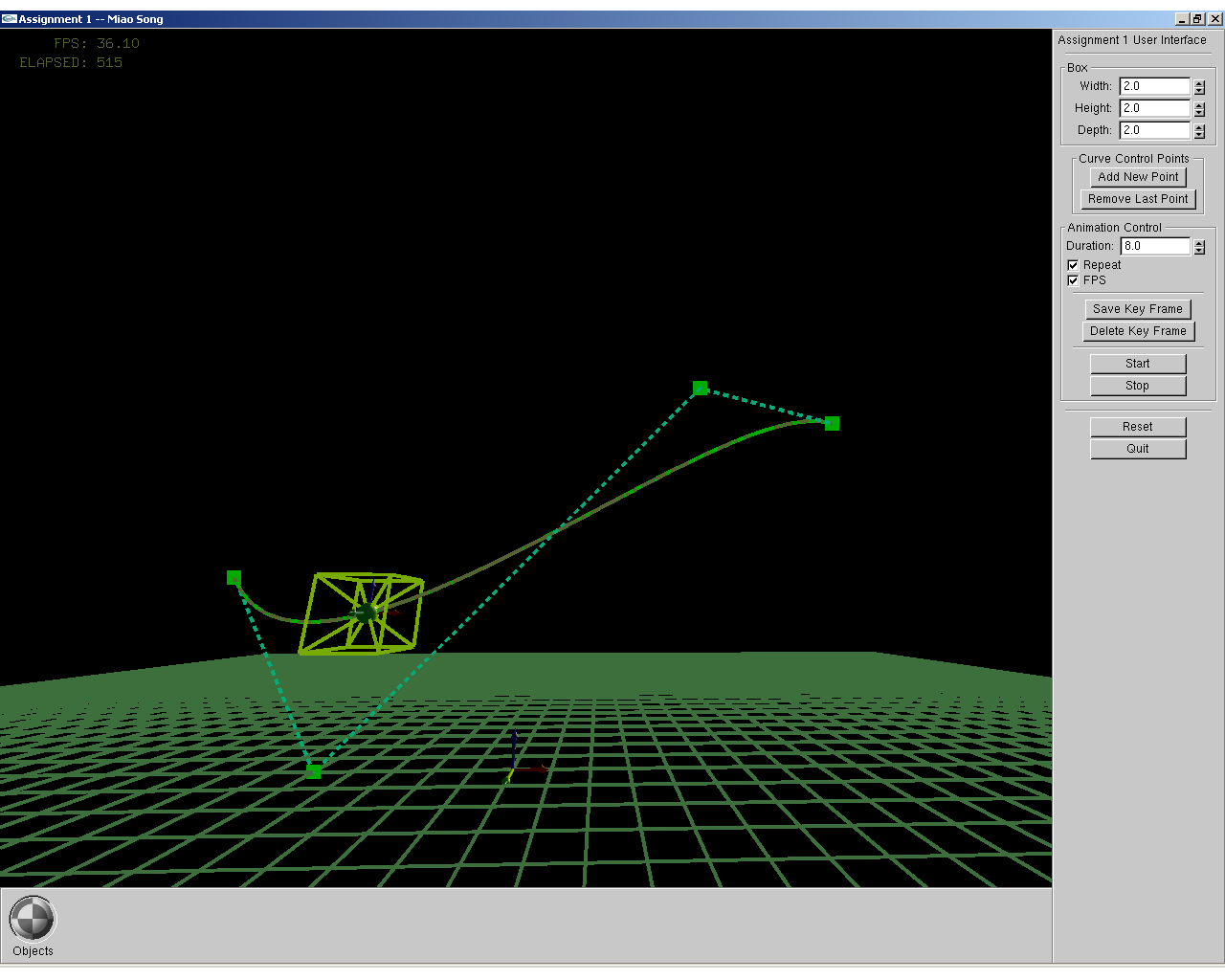}
	\caption{General Bezier-Curve Based Animation Screenshot}
	\label{fig:general-screenshot}
\end{figure}

In this image, this simulation displays the cubic Bezier curves for the control points. 
Four control points are required to specify the curve.
The implementation is in \api{Drawcurve()} function in the \file{drawTools.cpp} file.
The Bezier curve (or each curve segment) is only drawn when there are enough (4) points to have it done.

The user is able to specify the time the motion
should take from start to finish.
There is a {\glui} control for that. At the implementation level, the time is not in
true seconds, but allows to speed up or slow down the motion. 
It also depends on the frame rate of a particular environment, OS, hardware, etc.
The implementation is in \api{void GUI::guiContorollerCallback()} function in the \file{gui.cpp}
and \file{globals.cpp} files.

Another GUI rotation control at the bottom-left of the {\glut} window allows the user
to input rotations into the box object by manipulating an arcball control.
The control displays as a checkerboard-textured sphere, which the user manipulates directly.
The rotation control can be constrained to horizontal-only movement by holding the CTRL key, or to vertical
rotation by holding the ALT key.
Rotation controls can optionally keep spinning once the user releases the mouse button.
The local coordinate axes have been drawn with {\cugl}'s API.
The implementation is in \api{void GUI::guiContorollerCallback()} function in the \file{gui.cpp}.

The user also has a control of the camera that can look
around the animated scene using quaternions and the
camera as well as axes from the Concordia University Graphics Library ({\cugl})~\cite{cugl}.

The softbody's center point is following the curve in the animation while the
the physical based properties still in effect in the area of the softbody.
The end goal of this is to have the body follow the curve motion and react
accordingly. When different stiffness softbody objects placed on the
skeleton or otherwise character, the idea as they move with the character
attached by the attachment point, the animation takes places. This can
simulate the soft body parts of a running or walking human as well as
when humans breathe. The curve based animation is a testbed for a
periodical walk to see how the body behaves. We have not implemented
yet or attached any actual softbody to any skeleton or character yet
so this is deferred to the future work.

\paragraph{Feynman Algorithm.}
\label{sect:softbody-jellyfish-feynman}
 
We implement for comparative purposes the Feynman algorithm within a {\cpp}-based framework for the
softbody simulation.
To facilitate the comparison, we design
the timing measurements to be on the same hardware against that of 
Euler integrator in the softbody framework by varying different algorithm parameters.
Due to a relatively 
large number of such variations we implemented a GLUI-based user-interface to allow for much more finer 
control over the simulation process at real-time, which was lacking completely in the previous versions of 
the framework. The Feynman algorithm~\cite{physics-and-java} is still quite
efficient comparatively to Euler yet more accurate due to the step and a half
stepping. Our framework does not yet implement comprehensive statistic
measurements so we cannot provide actual numerical results, just visual
observations.
We show our current results based on the enhanced framework.

\longpaper
{
\paragraph{Modeling Uniform Softbody Sphere.}
\label{sect:softbody-jellyfish-modeling-uniform-sphere}

Modeling Uniform Softbody Sphere

{\todo}

\paragraph{Drag Animation.}
\label{sect:softbody-jellyfish-drag-animation}

Drag Animation

{\todo}

We simulate the mouse drag along the Bezier curve
by the attachment center point.
} % \longpaper

\begin{figure*}[htp!]
	\hrule\vskip4pt
	\begin{center}
	\includegraphics[width=\textwidth]{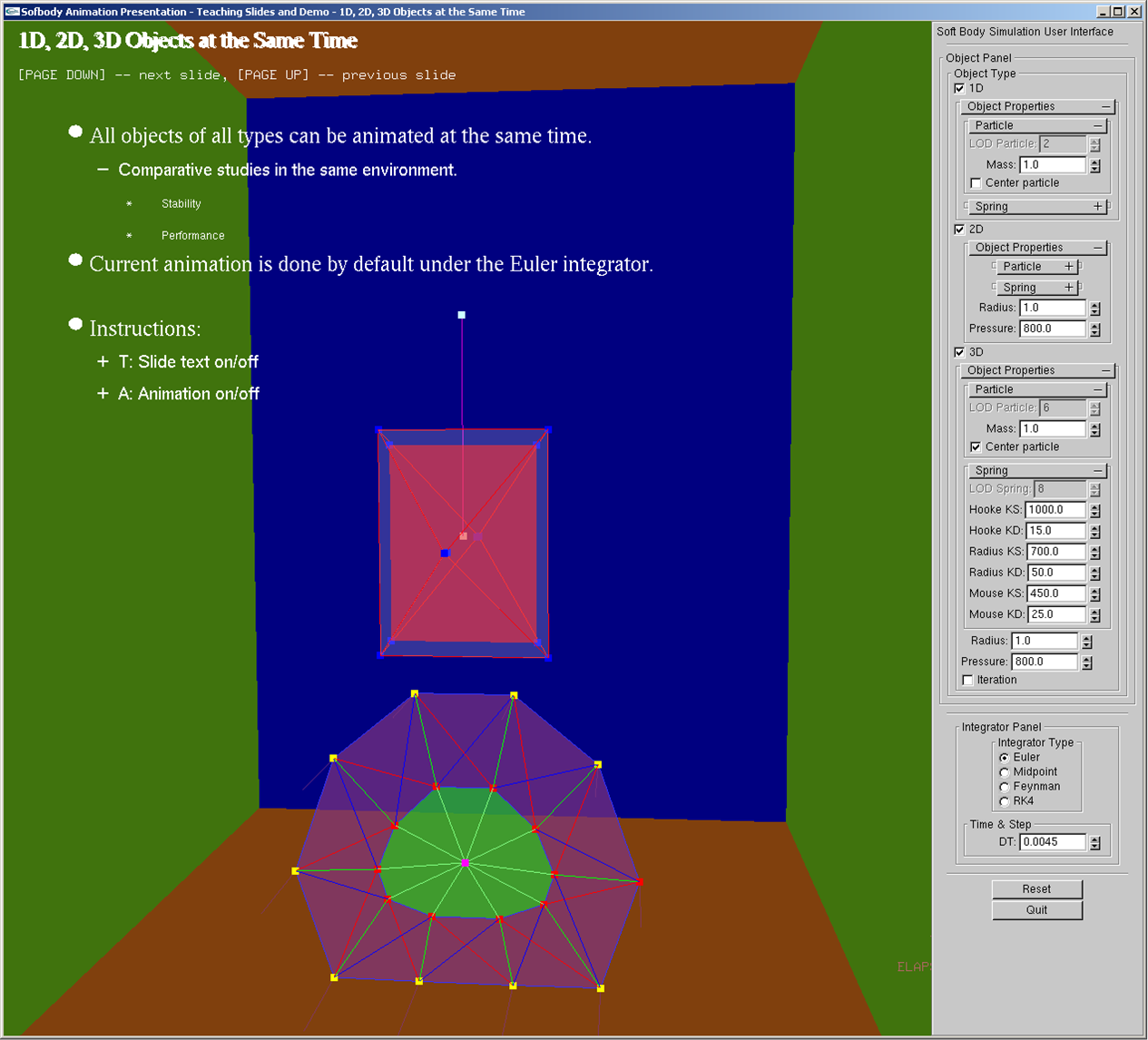}
	\caption{Three Types of Softbody Objects Simulated Together on a Single Slide Scene}
	\label{fig:Objects3}
	\end{center}
	\hrule\vskip4pt
\end{figure*}

\longpaper
{
The Softbody Simulation System's main goal is to provide real-time simulation of
a variety of softbody objects, founded in the core two- or three-
layer model for objects such as human and animal's soft parts and tissue,
and non-living soft objects, such as cloth, gel, liquid, and gas. 
Softbody simulation is a vast research topic and has a
long history in computer graphics.
The softbody system has gone through a number of iterations
in its design and development.
Initially it had limited user interface~\cite{msong-mcthesis-2007,softbody-framework-c3s2e08}.
Then the fine-grained to high-level level-of-detail (LOD) GLUI-based~\cite{gluiManual}
user interface has been added~\cite{softbody-lod-glui-cisse08},
GPU shading support was added~\cite{adv-rendering-animation-softbody-c3s2e09}
using the OpenGL Shading Language~\cite{rost2004}, a curve-based animation
was integrated, and software engineering re-design is constantly being applied.
The example of the common visual design of the LOD interactivity
interface is summarized in \xf{fig:Objects3}. The LOD components
are on the right-hand-side, expanded,
and the main simulation window is on the left (interactivity with
that window constitutes for now just the mouse drag and functional
keys).
Following the top-down approach configuration parameters, that
assume some defaults, were reflected in the GUI~\cite{softbody-opengl-slides}.
} % \longpaper

\begin{figure*}[htb!]
\hrule\vskip4pt
\begin{center}
	\subfigure[1D Elastic Object Simulation Slide]
	{\includegraphics[width=.3\slideimagewidth]{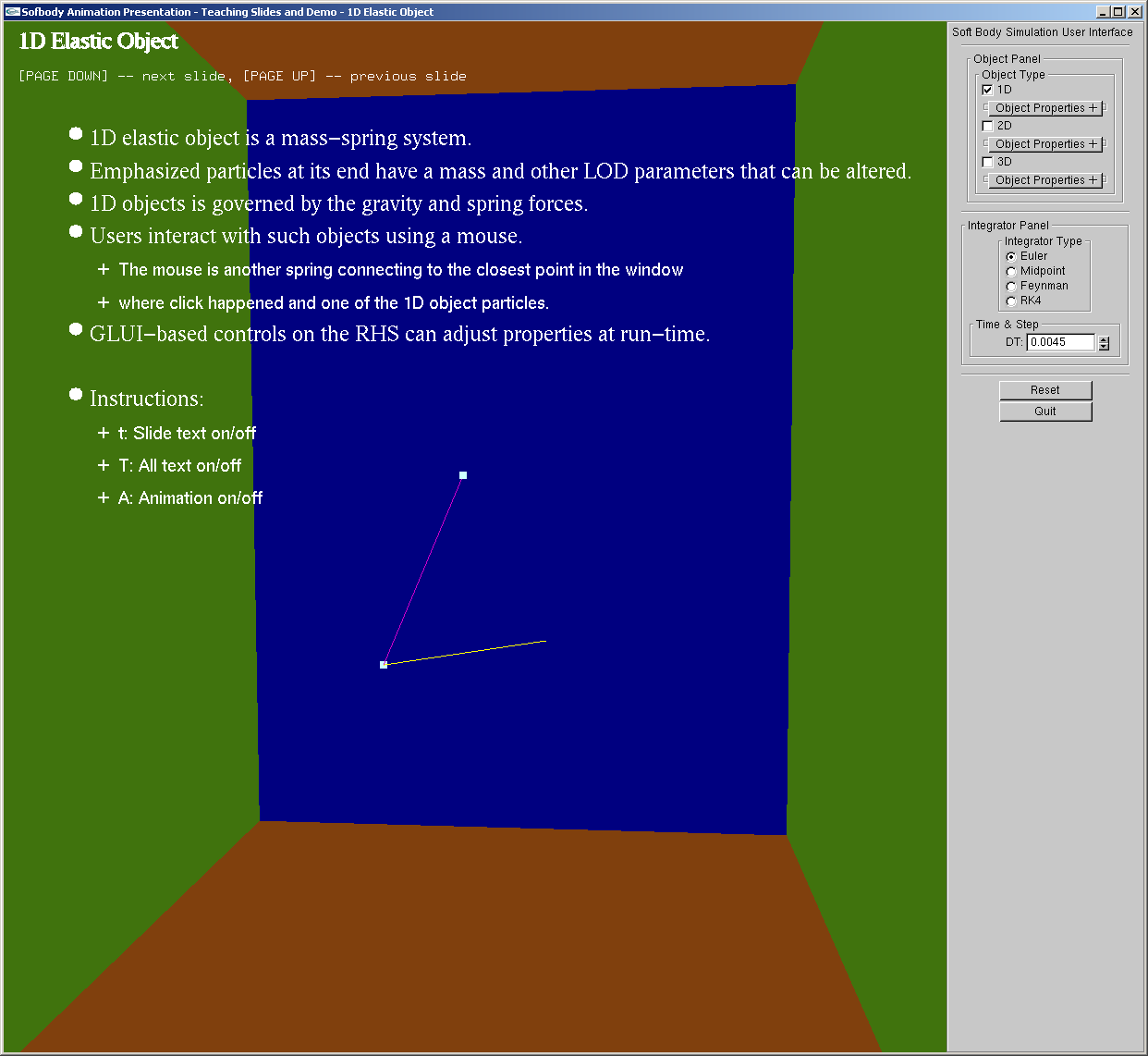}
	 \label{fig:1d-elastic-object}}
	\subfigure[2D Elastic Object Simulation Slide]
  {\includegraphics[width=.3\slideimagewidth]{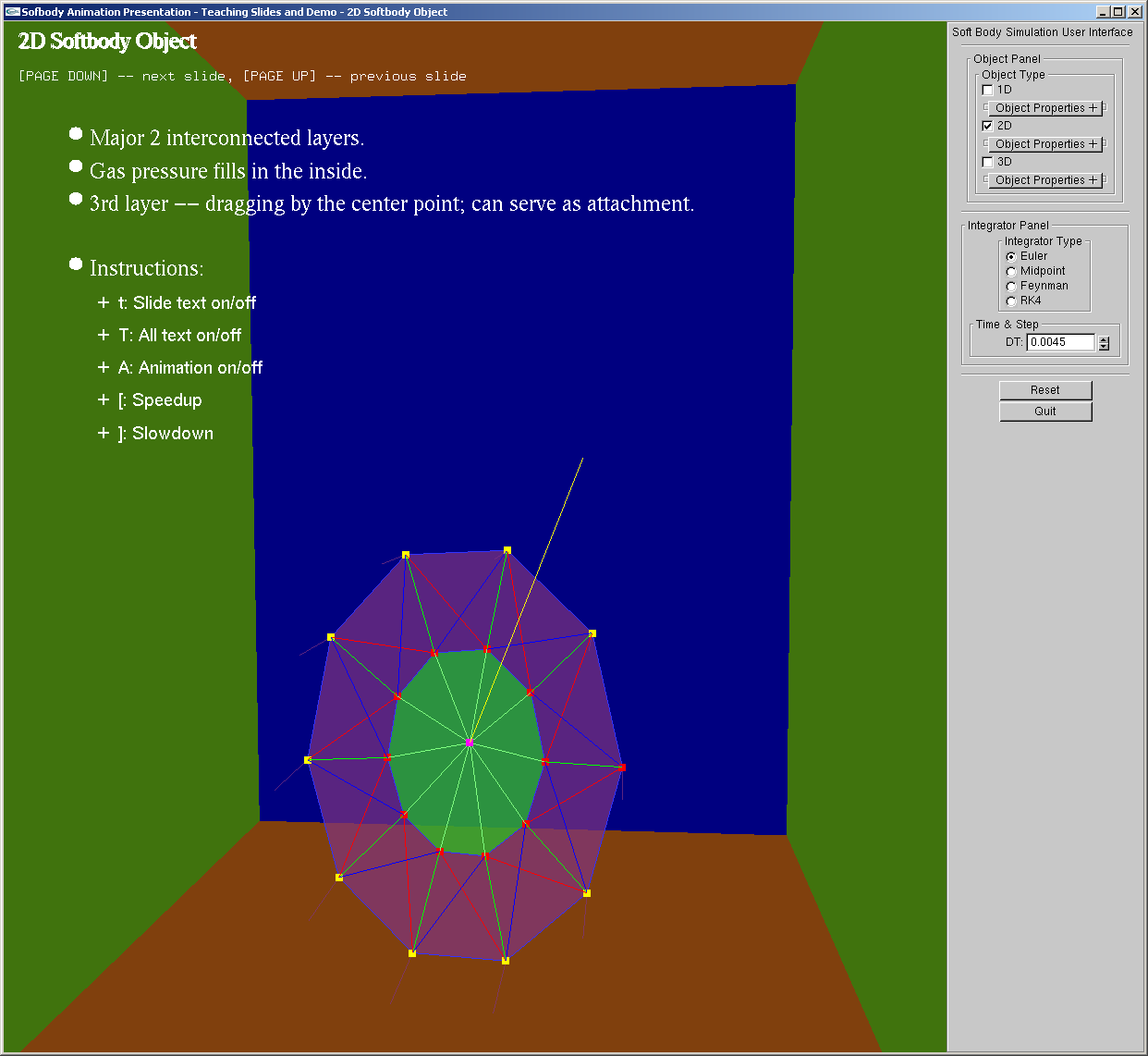}
	 \label{fig:2d-softbody-object}}
	\subfigure[3D Elastic Object Simulation Slide]
  {\includegraphics[width=.3\slideimagewidth]{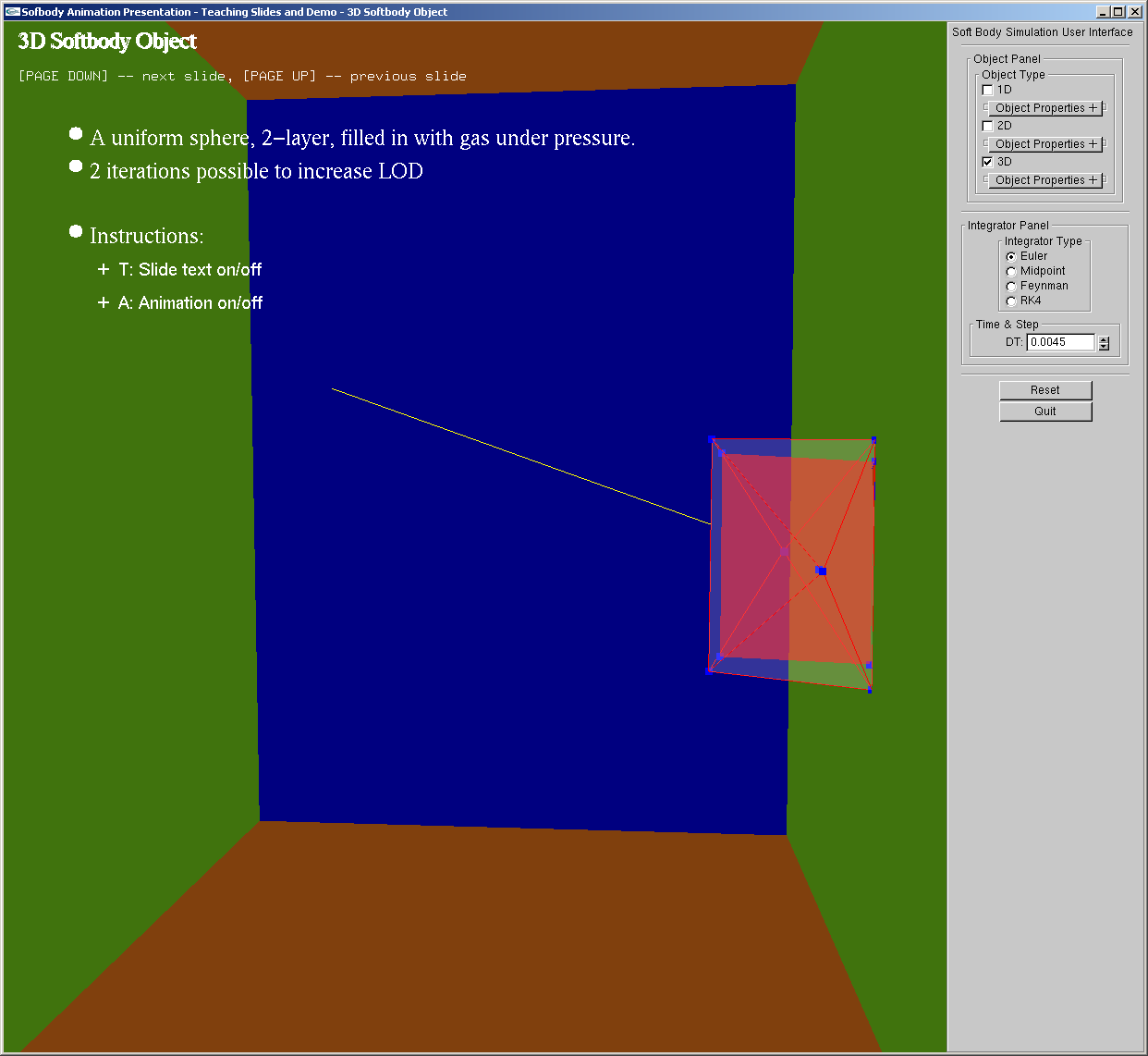}
	 \label{fig:3d-softbody-object}}
\caption{Simulation of Single 1D, 2D, and 3D Softbody Elastic Objects Slides}
\label{fig:object-dimensionality-slides}
\end{center}
\hrule\vskip4pt
\end{figure*}

\begin{figure*}[htb!]
\hrule\vskip4pt
\begin{center}
	\subfigure[Simulation with Euler Integrator Slide]
	{\includegraphics[width=.3\slideimagewidth]{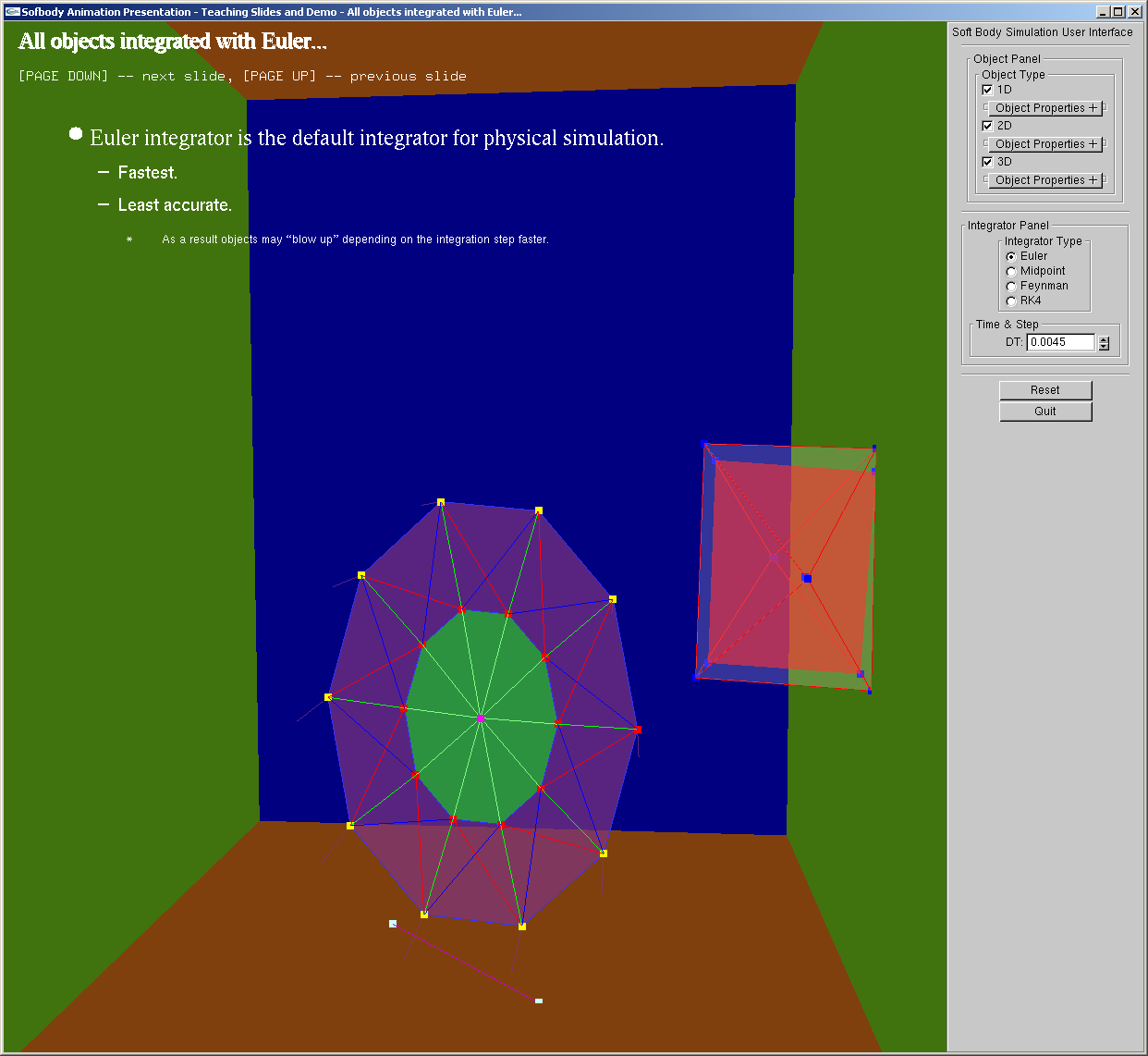}
	 \label{fig:all-objects-euler}}
	\subfigure[Simulation with Midpoint Integrator Slide]
  {\includegraphics[width=.3\slideimagewidth]{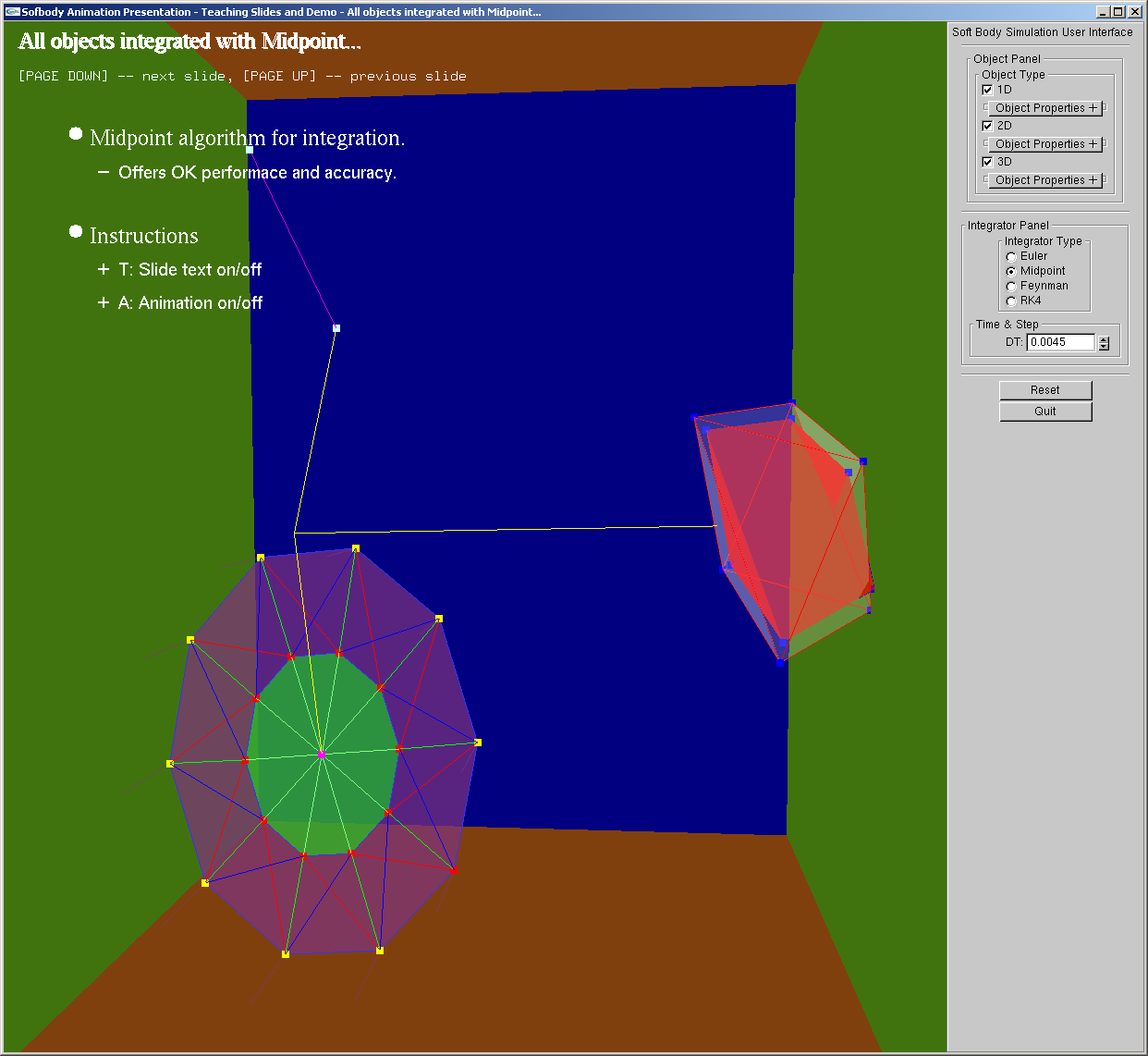}
	 \label{fig:all-objects-midpoint}}
	\subfigure[Simulation with Feynman Integrator Slide]
  {\includegraphics[width=.3\slideimagewidth]{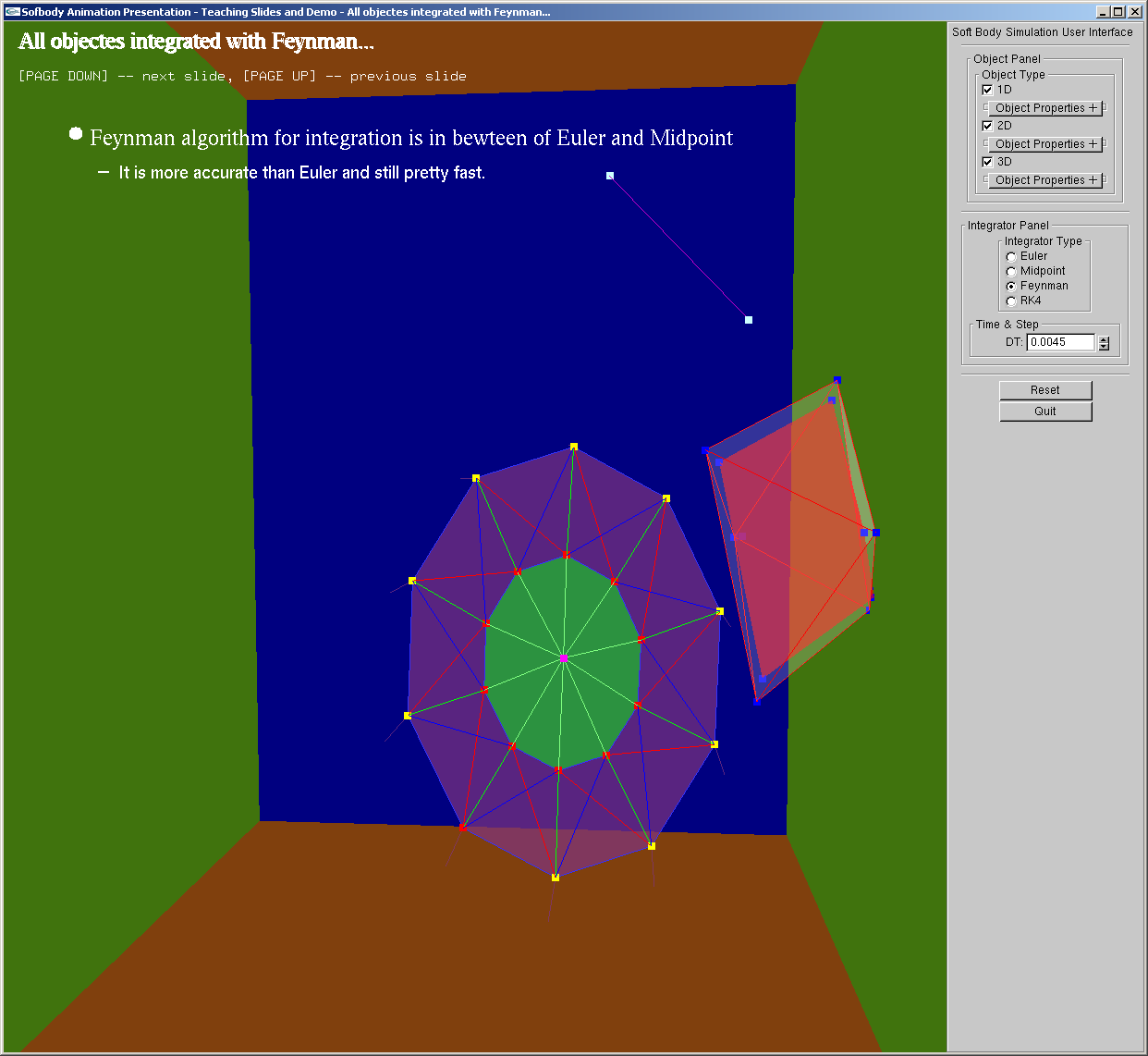}
	 \label{fig:all-objects-feynman}}
\caption{Simulation of all Softbody Object Types with Various Integrators Slides}
\label{fig:some-integrator-slides}
\end{center}
\hrule\vskip4pt
\end{figure*}

\longpaper
{
\newcommand{\softbodypresentationmain}{\file{SoftBodyPresentation.cpp}}
\newcommand{\sofbodysimulationmain}{\file{SoftbodySimulation.cpp}}
\newcommand{\sofbodysimulationslide}{\api{SoftbodySimulationSlide}}
\newcommand{\apimain}{\api{main()}}
\newcommand{\apianimate}{\api{animate()}}

The methodology consists of the design and implementation
modification required for the integration followed by making
the actual presentation slides. The source code
of the presentation is a part of the learning material
along the actual content of the material presented and
is prepared as such.
Separately, both frameworks and implementing systems define
the {\apimain} function, which cannot be included into
any of the libraries (both can be compiled into the library
files to be linked into other projects) because of the linker errors
when the object code from the two or more systems is
combined into a single executable.
We therefore started a new application with a new {\apimain},
the {\softbodypresentationmain}.
Additionally, both frameworks have to declare their own
namespaces, which both have not done in the past, similarly to CUGL~\cite{cugl} because
there are some common names of variables, classes,
or functions that clash on compilation. This is an overall improvement
not only for this work, but also for any similar type
of integration with other projects (cf. \xs{sect:future-work}).
Thus, we declared the namespaces \api{softbody:} and \api{slides:}
and move the clashing variables under those namespaces.
Most of the main code from {\sofbodysimulationmain} application
is encapsulated into a generic {\sofbodysimulationslide}
class that includes the default configuration of the
softbody simulation parameters~\cite{softbody-opengl-slides}
this class is inherited by the slides that do the actual simulation
of softbody objects.
Furthermore, the concrete slides that inherit from {\sofbodysimulationslide}
are broken down into some preset distinct configuration defaults and accompanying
tidgets. They override the {\apianimate} method (the ``idle'' function) as well as
the state LOD parameters per an example slide.

For the presentation in this work the demonstration slides currently include
the following:

\begin{enumerate}
\item
\api{TitleSlide} -- a typical title slide with the lecture/presentation title and the presenter information

\item
\api{TOCSlide} -- a tidget table of contents of the presentation

\item
\api{IntroductionSlide} -- a tidget introduction of the material

\item
\api{SoftbodySimulationSlide1D} -- a slide featuring the 1D elastic
object configured by default encased in the \api{ViewBox}, see \xf{fig:1d-elastic-object}.

\item
\api{SoftbodySimulationSlide2D} -- a slide featuring the 2D softbody object configured by default,
see \xf{fig:2d-softbody-object}.

\item
\api{SoftbodySimulationSlide3D} -- a slide featuring the 3D softbody object configured by default,
see \xf{fig:3d-softbody-object}.

\item
\api{SoftbodySimulationSlideAllD} -- a slide featuring all types of softbody objects configured by default,
as shown in \xf{fig:Objects3}, included into the slide environment.

\item
\api{SoftbodySimulationSlideAllEuler} -- all three objects configured by default to animate
under the Explicit Euler integrator, see \xf{fig:all-objects-euler}.

\item
\api{SoftbodySimulationSlideAllMidpoint} -- all three objects configured by default to animate
under the Midpoint integrator, see \xf{fig:all-objects-midpoint}.

\item
\api{SoftbodySimulationSlideAllFeynman} -- all three objects configured by default to
animate under the Feynman integrator, see \xf{fig:all-objects-feynman}.

\item
\api{SoftbodySimulationSlideAllRK4} -- all three objects configured by default to animate
under the RK4 integrator, see \xf{fig:all-objects-rk4}.

\item
\api{ShortcomingsSlide} -- a slide describing the limitations of the approach

\item
\api{ProjectedFeaturesSlide} -- a summary of some projected features for

\item
\api{ConclusionSlide} -- a preliminary conclusions slide

\item

\end{enumerate}

} % \longpaper

\begin{figure}[thb!]
	\hrule\vskip4pt
	\centering
	\includegraphics[width=\slideimagewidth]{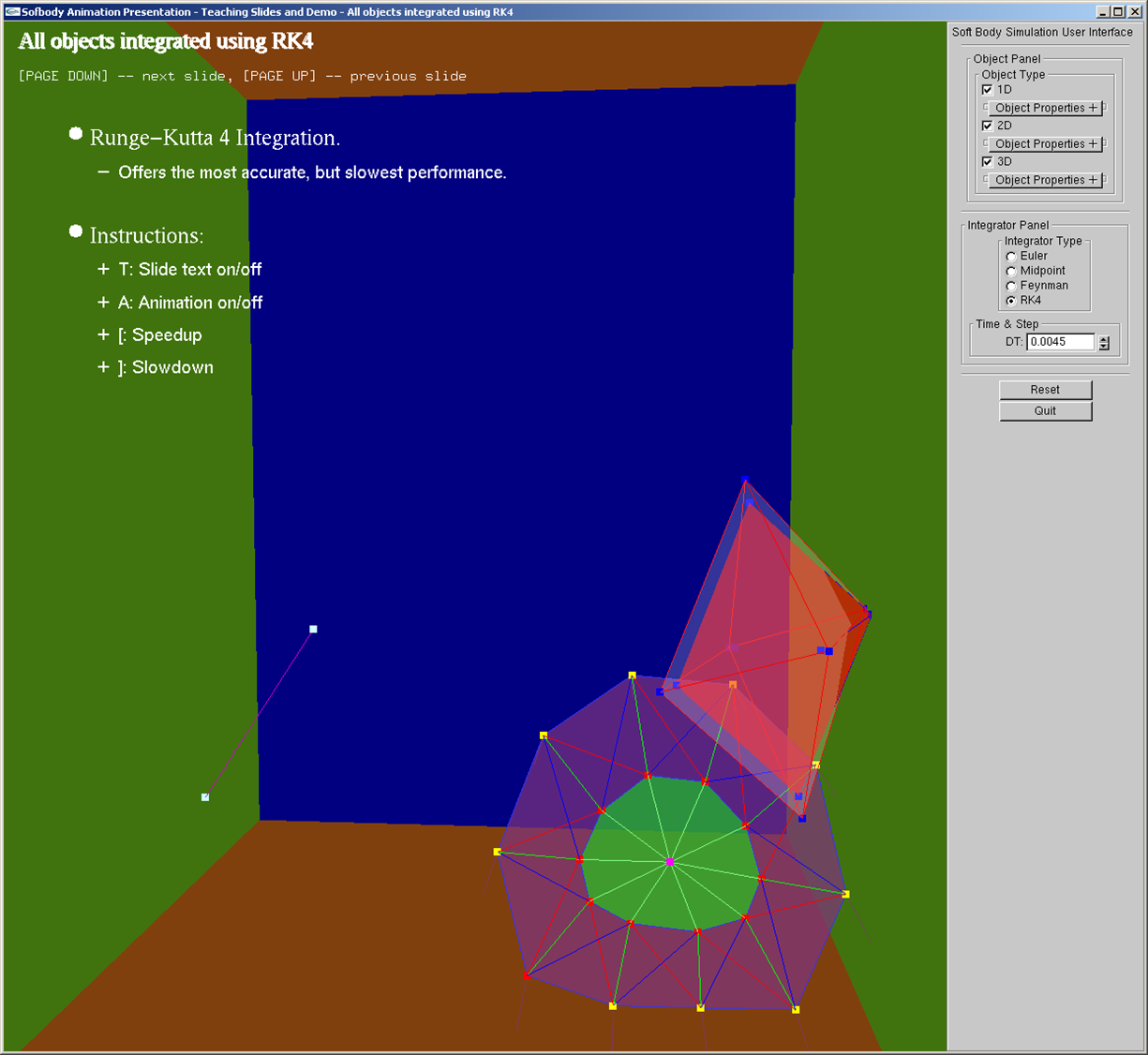}
	\caption{Simulation with Runge-Kutta 4 Integrator Slide}
	\label{fig:all-objects-rk4}
	\hrule\vskip4pt
\end{figure}

\longpaper
{

The core framework's design is centered around common dimensionality (1D, 2D, and 3D)
of graphical objects for simulation purposes, physics-based integrators,
and the user interaction component.
The \api{Integrator} API of the framework as of this writing is implemented by the well-known
Explicit Euler, Midpoint, Feynman, and Runge-Kutta~4 (RK4)-based integrators for their mutual
comparison of the run-time and accuracy.
The system is implemented using OpenGL~\cite{opengl,opengl-redbook}
and the {\cpp} programming language with
the object oriented programming paradigm~\cite{softbody-opengl-slides}.

This elastic object simulation system has been designed and implemented according
to the well known architectural pattern, the model-view-controller (MVC).
This pattern is ideal for real-time simulation because it simplifies
the dynamic tasks handling by separating data (the model) from user interface (the view).
Thus, the user's interaction with the software does not impact the data handling;
the data can be reorganized without changing the user interface.
The communication between the model and the view is done through the controller.
This also closely correlates to the OpenGL state machine, that is used as a core
library for the implementation~\cite{softbody-framework-c3s2e08,softbody-opengl-slides}.
} % \longpaper

\subsubsection{Advanced Softbody Rendering}
\label{sect:softbody-jellyfish-adv-rendering}

Following the description in \xs{sect:bg-advanced-rendering}, we apply
some of the techniques in here. Specifically, and LOD-related processing
and its user interface in \xs{sect:softbody-jellyfish-framework-ui},
{\gpu} shading in \xs{sect:softbody-jellyfish-gpu}\longpaper{, and
stereoscopic effects in \xs{sect:softbody-jellyfish-stereo}}.

\paragraph{LOD Interaction Interface.}
\label{sect:softbody-jellyfish-lod}
\label{sect:softbody-jellyfish-framework-ui}

The most important contribution of this work, aside from
greatly improving the usability of the simulation system
by scientists, is capture and comprehension of a hierarchy
of the LOD parameters, traditional and non-traditional.
For example, allowing arbitrary number of integration
algorithms constitute a run-time LOD sequence, which
would not normally be considered as LOD parameters, so
we introduce the higher-level LOD components, such as
algorithms, in addition to the more intuitive LOD
parameters like the number of geometric primitives
there are on the scene and so on~\cite{softbody-lod-glui-cisse08}.

We summarize an interactive {\glui}-based interface to the real-time
softbody simulation using {\opengl}. This interactivity focuses not
only the user being able to drag 1D-, 2D-, and 3D-deformable
elastic objects selectively or all at once, but also being able
to change at run-time various ``knobs'' and ``switches'' of the LOD %level-of-detail (LOD)
of the objects on the scene as well as their physically-based modeling
parameters (see, e.g., \xf{fig:object-dimensionality-slides},
\xf{fig:some-integrator-slides}, and \xf{fig:all-objects-rk4}).
We discuss the properties of such an interface in its
current iteration, advantages and disadvantages, and the main
contribution of this work~\cite{softbody-lod-glui-cisse08}.

In \xs{sect:bg-lod-techniques} we mentioned a number of LOD possibilities.
We implement some of them to a degree of the LOD levels in our softbody simulation system.
Most of these details are summarized in the corresponding
publication~\cite{softbody-lod-glui-cisse08} where we outline
the LOD hierarchy and its relationship to the corresponding
GLUI elements. The hierarchy of the LOD parameters varies
from the traditional geometry of the softbody objects, to
the physical simulation parameters, and all the way to
the integration algorithms used, where we consider
the algorithm complexity and accuracy to be an LOD
aspect along the traditional notions for the level
of detail.
As a part of the future work we plan
to extend and deepen further the important notion of
LOD according to the surveyed techniques for the
research and optimization purposes for larger
scale simulations~\cite{adv-rendering-animation-softbody-c3s2e09}.

The software design is centered around the notion of state.
The state is captured by a collection of variables of different
data types to hold the values of the simulation parameters.
The whole architecture follows the Model-View-Controller (MVC)~\cite{larmanUML}
design pattern, where the Model is the state of the geometry
and LOD parameters, View is the simulation window and the GUI
into the model details, and the controller that translates
users actions onto the GUI into the model changes, particularly
the LOD state~\cite{softbody-lod-glui-cisse08}.

The system state has to encode a variety of parameters 
mentioned earlier that are exposed to the interactive user at run-time. 
Our initial iteration of the visual design of the LOD interactivity
interface is summarized in \xf{fig:Objects3}. The LOD components
are on the right-hand-side (in their initial state, not expanded).
And the main simulation window is on the left (interactivity with
that window constitutes for now just the mouse drag and functional
keys).
Following the top-down approach we bring in more details of
configurable simulation parameters~\cite{softbody-lod-glui-cisse08}.

\paragraph{Adjustable Parameters.}

The adjustable LOD parameters can be categorized as dimensionality,
geometry, integration algorithms, force coefficients, and particle
mass~\cite{softbody-lod-glui-cisse08}.

\paragraph*{Dimensionality.}

Switching the 1-, 2-, and 3-dimensional objects present in the
simulation is simply done as checkboxes as shown in~\xf{fig:Dimension}.
The ``Object Properties'' buttons expand the higher-level representation
into the finer details of the objects of each dimension type~\cite{softbody-lod-glui-cisse08}.

The LOD on dimensionality is 1D, 2D, and 3D.
All three types of objects can be rendered on
the same scene at the same time. This is encoded
by the instances of state variables \api{object1D}
of type \api{Object1D}, \api{object2D} of type \api{Object2D},
and \api{object3D} of type \api{Object3D} as well as their corresponding
Boolean flags reflecting the status of whether top render
them or not. This is done to illustrate the object behavior
under the same simulation environment and how the objects
of different dimensionality respond to the same simulation
settings~\cite{softbody-lod-glui-cisse08}.

\paragraph*{Geometry.}

In 1D there are really no geometry changes. In 2D,
the geometry means number of springs comprising the
inner and outer layers of a 2D softbody. Increasing
the number of springs automatically increases the
number of the particles at the ends of those springs,
and brings a better stability to the object, but
naturally degrades in real-time performance.
In 
\xf{fig:LOD2d},
\xf{fig:LOD2d1},
\xf{fig:LOD2d2}, and
\xf{fig:LOD2d3} is an example of the 2D object
being increased from 10 to 18 springs, from the
UI and rendering of the object perspectives~\cite{softbody-lod-glui-cisse08}.

The geometry of increasing or decreasing of
the number of springs in 2D, as in
\xf{fig:LOD2d},
\xf{fig:LOD2d1},
\xf{fig:LOD2d2}, and
\xf{fig:LOD2d3} is not the same approach used
for the 3D objects in the original
framework. The 3D object is constructed
from an octahedron by an iteration of a 
subdivision procedure, as selectively
shown in
\xf{fig:LOD3d}
\xf{fig:LOD3d1},
\xf{fig:LOD3d2}, and
\xf{fig:LOD3d3}. The iteration increases
dramatically the number of geometrical
primitives and their interconnectivity
and the corresponding data structures, so
the efficiency of the real-time simulation
degrades exponentially with an iteration
as well as the detailed time step, which
can make the users wait on slower
hardware. Using the LOD GLUI components we
designed, a researcher can fine-tune the
optimal simulation parameters for the system
where the simulation is running. Since it is
a combination of multiple parameters, the complexity
between editing and seeing the result is
nearly immediate compared to the recompilation
effort required earlier~\cite{softbody-lod-glui-cisse08}.

\begin{figure*}[ht]
\hrule\vskip4pt
\begin{center}
	\subfigure[LOD 2D UI with 10 Springs]
	{\includegraphics[width=2in]{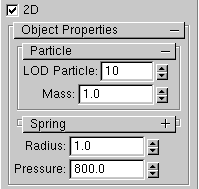}
	 \label{fig:LOD2d}}
	\hspace{.3in}
	\subfigure[LOD 2D Object with 10 Springs]
	{\includegraphics[width=2in]{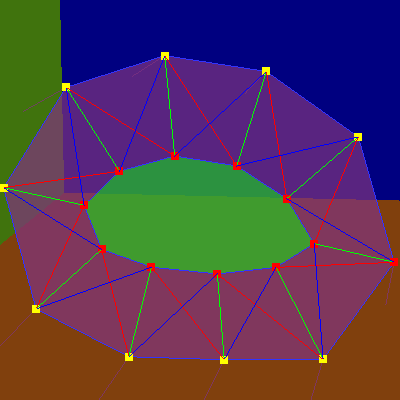}
	 \label{fig:LOD2d1}}
	\subfigure[LOD 2D UI with 18 Springs]
	{\includegraphics[width=2in]{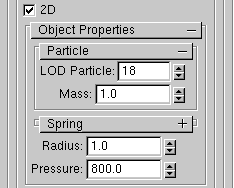}
	 \label{fig:LOD2d2}}
	\hspace{.3in}
	\subfigure[LOD 2D Object with 18 Springs]
	{\includegraphics[width=2in]{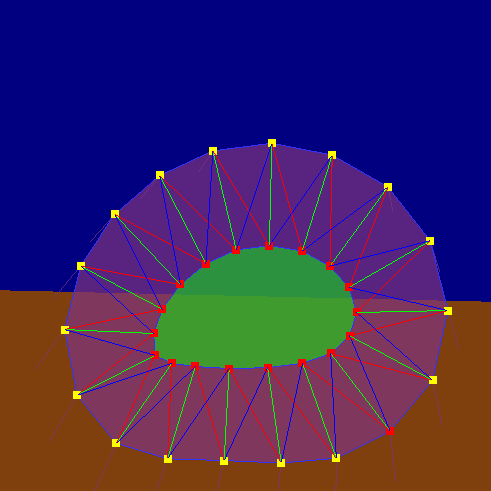}
	 \label{fig:LOD2d3}}
\caption{Some Examples of Various 2D LOD Settings}
\label{fig:2d-lod-examples}
\end{center}
\hrule\vskip4pt
\end{figure*}

\longpaper{IGNORED: \xf{fig:2d-lod-examples}}

\begin{figure*}[ht]
\hrule\vskip4pt
\begin{center}
	\subfigure[LOD 3D UI without Iteration]%
	{\includegraphics[width=2in]{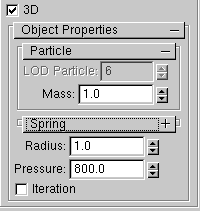}%
	 \label{fig:LOD3d}}%
	\hspace{.1in}%
	\subfigure[LOD 3D Octahedron]%
	{\includegraphics[width=2in]{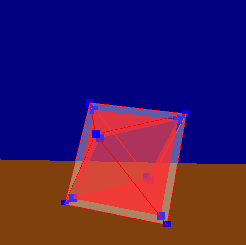}%
	 \label{fig:LOD3d1}}\\%
	\subfigure[LOD 3D UI with Iteration Once]%
	{\includegraphics[width=2in]{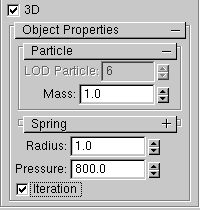}%
	 \label{fig:LOD3d2}}%
	\hspace{.1in}%
	\subfigure[LOD 3D Object with Iteration Once]%
	{\includegraphics[width=2in]{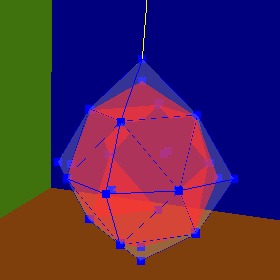}%
	 \label{fig:LOD3d3}}
\caption{Some Examples of Various 3D LOD Settings}
\label{fig:3d-lod-examples}
\end{center}
\hrule\vskip4pt
\end{figure*}

\longpaper{IGNORE: \xf{fig:3d-lod-examples}}

The 2D and 3D objects have an additional LOD parameter
related to their geometry complexity. In 2D the geometry LOD
specifies the number of particles the inner and outer layers
have and by extension the number of structural, radial, and
shear springs they are connected by. The 3D geometry LOD
is based on the way the sphere is built through a number
of iterations and the default set of points upon which
the triangular subdivision is performed on the initial
octahedron. The LOD here is encoded by the \api{iterations}
state variable~\cite{softbody-lod-glui-cisse08}.

\paragraph*{Integration Algorithm.}

Currently available integrator types are in \xf{fig:Integrator},
from the fastest, but least accurate and stable (Euler) to the
most accurate and stable (RK4)~\cite{msong-mcthesis-2007}.
The Softbody Simulation Framework allows addition of an arbitrary number
of implementations of various integrators that can be
compared in real-time with each others by switching from
one to another while the simulation is running~\cite{softbody-lod-glui-cisse08}.

\begin{figure*}[ht]
\hrule\vskip4pt
\begin{center}
	\subfigure[Object Type Dimensionality]%
	{\includegraphics[width=1.6in]{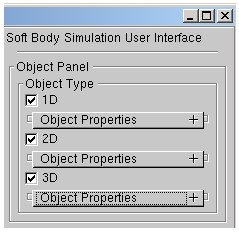}%
	 \label{fig:Dimension}}%
	\hspace{.3in}%
	\subfigure[Integrator Type LOD UI]%
	{\includegraphics[width=1.6in]{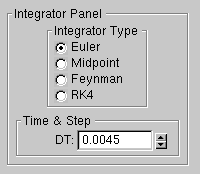}%
	 \label{fig:Integrator}}%
	\hspace{.3in}%
	\subfigure[Force Coefficients LOD UI]%
	{\includegraphics[width=1.6in]{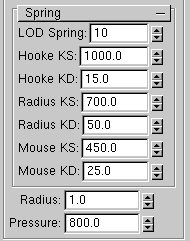}%
	 \label{fig:ForceCoefficient}}%
\caption{Some Examples of Various LOD Parameters}
\label{fig:lod-params}
\end{center}
\hrule\vskip4pt
\end{figure*}

\longpaper{IGNORED: \xf{fig:lod-params}}

The algorithmic LOD includes the selection of a physical
interpolation and approximation algorithms for the run-time
integration of the velocities and acceleration exhibited
by the particles based on physical laws, such as Newton's,
and the different types of integrators: Euler, mid-point,
Feynman, and RK4. These are the four presently implemented
integrators in the Integrators Framework of the system,
so the corresponding state variable is \api{integratorType}
that allows selecting any of the available integrators
and witness the effects of such selection at run-time.
The LOD aspect here goes about the simplicity and
run-time performance of an algorithm implementation
versus accuracy of approximation of the physical motion,
with the least accurate (but fastest) being the Euler's integrator,
and the most accurate (but slowest) being the RK4~\cite{softbody-lod-glui-cisse08}.

\paragraph*{Force Coefficients.}

In \xf{fig:ForceCoefficient}, we provide in the current form a way to
adjust the coefficients used in force accumulation calculations as
applied to each particle. KS and KD represent elasticity and damping
factors of different types of springs. The default Hooke parameter refers to
the structural springs comprising the perimeter layers of the object,
and then the radius springs, as well as the spring created by a mouse
drag of a nearest point on the object and the mouse pointer~\cite{softbody-lod-glui-cisse08}.

There are a number of forces that are taken into 
account in the simulation process. Each of them
corresponds to at least one state variable supporting
the simulation and the interface. The coefficients
correspond to Hooke's law forces on springs, gravity
force, drag force, gas pressure (for 2D and 3D enclosed
objects), and the collision response force. The
spring forces (structural on each layer, radial,
shear, and mouse drag) typically have the elasticity spring
coefficient, $KS$, and the damping force coefficient, $KD$.
The gravity force by default is $m \cdot g$, where both
the mass $m$ of a particle and $g$ can be varied
as LOD parameters~\cite{softbody-lod-glui-cisse08}.

\paragraph*{Particle Mass.}

Each object's properties have sub-properties (e.g., velocity, position, etc.) of comprising it
particles, including mass, as exemplified in
\xf{fig:LOD2d},
\xf{fig:LOD2d2},
\xf{fig:LOD3d}, and
\xf{fig:LOD3d2}.
We argue mass is an LOD parameter as each layer's particles of
the softbody object can get different masses (but uniform across
the layer), or even different masses for individual particles.
The latter case however does not scale well in the present
UI design, and has to be realized differently~\cite{softbody-lod-glui-cisse08}.
More specifically,
particle mass is another LOD parameter that affects
the simulation and can be tweaked through the interface.
Internally, the simulation allows every particle to have
its own mass (or at least two different layers in the
softbody can have two distinct particle weights), but
it is impractical to provide a GUI to set up each particle
with such a weight, however, in most simulations we are
dealing with the particles of uniform mass, so we can
allow resetting the mass of all particles with a single knob.
Thus, the default particle mass LOD parameter is mapped
to the \api{mass} state variable~\cite{softbody-lod-glui-cisse08}.

\paragraph{{\gpu} Shaders.}
\label{sect:softbody-jellyfish-gpu}
\label{sect:softbody-jellyfish-shader}

We created a small subframework for GPU shaders
to allow to work with the softbody simulation
system. The shaders themselves, the shader vertex
and fragment source text files come from various
open source vendors from their examples and
tutorials~\cite{glsl-clockworkcoders,opengl-codecolony,glsl-lighthouse3d,glsl-typhoonlabs,opengl-glsl-quick-guide,mokhov-initial-hair-modeling-project04}.
We implemented an abstract
ability to load and enable the shading of softbody
objects from either assembly shaders or {\glsl}
shaders. Our abstraction layer makes the softbody
system dependent only our own API, and we provide
the adapters to the code written by the other open
source developers. The class declarations are found in the
\apipackage{include/GPU} directory and called \api{Shader}
and \api{ShaderLoader} and their implementation is
in \apipackage{GPU}. There are their concrete implementation
classes such as \api{RGSMShaderAdapter} and \api{CWCShaderAdapter}
to map our functions to the loading and initialization functions
of the concrete shader loaders for the assembly (former)
and GLSL (latter) respectively. We tested some example
shader programs for their interesting material effects on
the softbody objects and the scene that worked.
Thus we succeeded to run the softbody system either
with the {\glsl} shaders to assembly vertex and fragment
shaders, but could not fully experiment with this
feature. However, it opened the doors for further
experiments and research and writing softbody-specific
shaders in the future work, including shader-based
texture mapping, stereoscopy, normal computation,
and others to speed things up. The screenshots
in \xf{fig:softbody-brick-assembly-shader} and
\xf{fig:softbody-green-glsl-shader} are simple
shading examples applied to softbody system
as a proof-of-concept.

\begin{figure}[htp!]
	\centering
	\includegraphics[width=\textwidth]{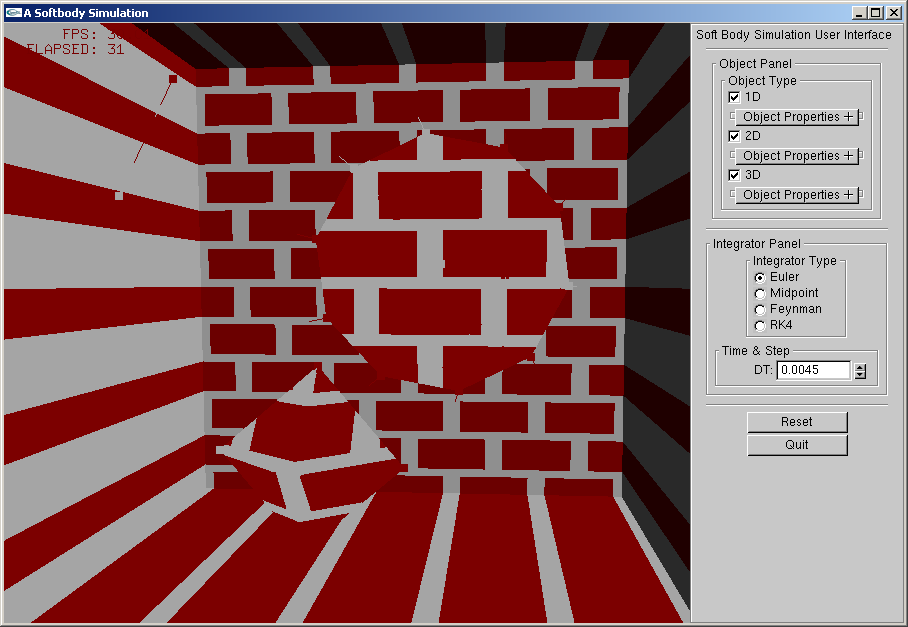}
	\caption{Simple Brick Assembly Shader Applied to the Softbody Simulation System}
	\label{fig:softbody-brick-assembly-shader}
\end{figure}

\begin{figure}[htp!]
	\centering
	\includegraphics[width=\textwidth]{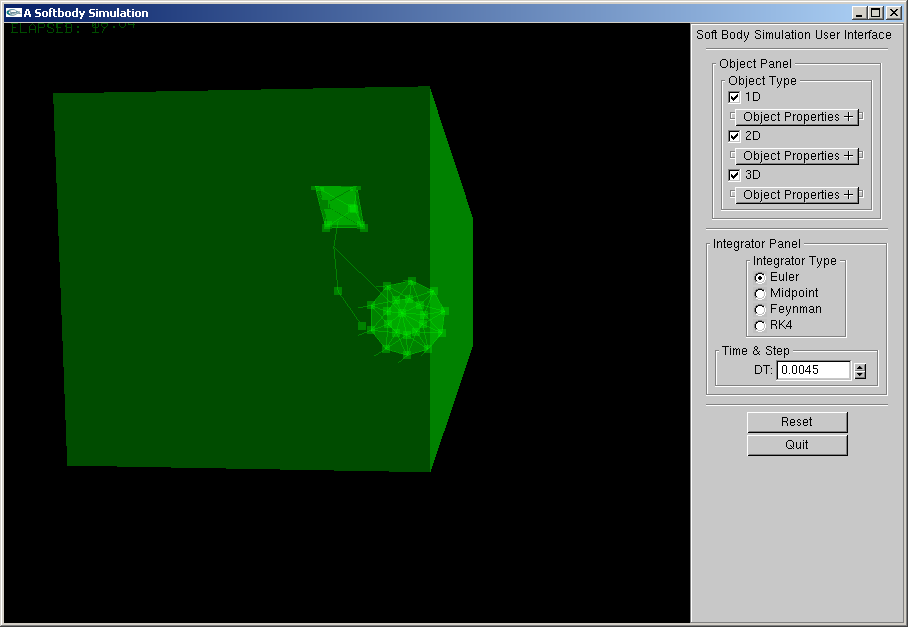}
	\caption{Simple GLSL Shader Applied to the Softbody Simulation System}
	\label{fig:softbody-green-glsl-shader}
\end{figure}

\longpaper
{
\paragraph{Stereoscopic Effects.}
\label{sect:softbody-jellyfish-stereo}

\cite{stereo3d}

A prototype two-layer elastic softbody object simulation
system was implemented using {\opengl} with multiple goals
in mind \cite{softbody-framework-c3s2e08,msong-mcthesis-2007}. One of the goals was modeling surgeon training
with haptic devices in a AR-type of system. Prior moving
to the use of haptic devices we need to define a notion
of softbody 3D-graphics based service. The system is
implemented on top of an extensible C++ framework with
recently added hooks to the LOD controls and the corresponding
GUI~\cite{softbody-lod-glui-cisse08}. Our goal in this work is to devise a methodology
of exposing the settings of the system in a form of a
library to allow its use as a service for a future
application of virtual surgical training.
\cite{adv-rendering-animation-softbody-c3s2e09}

When animators create stereoscopic scenes in 3D modeling and animation software, 
such as {\maya}~\cite{maya}, XSI~\cite{xsi}, 3DS Max~\cite{3dsmax} or Blender~\cite{blender},
they must render left and right eye views
and then import and composite them in another software to see the stereoscopic effect. 
Many current and recent tools offered little if any support in their visualization interfaces that allow 
a ``preview'' mode to check the stereo parameters 
before any rendering and compositing. An example of this problematic workflow 
was the production of ``Folding'',
an animated short film by Loader~\cite{loader-folding-movie} and
the feature film ``Journey To The Center Of The {Earth}''~\cite{jce-stereo-07}
both created using older Maya versions.

We propose a plug-in-based interface for {\maya} to
allow preview-based modeling and animation in stereo, using
a multi-view approach and an OpenGL-based core.
Our interface design is split into
Maya-independent OpenGL components that do not
require a stereo-capable graphics card (but
can work with one if available) as well as Maya-specific
use of this core, which is {\cpp}~\cite{complete-maya-programming-vol1} and {\mel}
(Maya Embedded Language)~\cite{mel-companion,mel-scripting,mel-camera-data-api,mel-interfaces,top2maya,complete-maya-programming-vol1,maya-ui-creation}.

Our design and development were done before Maya 2009 came out. Yet
when it came out with the built-in stereo support,
our advantage is that we still work with the previous
Mayas, open-source, suitable for integration into other projects, such as Blender
or ``pure'' {\C}/{\cpp} {\opengl}-based programs, an
we do not depend on the support of stereo buffers
provided by the graphics cards (but we also can take
advantage of one if there is any).

We had initially communicated with Autodesk,
the owners of Maya software, about collaboration but desisted in favour of
releasing the plug-in as open source. We opened a project on the
SourceForge website,
\url{http://sourceforge.net/projects/stereo3d/},
using a BSD (Berkeley Software Distribution) license.
A large portion of the code has been written using {\cpp}
and OpenGL (a Maya-independent core) with integration into the
Maya User Interface embedded language, mel, and a corresponding Maya
{\cpp} plug-in wrapper. By sharing this code, other developers may
improve it and/or apply it to plug-ins for other modeling, animation,
and game engine applications.

A second part of the plug-in tool generates a stereo camera rig using
either new or existing cameras. Our stereo camera rig is a hierarchy of
three cameras -- a central master control camera, and left and right
cameras for stereo viewing and rendering.

In this novel work we combine our previous results on the open-source 
stereoscopic plug-in framework in {\opengl} as well as real-time elastic 
Softbody Simulation System, and their presentation. We render the simulation 
in OpenGL in various stereoscopic techniques (parallel and toed-in anaglyphs, 
interlaced, graphics hardware-based, etc.) to measure their real-time 
overhead, responsiveness, in order to further their application to games and 
large 180-degree-screen virtual reality systems. In this research we work to 
overcome problems encountered during multi-view rendering of an existing 
OpenGL scene and color distortion. Another purpose of this work, aside from 
the stated research objectives, is to create a teaching material on 
stereoscopy and real-time simulation usable in a classroom via the OpenGL 
slides framework. Some preceding founding related work that we combine in 
here has been published separately: 

\url{http://doi.acm.org/10.1145/1670252.1670333}
\cite{stereo-plugin-interface}

\url{http://doi.acm.org/10.1145/1557626.1557647}
\cite{softbody-opengl-slides}

\url{http://doi.acm.org/10.1145/1370256.1370282}
\cite{softbody-framework-c3s2e08}

The design is based on the multi-view rendering
of a scene from two cameras corresponding to 
left and right eye positions. The two camera views are software-rendered,
either side by side or superimposed, into a single Maya view panel. 
These combined camera views 
are made stereoscopic by applying color masks for red-blue anaglyph
filtering, creating interlaced views, or compositing for use with a stereo graphics card. 
There are a significant number of viewing and camera parameters for the
stereoscopic visualization.
Modifying these variables  directly in the interface, 
records  changes on both the C++ and MEL sides 
thus altering the on-screen output in Maya at run-time.

The parameters include the interaxial distance between left and right cameras;
the pixel offset for post production determined by the image size and resolution
and the distance between the viewer and the image; and whether left and right
cameras are parallel or toed in.
These parameters work toward \emph{orthoscopic} (true-to-life) stereoscopy or they may be adjusted to
allow spatial distortion and experimentation.
The last parameter includes the debug mode to trace
the execution of the plug-in internals.

One of the design goals here is to maintain some
common Maya-interface compatibility at the MEL
level such that most people who are familiar with
Maya can get started right away. Thus, the views
and the menus should resemble those standard to
Maya, including most common menu items and default
actions. Users can model, move and animate objects
in the stereo view,  in wireframe or shaded mode, 
and create preview animations quickly without software rendering
and compositing in another software.

real-time modeling and previewing from the stereo
cameras in stereo can be done without any graphics card support.

\begin{figure}[htpb]%
	\centering
	\fbox{\includegraphics[width=\textwidth]{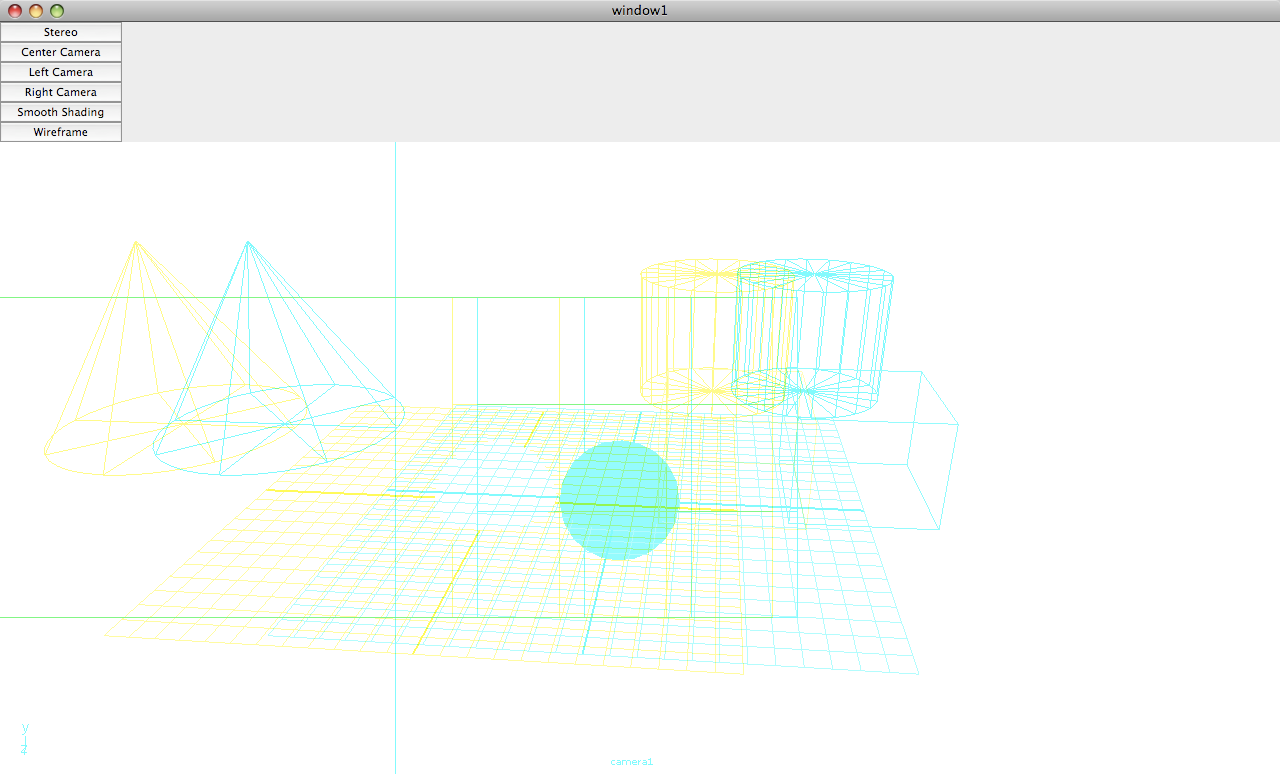}}
	\caption{Stereo multiview rendered together in Maya}%
	\label{fig:stereo-rendered-multiview}%
	\label{fig:stereo-window}
\end{figure}

\begin{figure*}[htb!]
	\centering
	\includegraphics[width=\textwidth]{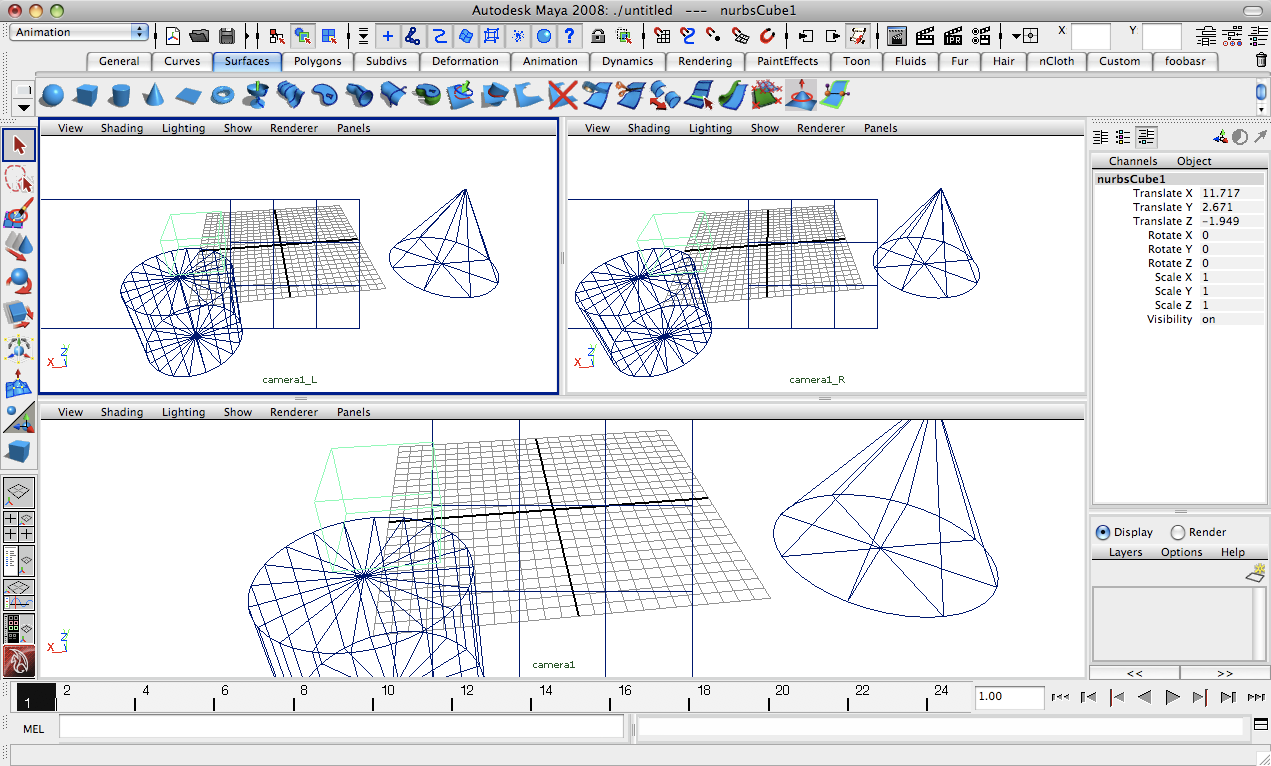}
	\caption{Stereoscopic plug-in setup in {\maya}}%
	\label{fig:stereo-modeling-views-maya}%
	\label{fig:maya-views}
\end{figure*}

IGNORED: \xf{fig:maya-views}, \xf{fig:stereo-modeling-views-maya}, \xf{fig:stereo-window}, \xf{fig:stereo-rendered-multiview}

We drew a lot of {\mel} programming tricks from a wide array of
references~\cite{mel-companion,mel-scripting,%
mel-camera-data-api,mel-interfaces,top2maya,complete-maya-programming-vol1,maya-ui-creation}.
We also consulted multiple resources for information about stereoscopy-related work in OpenGL,
C++ plug-ins for Maya, stereo graphics card manipulation and so
on~\cite{%
orthostereo,%
stereo-opengl-tutorial-gali,%
maya-api-mpxtools,%
maya-programming-code,%
maya-exporter,%
devkit-build-area-setup,%
rawkee,%
maya-plugins-highend3d,%
maya-mfncamera,%
lookat-vs-aim-constraint,%
uri-teaching-animation,%
pangeasoft-macosx-game-book,%
complete-maya-programming-vol2%
} in addition to those referenced earlier.
For historical background on stereoscopy and related
setups we examined another
subset of work cited here~\cite{%
retour-3d-stereo-07,%
about-stereo-cinema-49,%
media-art-in-3d,%
history-filmography-stereo-cinema-89,%
stereoscopy-where-79,%
stereo-foundations-82,%
stereoscopic-cinema-innovation-07,%
glasses-required-07,%
stereo-realist-manual,%
stereo-sue-06,%
binocular-vision-38,%
3d-filmmakers-3005%
}.

A number of equations have been built directly into our
stereo camera rig.  The user must select and input specific values --
output resolution, screen size, viewer distance, and interaxial distance. 
For parallel camera setups, the pixel offset required in post production
feeds back automatically and the camera lenses
are set to an orthoscopic (true-to-life) focal distance.  The user may adjust the
interaxial attribute to produce a deeper or shallower stereo effect without
altering the planar composition. A magnification attribute multiplies
the focal length to add spatial distortion. This gives an artist
considerably more control over the final output.

However the stereoscopic preview will not be accurate if the preview size 
and viewing proximity is different than the final output.
A stereo image projected onto a large screen
will look different than the same image displayed on a much smaller
computer screen. Thus stereo view plug-in makes some compensation for output size. It
generates a second set of offset cameras that are placed closer
together generating a shallower stereo image more suitable for display
on a monitor. The OpenGL either queries for or passes down from the
overlying plug-in layer the ratio of final output size to preview
output to calculate the distance between the secondary
offset cameras and performs the correct pixel offset normally done in
post-production.

The stereoscopic plug-in programmed for production in Maya does not
replace final output testing.  Hardware-rendered previews, which produce poor
color and image quality, are further degraded by the anaglyph
process. Furthermore, even though the plug-in compensates for screen size, it
cannot reproduce final viewing conditions.  Yet working in
real-time in an environment that replicates the final output
environment is rarely an option.  

However, these new stereo tools are very
useful for the real-time stereo preview mode. Any modeler or artist is no longer
compelled to work blindly on a flat screen and may develop a better
understanding of stereo space.  The camera rig when used properly
generates predictable stereoscopic illusions but is flexible enough to
produce intentional spatial distortion. Additional work is being done to
experiment with other stereo viewing methods such as the axis-aligned
frustum and interlaced stereo. Our plug-in framework is capable of
working with the set-ups where stereo-buffer graphics cards are not
available (which are the only target of similar tools these from Maya
devkit or elsewhere).
Such tools provide more power and control to the modeler, essential for the production
of meaningful and enduring work.
}

\subsubsection{Jellyfish Application of the {\softbodysys}}
\label{sect:softbody-simulation-subsystem}
\label{sect:softbody-jellyfish-in-jellyfish}

We overview our approach to implement a realtime simulation of a {\jellyfish}
using our physical based {\softbodysys} founded in laws of physics
and spring-mass systems, visualized in {\opengl}~\cite{jellyfish-c3s2e-2012}.
We outline our progress so far and the difficulties encountered
and ways we solve some and propose to solve some other difficulties
in this work.

We describe details of modeling and implementation of an
interactive work of the initial prototype of {\jellyfish}
using {\opengl} and the {\softbodysys} in \xs{sect:softbody-jellyfish-jelly-modeling}.
We review the animation and interaction aspects in \xs{sect:softbody-jellyfish-animation-interaction}
and lighting in \xs{sect:softbody-jellyfish-lighting}.

\paragraph{Jellyfish Softbody Modeling.}
\label{sect:softbody-subsys-modeling}
\label{sect:softbody-jellyfish-jelly-modeling}

The modeling for the {\jellyfish} softbody simulation includes environment (sea world),
2D jellyfish (see \xf{fig:2D-jellyfish} and \xf{fig:jellyfish-lights}),
and 3D jellyfish\longpaper{ (see \xf{fig:jelly-fish-model}, \xf{fig:3D-jellyfish-head-normals},
\xf{fig:jellyfish-particles-and-shell}, and related)}.
There is no 1D jellyfish.

The {\jellyfish} objects inherit from the {\softbodysys} \api{Object2D}
and \api{Object3D} classes, whose geometry was altered fit into the
approximation of the {\jellyfish} shape in terms of particles and the
corresponding connecting springs.

The tentacles framework is designed to provide 2D and 3D tentacles
of various configurations, animation, and LOD (\xs{sect:tentacles}).

\longpaper
{
The fluid simulation framework.
{\todo}
}

\begin{figure}[htp!]
	\centering
	\includegraphics[width=\textwidth]{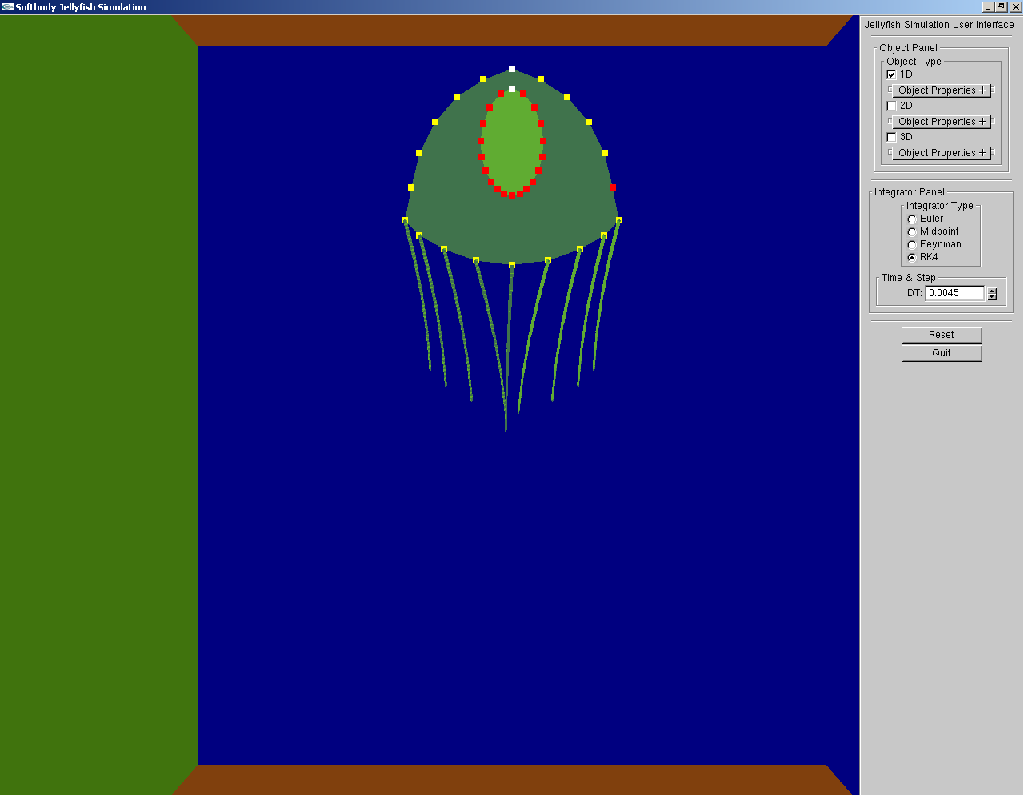}
	\caption{2D jellyfish}
	\label{fig:2D-jellyfish}
\end{figure}

\begin{figure*}[ht]
\hrule\vskip4pt
\begin{center}
	\subfigure[]
	{\includegraphics[height=2.6in]{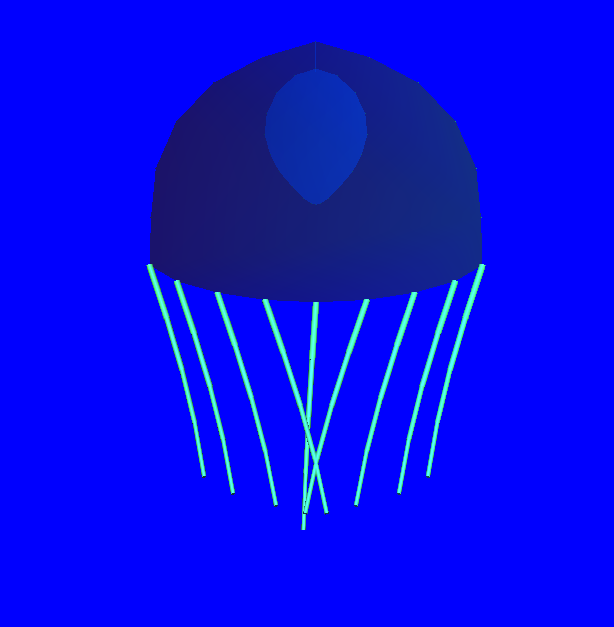}
	 \label{fig:jellyfish-lights-1}}
	\hspace{.3in}
	\subfigure[]
	{\includegraphics[height=2.6in]{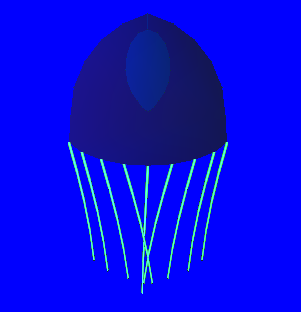}
	 \label{fig:jellyfish-lights-2}}
	\subfigure[]
	{\includegraphics[width=2.6in]{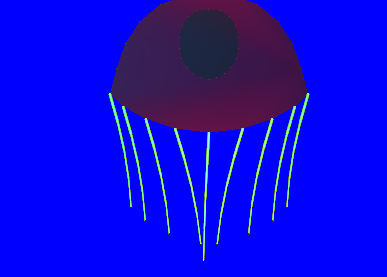}
	 \label{fig:jellyfish-lights-3}}
	\hspace{.3in}
	\subfigure[]
	{\includegraphics[width=2.6in]{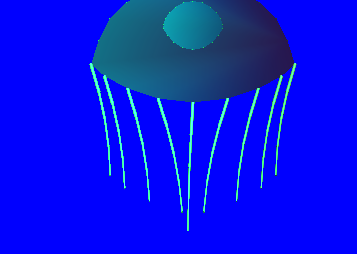}
	 \label{fig:jellyfish-lights-4}}
\caption{2D Jellyfish Lit Up with Lights}
\label{fig:jellyfish-lights}
\end{center}
\hrule\vskip4pt
\end{figure*}

\longpaper{IGNORED: \xf{fig:jellyfish-lights-1}, \xf{fig:jellyfish-lights-2}, \xf{fig:jellyfish-lights-3}, \xf{fig:jellyfish-lights-4}}

\longpaper
{
\begin{figure}[htpb]%
	\centering
	\includegraphics[width=\columnwidth]{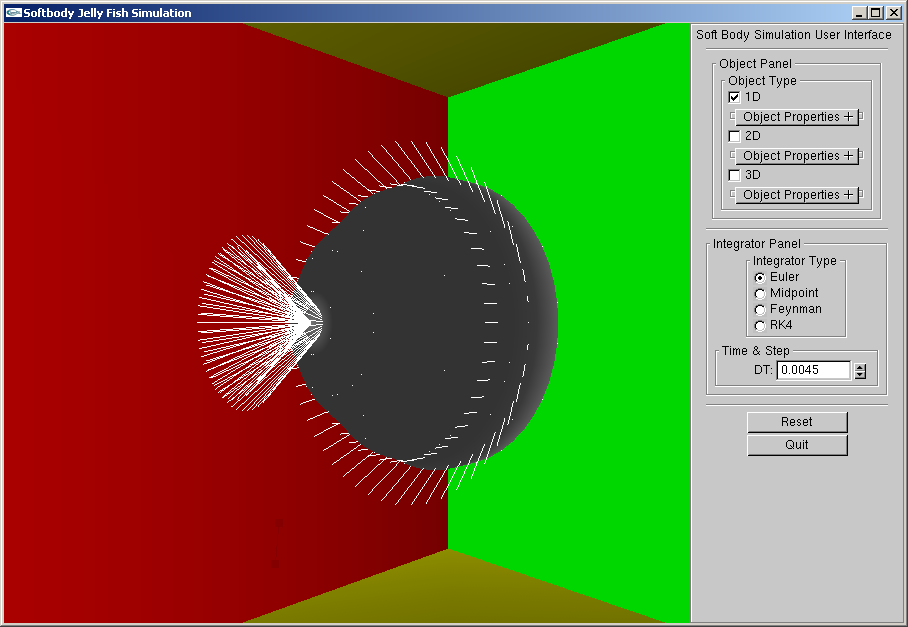}%
	\caption{3D Jellyfish Model}%
	\label{fig:jelly-fish-model}%
\end{figure}

\begin{figure}[htpb]%
	\centering
	\includegraphics[width=\textwidth]{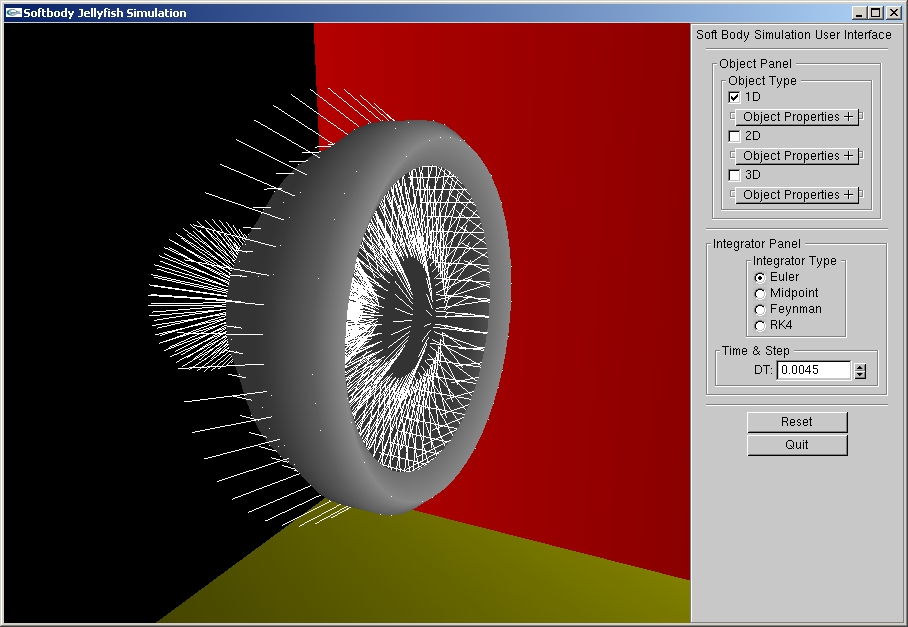}
	\caption{3D Jellyfish with Bell Normals}%
	\label{fig:jellyfish-head-normals}%
	\label{fig:3D-jellyfish-head-normals}
\end{figure}

\begin{figure}[htpb]%
	\includegraphics[width=\columnwidth]{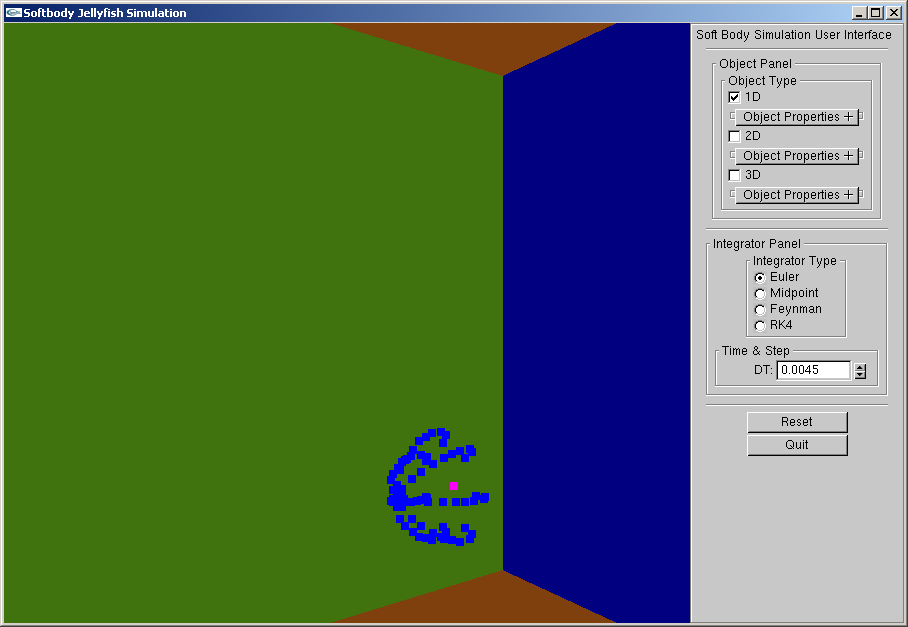}%
	\caption{Jellyfish Bell Individual Particles (Physical Based)}%
	\label{fig:jellyfish-head-individual-particles}%
\end{figure}

IGNORED: \xf{fig:jellyfish-head-individual-particles}

\begin{figure}[htpb]%
	\includegraphics[width=\columnwidth]{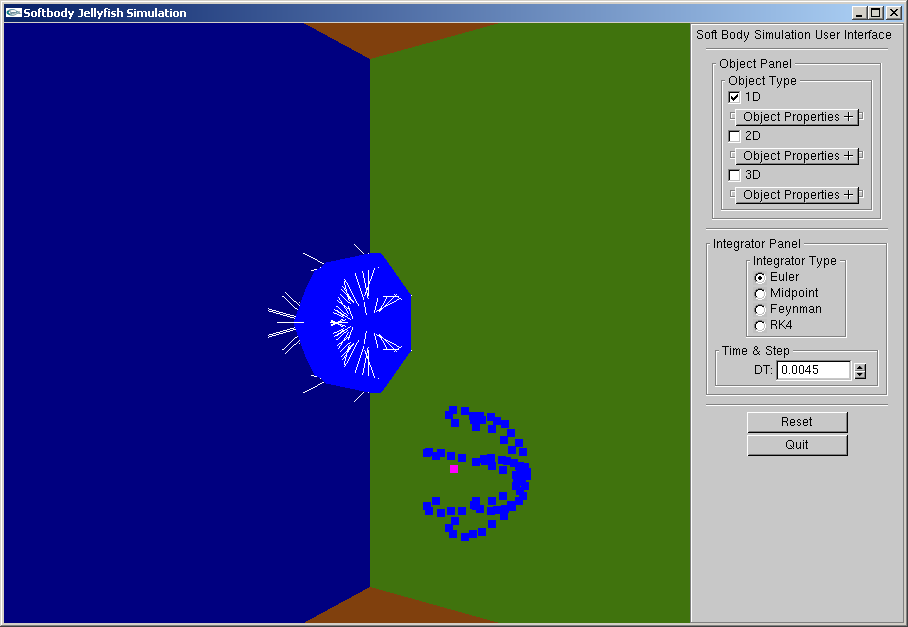}%
	\caption{Jellyfish Bell Individual Particles (Physical Based) and Shell (Same Scale)}%
	\label{fig:jellyfish-particles-and-shell}%
\end{figure}

\begin{figure}[htpb]%
	\includegraphics[width=\columnwidth]{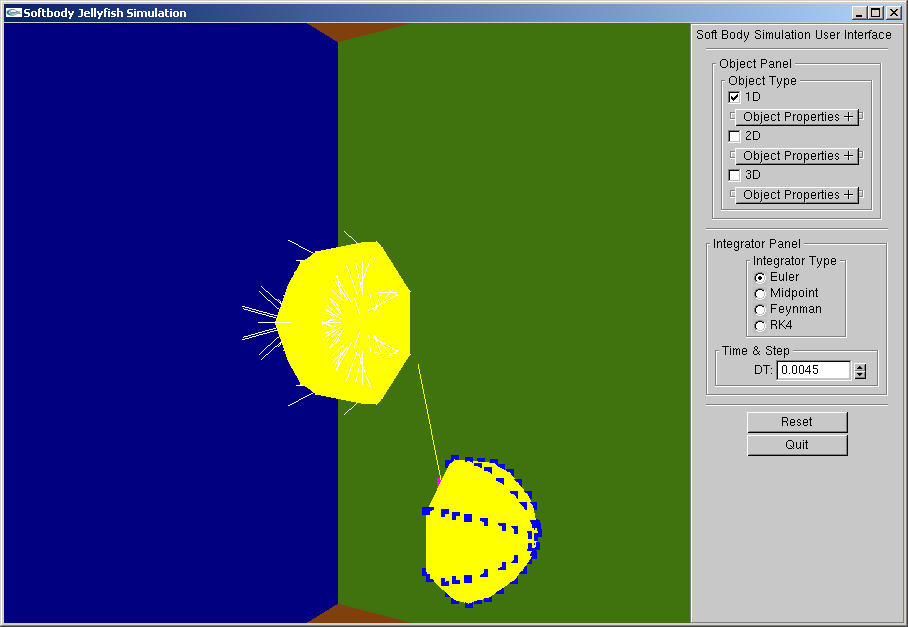}%
	\caption{Jellyfish Bell Static vs. Physical Based Faces}%
	\label{fig:jellyfish-static-vs-dynamic-faces}%
\end{figure}

\begin{figure}[htpb]%
	\includegraphics[width=\columnwidth]{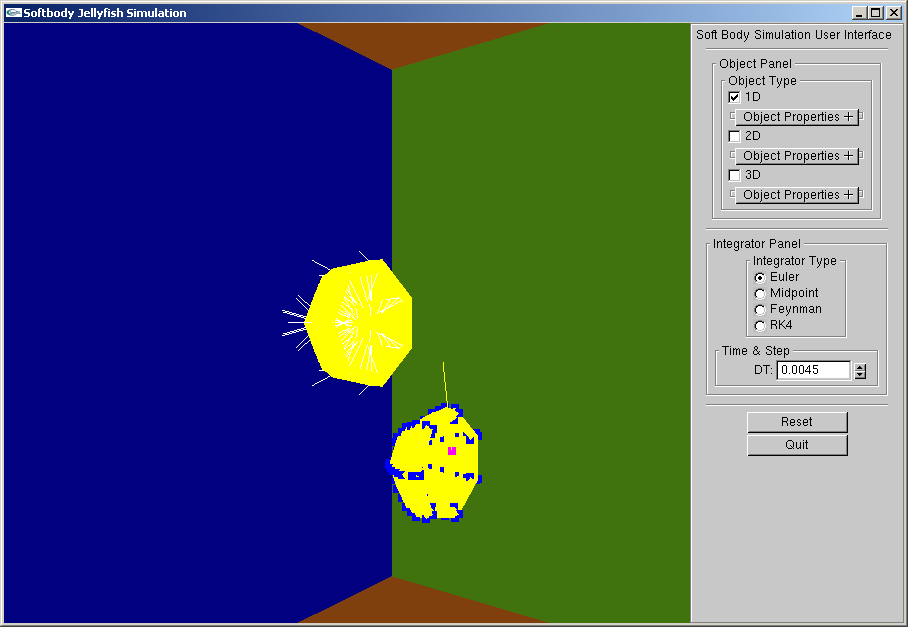}%
	\caption{Jellyfish Bell Static vs. Physical Based Faces In the Same Orientation}%
	\label{fig:jellyfish-static-vs-dynamic-faces-turned}%
\end{figure}

IGNORED: \xf{fig:jellyfish-static-vs-dynamic-faces}, \xf{fig:jellyfish-static-vs-dynamic-faces-turned}

\begin{figure}[htpb]%
	\includegraphics[width=\columnwidth]{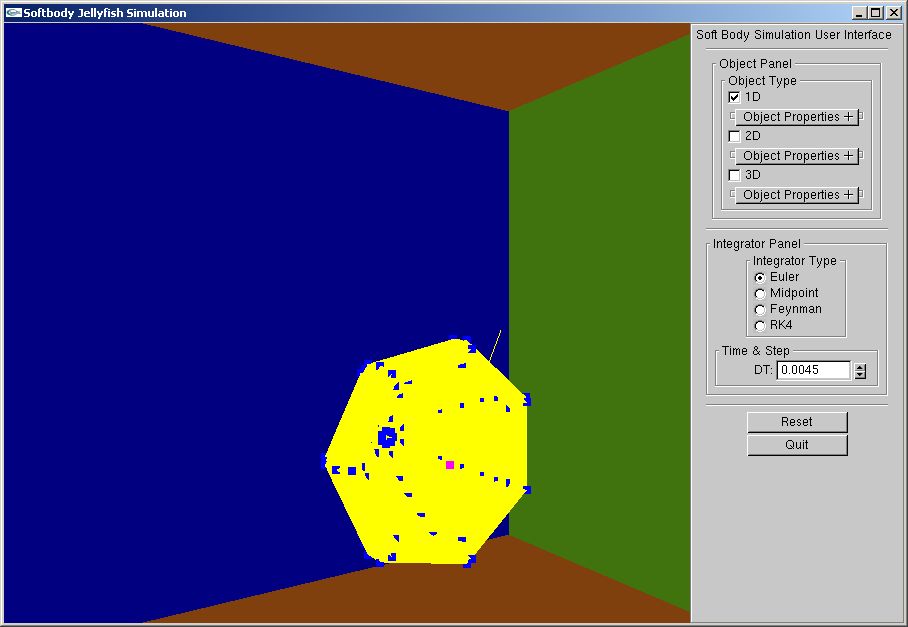}%
	\caption{Jellyfish with Physical Based Faces Scaled}%
	\label{fig:jellyfish-dynamic-turned-scaled}%
\end{figure}
} % end of \longpaper

\paragraph{2D.}
\label{sect:softbody-jellyfish-jelly-2d}

\longpaper
{
The 2D jellyfish model has been completed and 3D jellyfish is
partially done, see \xf{fig:2D-jellyfish} and \xf{fig:3D-jellyfish-head-normals}.
{\todo}
}

The 2D jellyfish (\xf{fig:2D-jellyfish}) is a structural modification of
a two-layered 2D softbody object with its spring-mass system for its bell. The
user retains the capability to ``drag'' it using the mouse, or it ``drags itself'' with the frontal
spring to simulate the real-time swimming realistically and let the physical 
laws in softbody do the rest. The tentacles (\xs{sect:tentacles}) are cylindrical joints individually
animated kinematically while attached to the bottom particles of the bell~\cite{jellyfish-c3s2e-2012}.

The dragging is done via the newly designed \api{Drag} object as a part
of the redesign and refactoring. In the nutshell, it simulates the user
pulling onto one particular of the particles of the softbody object
periodically and that causes the whole system to animate.

The pressure value on the enclosed {\jellyfish} is also set so
that it helps the animation of movement and ``breathing'' combined
with the drag.

\longpaper
{
\paragraph{3D.}
\label{sect:softbody-jellyfish-jelly-3d}

The 3D jellyfish bell is modeled using a Bezier curve revolved~\cite{cugl}
around an axis based on $12$ points and $51$ slices as shown in
\xf{fig:jellyfish-head-normals}.
At the current implementation this level of detail (LOD) turned out to be prohibitive on
a commodity hardware in terms of performance, so we reduced it to 7 slices.
Then, the points are
converted to physical based particles interconnected with a spring-mass
system of radial and shear springs to form a closed
softbody object (see \xf{fig:jellyfish-dynamic-turned-scaled}, \xf{fig:3D-jellyfish-head-normals}, and related figures).
The standard forces and pressure are
applied later to the object as a whole in the idle function and based
on the softbody simulation system.
The normals are set along with the texture coordinates for proper shading
and texture mapping.
While tentacles (not shown) are presently modeled using a
fast hair simulation environment \cite{mokhov-initial-hair-modeling-project04}
that models them either as lines or cones or cylinders depending on
the hardware at hand.
All this is to achieve real-time simulation~\cite{jellyfish-c3s2e-2012}.
}

\paragraph{Tentacles.}
\label{sect:tentacles}

True tentacles are rather complex to model and render in real-time for the animation
purposes as they would rely on complex modeling and animation inspired from the
biomechanical modeling and description of {\jellyfish}
locomotion~\cite{bio-mechs-jellyfish-swimming-2002,synthetic-engineered-jellyfish}
and fluid dynamics.
Instead we build upon simplified models that are attached to and depended on the
{\jellyfish} bell softbody design or have an independent own proper motion.
We also define \apipackage{tentacles} as a mini-framework, and provide incrementally
different implementations of it, from simpler and less accurate, but fastest,
to more realistic and detailed, suitable for PoC demonstrations and adaptive
LOD depending on the hardware available where the program is run~\cite{jellyfish-c3s2e-2012}.

We produced an implementation based on inverse kinematics (done in 2D at the present) with
the tentacles as cylindrical joints individually animated kinematically as \api{Linkage}~\cite{PG021,PG022,cugl}
while attached to the bottom particles of the bell~\cite{jellyfish-c3s2e-2012}. The other
implementations in the work involve the use of the softbody elastic springs,
or the latter two mixed to some degree with the hair/thread simulation from
another project~\cite{jellyfish-c3s2e-2012}.

\longpaper
{
The tentacles of a 3D jellyfish, which have not been fully completed yet,
are rather complex to model and render in realtime for the animation
purposes. We also define this as a ``mini framework'', and provide incrementally
different implementations of it, from simpler and less accurate, but fastest,
to more realistic and detailed, suitable for PoC demonstrations and adaptive
LOD depending on the hardware available where the demo is ran.
The first implementation of the tentacles is provided by the realtime 
implementation of the earlier project by 
Mokhov~\cite{mokhov-initial-hair-modeling-project04} from thread-like
objects simulation that includes, hair, grass, etc. It is simple and fast,
allowing connected animated threads via meshes of line, triangle and
cylinder types (the latter allow texture mapping for visual realism).
This is the first instance of the tentacles mini-framework implementation.
We are considering other implementations as they become available, such
as based on inverse kinematics, softbody, or the latter two mixed to
some degree with the hair simulation.

The first implementation of the tentacles is provided by the realtime 
realization of the earlier project by Mokhov~\cite{mokhov-initial-hair-modeling-project04} from thread-like
objects simulation that include hair, grass, etc. It is simple and fast,
allowing connected animated threads via meshes of line, triangle, and
cylinder types (the latter allow texture mapping for visual realism).
This is the first instance of the tentacles mini-framework implementation~\cite{jellyfish-c3s2e-2012}.
}

\paragraph{Environment.}
\label{sect:softbody-jellyfish-env}

An undersea project~\cite{real-time-sea-world-2003} would serve
as an ultimate environment for this interactive simulation;
which is currently abstracted by a partially translucent ``aquarium box''.
Presently, this box is waterless environment with gravity~\cite{jellyfish-c3s2e-2012}.
The undersea scene will include some seaweed, fish, dolphin,
and other sea creatures, as, e.g., modeled in \animtitle{Spectacle}~\cite{spectacle-anim-film-2003}.

For a blackbox installation, the back wall or all walls, ceiling and floor in the
box will have an AVI movie played back the author shot at a Vancouver zoo of the
real jellyfish in the background.
The worlds framework design is detailed further in the next section.

\paragraph{The Worlds Framework.}
\label{sect:softbody-jellyfish-worlds}

The system has been extended to allow switching between different scenery worlds
in the simulation relatively easier (before it was nearly impossible). The default \api{ViewSpace},
which was an open box became a framework of worlds. It's default implementation
got split out into the \api{worlds::Box} while defining the initial API that
could be used to implement other worlds. The second world being added to the
system using this new API is the previously mentioned underwater sea world conceived by Song {\etal} from
an earlier project~\cite{real-time-sea-world-2003}, which is a more natural aesthetic
environment for a {\jellyfish} than a testing box~\cite{jellyfish-c3s2e-2012}.

At the moment, switching between the worlds is done in the main application
code by assigning one view space instance or another via `W'. A {\glui} option for this is
planned in one of the several revisions~\cite{jellyfish-c3s2e-2012}.

\paragraph{Animation and Interaction.}
\label{sect:jellyfish-animation}
\label{sect:jellyfish-interaction}
\label{sect:softbody-jellyfish-animation-interaction}

The animation in this project~\cite{jellyfish-c3s2e-2012} is primarily physical based and
is principally done through the {\softbodysys} and its framework
detailed earlier through the chapter and the previous work~\cite{softbody-framework-c3s2e08}.
The {\jellyfish} is composed of the softbody bell and tentacles
with the bell having several attachment particles arranged in a circular
fashion. The user ``drags'' the jellyfish by those points.
If no interaction is taking place; and self-dragging swimming is not strong enough,
the jellyfish ``falls gently down''
on the floor and gets ``squashed'' there by ``bouncing off'' two or three times
(since there is no physical water environment present yet making the jellyfish
being effectively let go and drop on the floor) just as a plain softbody object would.
As before, the interaction UI on the right hand side allows selection
of different integration algorithms, time step and level of detail
to alter the animation in real-time to select most optimal parameters
for the simulation. Haptic and motion-capture based interaction
are partially implemented as well~\cite{jellyfish-c3s2e-2012}.

\longpaper
{
Tentacle animation animation is ...
{\todo}
}

\paragraph*{{\falcon} Haptics Softbody Interface.}
\label{sect:softbody-jellyfish-haptics}

\begin{figure}[hptb!]%
	\centering
	\includegraphics[width=.8\columnwidth]{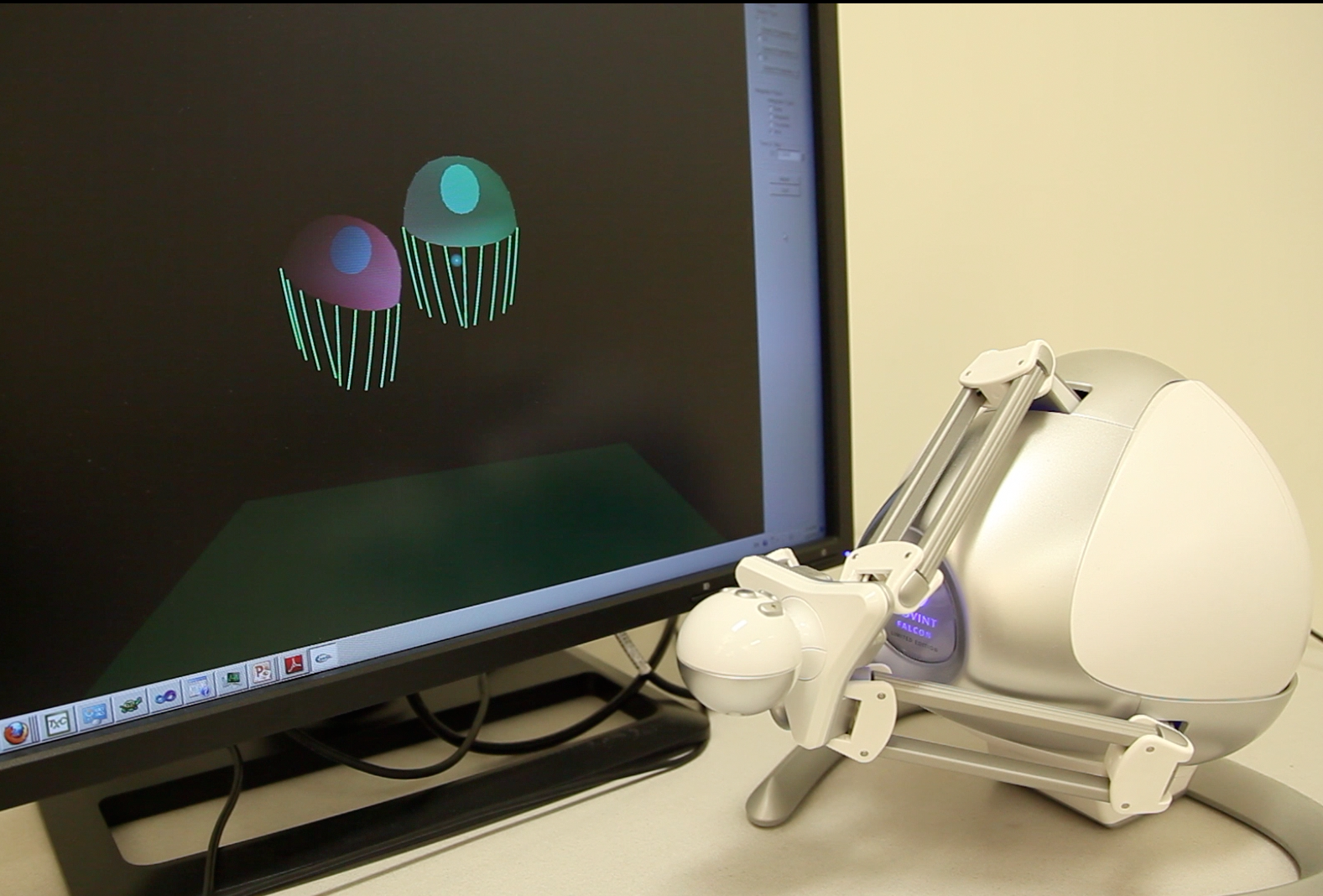}%
	\caption{Jellyfish Connected with Haptics Device}%
	\label{fig:jellyfish-haptics}%
\end{figure}

Here we detail our realization-specific information about the interaction
aspects using a Novint {\falcon} haptic device.
For the background information on {\falcon} haptics see \xs{sect:bg-falcon-haptics}
and \xs{sect:softbody-jellyfish-api-falcon}.

The link between the \api{HapticsClass}~\cite{falcon-sdk} and the {\softbodysys}
is via \api{Object2D}. \api{Object2DJellyfish} and \api{Object3DJellyfish} are
various degree descendants of \api{Object2D} and all can be passed on via 
\api{HapticsClass::init()} at the initialization time.

The subsequent interaction links the button on {\falcon} to the \api{Drag} spring,
same as the mouse, and the virtual haptic ``cursor'', visualized as a sphere in
the \api{BoxWorld}. When the cursor and the {\jellyfish} collide (sphere-softbody-particle
intersection test) the spring-mass forces trigger the servo motors within the device
to partially block it and create a drag force pulling or pushing onto the {\jellyfish}
where the resistance can be felt. In other way, the servo cursor moves along with the
{\jellyfish} and when it's at the bottom, the interfacing user can feel the mass
of it.
In \xf{fig:jellyfish-haptics} is the illustration of two {\jellyfish}, one of which
is interfaced with {\falcon} by mapping the particle forces to the servo motors
and observing the synchronized motion effect mapped exaggerated ``breathing''
felt through the haptics.

\paragraph{Lighting.}
\label{sect:softbody-jellyfish-lighting}

Currently, three moving spotlights light up the scene as if it's undersea highlighting
portions of the projections on the blackbox walls simulating
sun blinking and glints.

\longpaper{{\todo}}

\chapter{Tangible Memories in Interactive Documentary}
\label{chapt:interactive-docu}

Linear documentary story telling has its own limitations and every documented story
becomes an event of the static past after being first released.
What makes an interactive installation documentary piece different from a
traditional documentary is the aspect of interactivity itself
and the audience being able to compose their preferred pathways and storylines selectively,
bidirectionally, with the ability to go into details of interesting points to them unlike
a linear passive progression. In the advanced way, the interaction can be also
immersive and bidirectional with haptics and virtual reality techniques.

Why is it important to introduce interactive
media technology to traditional documentary film making?
The research could be a long journey as witnessed by some of
the items in \xs{sect:interactive-media-docu-background},
but we take a small step at a time.

\begin{quote}\emph{``Mostly, the memories are like little pictures in my head,
like float around in the bubbles... So, when I want to see them,
I search all those bubbles... Then I go into one, I can remember them...''}
\end{quote}
That is how memories are for the little girl in the award-winning short documentary film
\docutitle{I Still Remember}.

We would like to extend this idea to a real touchable
interactive installation. The prototype is augments the
short film with interaction and projection in the form of the \emph{memory bubbles}.

In this chapter we further review the methodology used in \xs{sect:interactive-docu-methodology},
the artistic and software design and implementation in \xs{sect:interactive-docu-design-impl},
and finally the concluding summary in \xs{sect:interactive-docu-summary}.

\section{Methodology}
\label{sect:interactive-docu-methodology}

In general, it is difficult to make a traditional documentary film interactive.
How the computer graphics techniques could possibly be applied to and combined with cinema and
interactive media to me is also a very exciting
topic~\cite{haptics-cinema-future-grapp09,role-cg-docu-film-prod-2009}.
I explore this topic in detail further on by stating general goal, conceptual design,
and my proof-of-concept realization of it.

\subsection{Goal}

First of all, the film elements in interactive documentary should be
controlled by real-time
conditions as well as algorithmically controlled behaviors. 
Then, enhanced with haptic devices and stereoscopic effects, the feeling of the interaction
will consume the audience to be an integral part of the storyline in the movie, which
is a lot much more than a game. And the perceptual experiences can be different each
time the movie is played. Moreover, I wish I could cross over another barrier in 
traditional documentary film: that the cinema screen is flat, a piece of cloth. 
The documentary film could have depth, which would be consciously or unconsciously
associated with audience behaviors.

\subsection{Proof of Concept}
\label{sect:interactive-docu-poc}

\begin{figure*}[htpb!]%
	\centering
	\includegraphics[width=\textwidth]{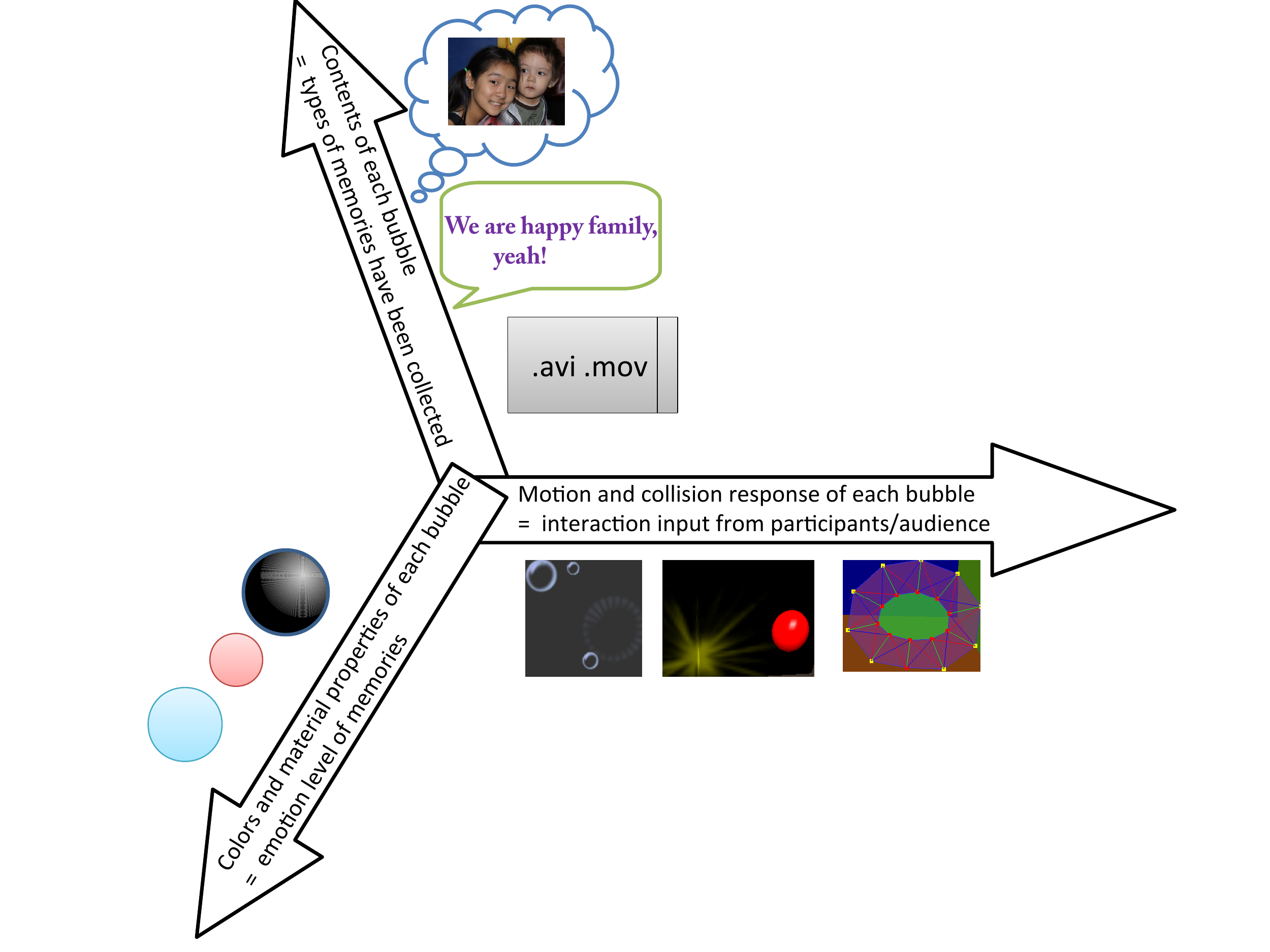}%
	\caption{Overall Interactive Documentary Process}%
	\label{fig:process}%
\end{figure*}

In the interactive documentary installation, the \docutitle{Tangible Memories},
I first made the \docutitle{I Still Remember}
documentary's~\cite{i-still-remember-opengl-remake-2011}
floating memory bubbles interactive with the audience's participation using
a near day-to-day {\opengl}.

\begin{figure*}[htpb!]%
	\centering
	\includegraphics[width=\textwidth]{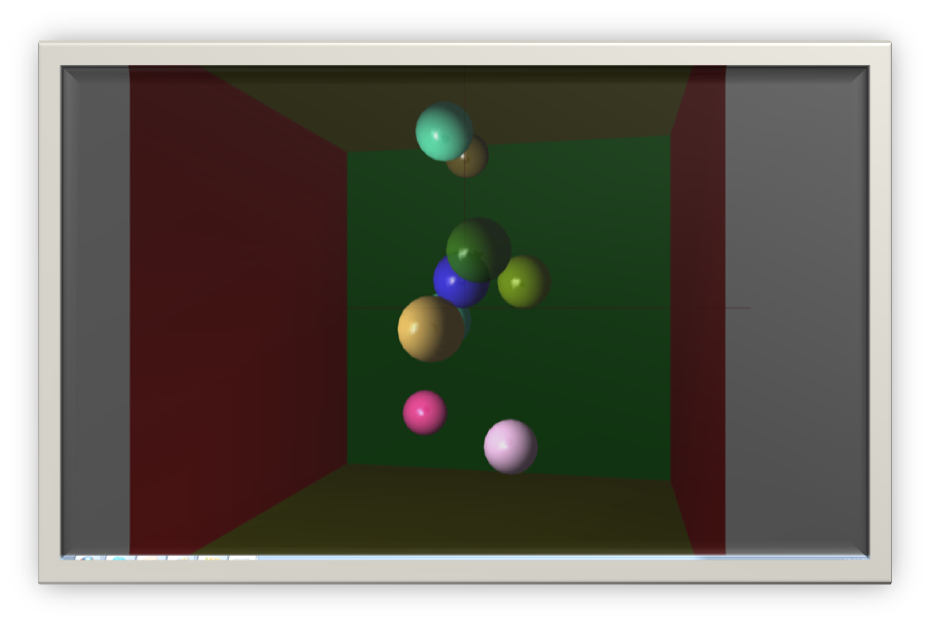}%
	\caption{Simple OpenGL Bubbles Prototype Sample}%
	\label{fig:bubbles-in-box}%
\end{figure*}

By porting some of the of the \docutitle{Tangible Memories} system
with later enhancements using
an inexpensive MoCap system, such as {\kinect} (with {\xna}),
we could achieve the real interaction for documentary perception experience.

The virtual memory bubbles are projected
in a physical environment, such as walls and ceilings with multi-projectors.
Audience will not sit and passively watch the documentary film,
instead, they need to use their body movement or voice to advance the dynamic scenarios
of the story or stories.

The resulting ``artistic high-level algorithm'''s pseudo-description, with the process
and basic design, are conceptually illustrated in \xf{fig:process}, \xf{fig:bubbles-in-box},
and in \xg{algo:conceptual-process-bubbles} respectively.

\begin{algorithm}[ht!]
\hrule\vskip4pt
 The media represent memories could be home
 videos, photos, audio, and animated text (e.g. as in \cite{cityspeak})\;

 Model each bubble object in the first iteration
 as a translucent sphere with a flat polygon inside,
 procedurally\;
 \Begin
 {
   Each bubble is defined by the property of the contained
   media and the outer color. Of course, the bubbles 
   also have a radius $r$ and a position $(x,y,z)$\;
   
   Each bubble has a unique color, which represents the emotion level of memories,
   such as lightheartedness and sadness or rainbow-like colors or ease of calling out\;

   The media polygon is a 2-sided quad, onto
   which a texture photograph can be mapped, or a
   text item with or without animation, or more generally
   a video clip\;

   \tcp{A more advanced bubble modeling under consideration
   is the utilization of the softbody objects instead of
   plain spheres to add some realism and tangibility}
 }

 The audience is allowed to interact with the bubbles, and the bubbles' motion
 and collision response are to be based on the behavior and gestures of the audience\;
 
 \tcp{The effects of bubbles could be soap bubble bursting, softbody dynamics,
 and rigid body sparkling collision}

\hrule\vskip4pt
\caption{``Artistic High-Level Algorithm'' for \docutitle{Tangible Memories}}
\label{algo:conceptual-process-bubbles}
\end{algorithm}

There are two runnable proof-of-concept programs we have developed in {\opengl} and {\xna} to demonstrate
how the 3D bubbles float randomly in the space. Each of them contains
a piece of memory, and will project the bubbles on the environment surfaces,
such as walls, ceiling, or floor.

\section{Design and Implementation}
\label{sect:interactive-docu-design-impl}

The design and implementation of the interactive documentary work focus on
several key aspects: modeling, animation, and interaction. These aspects
are supported by specific technologies used. In \xs{sect:interactive-docu-conceptual-design}
we begin the conceptual overview of the work as it was being designed.
Then the realization details are presented in \xs{sect:interactive-docu-implementation}.

\subsection{Conceptual Design}
\label{sect:interactive-docu-conceptual-design}

\begin{figure*}[htpb]%
	\centering
	\includegraphics[width=\textwidth]{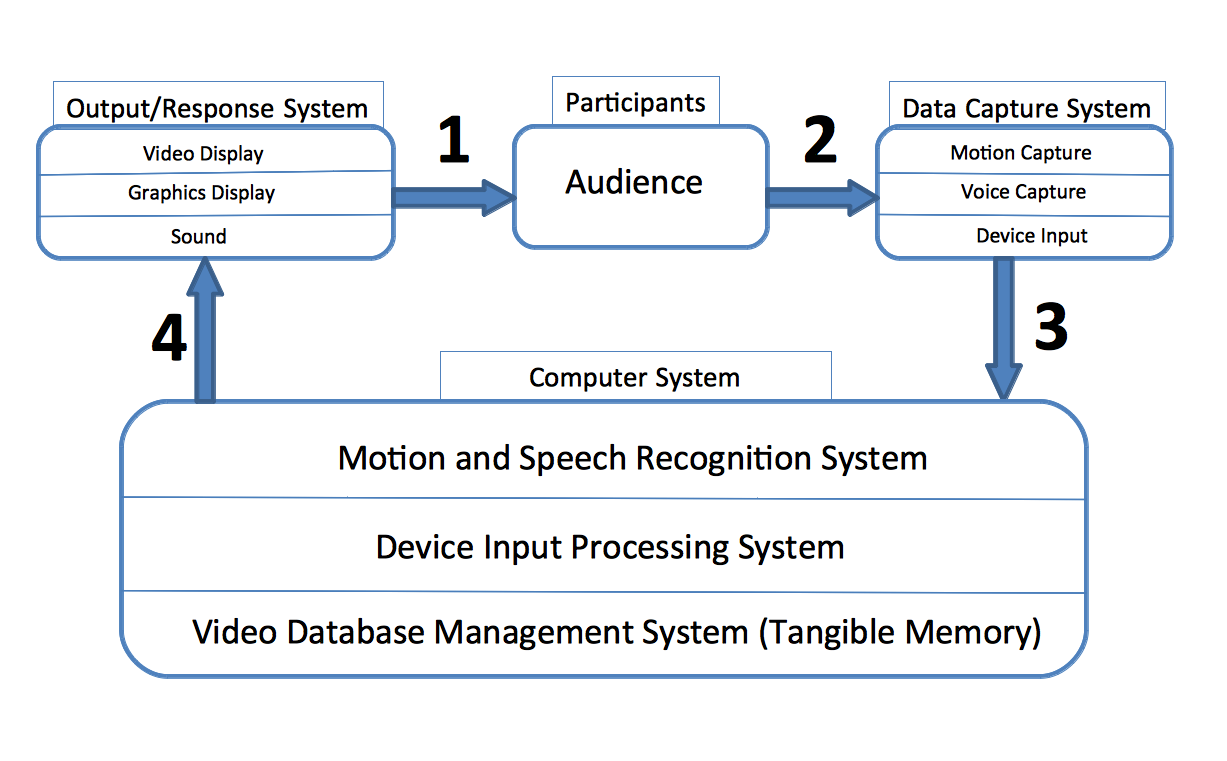}%
	\caption{Conceptual Design of an Interactive Documentary Installation}%
	\label{fig:ConceptualDesign-documentary}%
\end{figure*}

As mentioned,
the \docutitle{Tangible Memories} prototype is the augmented \docutitle{I Still Remember}
short film with interaction and projection of the ``memory bubbles''
in order to visualize how memories are represented by the ten-year-old girl's recollection
(see \xs{sect:interactive-docu-poc}, specifically \xf{fig:process} and \xf{fig:bubbles-in-box})
including additional footage not originally present in the linear \docutitle{I Still Remember}
and the ability to expand it arbitrarily further in the future.

The system's conceptual design is presented in \xf{fig:ConceptualDesign-documentary},
which is an evolved version of the design from \xc{chapt:softbody-objects} (following
the same core principles). The participants, the interactive audience, are central
to interact with the installation providing motion, voice, and other captured data
as an input to the computer system that processes the supplied input and produces
a real-time graphical and audio feedback to the interacting audience.

\subsection{Implementation}
\label{sect:interactive-docu-implementation}

There are two implementations: the first is in {\opengl}~\cite{i-still-remember-opengl-remake-2011}
(primarily with mouse and keyboard interaction only; and haptics planned)
and the newer one is in {\xna} ({\kinect}, mouse, voice, and keyboard interaction
and higher quality footage). We further detail the realization details
of the two versions beginning with the API in \xs{sect:interactive-docu-api},
modeling in \xs{sect:tangible-memory-modeling}, animation in \xs{sect:interactive-docu-animation},
content (\xs{sect:interactive-docu-media-content}),
the interaction aspects (\xs{sect:interactive-docu-interaction}), and
projection in \xs{sect:interactive-docu-projection}.

\subsubsection{APIs}
\label{sect:interactive-docu-api}

We describe some specific APIs that are dominant in the realization of
the interactive documentary proof-of-concept work.

\paragraph{{\opengl} and {\glut}.}
\label{sect:impl-opengl-glut}

The general {\opengl} and {\glut} concepts are described in \xs{sect:bg-opengl}.
A {\glut} window is created with registered callbacks for idle animation, mouse and keyboard
handling along side with a {\glui}~\cite{gluiManual} interface for more control
and based on earlier successful {\glut} examples~\cite{opengl-glut-examples}.

With {\glut}, we set up our handling for the \api{Mouse()} and \api{Motion()} callbacks
with \api{glutMouseFunc()} and \api{glutMotionFunc()} for mouse-based tracking and clicking
for interaction purposes to estimate a 2D click in the {\glut} window to the nearest bubble
in the 3D perspective environment. Our \api{Keyboard()} and \api{SpecialKeys()} callbacks registered with
\api{glutKeyboardFunc()} and \api{glutSpecialFunc()} respectively enable the keyboard-based
interaction (see \xs{sect:tangible-memories-opengl-glut-interaction}. The rendering
and timer-based updates callbacks \api{Display()} and \api{Update()} are registered
with the {\glut}'s functions \api{glutDisplayFunc()} and \api{glutTimerFunc()}.

\paragraph{{\xna} and {\hlsl}.}
\label{sect:impl-xna}

The {\xna} framework provides a collection of relatively easy to use
APIs for fast game development in {\csharp} (and other Microsoft .NET languages)
for various Microsoft platforms. We use {\xna} and .NET APIs for speech processing in a relationship
with {\kinect} mentioned in \xs{sect:impl-kinect} as well as to manage
the various media content. In particular, loading the media such as
photographic images and video playback within an interactive scene
and the corresponding keyboard and mouse controls.

The specific API is used for video playback control and getting
movie frames as textures. This is achieved by the provided \api{VideoPlayer}
and \api{Video} classes. The \api{Microsoft.Xna.Framework.Content} framework
has a variety of importers for media types (e.g. \file{.wmv}, \file{.png})
that pre-compile the media into the usable format by the rest of the {\xna}
components, such as \api{VideoPlayer} that streams the frames as textures
into the running program and allows for the user control (pausing, resuming,
stopping, etc.).

\api{SpriteBatch}, \api{Texture2D}, \api{SpriteFont}, and the likes are
helpful and commonly used classes in {\xna} to manage the collection and
representation of 2D graphics to be rendered in the pipeline in a certain
order. The \api{Model} class generally represents a 3D mesh. The \api{Effect}
class allows for shader-based effects and the likes to be applied to the
models. There are various supporting data types and structures, such
as \api{Vector2}, \api{Vector3}, \api{Rectangle} etc., that are commonly used by the various
XNA components.

The \api{Game} API provides methods for initializing and loading/unloading the media
content, provides a control loop, make idle and otherwise updates and drawing,
a somewhat similar to {\opengl}/{\glut} pipeline and callbacks.

For the video playback we used the {\xna} \api{VideoPlayback} 4.0 sample of how to play
frames and control the \api{VideoPlayer} and \api{Video} stream from Microsoft
under their Permissive License (Ms-PL).

\paragraph{{\kinect} and Speech SDK.}
\label{sect:impl-kinect}

We rely on a number Kinect API functions to make a rapid proof-of-concept
demonstration of the interactive \docutitle{I Still Remember} memory bubbles remake
as \docutitle{Tangible Memories}~\cite{tangible-memory-2011} from {\opengl} primarily because {\kinect} API~\cite{kinect-ms-sdk}
originally offered built-in skeleton tracking and associated speech components and a number of code samples (see \xs{sect:bg-code-samples})
of how to use those functions. The API also offers the integration with the {\xna} framework.
(The more recent {\kinect} SDK starting from 1.5 includes {\cpp} samples now as well among other new features.)

We use the \apipackage{Microsoft.Kinect} framework~\cite{microsoft-kinect-namespace} to obtain references to class instances like \api{KinectSensor},
and its API to access the skeleton stream frames (the \api{KinectSensor.SkeletonStream} class), from which
\api{Skeleton}, \api{Joint}, and \api{JointType} data are received, along side with
the \api{ColorStream} and \api{ColorImageFrame} as an input from the normal camera.

\api{KinectAudioSource}, \api{KinectSensor.AudioSource}, combined with
the API from \apipackage{Microsoft.Speech.Recognition} framework and its Speech SDK
(a part of the .NET Framework 4)~\cite{system-speech-recognition-namespace}
to allow for the language-dependent speech dictionary recognition to offer voice based command
interaction with the system.

\api{KinectStatus} provides information about the sensor state and availability.

We used an example implementation as a source of reference and inspiration
for Kinect skeleton tracking, including the hand joints, and rendering the
overall input video stream. The Kinect API reference, examples, speech
recognition come from the Microsoft Kinect for Windows SDK Sample Browser
from~\cite{kinect-fundamentals-skeleton-tracking,kinect-ms-sdk}.

The other details of the use of the Kinect API in \xs{sect:tangible-memories-kinect-control}.

\subsubsection{Modeling}
\label{sect:tangible-memory-modeling}

We describe the modeling for both {\opengl} and {\xna} versions.
This includes the memory bubbles themselves, media texture polygons, and the
simple environment they float in.

\paragraph{Modeling in {\opengl} and {\glut}.}

\begin{figure*}[htpb!]%
	\centering
	\includegraphics[width=\textwidth]{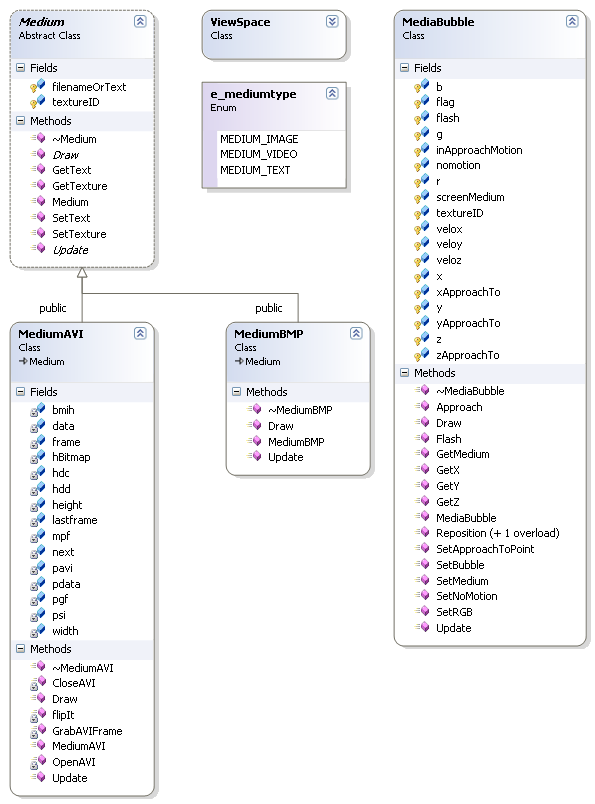}%
	\caption{Class Diagram Model of the Software Side of the {\opengl} Installation}%
	\label{fig:i-still-remember-class-diagram}%
\end{figure*}

Modeling is done as per the related algorithm steps in \xg{algo:conceptual-process-bubbles},
all of which are done using {\opengl}.
As stated, each bubble is defined by the property of the contained
media (\api{screenMedium}) and the outer color $(r,g,b)$. Naturally, the bubbles 
also have a radius $r$, a position $(x,y,z)$, and velocity
$\overrightarrow{vel}=(vel_{x}, vel_{y}, vel_{z})$~\cite{i-still-remember-opengl-remake-2011}.

The bubbles float in a confined rectangular space, a world ``box'',
designed for multi-projecting on different surfaces in an
installation environment of a blackbox or space alike.

In \xf{fig:i-still-remember-class-diagram} is a software component
model for the {\opengl} installation detailing the system design.
The \api{MediaBubble} class is a representation of a memory bubble
with a media of one of three types of image, video, or text. It contains
the properties described above as well as animation-related properties
for the media display or playback as to where to approach to in the virtual space, flash up
to indicate selection by the user, or return back to the bubble pool. It contains the
\api{screenMedium} instance reference to a particular media item
contained within: \api{Medium}. 

The \api{Medium} is a generic media class that provides a common API
for media items to be rendered within media bubbles. It provides for getting
and setting the actual media as well as \api{Draw()} and \api{Update()} calls
for rendering and animation to be obligatorily overridden by concrete media
class implementations.

The \api{MediumBMP} and \api{MediumAVI} classes are such implementations
in that they concretely implement the media-specific rendering functionality.
The former delegates the work to a \api{BMPLoader} to load and store a
single texture in the \file{.bmp} format, whereas the latter wraps and invokes the
AVI loader and playback example~\cite{nehe-lesson-35-avi} for playback of the
AVI-formatted video content. Each media bubble can contain any of these.

\api{ViewSpace} provides the updatable world model/environment for modeling of the
documentary display on screen, debugging in a color box, or a projection in
the blackbox. A light source is also modeled within \api{ViewSpace} to make the
bubbles shiny.

\paragraph{Modeling in {\xna} and {\hlsl}.}
\label{sect:interactive-docu-modeling-xna}

\begin{figure*}[htpb!]%
	\centering
	\includegraphics[width=\textwidth]{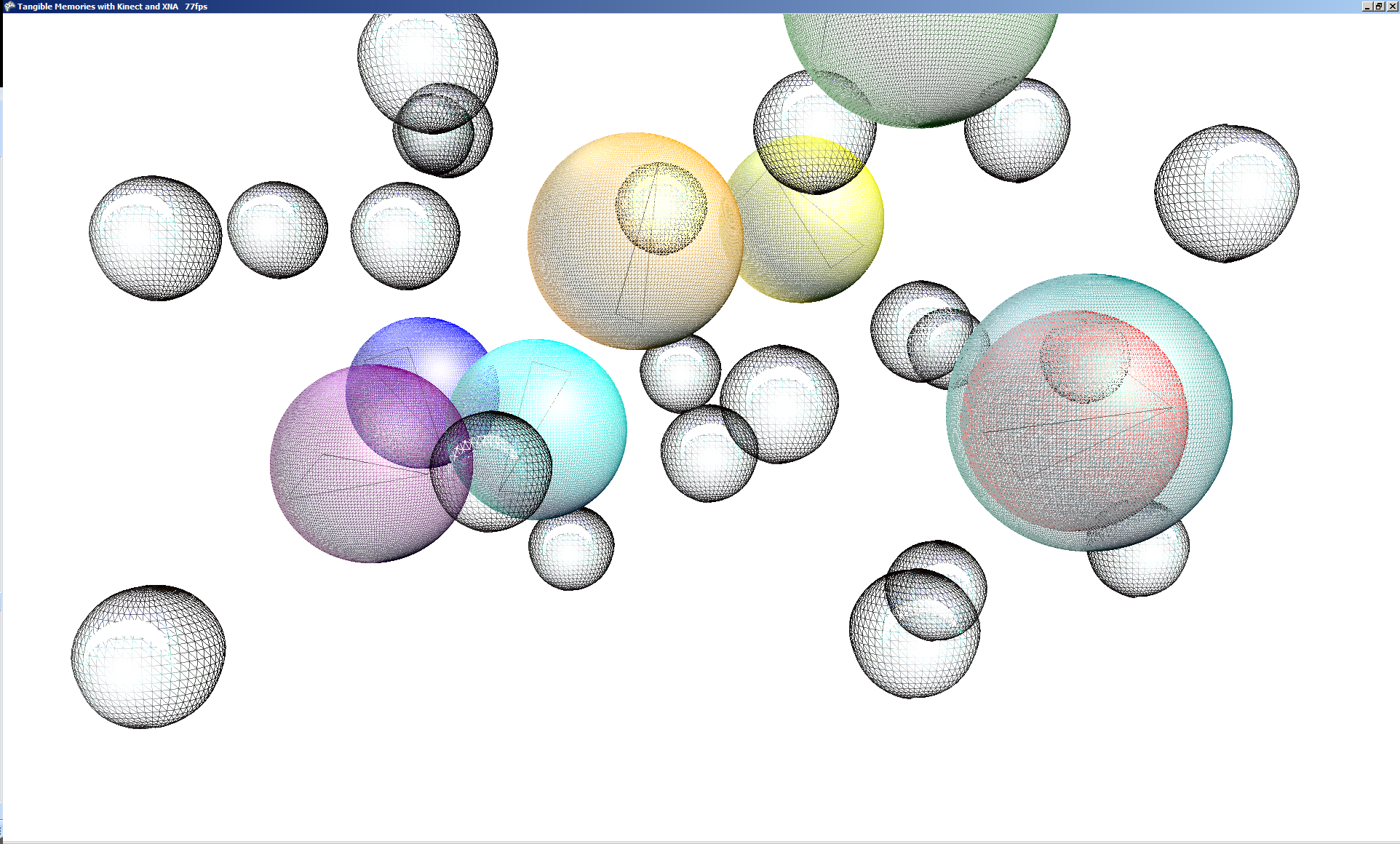}%
	\caption{Modeling of the Fancy and Media Memory Bubbles in {\xna}}%
	\label{fig:tangible-kinect-2-white-wireframe-model}%
\end{figure*}

The classical performance measuring sample bubbles~\cite{xna-performance-sample}
are computationally modeled sphere meshes with a programmatically specified LOD.
Fancy soap animated bubbles~\cite{fancy-bubble-sample-xna3} have their spherical
mesh stored as a \file{sphere.X} model coupled with the \file{Bubble.fx} {\hlsl} shader
file that does all the texture mapping, pixel/vertex shading, and normal alterations.
The wireframe mesh of the two types of bubbles is illustrated in
\xf{fig:tangible-kinect-2-white-wireframe-model}.

The fancy bubble model has an environment map of using the \api{RenderTargetCube} {\xna}
API to selectively map either a skybox, or video faces (real-time or pre-recorded)
and other animations. The ground floor (if shown) as well as the projected
media inside the rainbow color bubbles are a simple flat square modeled as two
rectangles from \file{Ground.x}~\cite{xna-performance-sample}
(see \xf{fig:bubble-installation-2} on \xp{fig:bubble-installation-2}, for example).

\subsubsection{Animation}
\label{sect:interactive-docu-animation}

Some of the animation aspects in this documentary are quite simple
and through that simplicity the main message is unobscured. However,
there are some fancier animation features added to illustrate various
attractive capabilities.

\paragraph{{\opengl}.}

The animation in the {\opengl} version is simple bouncing off the walls media
bubbles in a box world (which is removable for projection purposes). Once
a bubble is called upon, it changes its pseudo random bouncing trajectory
towards the viewer to occupy the majority of the visible space. Then it is
sent back when the user is done watching.

\begin{figure*}[htpb!]%
	\centering
	\includegraphics[width=\textwidth]{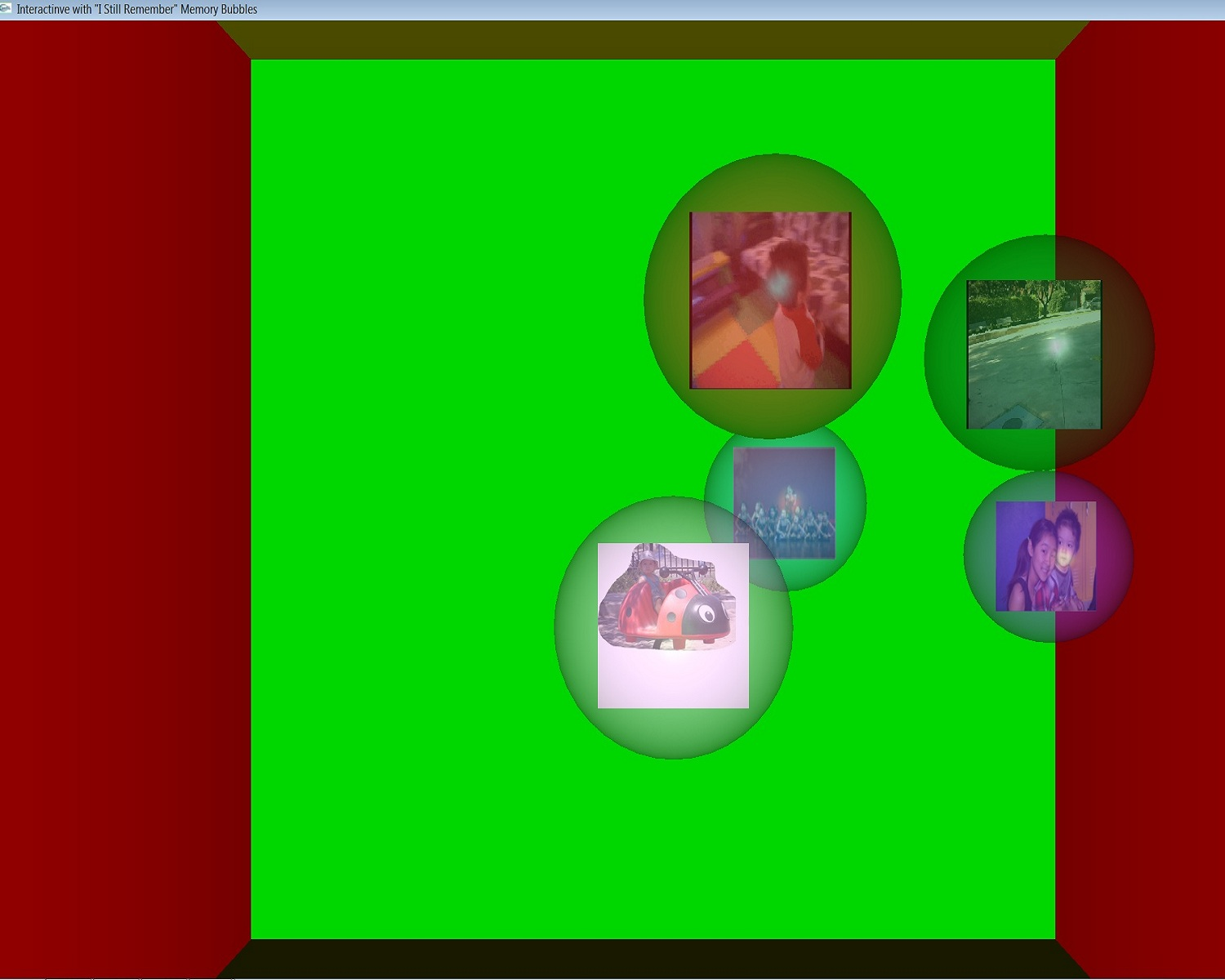}%
	\caption{Bubbles Float with the Screen Media Inside}%
	\label{fig:bubbles-float-example}%
\end{figure*}

\begin{figure*}[htpb!]%
	\centering
	\includegraphics[width=\textwidth]{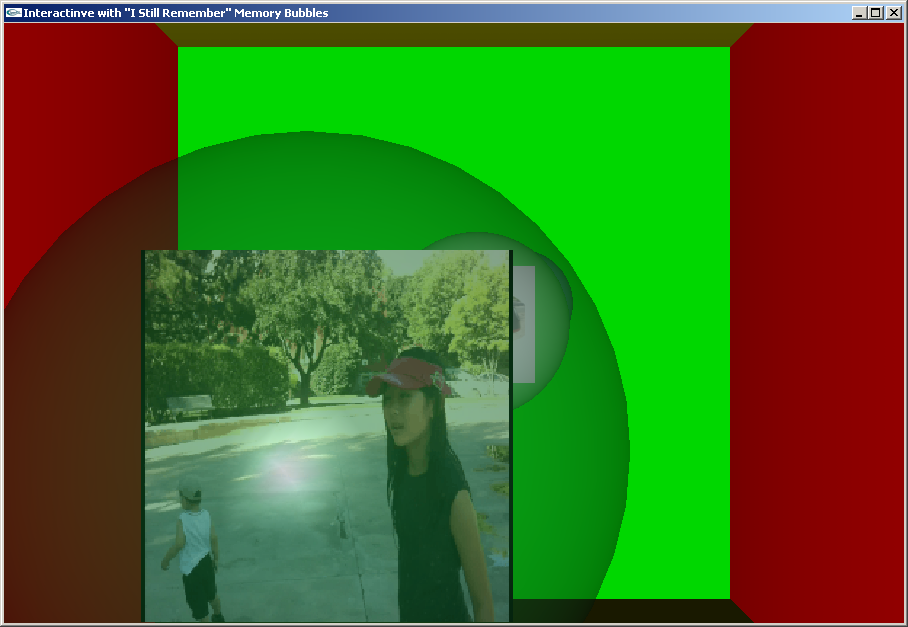}
	\caption{A Close Up Example of a Selected Bubble with a Video Content Playing}
	\label{fig:selected-bubble-zoom-in}
\end{figure*}

\begin{figure*}[htpb!]%
	\centering
	\includegraphics[width=\textwidth]{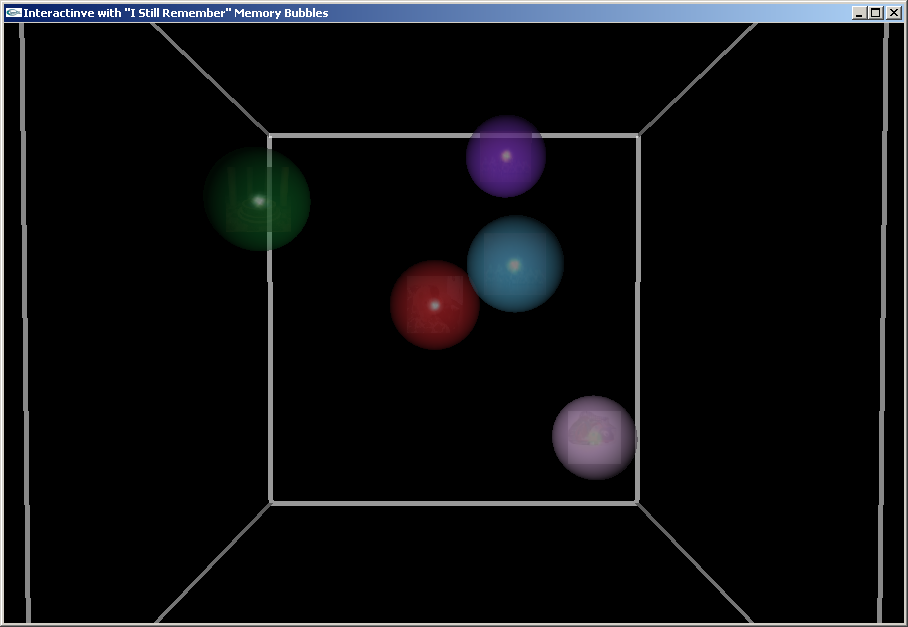}
	\caption{{\opengl} \docutitle{Tangible Memories} in the Black Background for Blackbox Projection}
	\label{fig:bubbles-opengl-blackbox-1}
\end{figure*}

\begin{figure*}[htpb!]%
	\centering
	\includegraphics[width=\textwidth]{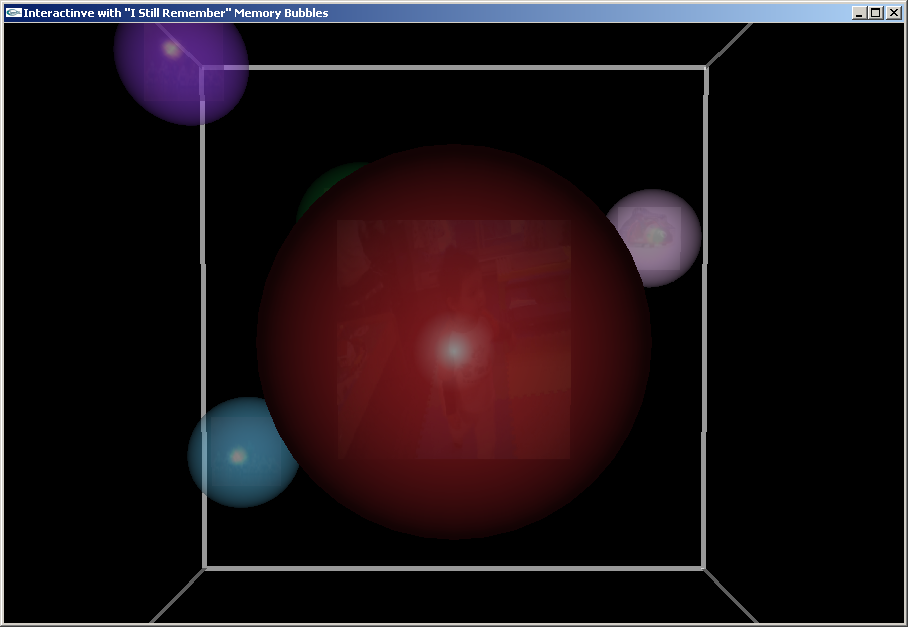}
	\caption{{\opengl} \docutitle{Tangible Memories} in the Black Background with the Red Bubble Called Out}
	\label{fig:bubbles-opengl-blackbox-2}
\end{figure*}

More specifically, the special animation exists when a bubble is selected (\api{selected} is
\api{true}) by whatever means (mouse click or keyboard): it flashes, then 
it changes its trajectory to approach the near plane, towards the default
viewing position of the synthetic camera ($(x_{ApproachTo},y_{ApproachTo},z_{ApproachTo})$
in \xf{fig:i-still-remember-class-diagram}) such that the 
media in the bubble comes into full view and fills up the screen. When the 
selected bubble is pushed back (by the same means), it returns to the pool of 
the bouncing bubbles.

The AVI playback~\cite{nehe-lesson-35-avi} animation of the
video media is done by swapping textures, the movie frames, on the media
polygons. The video media is constantly played in a loop.

The animation sequence is driven via the \api{Update()} {\glut} callback
set through \api{glutTimerFunc()} programmed to update every 20ms. This allows
for balanced consistent animation speeds across different hardware.

\paragraph{{\xna}.}

The basic core animation of the memory bubbles in the {\xna} version is
very similar to that of {\opengl}. Additional optional collision detection is
performed among the bouncing bubbles themselves~\cite{xna-performance-sample}. A large number of
bubbles are possible, but the zoom-in playback animation is done only for the
rainbow-like color bubbles. The media playback is animated only when
it is called upon and is done similarly as in the {\opengl} version
by texture-mapping the video frames one at a time.

Additionally, the fancy bubbles that mingle among the rainbow color ones
are animated as floating soap bubbles adapted from~\cite{fancy-bubble-sample-xna3}.
That animation can be altered by playing various video feed, real-time,
or pre-recorded textured on the bubble as it moves and shakes (this feature is
explored more in detail later in this chapter and subsequently in \xc{chapt:illimitable-space}). 
The waving and shaking effect can also be altered by
the beat of a music tune being played back along with the use of
the BASS.NET~\cite{bass-audio-library-dot-net} library\index{Libraries!BASS!BASS.NET}\index{API!BASS.NET}.
This wavy \api{Effect} is primarily implemented in the {\hlsl} vertex and pixel shaders~\cite{fancy-bubble-sample-xna3}
that alter the bubble's mesh and perturb normals and at the same time re-applying
textures (environment, skybox, or otherwise) to the new coordinates. These fancy
soap bubbles were augmented to collide like the classical media bubbles do with all the
other bubbles in their environment.

The media plane inside the media bubbles is animated via simple matrix
transformations to rotate the ground plane around $x$, $y$, and $z$
axes by the ``padded'' \api{wave} value---this also allows for the
music-based animation to rotate them along with a beat of a tune if
a tune playback is enabled.

Hand-based animation is also there to partially rotate two fancy bubbles that
can also be ``held'' with the left and right hands via {\kinect} skeleton tracking.
This aspect, however, goes beyond \docutitle{Tangible Memories} is more about
an interactive performance described in \xc{chapt:illimitable-space}.

The animation sequence is driven from the {\xna} \api{Update(GameTime)} callback
of the \api{Game}-derived class (see \xg{algo:kinect-audio-recognition-bubbles})
that takes into the account elapsed time so the animation could be adjusted to
the frame rate of the slower or faster hardware.

\subsubsection{Media Content}
\label{sect:interactive-docu-media-content}

The media bubbles are generically designed to contain the main three types of
visual media---a photograph/picture, text, and video. All three can also be
animated as desired and augmented with sound (especially the video content).
There are some minor differences at present in the two installations produced,
but we list the content contained in the bubbles. All the content is of our own
production and the audience is invited to run and interact with the installations
for themselves and explore each bubble.

\paragraph{{\opengl} Installation.}

Below is the {\opengl}'s installation bubble contents as it was captured at some point in time.
The bubble colors are described, but are obviously the best perceived in the electronic
form or a color printout than a black-and-white printout.
The media contents in this version are picked randomly shot by home digital camera
for prototyping purposes, such as:

\begin{enumerate}

	\item
The green-shaded bubble has a short video of the kids playing in Montreal Westmount
park's fountain in 2009.

	\item
The red-shaded bubble has a another short video of my son in 2009
wandering at home.

	\item
The light blue contains a photograph of my daughter with her dance school
at a Chinese New Year performance in 2006.

	\item
The white bubble has a photograph of my son on the ladybug buggy in a park.

	\item
The purple bubble has a photograph of the kids at home.
\end{enumerate}

\paragraph{{\xna} Installation.}

The media contents in this installation are more of professional making and complete.
Most of the footage is shot with a professional HD camera.

\begin{enumerate}

	\item
The orange bubble contains the award-winning 13-minute \docutitle{I Still Remember} with the interview
of my daughter shot in the Hexagram blackbox; the work that prompted this
whole project of the interactive documentary presented in this
chapter~\cite{song-still-remember-movie-bjiff2011,i-still-remember-opengl-remake-2011}.

	\item
The violet/purple bubble has the footage of the ceremony of \docutitle{I Still Remember}
receiving the Best Short Documentary Award at the 1st BJIFF (Beijing International Film Festival)
at the closing ceremony.

	\item
The yellow bubble is the collage of VWIFF (Vancouver Women in Films Festival), at which
\docutitle{I Still Remember} also was officially screened.

	\item
The blue bubble has the interview by the BTV (Beijing Television) \emph{View China} Program
about how \docutitle{I Still Remember} was made.

	\item
The red bubble contains a short mini-documentary \docutitle{St-Valentine},
in which children were preparing the Valentine candies and cards for
their classmates and teachers. They are also the real scenes and life behind
\docutitle{I Still Remember}. \docutitle{St-Valentine} was edited by my daughter.

	\item
The light blue / cyan bubble contains a short 3-minute documentary my
daughter did about her little brother---\docutitle{My Little Brother} that
got selected as one of the top 100 international films and was screened
at the Young Cuts 2012 film festival taking place at Concordia in October 2012.

	\item
The green bubble features the \animtitle{Spectacle} animation I did
in {\maya}~\cite{spectacle-anim-film-2003}.

	\item
The teal bubble, which is not a part of the 7 rainbow-like color bubbles,
has a real-time video feed from the {\kinect}'s color stream camera itself.
\end{enumerate}

\subsubsection{Interaction}
\label{sect:interactive-docu-interaction}

We describe the implementation of the interaction and the user control of
the work for both versions ({\opengl} in \xs{sect:tangible-memories-opengl-glut-interaction}
and {\xna}/{\kinect} in \xs{sect:tangible-memories-kinect-control})
and their particularities.

Similarly to the followup chapter's recorded clips and the discussion on the generalization
of the proposed system \xs{sect:tangible-interaction-animation}, we record the interaction
example Clip 4---voice control of the bubbles (with a person calling out
the bubbles, etc.) along with the narration about multidisciplinary the research
recorded in the whitebox production room.

\begin{figure*}[htpb!]%
	\centering
	\includegraphics[width=.9\textwidth]{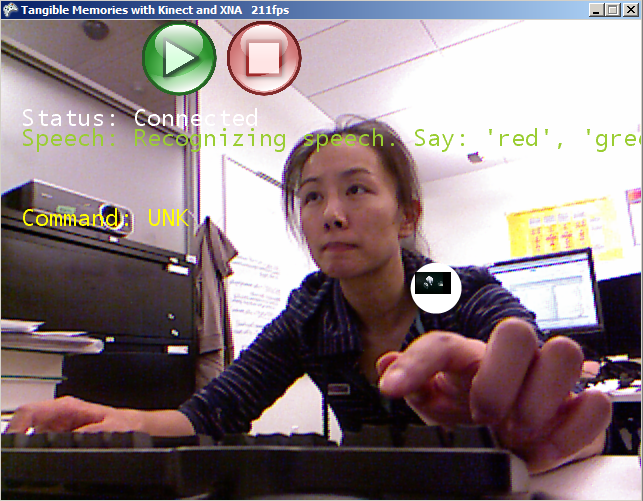}%
	\caption{Interaction in Debug Mode with Speech and Keyboard On}%
	\label{fig:tangible-kinect-1}%
\end{figure*}

\begin{figure*}[htpb!]%
	\centering
	\includegraphics[width=\textwidth]{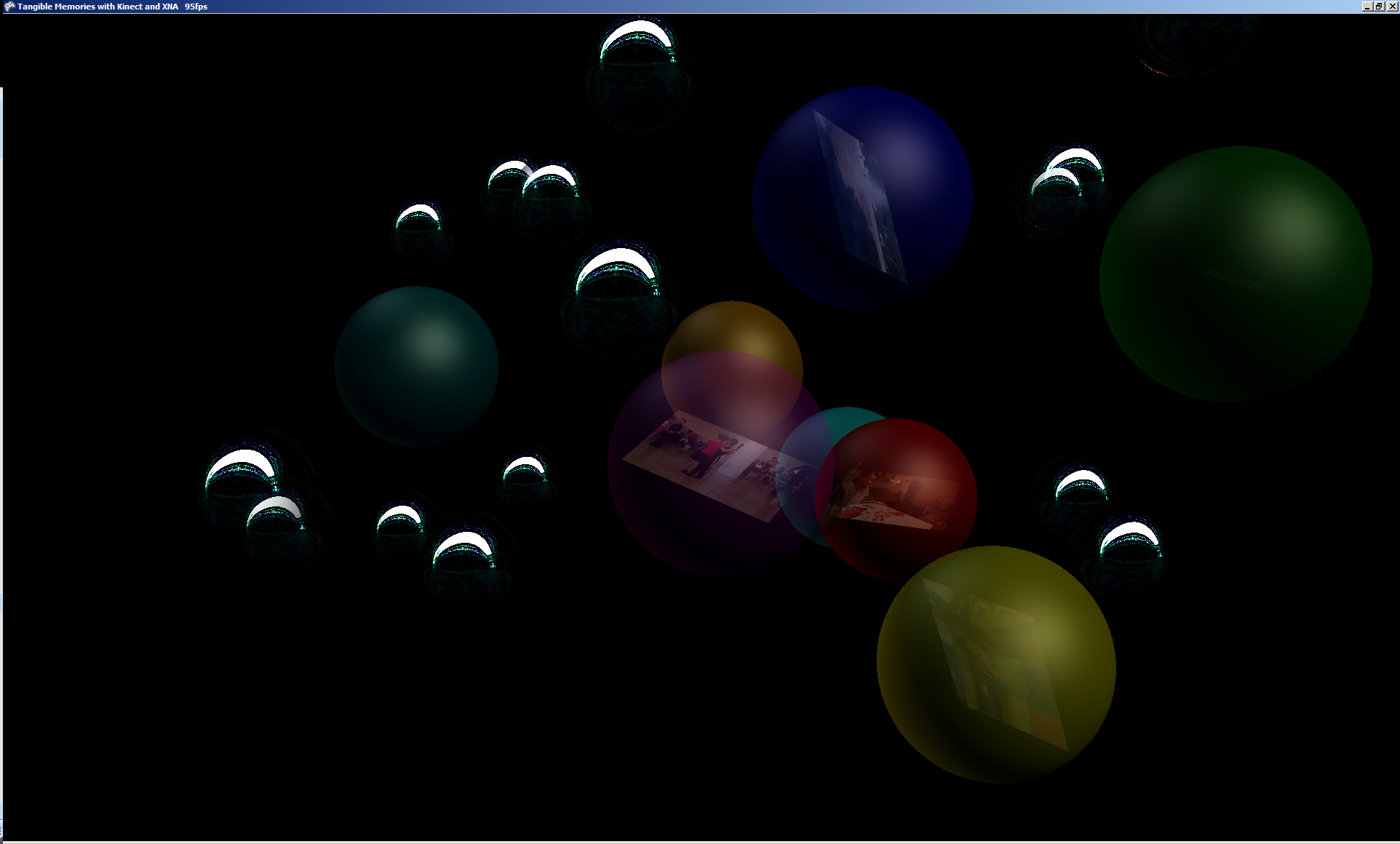}%
	\caption{Fully Shaded {\xna} Bubbles on Black Background}%
	\label{fig:tangible-kinect-2}%
\end{figure*}

\begin{figure*}[htpb!]%
	\centering
	\includegraphics[width=\textwidth]{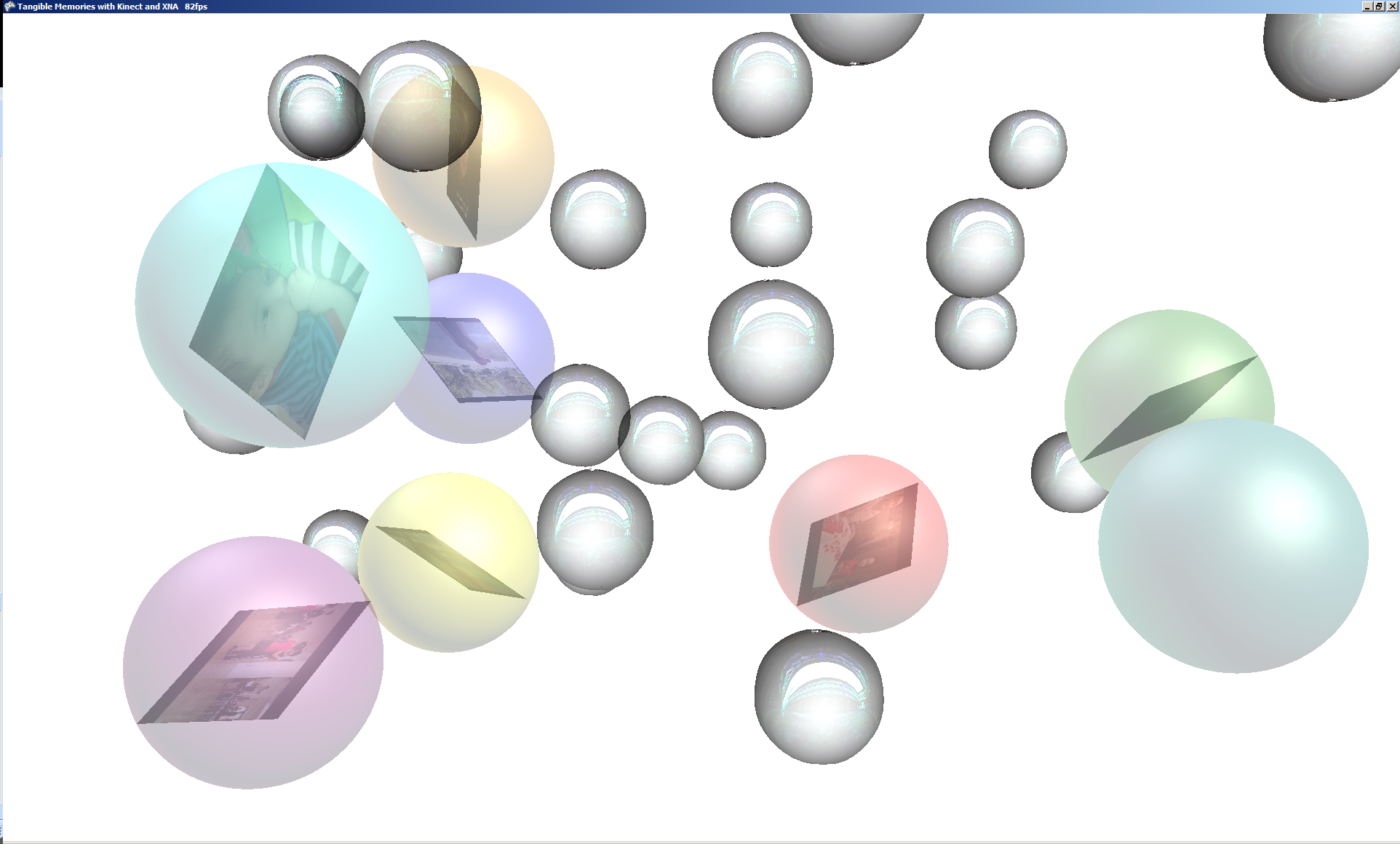}%
	\caption{Fully Shaded {\xna} Bubbles on White Background}%
	\label{fig:tangible-kinect-2-white}%
\end{figure*}

\begin{figure*}[htpb!]%
	\centering
	\includegraphics[width=\textwidth]{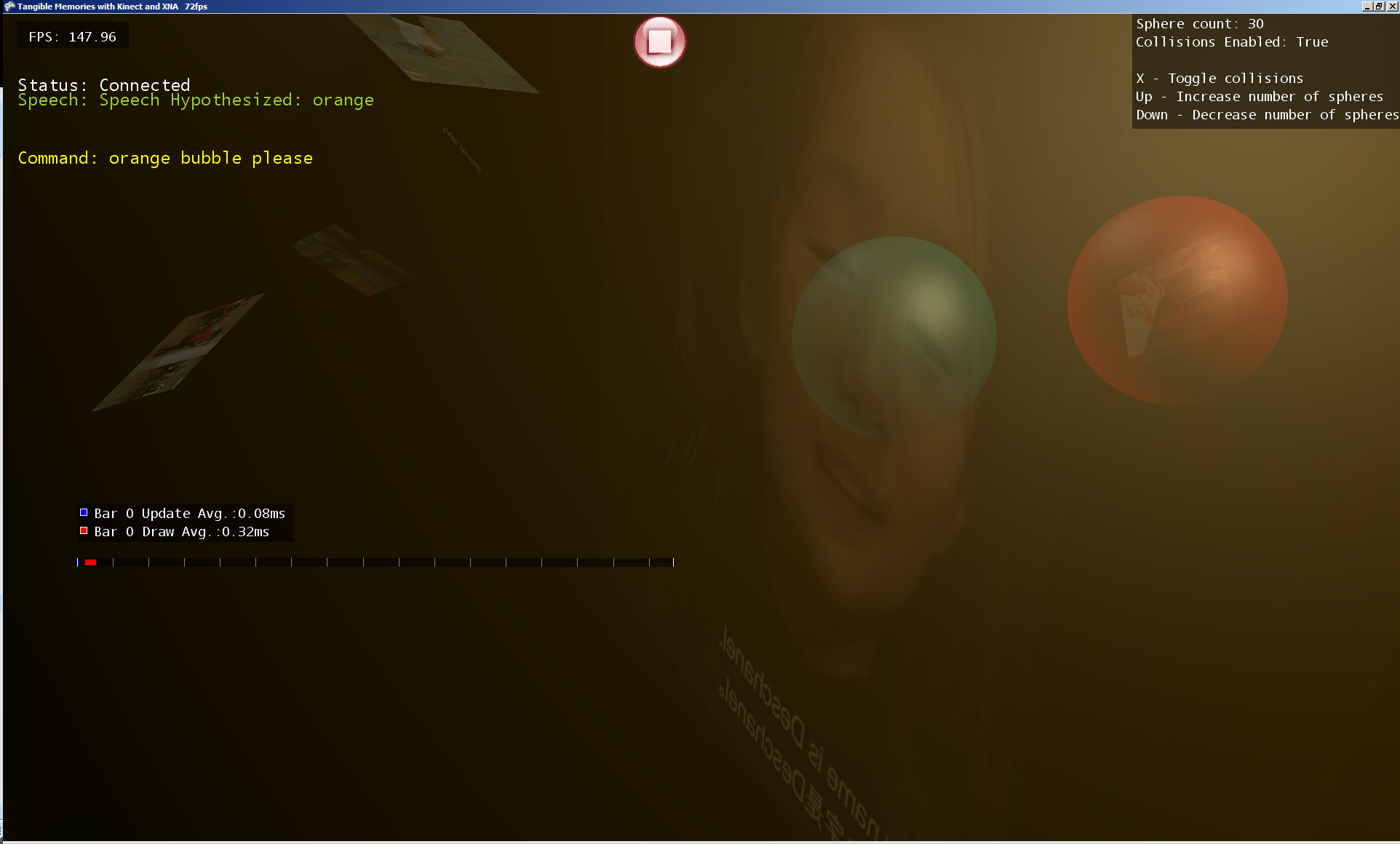}%
	\caption{Calling out an Orange Bubble by Voice Containing the \docutitle{I Still Remember} Documentary Playing 
	(Debug Mode, Black Background)}%
	\label{fig:tangible-kinect-3}%
\end{figure*}

\begin{figure*}[htpb!]%
	\centering
	\includegraphics[width=.9\textwidth]{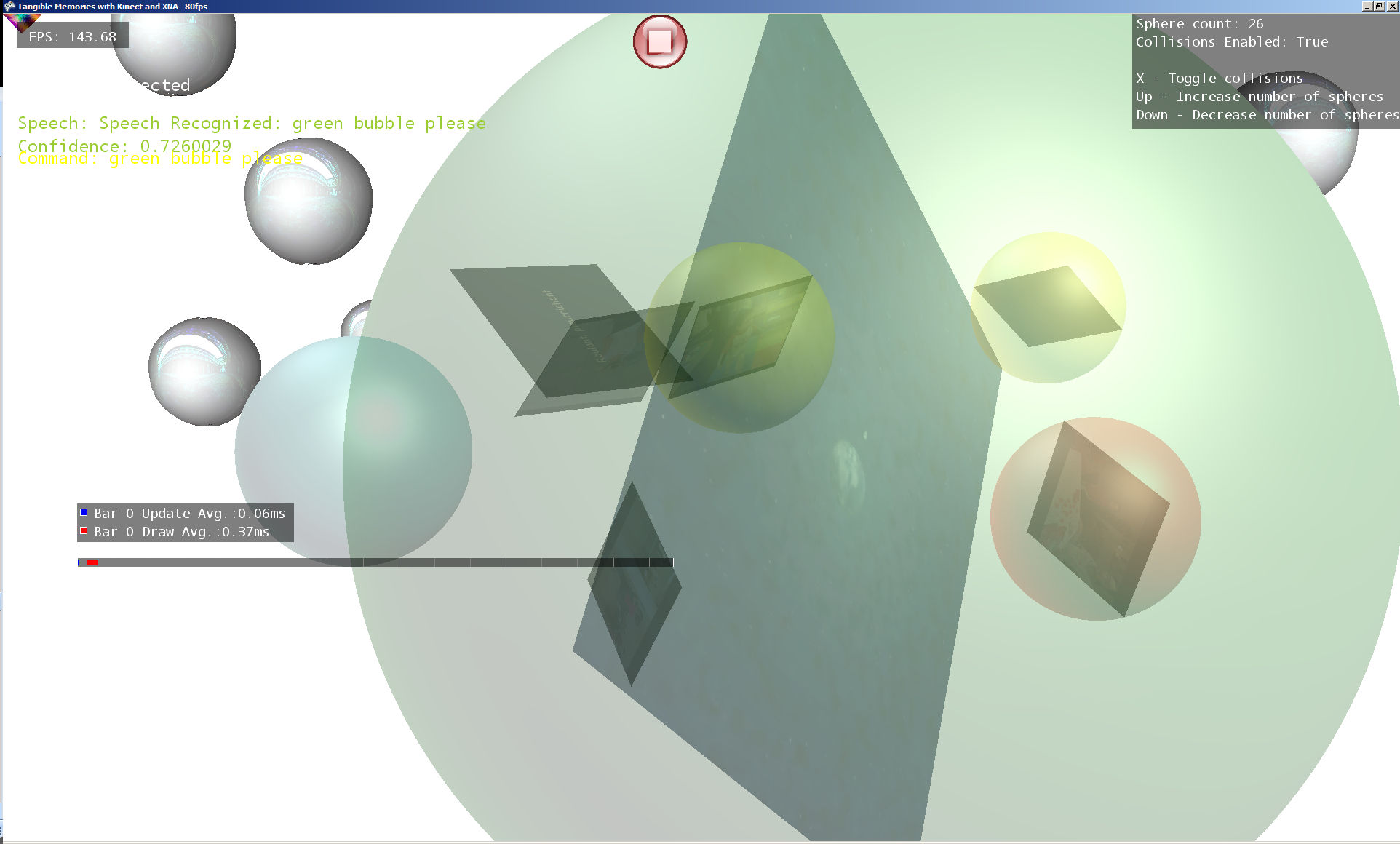}%
	\caption{Calling out a Green Bubble by Voice Containing the \animtitle{Spectacle} Animation Playing 
	(Debug Mode, White Background)}%
	\label{fig:tangible-kinect-3-white}%
\end{figure*}

\begin{figure*}[htpb!]%
	\centering
	\includegraphics[width=.9\textwidth]{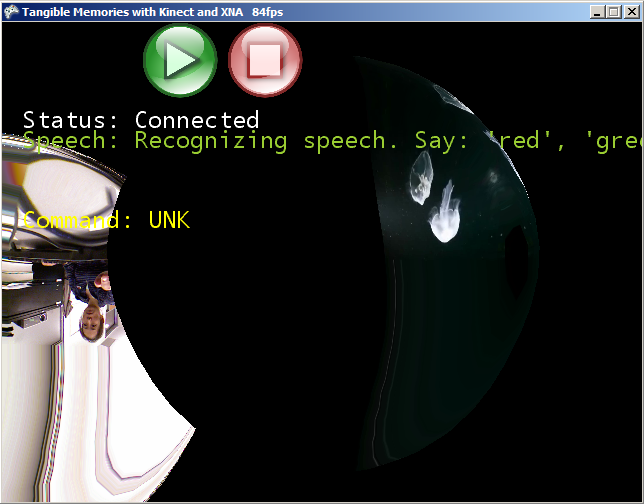}%
	\caption{Two Hand-Controlled Fancy Bubbles With Read and Recorded Video Feed in Debug Mode}%
	\label{fig:tangible-kinect-4}%
\end{figure*}

\paragraph{{\opengl} Version.}
\label{sect:tangible-memories-opengl-glut-interaction}

The present level of interactivity (shown, e.g., in \xf{fig:bubbles-float-example}) by the audience is
via simple mouse clicks in the 2D coordinates, which are
then translated to the nearest bubble, which is then
``selected'' and zooms in slowly onto the viewer
with its content gradually taking over the full screen.
In the bubble the corresponding media content (photograph, text, video clip)
is rendered (see \xf{fig:selected-bubble-zoom-in}).
Clicking the zoomed-in content makes it recede back to the
bubble pool of memories~\cite{i-still-remember-opengl-remake-2011}.
A similar effect can be achieved by the assigned keys `1'--`5' for quick
and precise testing and showcasing. `W' can also switch the box
environment from the color (\xf{fig:bubbles-float-example} and \xf{fig:selected-bubble-zoom-in})
to black box for projecting (\xf{fig:bubbles-opengl-blackbox-1} and
\xf{fig:bubbles-opengl-blackbox-2}).

More advanced methods of interaction in the works are described in
\xs{sect:interactive-docu-future-directions}.

\paragraph{Kinect Skeleton and Audio Control.}
\label{sect:tangible-memories-kinect-control}

\begin{algorithm}[ht!]
\tiny
\hrule\vskip4pt

\api{Program.Main()}\;
\Begin
{
	\api{Game.Run()}\;
	\Begin
	{
		\api{Initialize()}\;
		\Begin
		{
			Register Kinect observer/listener method of status change to be
			\api{KinectSensor.KinectSensors.StatusChanged} = \api{KinectSensors\_StatusChanged()}\;
			
			\api{DiscoverKinectSensor()}\;
			\Begin
			{
				\lForEach{\api{KinectSensor} sensor $\in$ \api{KinectSensor.KinectSensors}}
				{
					\If{sensor.Status == \api{KinectStatus.Connected}}
					{
						\tcp{Found one, set our sensor to this}
						kinectSensor = sensor\;
						break\;
					}
					\tcp{Init the found and connected device}
					\If{kinectSensor.Status == \api{KinectStatus.Connected}}
					{
						InitializeKinect()\;
						\Begin
						{
							\tcp{Color stream}
							kinectSensor.ColorStream.Enable(ColorImageFormat.RgbResolution640x480Fps30)\;
							kinectSensor.ColorFrameReady = kinectSensor\_ColorFrameReady\;

							\tcp{Skeleton Stream}
							kinectSensor.SkeletonStream.Enable
							(
									new TransformSmoothParameters()
									{
											Smoothing = 0.5f,
											Correction = 0.5f,
											Prediction = 0.5f,
											JitterRadius = 0.05f,
											MaxDeviationRadius = 0.04f
									}
							)\;

							kinectSensor.SkeletonFrameReady = kinectSensor\_SkeletonFrameReady\;
							kinectSensor.Start()\;
							Speech.Program.StartKinectAudioProcessing(kinectSensor)\;
						}
					}
				}
			}
		}
		
		\tcp{Loads media files: images, videos, models, effects}
		\api{LoadContent()}\;

		\tcp{Updates within a loop like idle}
		\api{Update(GameTime)}\;
		\Begin
		{
			Process player states to stop/resume\;
			Process call back keys\;
			Update skeleton/bubble capture condition\;
			\tcp{The positional skeletal tracking and voice streams are updated
			by the registered Kinect callbacks in the \api{InitializeKinect()}
			method where the variables for joints of the left and right hands
			are tracked. Likewise, asynchronous audio input source from Kinect
			is tracked by a speech recognized initialized there and a global
			variables set with the states that are used here.}
			Update video playback condition of a captured bubble (by gesture of a left hand
			in predefined spots or voice commands ``stop'' or ``play'')\;
			Update free-floating memory bubbles position\;
		}
		
		\tcp{Renders content}
		\api{Draw(GameTime)}\;
		\Begin
		{
			Update free-floating memory bubbles positions\;
			\tcp{The playback is stopped when bubbles freely float for speed}
			Acquire and render video frames as textures for close-up video playback of
			a captured bubble\;
			Update and render left capture hand for play and stop buttons\;
		}
		
		\tcp{Unloads/disposes of contend upon exiting the loop}
		\api{UnloadContent()}\;
	}
}

\hrule\vskip4pt
\normalsize
\caption{{\kinect}/{\xna} Interaction Algorithm}
\label{algo:kinect-audio-recognition-bubbles}
\end{algorithm}

In \xg{algo:kinect-audio-recognition-bubbles} is the complete overview of the
algorithm how {\kinect}- and {\xna}-based interaction are performed (that goes
beyond the interactive documentary concept and into the interactive performance,
a topic of \xc{chapt:illimitable-space}).
The details of the keyboard, gesture, and speech based specific commands
for interaction are listed in \xs{appdx:kinect-interaction}.

\begin{figure*}[htpb!]%
	\centering
	\includegraphics[width=\textwidth]{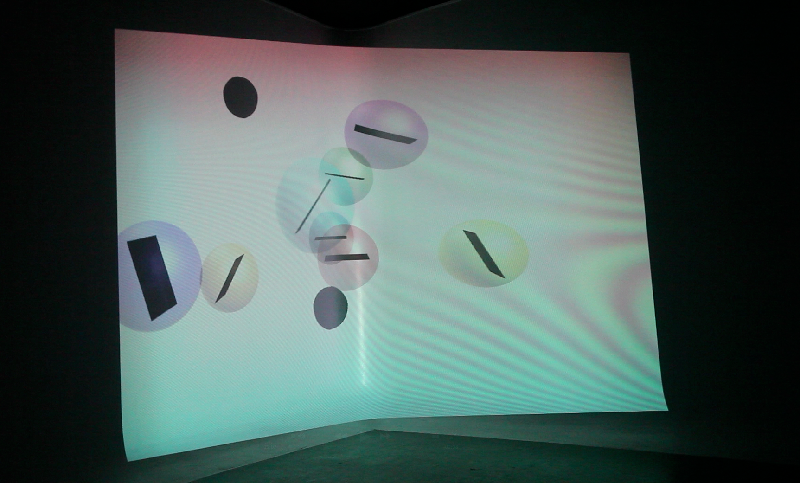}
	\caption{Projection of \docutitle{Tangible Memories} in the Production Room Corner}
	\label{fig:bubble-installation-1}
\end{figure*}

\begin{figure*}[htpb!]%
	\centering
	\includegraphics[width=\textwidth]{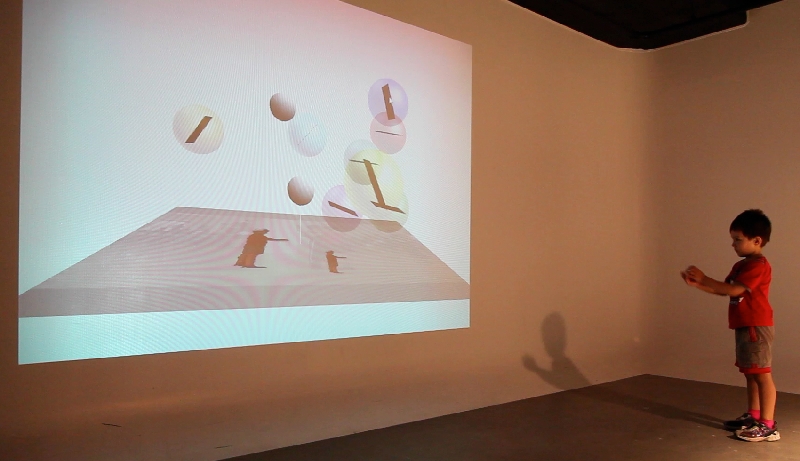}
	\caption{Audience Interacts with the Projection of \docutitle{Tangible Memories} in the Production Room}
	\label{fig:bubble-installation-2}
\end{figure*}

In \xs{sect:impl-kinect} some audio and {\kinect} APIs that we use are mentioned.
Specifically, we capture the audio stream and run \api{Microsoft.Speech}'s
recognizer for the speech-based interaction. We define a dictionary of
words and phrases (see \xs{sect:kinect-voice-commands}) using the English grammar instance
(which we plan to try with other languages).

Skeleton tracking provided by the {\kinect} API allows tracking joints,
and, in our case, specifically left and right hands to hold on to or
bounce off the bubbles in this installation instance.

In \xf{fig:tangible-kinect-1}, \xf{fig:tangible-kinect-2},
\xf{fig:tangible-kinect-3}, and \xf{fig:tangible-kinect-4} are
some screenshot results of the prototype {\xna}-based installation
with {\kinect} and speech recognition.

The bubbles can be called out by their color or by keys (see the
\xg{algo:kinect-user-docu-keyboard-setup} and \xc{appdx:kinect-interaction}
for more specific details on that aspect).

\begin{algorithm}[ht!]
\hrule\vskip4pt
\Begin
{
	\tcp{By default}
	Music off\;
	Enable speech by default (bug: still need to press `T' twice)\;

	\tcp{Start as normal with bubbles}
	\Begin
	{
		Show full screen (press `F' and then double-click the title bar)\;
		More or less fancy bubbles\;
		\Begin
		{
			By keys ``$\uparrow$''/``$\downarrow$''\;
			By speech saying ``more bubbles'', ``less bubbles''\;
		}
		``PgDown''/``PgUp'' zoom in/out for the best fitting projection\;
		Call out the media bubbles\;
		\Begin
		{
			\tcp{See the color list in the keys list}
			By voice using the bubble color\;
			By key\;
			\Begin
			{
				``F4'' -- simulate ``red'' command for testing\;
				``F5'' -- simulate ``orange'' command for testing\;
				``F6'' -- simulate ``yellow'' command for testing\;
				``F7'' -- simulate ``green'' command for testing\;
				``F8'' -- simulate ``cyan'', ``light blue'' command for testing\;
				``F9'' -- simulate ``blue'' command for testing\;
				``F10'' -- simulate ``violet'', ``purple'' command for testing\;
			}
		}
	}
}
\hrule\vskip4pt
\caption{Configuration and Setup Interaction}
\label{algo:kinect-user-docu-keyboard-setup}
\end{algorithm}

\subsubsection{Projection}
\label{sect:interactive-docu-projection}

The installation is projected in a blackbox
or an enclosed production room alike with a full-screen
mode on with either black or white background. On the low-end the inexpensive
projectors are used with standard VGA-based $1024 \times 768$ projected image. At the
higher end the HD projection for large spaces can be used.

In \xf{fig:bubble-installation-1} and \xf{fig:bubble-installation-2} are examples
of a medium scale projected interactive installation of the {\xna} \docutitle{Tangible Memories}
inside a production ``whitebox'' room with a white background and the ground floor
rendered with green screen capture techniques and extraction of audience interacting with
the bubbles.

\chapter{Illimitable Space in Responsive Theatre}
\label{chapt:illimitable-space}

\newcommand{\theatreelement}[1]{\emph{#1}\index{Theatre elements!#1}}

Theatre is not only the nexus where computer technology
and artistic creation meet, but also the most powerful artistic
form in itself. Live performance includes all the possible elements,
lighting, fashion, video, music, storytelling, painting,
sculpture, computer graphics and animation, mobile media,
whatever term one could think about: all could be integrated into theatre performance.
However, I would still use Aristotle's six essential elements of drama~\cite{poetics-aristotle,wiki:poetics-aristotle},
\theatreelement{Spectacle}, \theatreelement{Character}, \theatreelement{Plot},
\theatreelement{Diction}, \theatreelement{Melody}, and \theatreelement{Thought},
to guide this theatre work innovated with the new technology and thoughts.
We map these elements to the contributions in this work in \xs{sect:tangible-interaction-theatre-elements}.
In fact, it could be the backbone carrier of different media types,
which binds science and arts elements.

In \xs{sect:illimitable-space-methodology} we further describe the
methodology behind our specific approach, then we delve into the
details of the realization of the installation in \xs{sect:illimitable-space-design-impl},
and we summarize our findings in \xs{sect:illimitable-space-summary}.

\section{Methodology}
\label{sect:illimitable-space-methodology}

The core ideas behind the interactive performance include a {\kinect}-driven,
real-time CG animation with the audience and participants interacting with
the system (and potentially each other).

There are four primary examples in this work described in \xs{sect:illimitable-space-poc}.
However, the possibilities with the system prototype are not limited at all
to just these specific four elements and forms and it can be reconfigured
for a large number of interactive performance scenarios.

Thus, we state the goal of the methodology in \xs{sect:illimitable-space-goal},
followed by the illustration of the proof-of-concept realization of this goal
in \xs{sect:illimitable-space-poc}.

\subsection{Goal}
\label{sect:illimitable-space-goal}

First of all, traditional theatre is limited within three walls, a ceiling, and a floor.
I intend to break this standard rule to free actors and audience
in a dynamic illimitable space.
Second, the conventional relationship between the audience and actors,
spectating vs. performing, needs to be altered;
instead, audience will also actively participate the \theatreelement{Plot}
and be a part of the performance. 

\subsection{Proof of Concept}
\label{sect:illimitable-space-poc}

The presented proof of concept (PoC) is illustrated in the follow up stills
of a test performance production in a Hexagram Concordia video production room. The illustrations,
in \xf{fig:dance-kinect}, \xf{fig:watersleeve-beginning-real-1-1}
\xf{fig:watersleeve-real-2-1}, \xf{fig:sleeve-1}, and \xf{fig:sleeve-2},
show the perception of depth, performance, and real-time visualization
of sound (in this example, we visualize the music background).
This is the first proof-of-concept realization of the
installation that has vast possibilities to be used in many performance
art aspects, some of which we will explore in the future directions and work
in \xs{sect:tangible-interaction-future-directions}.

The first is the dancer with the three graphical figures of her performing and animated
at the same time: the depth shadow, greenscreen background replacement and real-color
image, and a ``stick-man'' skeleton (see \xf{fig:dibubu-four-dancers-1} and \xf{fig:dibubu-four-dancers-2}).
All are founding core techniques that can be
augmented and wrapped with various extra features available in today's technologies
such as ``avateering''~\cite{kinect-ms-sdk}\index{avateering} (let's call it \emph{digital puppetry}\index{digital puppetry}),
depth and skeleton based interaction with the virtual elements, and art. This
example~\cite{song-phd-trio-dance-kinect-video-2012}\footnote{\url{https://vimeo.com/49682696}}
combines some of the elements from the subsequent examples that were incrementally built up.

A real-time green screen mapped audience is added to a pre-recorded video footage
(see \xf{fig:virtual-audience-dancers-1} and \xf{fig:virtual-audience-dancers-2}), in the effect
to multiply the audience, with various sizes, distances, depths, and potentially projected onto different scene
elements~\cite{song-phd-timmy-greenscreen-kinect-video-2012}\footnote{\url{https://vimeo.com/50069419}}.

Water-sleeve Chinese dance with music visualization blended together is the third example
of the real-time visualization of the rendered depth in rainbow-like colors,
(e.g. shown in \xf{fig:dance-kinect}, \xf{fig:sleeve-1}, \xf{fig:sleeve-2}, and related).
This is the classical example of working with the depth stream and real-time music
visualization~\cite{song-phd-depth-water-sleeve-dance-kinect-video-2012}\footnote{\url{https://vimeo.com/49399617}}.

Real-time interaction and video feed can be projected onto a multitude
of virtual and real surfaces. In the fourth example, we re-use the augmented fancy soap bubble
design~\cite{fancy-bubble-sample-xna3} from \xc{chapt:interactive-docu}, to project
the depth stream onto it while the bubble's wave animation and rotation is affected
by the beat of the music played as well as waving hands (gestures) to rotate it. A 6-plane render target projection
is mapped onto the bubble's sphere; each face can contain a real-time video feed, a stored video,
a static texture, an animation, music visualization, green screen effects, or all of the above blended
based on the desired configuration of the installation artist. See \xf{fig:timmy-fancy-bubbles-1}
and \xf{fig:timmy-fancy-bubbles-2} for
examples~\cite{song-phd-memory-bubble-video-2012}\footnote{\url{https://vimeo.com/51329588}}.

\section{Design and Implementation}
\label{sect:illimitable-space-design-impl}

The design and realization of this installation centers around the notion
of special data streams provided by the {\kinect} sensor and its drivers~\cite{kinect-ms-sdk,kinect-ms-toolkit}.
These specifically include in our case the color stream, depth stream, skeleton
stream, and audio stream (source). The streams are provided by the sensor's driver to
the installation application in real time with the corresponding API
(see \xs{sect:illimitable-api}).

We combine color and depth streams to form a single texture frame by frame,
which is used in a green-screen like processing regardless the background,
or rainbow-color the depth layers that are truly representing the perception
of depth (see, e.g., \xf{fig:dance-kinect}).

Another important aspect of the installation is the sound (see \xs{sect:illimitable-lighting-audio}).
A melody playback via the {\bassaudiolib} library~\cite{bass-audio-library-dot-net} is
translated into the visual representation and blended into the
depth/color images. It also can drive the animation of a fancy
soap bubble presented earlier in \xc{chapt:interactive-docu} for its
music-driven wave pattern.

The skeleton stream gives us a reference point of the performer
presence in the scene. These data are important for the gesture-based
interaction aspects (see \xs{sect:tangible-interaction-animation}).

We begin the review of the conceptual design in \xs{sect:illimitable-space-conceptual-design}
and the associated challenges in \xs{sect:illimitable-space-challenges}. We then present
a detailed overview of the technical production and realization of the
work in \xs{sect:illimitable-space-impl}.

\subsection{Conceptual Design}
\label{sect:illimitable-space-conceptual-design}

The logical pipeline of the performance installation is depicted in
\xf{fig:ConceptualDesign-theatre},
which combines computer graphics,
documentary montage, interactive media, and theatre performance
into one artistic piece, which transitions through
four primary states:

\begin{figure*}[htpb]%
	\centering
	\includegraphics[width=\textwidth]{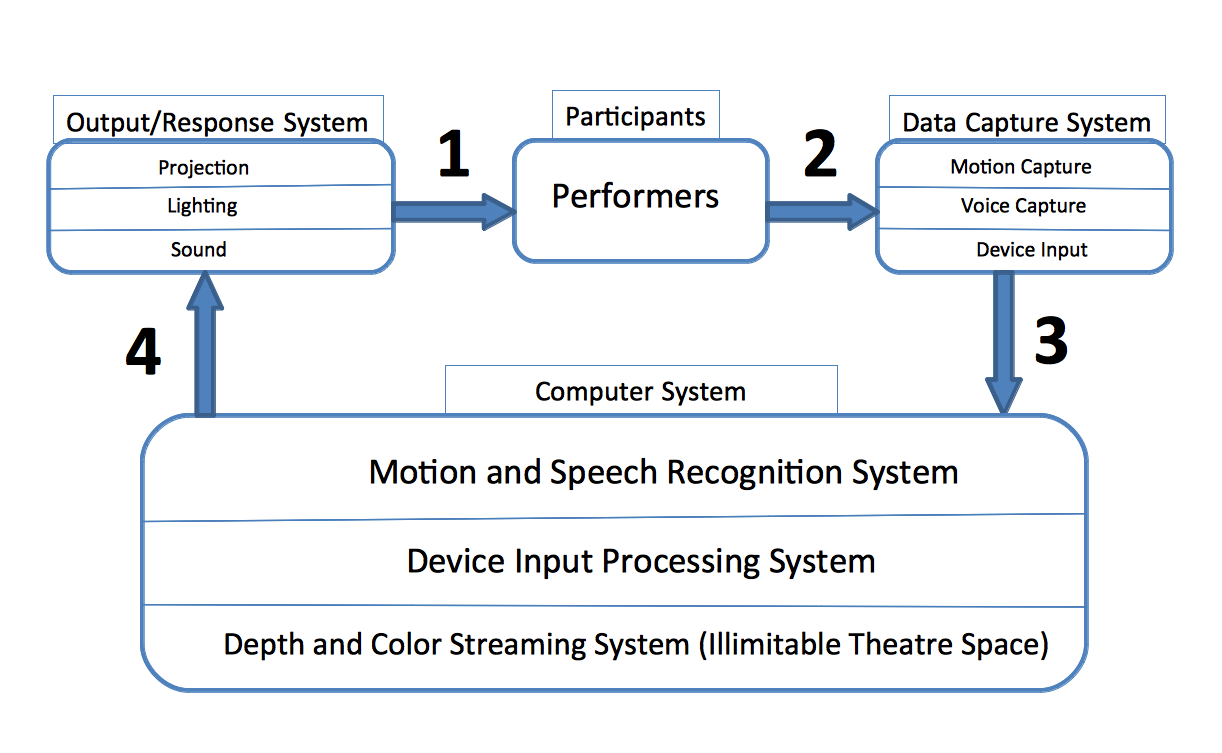}%
	\caption{Conceptual Design of an Interactive Performance Installation}%
	\label{fig:ConceptualDesign-theatre}%
\end{figure*}

\begin{itemize}
\item
State 1:
In a makeshift initial theatrical space, such as blackbox, where lighting, projection, sound
props, and related facilities are properly set up for actors' performance, audience's participation,
and other user group's interaction devices.

\item
State 2:
Participants, including actors, audience, and other users
could freely move in the physical space. They could move their body in different postures
with different gestures, talk or sing, and could play
musical instruments, or touch haptic-enabled devices.
The state of human being's performance,
such as motion of actors' body movements,
audience's voices and the forces of users applied through the haptic devices,
are captured through the integrated Data Capture System to generate
a response.

\item
State 3:
The captured data from the theatrical space are transferred to the
PACS (Performance Assistant Computer System) to be analyzed.
The computer system mainly contains three big components:
recognition, computing, and media output.
The motion recognition system, voice recognition system,
and device processing system are filtering, parsing,
and sorting motion, audio, and device/sensor input streams
captured into system by various capture devices.
The computer system then computes and simulates
physical based computer graphics and dynamic audio based on
the mathematical and physical based algorithms,
and fetch proper video footage from a video database management
system following some querying algorithm.

\item
State 4:
The outputting system then reconfigures the theatrical technologies
afterward based on the computation results and state updates.
The outputting system then re-renders the newly computed computer
generated imagery onto projectors, plays generated audio signals through speakers,
and even controls dynamic lighting (interacted with originally
as a result of participants' actions). 
\end{itemize}

In fact, the presented abstract system is applicable and can be generalized
to incorporate all the presented concrete simulation and animated systems
so far: the
Softbody Simulation (see \xc{chapt:softbody-jellyfish}),
Tangible Memories (see \xc{chapt:interactive-docu}),
and the Illimitable Space performance presented in this chapter.

\begin{figure*}[htpb!]%
	\centering
	\includegraphics[width=\textwidth]{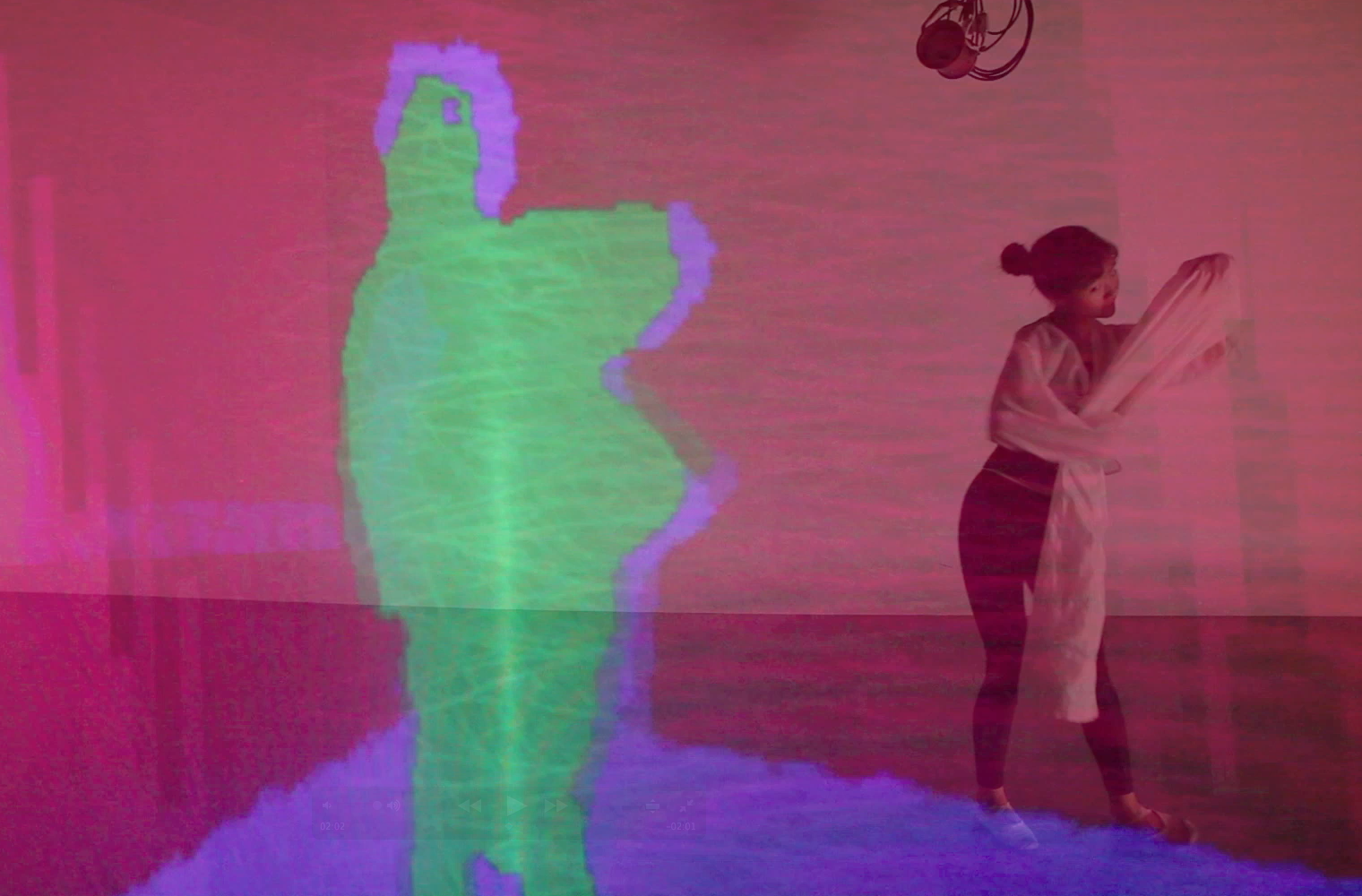}%
	\caption{Depth-Based Dance Performance Capture and Visualization}%
	\label{fig:dance-kinect}%
\end{figure*}

\subsection{Challenges}
\label{sect:illimitable-space-challenges}

The core posed challenges presented in this work have the same constraint:

\begin{itemize}
	\item Realtime sound/music
	\item Realtime graphics/video effects
	\item Realtime CG objects physical based simulation/digital puppet theatre
	\item Live performance
\end{itemize}

That is being live and real-time--i.e. happening \emph{now} and at the
same time as opposed to a pre-recorded post-production piece. The hardware
and software technology we use today allow us, inexpensively, to meet
this hard constraint.

\subsection{Implementation and Production}
\label{sect:illimitable-space-impl}

The implementation, realization, and production of the interactive performance
involves the use of the technological innovations in terms of software and hardware
presented earlier as well as traditional dance, and documentary art forms.
This corresponds to the software APIs used in \xs{sect:illimitable-api},
modeling details in \xs{sect:illimitable-modeling}, projection aspects described
in \xs{sect:illimitable-space-projection}, details on the lighting setup
and audio environment are described in \xs{sect:illimitable-lighting-audio},
the live performance aspects weaved into the installation as well as some
preliminary proof-of-concept stills are rendered in \xs{sect:illimitable-live-performance},
then how the performance interaction and the real-time animation with it are
detailed in \xs{sect:tangible-interaction-animation}, and, then, we review
how that maps back to the
most prominent
theatre elements in \xs{sect:tangible-interaction-theatre-elements}.

\subsubsection{APIs}
\label{sect:illimitable-api}

We rely on a number of the APIs introduced and detailed earlier
in part or in detail in \xs{sect:bg-kinect} and \xs{sect:interactive-docu-design-impl}
as the technologies used are integrated together in the same application, so they can be used for different
installation types, including the theatrical performance or interactive documentary,
but the emphasis shifts between different aspects of the implementation can be
configured for the desired installation type and environment.

\paragraph{{\xna}.}
\label{sect:illimitable-api-xna}

A lot of the {\xna} API described earlier in \xs{sect:impl-xna} is directly
applicable in this installation as well as the two share a lot of elements in common and the
illimitable space installation is in fact a generalization of all the
approaches presented here.

The additional API is used more in this installation is the \api{RenderTargetCube}
from \apipackage{Microsoft.Xna.Framework.Graphics}
for the dynamically updated sphere mapping onto the fancy soap bubble model with the
desired media (instead of the default static \api{TextureCube}).
Each face in the cube is an updatable \api{Texture2D} allowing us
to project onto the bubble any kind of video playback, animations, visualizations,
etc.

A \api{Model} instance is loaded with \file{Ground.x}
for the primary ground projection (``ground'' here denotes a flat
surface that is at the ``bottom'' of the virtual environment in the installation by default,
but technically can be projected anywhere, e.g. such as walls, ceiling, or media bubble's content).
The \api{Effect} instance is loaded with \file{Bubble.fx}~\cite{fancy-bubble-sample-xna3} as before
for the {\hlsl} fancy soap bubble shaders (a single pass
collection of the vertex and pixel shader to map the glow texture to the cube
map onto the bubbly sphere while deforming its mesh).

The \api{RenderTarget2D} is used for blended textures used in the sound visualization
projected onto the ground. There are targets each having \api{DepthFormat.Depth24}
color bits and render the sound frequency and wave coefficients as rotating amplitude curves
and spectrum bands blended into the same \api{SpriteBatch} with the alpha channel enabled
for transparency and without erasing the previous content for the curves~\cite{talkshowhost-xna}
(see \xs{sect:illimitable-api-bass} for more details). We subsequently blend this visualization
with the depth, skeleton, and green-screen color images to form a final image, such as,
e.g., in \xf{fig:dibubu-four-dancers-1} and \xf{fig:dibubu-four-dancers-2}.

\paragraph{{\kinect}.}
\label{sect:illimitable-api-kinect}

A lot of the {\kinect} SDK API described in \xs{sect:impl-kinect} is also
applicable here. Additional elements in this installation include the
more extensive use of the \apipackage{Microsoft.Kinect.KinectSensor}'s \api{DepthStream}, \api{SkeletonStream},
and \api{AudioStream} APIs. They help to realize the methodology and conceptual design
described in this chapter. The algorithm presented earlier
in \xl{algo:kinect-audio-recognition-bubbles} is applicable here as well
with the extensions mentioned: just extra paths that are configured to invoke
the processing of the three additional streams, mapping their content
to color and visualization, and finally rendering them.

Most of the relevant API comes from the \apipackage{Microsoft.Kinect} package.

First, we work with the visible spectrum camera and its color stream. We
capture each \api{ColorImageFrame} with the default format of \api{ColorImageFormat}
set to \api{RgbResolution640x480Fps30}, which is a reasonable compromise.
Then we maintain a \api{ColorImagePoint} array of color coordinates for maintaining the
depth-to-color mapping for the depth visualization. The \api{KinectSensor}'s API
provides a helper method to that effect of \api{KinectSensor.MapDepthFrameToColorFrame()}.
The majority of this color stream API invocation comes from the callback handler called
\api{kinectSensor\_ColorFrameReady()}, which in addition to what was described does
the greenscreen processing/abstraction by capturing the depth data of the object in motion
in the {\kinect}'s field of view, that becomes tracked and the corresponding set of indices
are mapped into the color frame's image and the rest of the pixels are set to transparent
as in the green screen example~\cite{kinect-ms-toolkit}.

Similarly to the color stream frames, each \api{DepthImageFrame} is acquired by our
callback handler \api{kinectSensor\_DepthFrameReady()} at the \api{DepthImageFormat} of
\api{Resolution320x240Fps30}. Each depth value is equally-spaced in the spectrum beween
the nearest and farthest values to map to the near-rainbow color layers with the red 
begin the farthest and violet the nearest layers (see e.g. \xf{fig:sleeve-1} and \xf{fig:sleeve-2}).

The \api{KinectSensor.AudioSource} serves as the audio stream input, typically however
enabled when the speech recognition is on (see \xs{sect:impl-kinect}). The audio stream
processing from an \file{.mp3} file for visual music feedback is described in \xs{sect:illimitable-api-bass}.

We also receive in the callback handler \api{kinectSensor\_SkeletonFrameReady()}
\api{SkeletonFrame}s with the \api{Skeleton} data that allow us
to track \api{Skeleton.Joints}. These are used in \api{SkeletonPoint}s
together with the corresponding method called \api{KinectSensor.MapSkeletonPointToColor()}~\cite{kinect-fundamentals-skeleton-tracking,kinect-ms-sdk}
both to visualize the skeleton as an element of debugging or performance (e.g., see \xf{fig:dibubu-four-dancers-1}
and \xf{fig:dibubu-four-dancers-2}) and to use this information to interact with the virtual objects on the scene,
e.g., bouncing things off (\api{Skeleton.Joints} gives us a lot of elements of a tracked skeleton
encoded in \api{JointType}, e.g. \api{JointType.Head}, \api{JointType.HandRight},
\api{JointType.FootLeft}, \api{JointType.AnkleRight}, \api{JointType.HipCenter}, and many others)
or rotating the objects by hand-waving (see, e.g., \xf{fig:timmy-fancy-bubbles-1}
and \xf{fig:timmy-fancy-bubbles-1}).

\paragraph{BASS.NET.}
\label{sect:illimitable-api-bass}

The .NET wrapper~\cite{bass-audio-library-dot-net} of
{\bassaudiolib}~\cite{bass-audio-library} offers API to tap into the audio stream's properties
including the beat via frequency characteristics. This information is used as an input to drive the animation
of the music visualization on the projection plane as well as the fancy soap bubble
rotation and wavy effect (e.g. \xf{fig:fancy-bubble-1} and \xf{fig:fancy-bubble-2}).
(Additionally, in the \docutitle{Tangible Memories} case the bubbles' random floating motion
can be affected by the music being played back at the time.)

We have integrated the music processing referenced from \api{TalkShowHostXNA}~\cite{talkshowhost-xna}
that relies on {\bassaudiolib}. The library is loaded, setup, and configured to playback \file{.mp3}
files.
More specifically,
from the \apipackage{Un4seen.Bass} package \api{BassNet} is loaded and registered (\api{BassNet.Registration()}
and the \api{Bass} class is the one that provides the functions needed
to playback \file{.mp3} file and access its properties during the playback.

\api{Bass.BASS\_Init()} initializes the library to sample at 44.1kHz.
Upon success, the library loads the specified file and creates a stream out of it and
returns a handle to it using the \api{Bass.BASS\_StreamCreateFile()}
call. At the same time it starts the playback with \api{Bass.BASS\_ChannelPlay()}.

The \api{Bass.BASS\_ChannelGetData()} call acquires the spectral characteristics of the sound
being played (usually a song, and more specifically for the Chines water sleeve dance performance
it is called \worktitle{Song}{Pipa Song} in Pinyin (``pipa'' is ``lute'')).
The \api{fltRotationWaves1} scalar variable is constantly updated in \api{TalkShowHostXNA} via
the fast Fourier transform (FFT)~\cite{shaughnessy2000} analysis to compute the 2048 frequency
coefficients and their peaks during the playback. We use the value of that variable to
update our \api{wave} variable that controls the fancy soap bubble wavy effect as well as
its rotation with the music beat.

The channel spectra is used to draw the spectral bars at the bottom-left and upper-right
mirror corner roughly equally distributed over the spectral coefficients. At the same time
the two wave patterns are drawn using different colors, remembering the previous state
and translating it sideways so it gradually fades away from the same FFT data combining
certain frequencies per pixels and approximating the lines between the two adjacent
point pixels.

Finally, the clean up BASS API includes functions
\api{Bass.BASS\_ChannelStop()},
\api{Bass.BASS\_StreamFree()},
\api{Bass.BASS\_Stop()}, and
\api{Bass.BASS\_Free()} to stop the playback and free resources.

\subsubsection{Modeling}
\label{sect:illimitable-modeling}

The modeling in this case is often very dynamic and is dependent
on the data streams from {\kinect}. There are also static elements
modeled in the installation. Otherwise, the modeling is very
simple in many aspects making it perform robustly and responsively in real-time.

\paragraph{Environment.}

The virtual environment gives the audience realistic
experience. %in the seasonal change.
This is to be done by either using video footage or
rendering synthetic 3D models to build the virtual environment
for the surrounding walls.
We also create a virtual skybox on the ceilings
and computer generated textures on the floor.

\paragraph{Media Rendering.}

Most of the media is rendered into a rectangular model of a polygon
consisting of two triangles or a render cube consisting of 6 of such
polygons. They have texture objects associated with them to which the
dynamic texture data are written. The alpha transparency for certain \api{Color}
pixels is used to blend multiple textures rendered onto the same polygon.
The soap bubble model and the polygon model are the same as
in \xs{sect:interactive-docu-modeling-xna}.

\subsubsection{Projection}
\label{sect:illimitable-space-projection}

A similar setup as in \xs{sect:interactive-docu-projection} can be and is used
in here in the basic installation setup and requirements.
The examples of the PoC projection are in
\xf{fig:water-sleeves-beginning},
\xf{fig:water-sleeves-figure-parenthesis},
\xf{fig:timmy-fancy-bubbles-1},
\xf{fig:timmy-fancy-bubbles-2},
\xf{fig:dibubu-four-dancers-1}, and
\xf{fig:dibubu-four-dancers-2}.
The future augmentation of the projection possibilities are further described in
\xs{sect:illimitable-space-projection-future}.

\begin{figure}[htpb!]
\centering
\subfigure[The Dancer is Ready to Begin Her Water Sleeve Dance]{
\includegraphics[width=.38\textwidth]{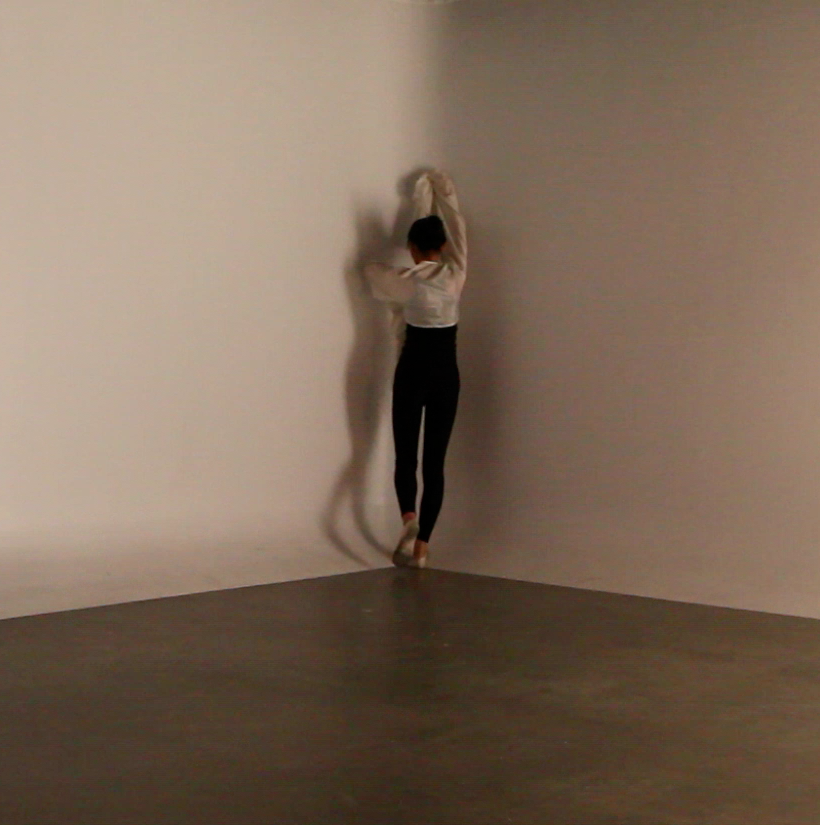}
\label{fig:watersleeve-beginning-real-1-1}
}
\subfigure[Depth View of the Same Scene; %
the Dancer is Not Detectable and Will Emerge Into the Scene. %
Music Visualization is Playing On Top Of the Depth Image.]{
\includegraphics[width=.52\textwidth]{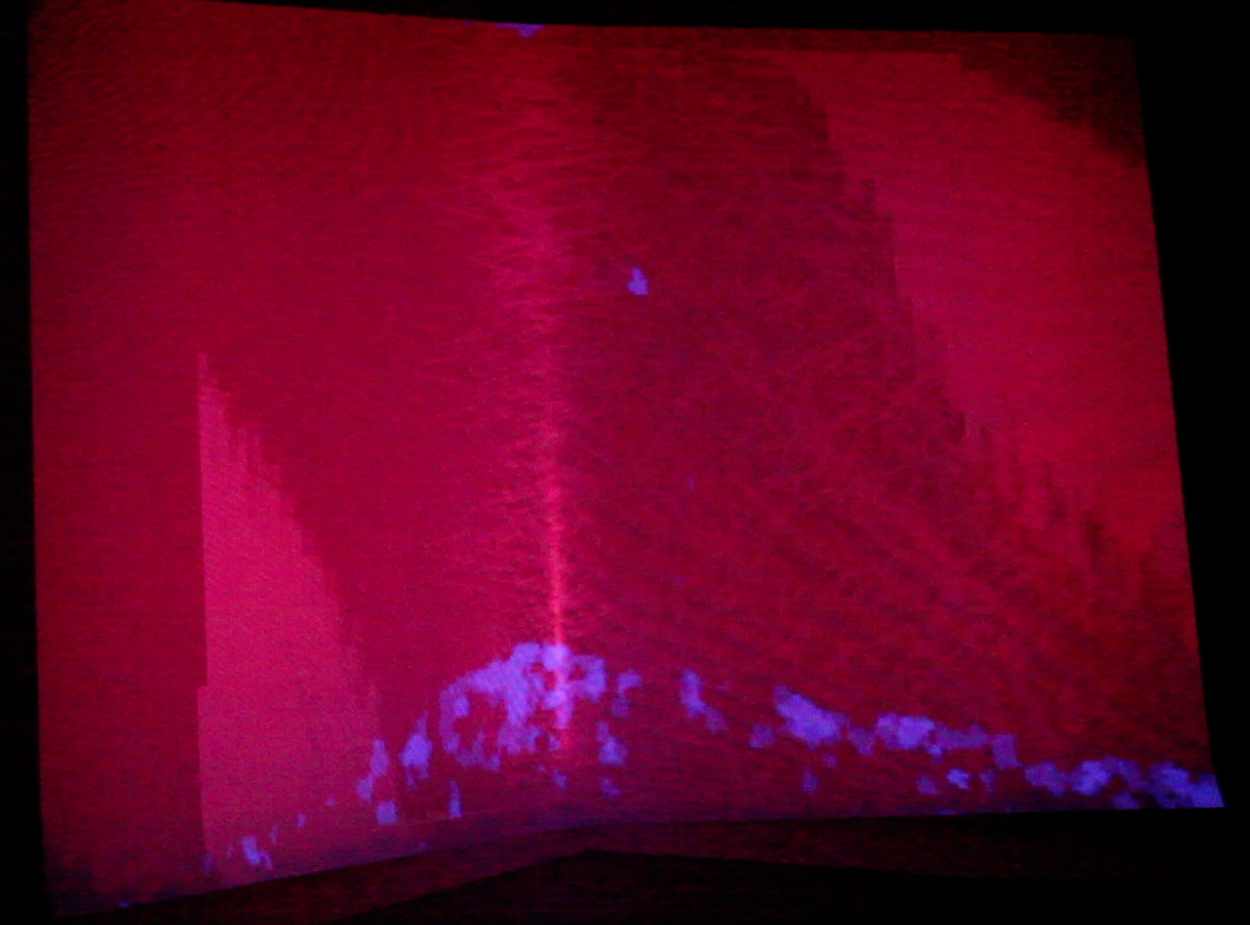}
\label{fig:watersleeve-beginning-depth-1-2}
}
\caption
	[Beginning of the Water Sleeve Dance by a Dancer (Real and Depth)]
	{Beginning of the Water Sleeve Dance by a Dancer (Real and Depth)}
\label{fig:water-sleeves-beginning}
\end{figure}

\begin{figure}[htpb!]
\centering
\subfigure[The Dancer is in the Middle of a Figure in Water Sleeve Dance]{
\includegraphics[width=.38\textwidth]{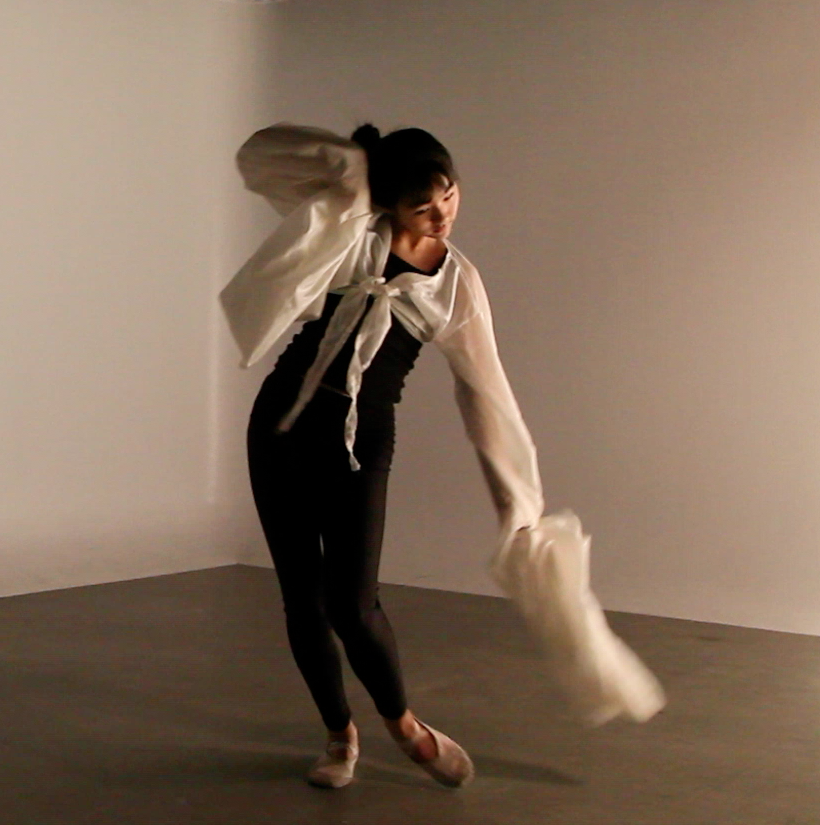}
\label{fig:watersleeve-real-2-1}
}
\subfigure[Depth View of the Same Scene; %
the Dancer is Detectable and Rendered with Different Depth Color Layers. %
Music Visualization is Playing On Top Of the Depth Image.]{
\includegraphics[width=.52\textwidth]{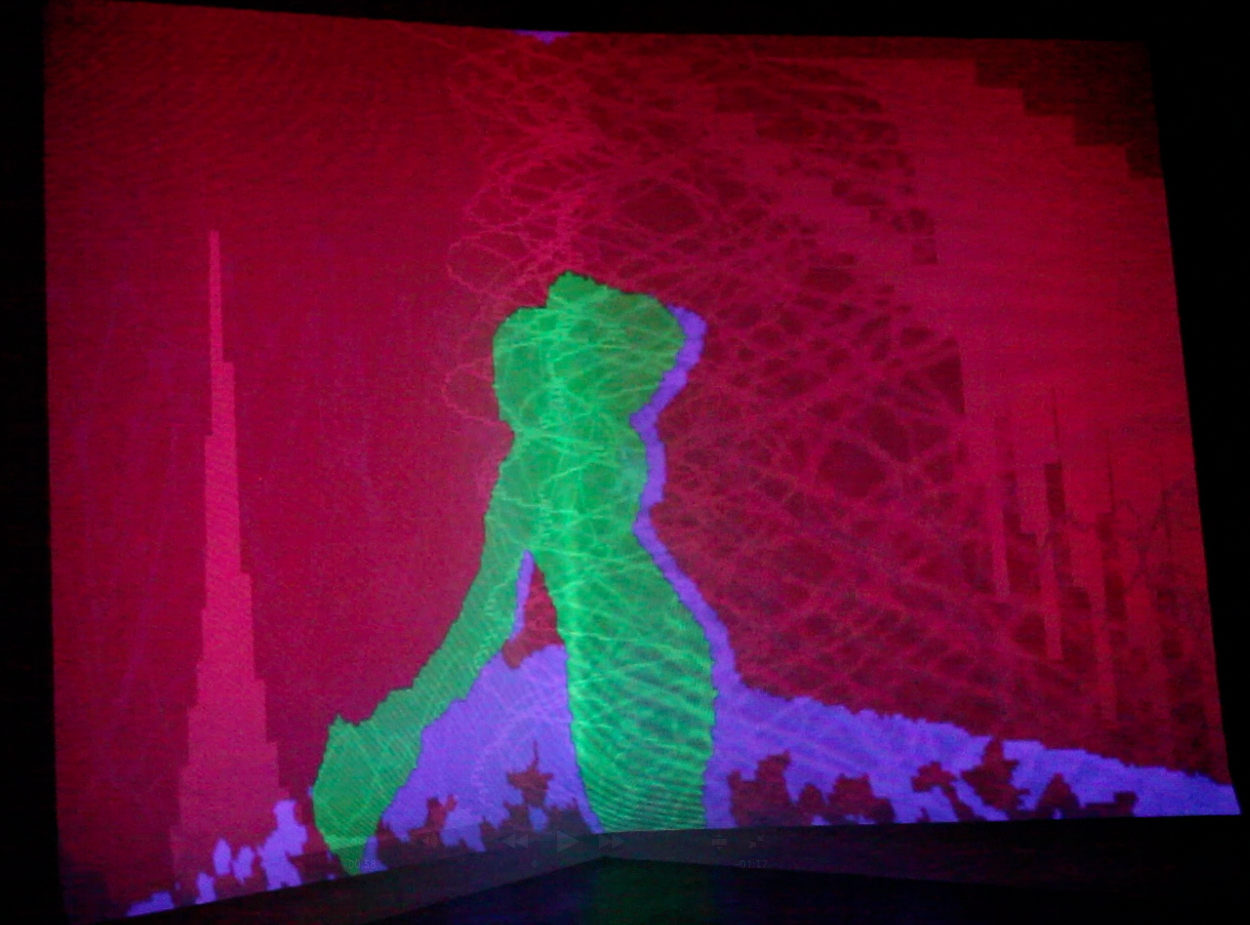}
\label{fig:watersleeve-depth-2-2}
}
\caption
	[A Figure in the Water Sleeve Dance Performance by a Dancer (Real and Depth)]
	{A Figure in the Water Sleeve Dance Performance by a Dancer (Real and Depth)}
\label{fig:water-sleeves-figure-parenthesis}
\end{figure}

\subsubsection{Lighting and Audio}
\label{sect:illimitable-lighting-audio}

Lighting is static for now (but is planned to be dynamic per \xs{sect:illimitable-lighting-audio-future}
using audio and motions as an input). It is setup in such a way using production
room spotlights as to highlight the performance area, but not directly at
the performer in case of smaller spaces, but toward the edge, such that the
performer and audience are not blinded and the colors captured by {\kinect}
for the color stream are not oversaturated. The area has to be lit just
enough for visibility and good quality video capture. There are almost no
spotlights directed to a nearby area in the production environment so as to
keep it dark enough for the projection to be the most crisp. Combinations
exist with almost no lighting as well as some lighting cast into the
projection and performance areas when the two are overlapping or merged.
Some of these are exemplified in the stills in this chapter. In the general
case the {\kinect}'s color stream should not be capturing projected performance
except the specific case where it is desired for additional add-on effects
of ``infinite depth''.

The sound in the installation configuration that does affect the visualization
or the animation of the projected fancy bubble comes from usually an \file{.mp3}
file read and played back via the BASS.NET library. The frequency patterns of
the beat are read and fed to the animation and visualization variables. This mode
of operation is primarily for a dance-like performance.

There is a known issue where the sound playback
may cause problems when the speech recognition is enabled at the same time.
That is when either a video playback with the sound on or a melody have speech in them resembling
the dictionary of words we used primarily in the interactive documentary
configuration (to call out bubbles, etc., see \xa{sect:kinect-voice-commands})
are detected, the system may accidentally begin a playback of that media content;
therefore, the speech recognition is usually disabled
in this installation when the audio playback is on.

\subsubsection{Live Performance}
\label{sect:illimitable-live-performance}

A live performance could be a play, a concert, or a dance show performed in front of audience.
The fascination of live performance is that audience could direct interact and connect with actors,
and it is the best way to engage audience with the performance and give actors emotional response.
It is the thrilling magic moment co-created by actors and audience.

Live performance in this installation, has even more advanced meaning of co-creation and interaction
between actors and audience.

The main disadvantages of the great recent works that involve performance and technology, e.g.
by Hollogne, Tsakos, and Puma~\cite{natasha-tsakos,marc-hollogne,puma-lift-commercial,puma-lift-behind-scenes}
is that, for the scenarios involving technology, the actors must rehearse, practice, and synchronize their
activities with the digital and cinematic pre-recorded and CG video replay, which is usually a pre-filmed
or an offline animated piece that is well timed with the dialog and actions of the on-stage actors. In my
opinion, this staticness unnecessarily constraints the actors and denies them extra dynamism and creativity
and instead of focusing on and immersing entirely into the \theatreelement{Character} the role
they are playing, they should also get distracted to recall at which time and place to synchronize
with the new media.
\textbf{The proposed work here is to remove such constraints and limitations and let
the actors fully embrace their performance while the technology dynamically would adjust and respond to the
actors' actions and \theatreelement{Diction} (speech). Thus, the actors are liberated from the
necessity of synchrony with technology.}

Let's take long sleeve dance performance for example: a ghost character could be visible/hidden
at different distances to the audience on the stage without the need to rehearse every
time for accurate timing and positioning (see, e.g., \xf{fig:water-sleeves-beginning}
and \xf{fig:water-sleeves-figure-parenthesis}).

In the fancy soap bubble example (\xf{fig:timmy-fancy-bubbles-1} and \xf{fig:timmy-fancy-bubbles-2}),
the audience could be augmented to use their gestures to alter (perhaps just slightly) the \theatreelement{Plot} of story
from a pre-scripted version possibly influencing the
actors who may change their decision in the performance based on the audience choices and inputs
(a traditional puppet show for children in Sherbrooke's cultural festival in 2012 has done something similar
and in some cases the puppets asked for children's advice and acted upon it in some parts of the
\theatreelement{Plot}).

In the virtual audience example (see \xf{fig:virtual-audience-dancers-1} and \xf{fig:virtual-audience-dancers-2}),
the audience may be captured to be projected onto the stage, and actors are part of
the virtual world as well in real-time, unlike Hollogne's.

There is more depth and connection of the augmented performance between the real, physical world
and the virtual world. Using a skeleton stream delivered in real-time, four dancers, actors
could drive in real-time a team of digital puppets (avatars) instead of one-to-one performance.

\begin{figure*}[htpb!]%
	\centering
	\includegraphics[width=\textwidth]{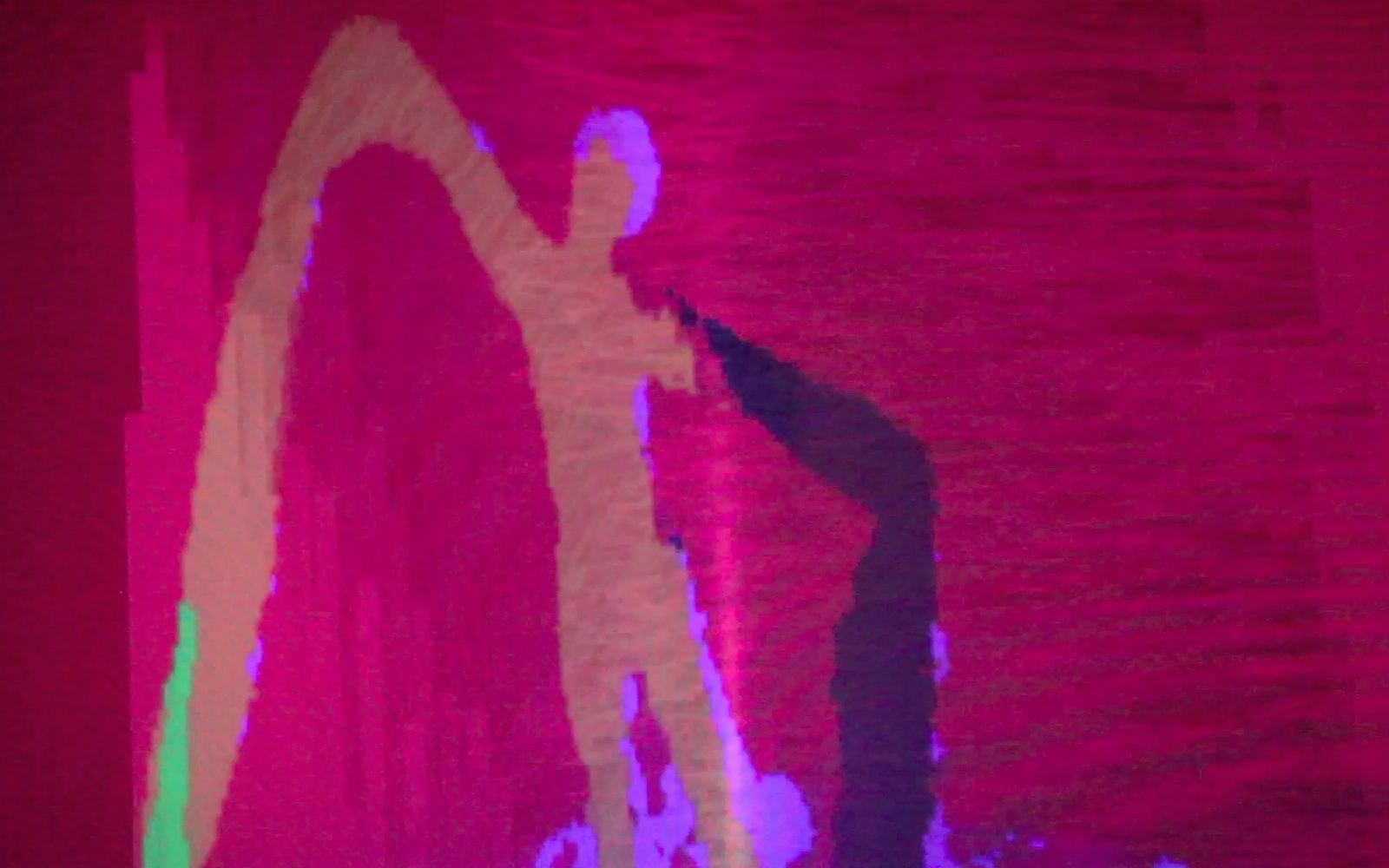}%
	\caption{Depth-Based Dance Performance Capture and Visualization of Long Sleeves 1}%
	\label{fig:sleeve-1}%
\end{figure*}

\begin{figure*}[htpb!]%
	\centering
	\includegraphics[width=\textwidth]{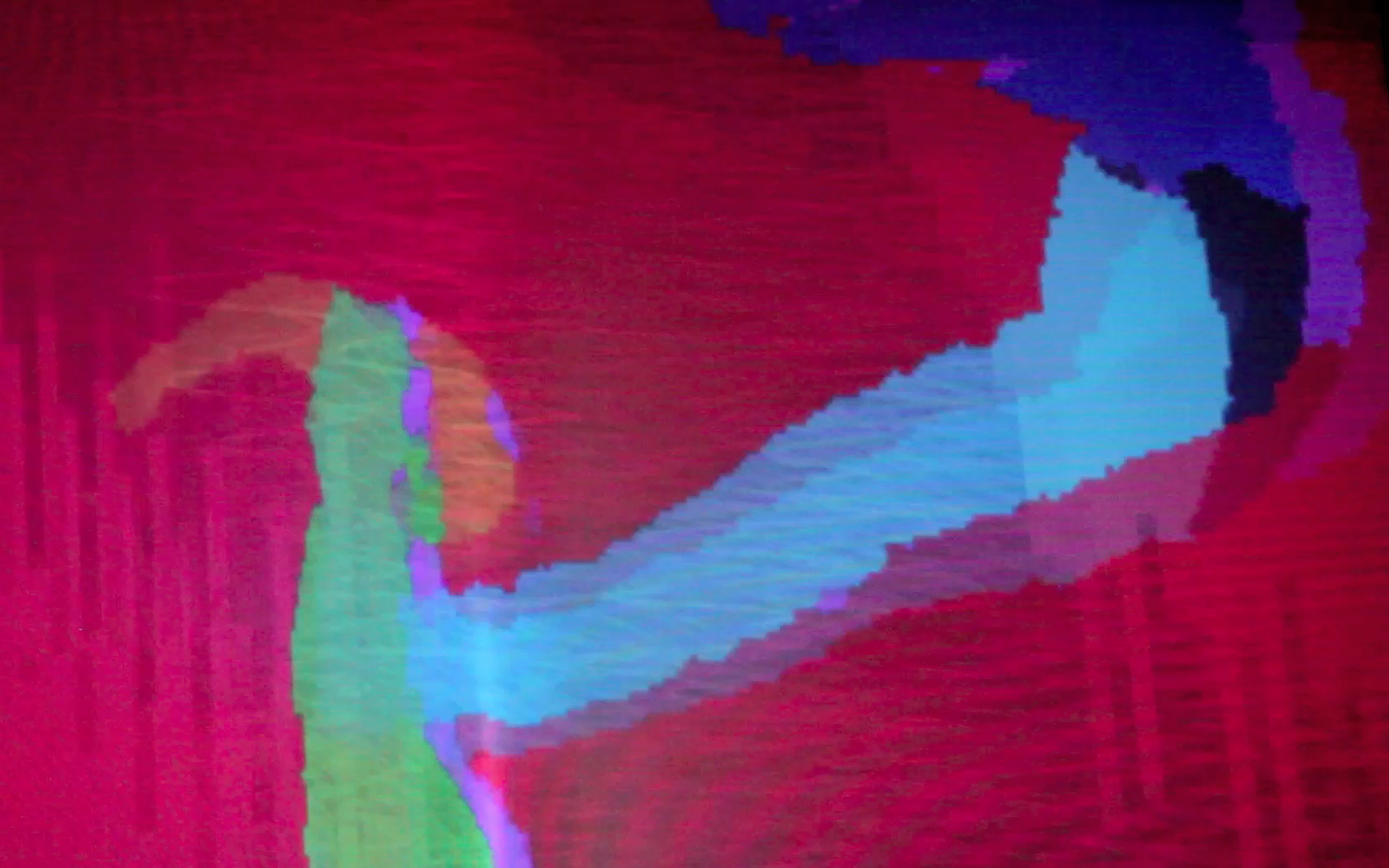}%
	\caption{Depth-Based Dance Performance Capture and Visualization of Long Sleeves 2}%
	\label{fig:sleeve-2}%
\end{figure*}

\begin{figure}[htpb!]
\centering
\subfigure[]{
\includegraphics[width=.47\textwidth]{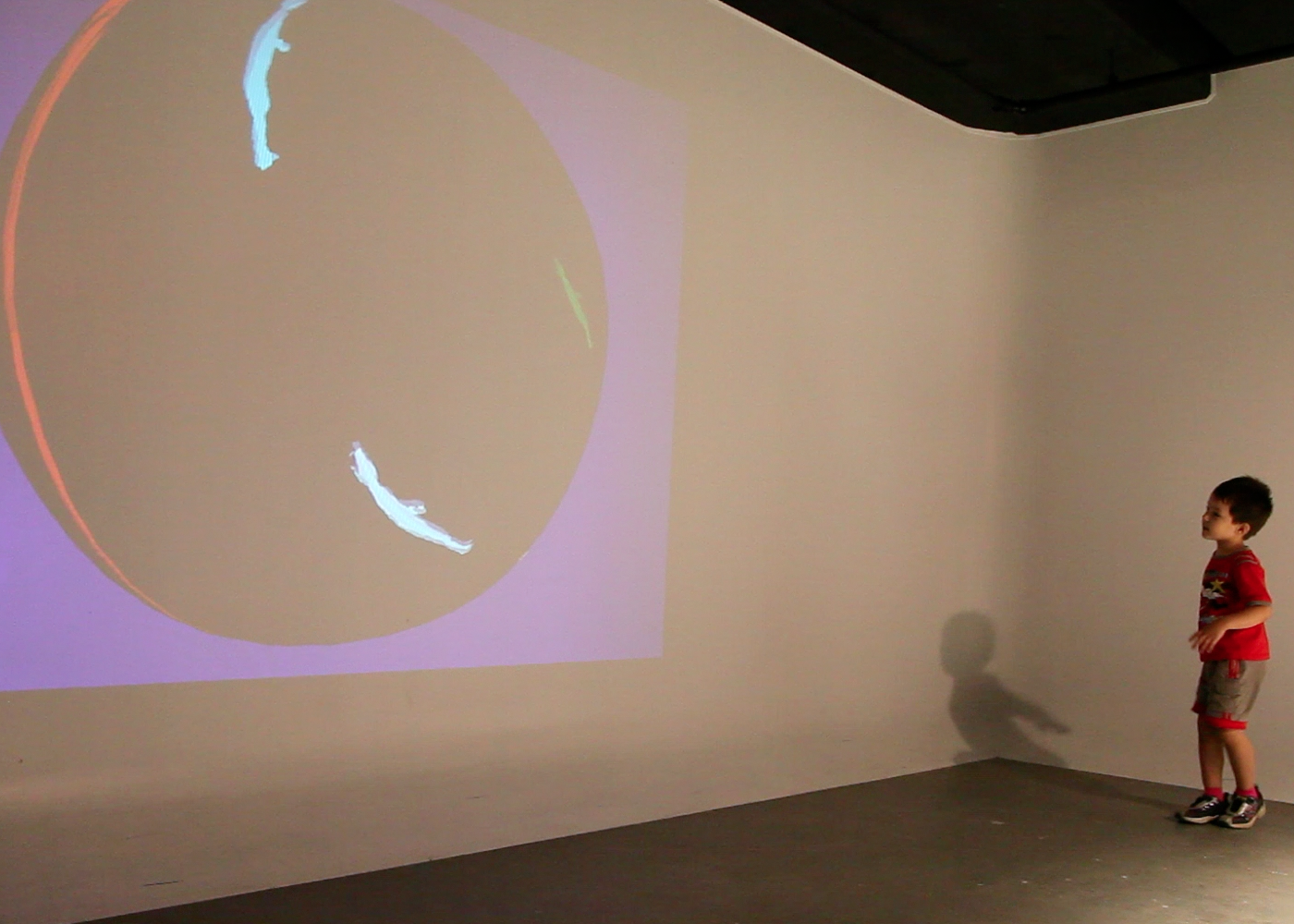}
\label{fig:fancy-bubble-1}
}
\subfigure[]{
\includegraphics[width=.47\textwidth]{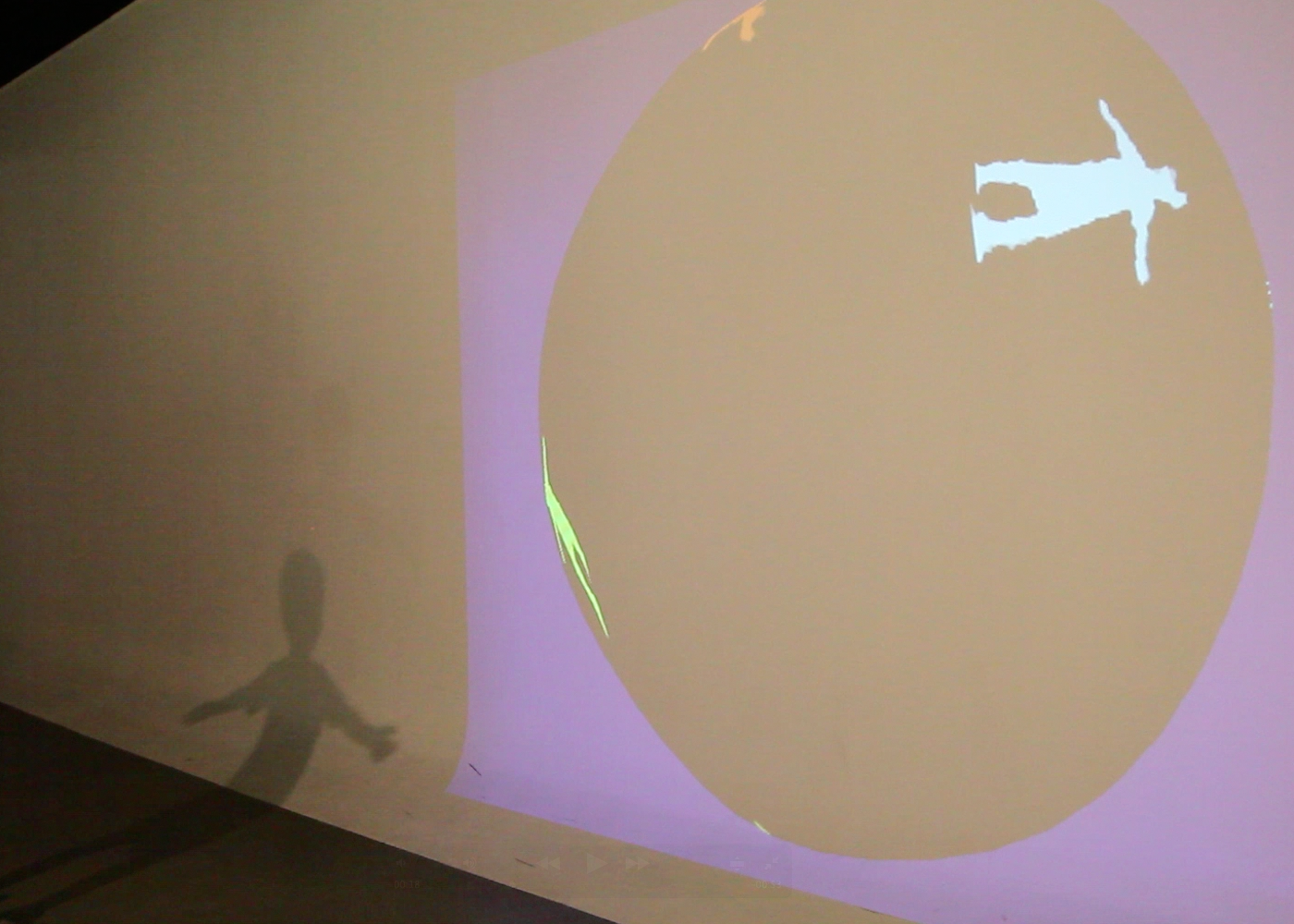}
\label{fig:fancy-bubble-2}
}
\caption
	[Fancy Soap Bubble With Depth Green Screened Audience Images Mapped Onto Its Surface Animated by a Melody]
	{Fancy Soap Bubble With Depth Green Screened Audience Images Mapped Onto Its Surface Animated by a Melody}
\label{fig:timmy-fancy-bubbles-1}
\end{figure}

\begin{figure}[htpb!]
\centering
\subfigure[]{
\includegraphics[width=.47\textwidth]{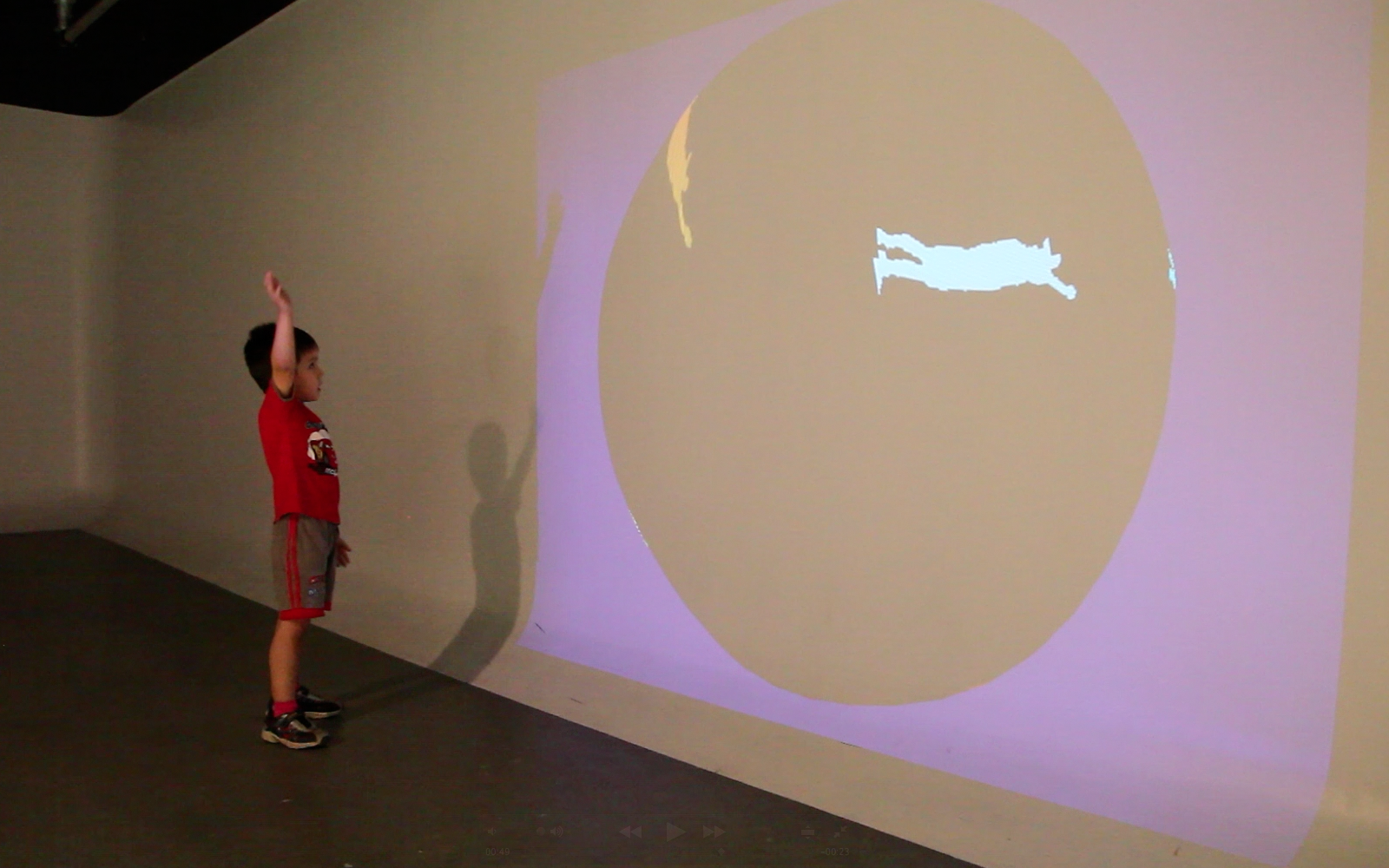}
\label{fig:fancy-bubble-3}
}
\subfigure[]{
\includegraphics[width=.47\textwidth]{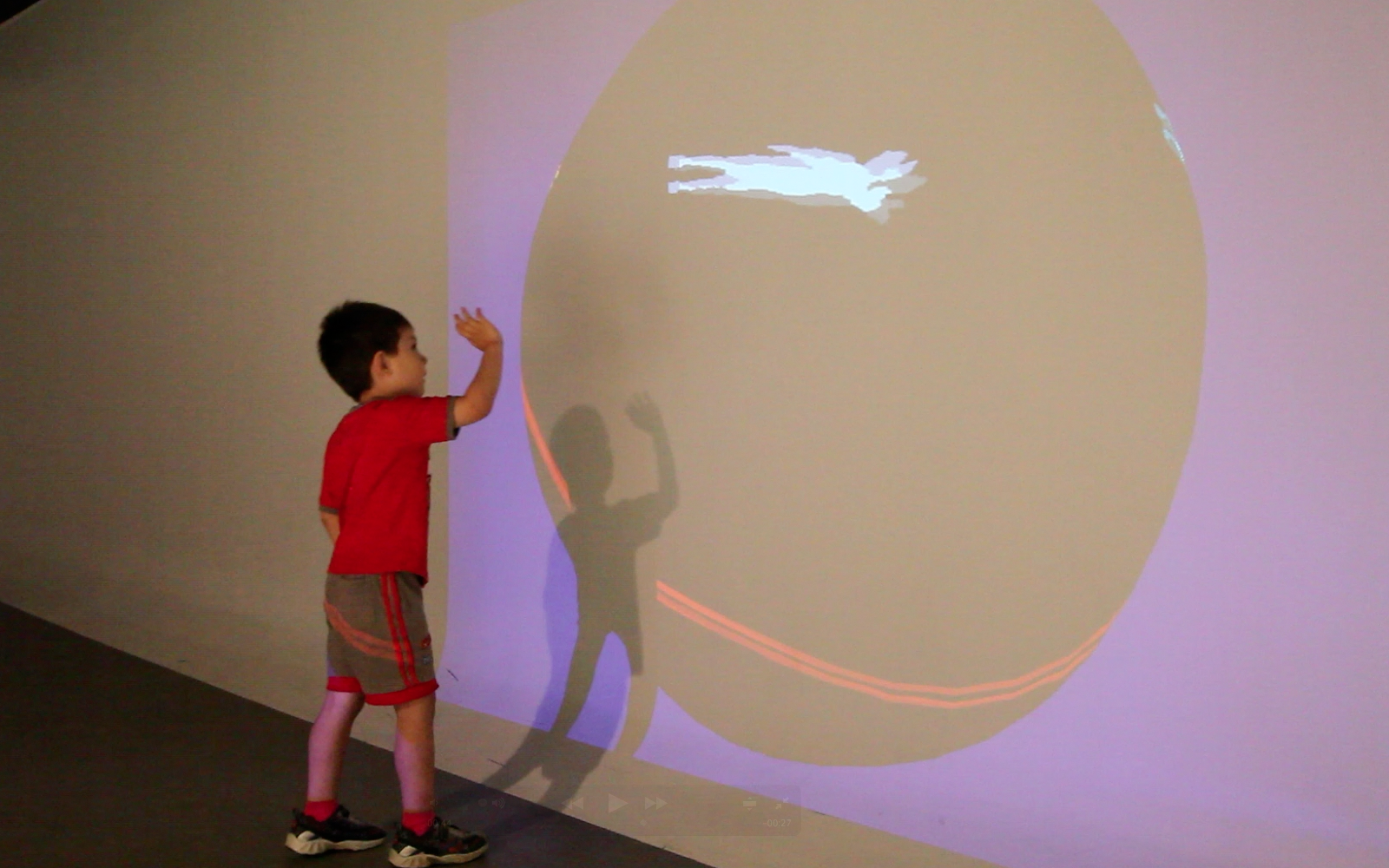}
\label{fig:fancy-bubble-4}
}
\caption
	[Fancy Soap Bubble With Depth Green Screened Audience Images Mapped Onto Its Surface Animated by Hand-Waving]
	{Fancy Soap Bubble With Depth Green Screened Audience Images Mapped Onto Its Surface Animated by Hand-Waving}
\label{fig:timmy-fancy-bubbles-2}
\end{figure}

\begin{figure}[htpb!]
\centering
\subfigure[]{
\includegraphics[width=.47\textwidth]{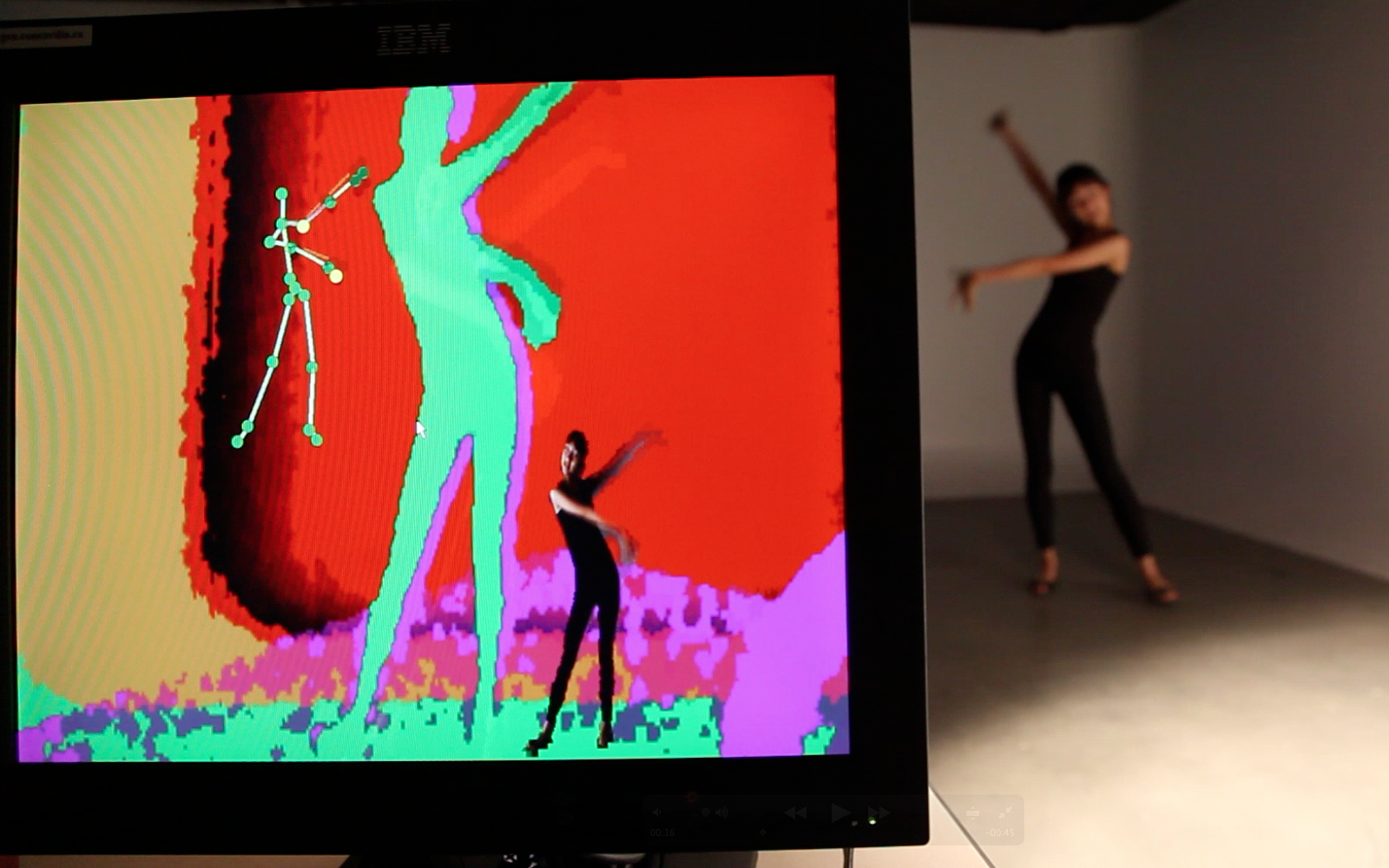}
\label{fig:four-dancers-1}
}
\subfigure[]{
\includegraphics[width=.47\textwidth]{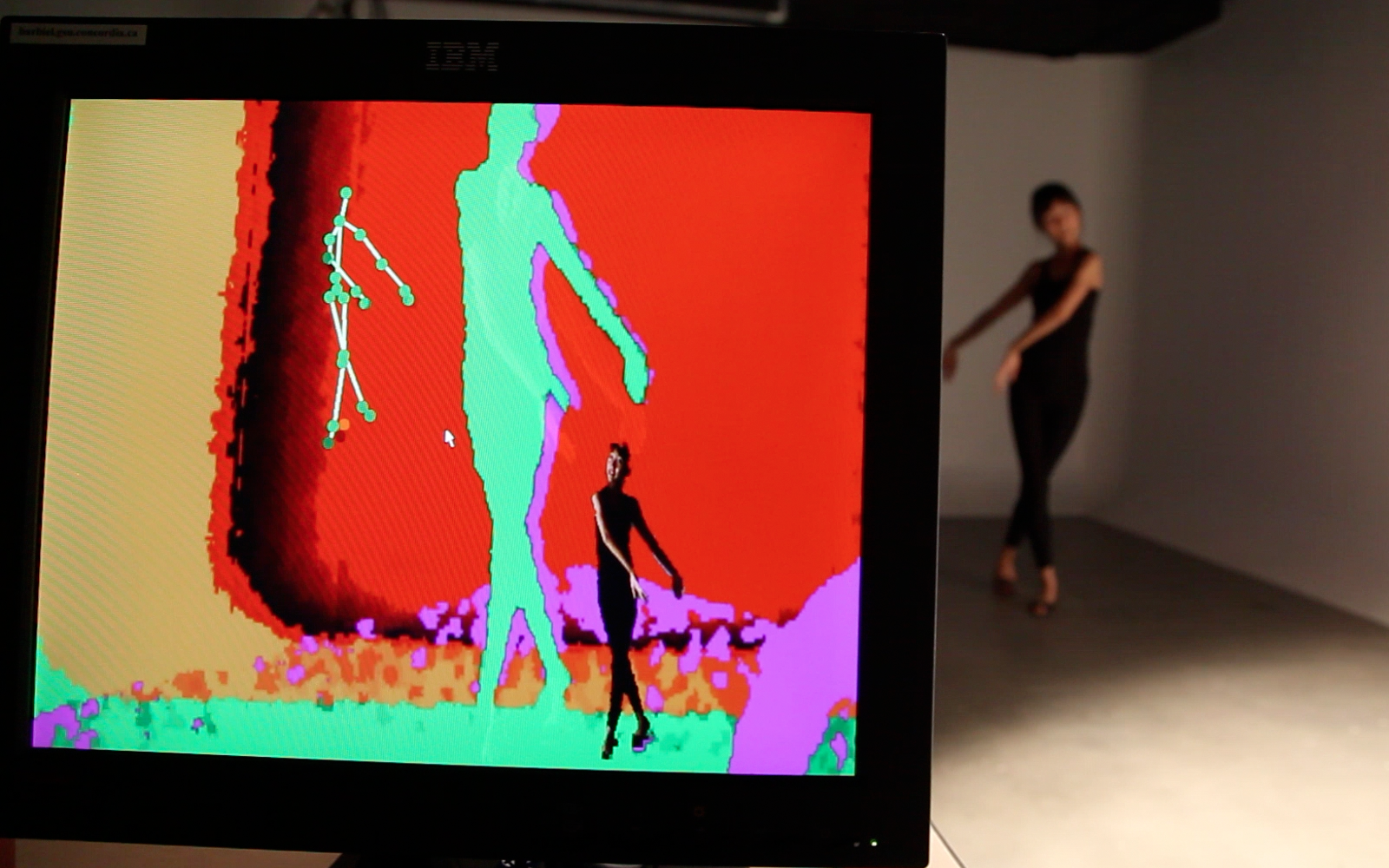}
\label{fig:four-dancers-2}
}
\caption
	[Four Dancers: Real, Depth, Greenscreened, and Skeleton]
	{Four Dancers: Real, Depth, Greenscreened, and Skeleton}
\label{fig:dibubu-four-dancers-1}
\end{figure}

\begin{figure}[htpb!]
\centering
\subfigure[]{
\includegraphics[width=.47\textwidth]{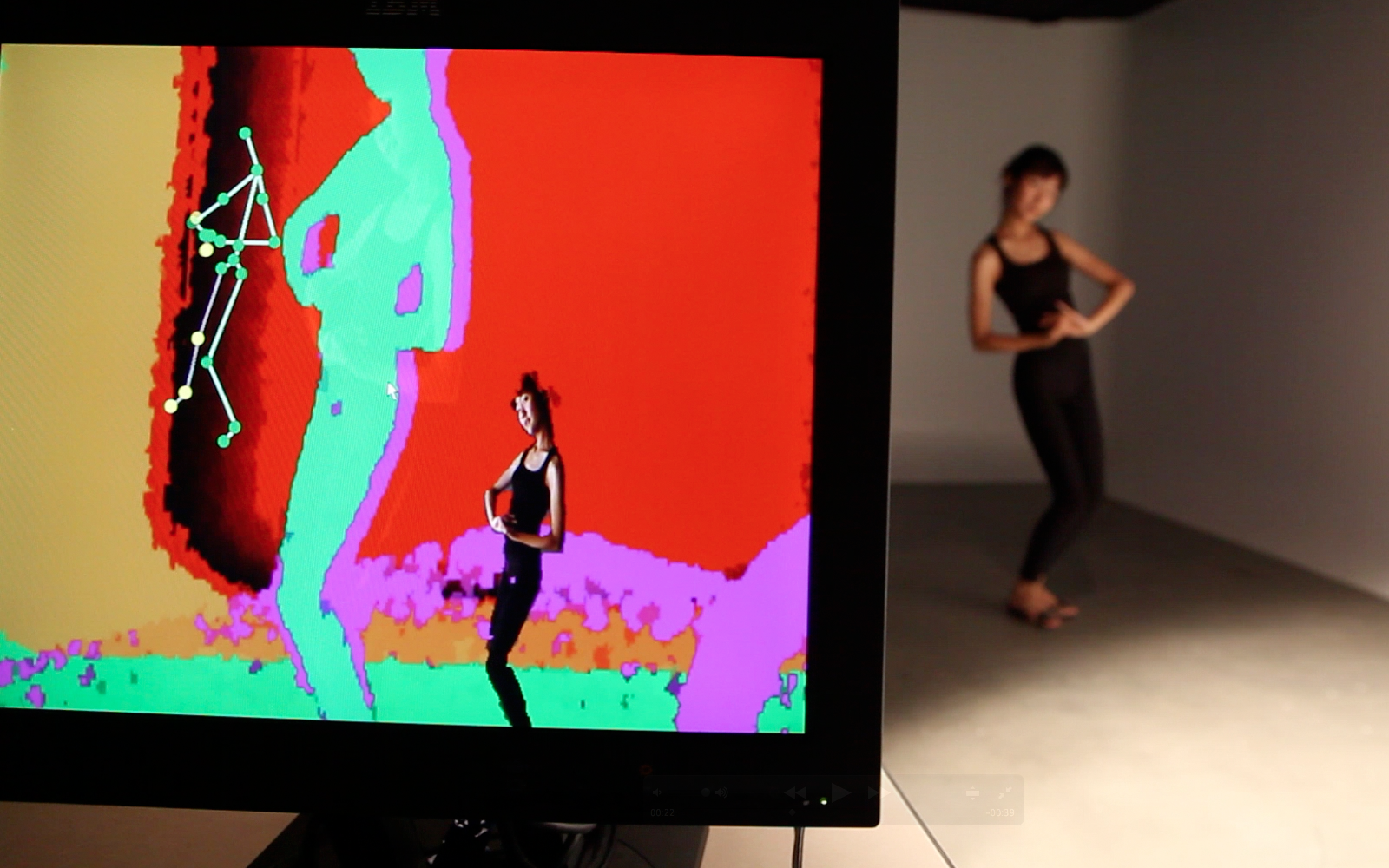}
\label{fig:four-dancers-3}
}
\subfigure[]{
\includegraphics[width=.47\textwidth]{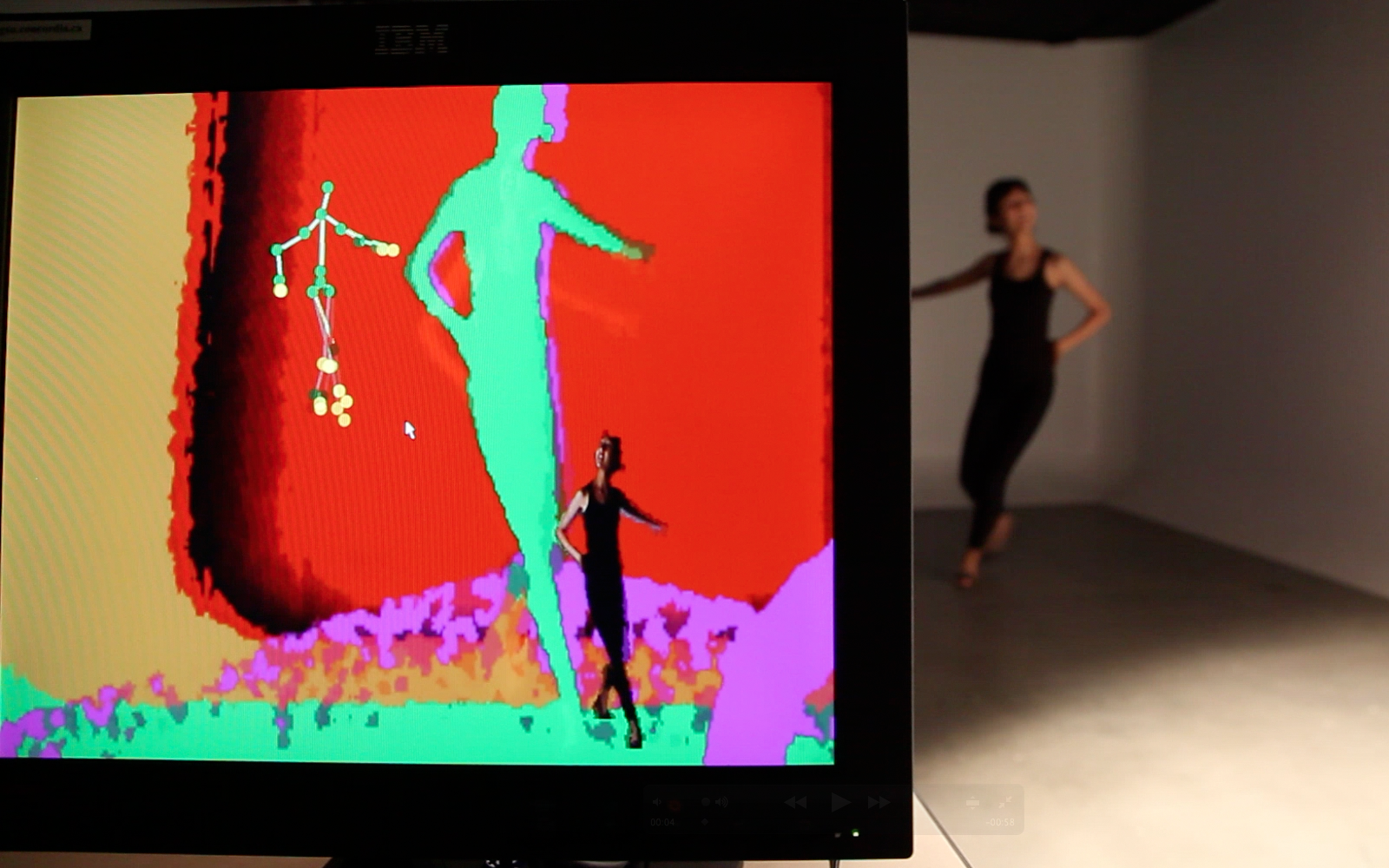}
\label{fig:four-dancers-4}
}
\caption
	[Four Dancers Again: Real, Depth, Greenscreened, and Skeleton]
	{Four Dancers Again: Real, Depth, Greenscreened, and Skeleton}
\label{fig:dibubu-four-dancers-2}
\end{figure}

\begin{figure}[htpb!]
\centering
\subfigure[]{
\includegraphics[width=.47\textwidth]{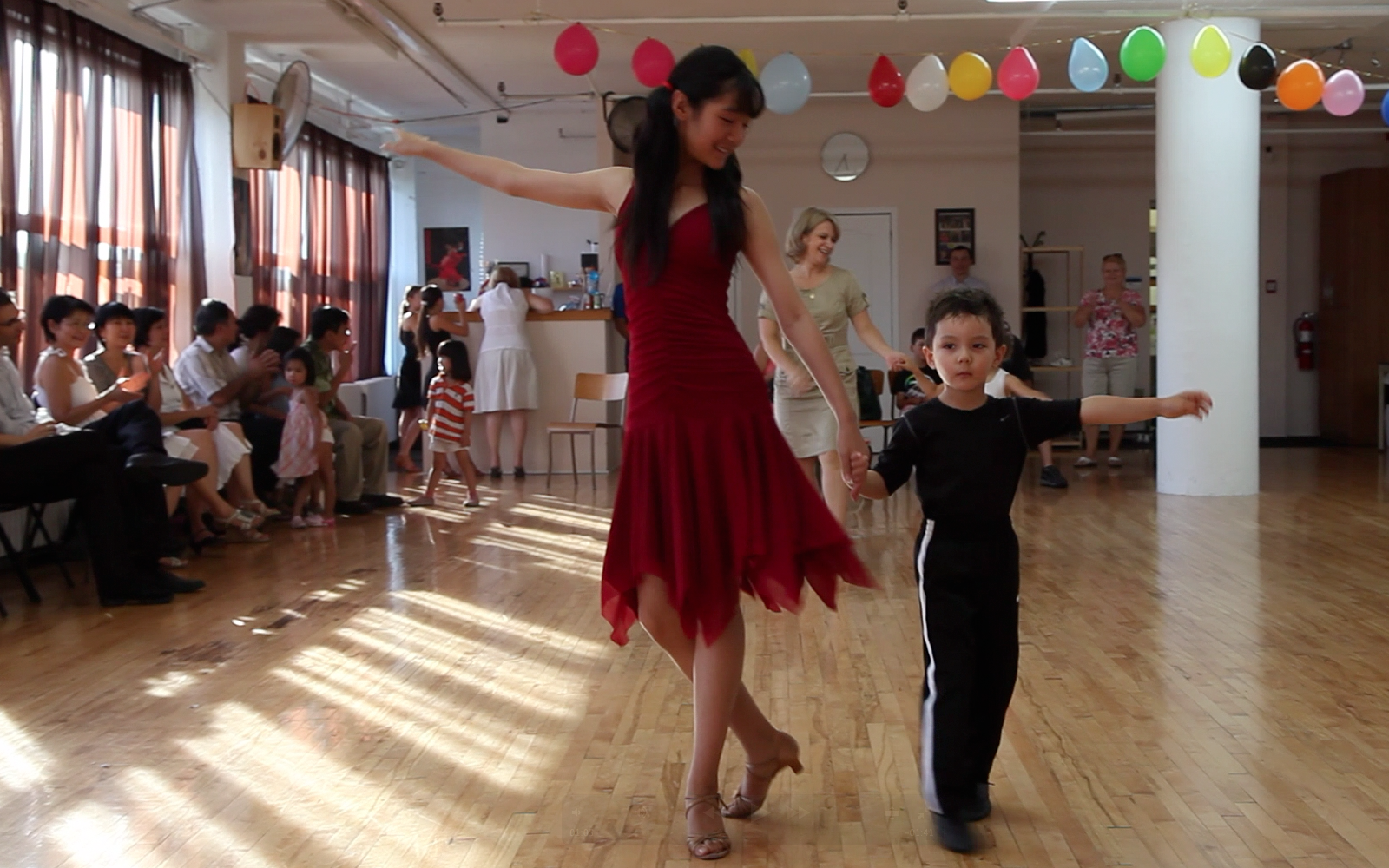}
\label{fig:virtual-audience-1}
}
\subfigure[]{
\includegraphics[width=.47\textwidth]{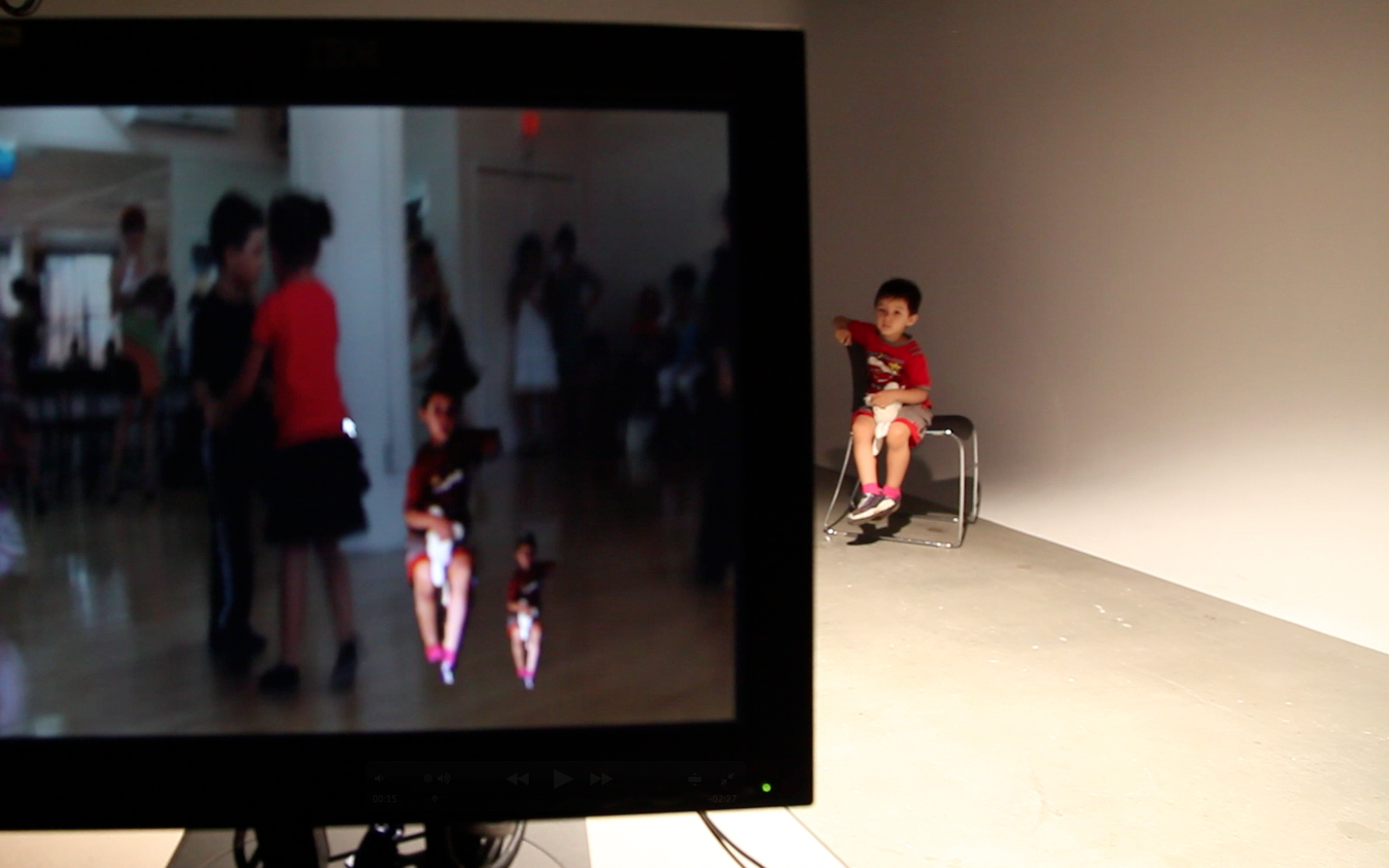}
\label{fig:virtual-audience-2}
}
\caption
	[Recorded Real Event Performance with Real-Time Additional Virtual Audience]
	{Recorded Real Event Performance with Real-Time Additional Virtual Audience}
\label{fig:virtual-audience-dancers-1}
\end{figure}

\begin{figure}[htpb!]
\centering
\subfigure[]{
\includegraphics[width=.47\textwidth]{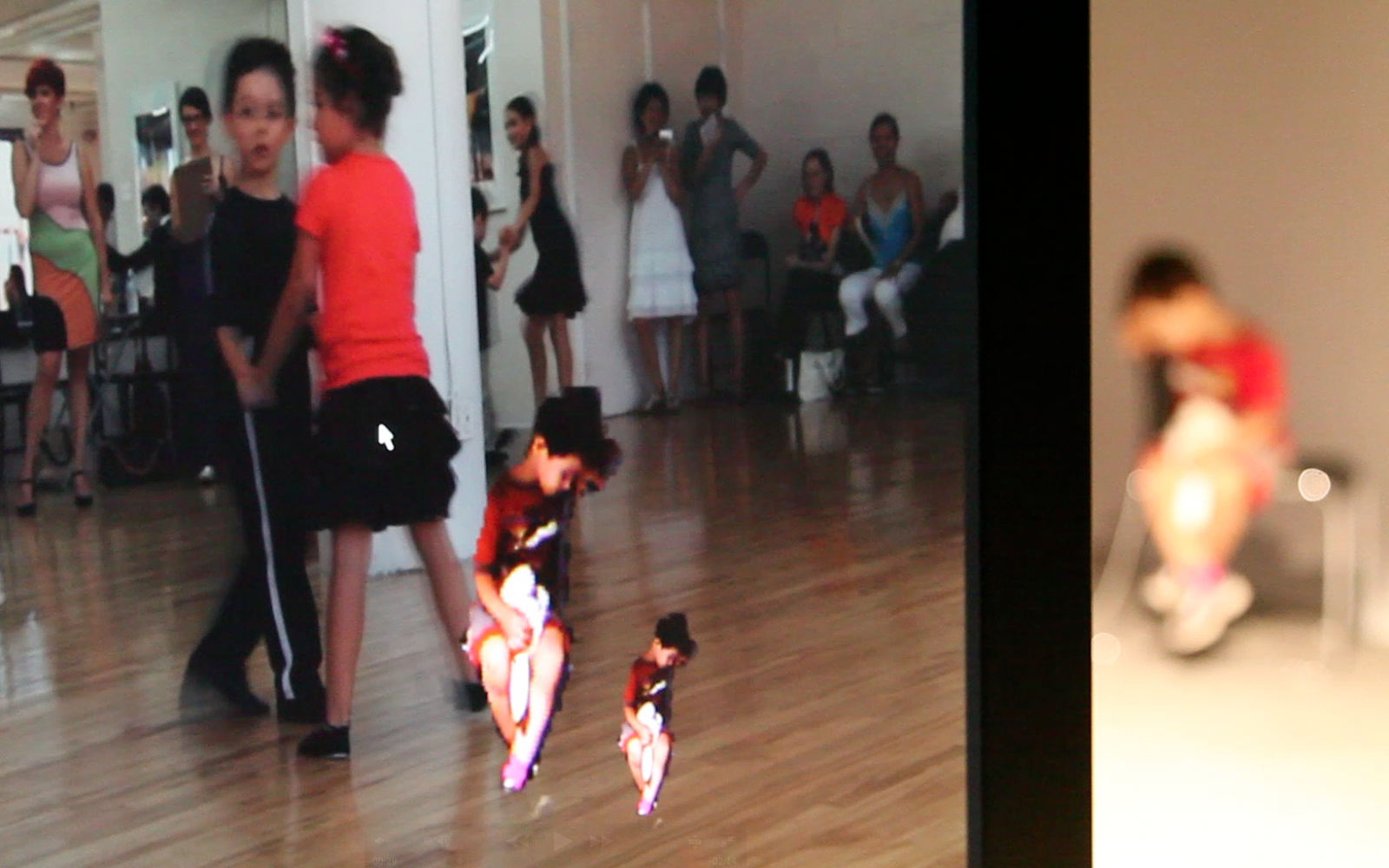}
\label{fig:virtual-audience-3}
}
\subfigure[]{
\includegraphics[width=.47\textwidth]{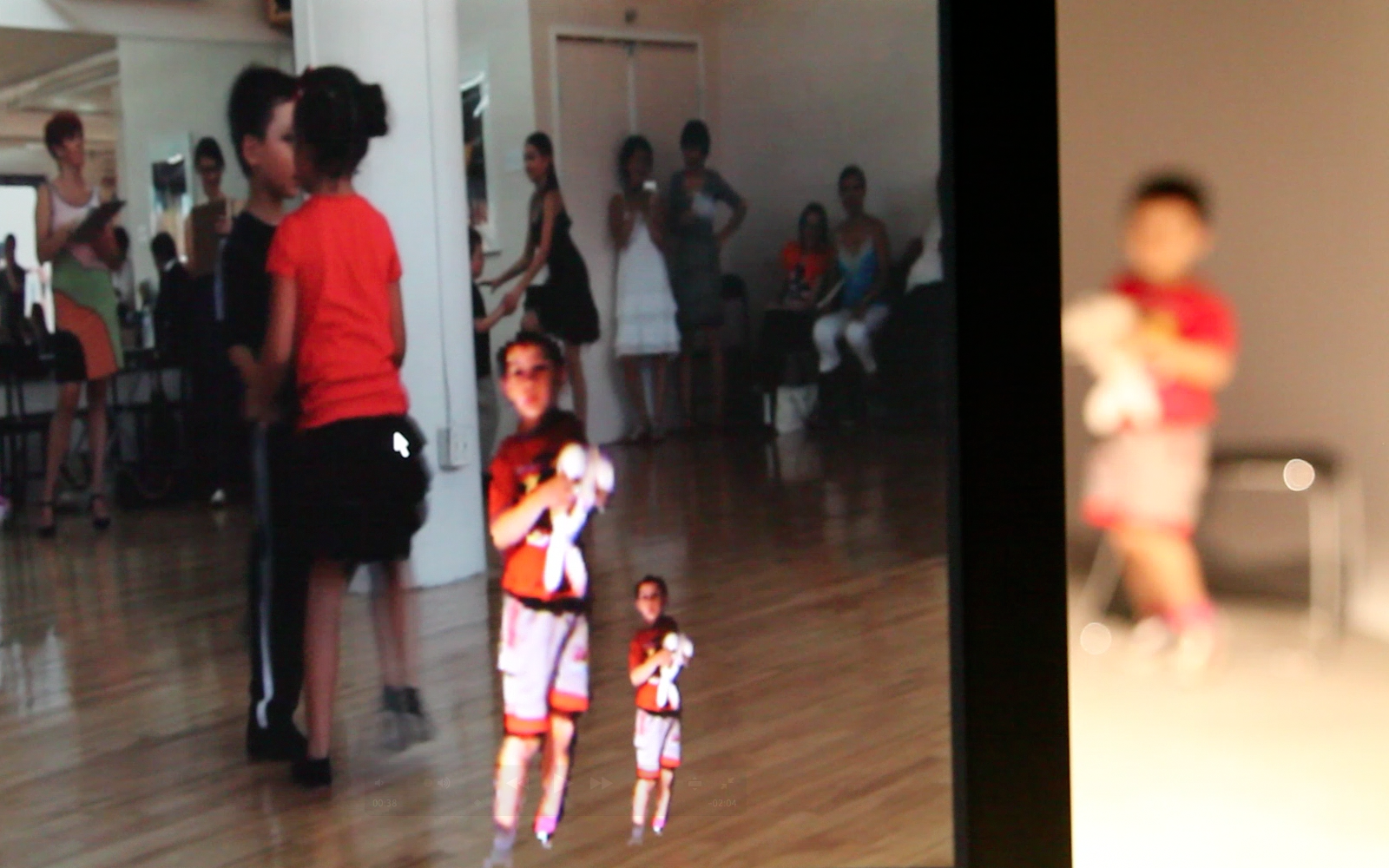}
\label{fig:virtual-audience-4}
}
\caption
	[Recorded Real Event Performance with Real-Time Additional Virtual Audience 2]
	{Recorded Real Event Performance with Real-Time Additional Virtual Audience 2}
\label{fig:virtual-audience-dancers-2}
\end{figure}

\subsubsection{Animation and Interaction}
\label{sect:tangible-interaction-animation}

Devices, such as {\kinect}, camera, and microphones
could capture the participants' acting data
so that they could interact with 
the CG graphics and their animations.
When there are no interruptions from people, 
the CG graphics, lighting, and audio have their own
initial default states and perform based on the default scenarios.

As mentioned earlier the interactive performance is the nexus of many things
that come together all related to animation and its dynamic real-time alteration
in the computer graphics world, such as music control of a big fancy soap bubble
and small bubbles, green screen and ground projection, motion-tracked skeleton,
gestures with a gentle swipe of the fancy soap bubble.

In \xl{algo:illimitable-space-installation-demo} is the interaction
sequence setup for this installation to get the desired configuration
with or without a soundtrack. There it's primarily a keyboard-based
interaction to set things up. After that, the details related to
depth, audio, and skeleton streams come into play.

The proof of concept illustration of the presented interaction
is depicted in the following clips, some of which were previously
mentioned in \xs{sect:illimitable-space-poc}:

\begin{itemize}
	\item 
Clip 1
  --- fancy soap bubble ``dancing'' with music and audience ``green-screen"
    color approach to observe and swipe. It encompasses the {\hlsl} shader
    deforming the soap bubble, while the textures onto it are dynamically
    mapped from various streams. The waving pattern is fed to the shader
    from the melody playback into the \api{wave} variable deforming the
    soap bubble model. It is also effected by the hand gestures
    gently swiping rotating the bubble (see, e.g., \xf{fig:timmy-fancy-bubbles-1}).

	\item 
Clip 2
  --- long sleeve dance with the depth camera and music visualization
    on the ground floor~\cite{song-phd-depth-water-sleeve-dance-kinect-video-2012}.
    This is becoming the classical performance
    piece that can be combined with all the others. The graphical
    imagery is affected by the data from the {\kinect}'s depth
    stream and the audio stream from the melody playback via BASS.
    See \xf{fig:sleeve-1} and \xf{fig:sleeve-2} as the examples
    of the projected stills from the clip.

	\item
Clip 3
   --- a recorded children dance performance video with one of them
   a real-time audience multiplied in the space and greenscreen-projected
   on to the video feed~\cite{song-phd-timmy-greenscreen-kinect-video-2012}
   (see \xf{fig:virtual-audience-dancers-1} and \xf{fig:virtual-audience-dancers-2}).

\end{itemize}

\begin{algorithm}[ht!]
\hrule\vskip4pt
	\Begin
	{
		Toggle white/black background for white walls or blackbox with `W';
		Enable ground projection with `G'\;
		Switch to the projected mode, `P'\;
		Take off the fancy projected bubble, `3'\;
		Revert to the frontal view with `P'\;
		Enable main melody playback with `M'\;
		Set back to the projection mode, `P'\;
		Adjust the zoom as necessary via the field-of-view angle $\lambda$ with `PgUp', `PgDown'\;
		\tcp{The installation support is running now; time to act and perform.}
	}
\hrule\vskip4pt
\caption{High-Level Interaction Algorithm for \worktitle{Illimitable Space}{Installation}}
\label{algo:illimitable-space-installation-demo}
\end{algorithm}

\chapter{Conclusion}
\index{Conclusion}
\label{chapt:conclusion}

In this chapter,
I gather together all the thoughts on the achievements, contributions, their
advantages and limitations at the time of this writing and the rich possibilities
and future directions in the follow up work, academically and commercially.
The chapter reviews the rationale and findings of this research (\xs{sect:conclusion-rationale-findings}),
multi-\index{research!multidisciplinary}, inter-\index{research!interdisciplinary}, and
trans-disciplinary\index{research!transdisciplinary} research realities (\xs{sect:conclusion-ndisciplinary-realities}),
consolidated contributions overview (\xs{sect:conclusion-contributions}), followed by
the three summaries of main themes of this thesis
(in \xs{sect:softbody-jellyfish-summary}, \xs{sect:interactive-docu-summary}, and \xs{sect:illimitable-space-summary}
respectively).

\section{Rationale and Findings}
\label{sect:conclusion-rationale-findings}
\index{rationale}
\index{findings}

The proposed research, scoped as
\textbf{{\myphdthesistitle}},
covers in the reverse order arriving at:

\begin{enumerate}
\item
theatre arts and the performance,
\item
theoretical and practical experience in film/video production,
\item
extensive 3D computer graphics knowledge and programming skills, and
\item
experience in the new media field.
\end{enumerate}

\noindent
Any of the above mentioned research directions is a very wide field of research and creation.
I have also discovered that the deeper research I do,
the more questions I encounter (and am eager to find answers to).
I really need to force myself sometimes
to keep the focus on several findings
and constrain myself to thinking rationally, 
which is often quite conflicting with my artistic creativity.

The research findings, positions, and installations argued throughout
the thesis indicate the following statements are \emph{true}:

\begin{itemize}
\item
Virtual reality, computer generated images, and new digital media technology
will not destroy 
the live performance of traditional theatre and the aesthetics
of documentary film, instead, the technology will enhance and
improve their representation.

\item
The boundaries among different art forms, such as film, theatre,
computational arts, tend to be blurrier and blurrier
because people are more interested in innovative combinations of
various art forms.
\end{itemize}

\noindent
In short, my position is:
\begin{quote}
\hspace{4cm}{\em Technology is essential; art is eternal.}
\end{quote}

\longpaper
{
\noindent
I discuss more of the personal opinion statement in \xc{chapt:discussion}.
}

\section{Multi-, Inter-, and Transdisciplinary Research Realities}
\label{sect:conclusion-ndisciplinary-realities}
\index{research!multidisciplinary}
\index{research!interdisciplinary}
\index{research!transdisciplinary}

After many years of knowledge, practices and leadership accumulated in various projects, 
I feel I have at last arrived at a very productive stage on emergence of 
traditional art forms combined with the rapidly developing technology. 

Moreover, after reading some of the literature, such as
Artaud's \emph{Theatre of Cruelty}\index{Theatre of Cruelty}~\cite{theatre-its-double-1970},
Grotowski's \emph{Poor Theatre}\index{Poor Theatre}~\cite{poor-theatre-1968}, and
Brook's \emph{Holy Theatre}\index{Holy Theatre}~\cite{empty-space-1968}, I am 
astonished that their ideas are still quite relevant to today's modern 
theatre and its performance, some of them could even continue leading the 
direction in today's new media and technology era.

Originally, I came to study computer science from a pure artistic background.
After ten years of extremely intensive training in science and computers,
at some period of time, I
partially lost connection to
my creative abilities as an artist.
Now, I am able to regain it all back and to combine the power of both the artistic and scientific backgrounds.
The whole process is such a painful experience!
More than this, only considering the artistic background,
there are significant differences between a theatre actor to a documentary
filmmaker and a director, and to a media artist; let alone scientists.

The unique experience and this project are a worthy undertaking given a good amount of time and life
to analyze and document, with a group of academic and artistic professionals and experts to supervise
and contribute.
Hopefully, the thesis itself will be a good contribution to the new students
and researchers who would proceed a multidisciplinary or interdisciplinary research.

\section{Contributions Overview}
\label{sect:conclusion-contributions}
\index{Contributions!Overview}

Here is a consolidated
short summary of the work accomplished as a part of or alongside the
research and creation, in terms of publications of the related work, design and implementation,
exhibitions, installations, performances, and other showings.
\index{research!contributions}
\index{research!publications}
\index{research!shows}
\index{research!installations}

While working on the research and analysis detailed in this thesis,
as well as the corresponding creation, design and implementation we produced
or are in the process of realization of the following works:

\begin{NoHyper}
\small
\begin{itemize}

	\item \bibentry{jellyfish-c3s2e-2012}

	\item \bibentry{i-still-remember-opengl-remake-2011}
	\item \bibentry{jellyfish-grand-2011}
	\item \bibentry{tangible-memory-2011}
	\item \bibentry{song-still-remember-movie-bjiff2011}
	\item \bibentry{water-ink-animation-film-2011}

	\item \bibentry{presentation-hexagram-2010}
  \item \bibentry{vr-medical-research-docu}
	\item \bibentry{soen-spec-cg-simulation-systems}
	\item \bibentry{song-still-remember-movie}
  \item \bibentry{msong-mcthesis-book-2010}
  \item \bibentry{vr-poster-ecsga10}
  \item \bibentry{softbody-poster-ecsga10}
  \item \bibentry{pain-performance-vr-pmmIII}

	\item \bibentry{marf-writer-ident}
	\item \bibentry{adv-rendering-animation-softbody-c3s2e09}
	\item \bibentry{haptics-cinema-future-grapp09}
	\item \bibentry{opengl-slides-grapp09}
	\item \bibentry{stereo-plugin-interface}
	\item \bibentry{softbody-opengl-slides}
	\item \bibentry{softbody-teaching-opengl-slides}
	\item \bibentry{role-cg-docu-film-prod-2009}
	\item \bibentry{role-cg-docu-film-presentation-2009}
	\item \bibentry{feynman-euler-softbody-glui}
	\item \bibentry{abra-cslp-video-2009}

	\item \bibentry{stereo3d}
	\item \bibentry{softbody-lod-glui-cisse08}
	\item \bibentry{opengl-slides-cisse08}
  \item \bibentry{softbody-framework-c3s2e08}

	\item \bibentry{her-story-miao-2007} (in \cite{her-story-2007})
	\item \bibentry{msong-mcthesis-2007}

	\item \bibentry{real-time-sea-world-2003}
	\item \bibentry{spectacle-anim-film-2003}
	\item \bibentry{tangram-flash-game-2003}
\end{itemize}
\normalsize
\end{NoHyper}

\section{Softbody, Jellyfish, and CG Summary}
\label{sect:softbody-jellyfish-summary}
\index{Contributions!Softbody Simulation}
\index{Contributions!Jellyfish Simulation}
\index{Contributions!CG}

Through the study of the particular case of the {\softbodysys}
and its evolution, we explicitly set a collection of requirements for similar interactive
computer graphics simulation systems. We hope that this set of requirements
can be used and expanded on by researchers in the field and will mature to be
a reference standard for similar systems that will be created
and revised accordingly~\cite{soen-spec-cg-simulation-systems}.

We have encountered some integration difficulties
due to the frameworks' original design and implementation
considerations discussed~\cite{soen-spec-cg-simulation-systems,softbody-teaching-opengl-slides}.

Physical based softbody simulation is still a very hot
topic in computer graphics~\cite{vega-deformable-lib}, especially, with realtime animated
feature, such as {\jellyfish}, to my knowledge based on own recent survey, there have not been an algorithm
or an application published yet.
Moreover, we integrate a number of technologies for the interactive user-controlled,
or, rather, user-assisted control of a swimming {\jellyfish}~\cite{jellyfish-grand-2011}.
\longpaper
{
In the foundation
of this interactive installation are a number of hardware and software
platforms.
The hardware includes the Kinect platform from Microsoft that
allows for the sensory inputs along with the PC rendering equipment
running either MacOS~X or Windows with OpenGL drivers, and the Jitter/MaxMSP~\cite{maxmsp,jitter}
setup. 
}

We further discuss the limitations of the proposed
approach as well as the future (and ongoing) work
on this and the related projects.

\subsection{Softbody, Jellyfish, and CG Contributions}
\label{sect:conclusion-softbody-jelly-fish-contributions}
\index{Contributions!Softbody Simulation}
\index{Contributions!Jellyfish Simulation}
\index{Contributions!CG}

To a various degree of completeness in the implementation, the contributions
include the following:

\begin{itemize}

\item
\index{Contributions!Softbody Simulation!Center particle}
We have modeled a third layer in the softbody objects
consisting of a single center particle connected by the
radius springs to the inner layer particles for
for 2D and 3D softbody objects. The particle
serves conveniently as the single attachment point,
and, is, therefore, easy to use instead of attaching
multiple points of the second layer. It also
adds stability to the softbody objects at the smaller
time steps in Euler and Feynman algorithms making
them more usable\longpaper{ while retaining the real-time
simulation speed where RK4 is still visibly slower
on commodity computers}.
This setup provided the means
of attachment of the softbody objects on
``hardbody'' ones or animate along the curve~\cite{adv-rendering-animation-softbody-c3s2e09}.

\item
\index{Contributions!Softbody Simulation!LOD}
The {\softbodysys} LOD aspect got more prominent and visible
with the {\glui} interface, especially for the LOD
hierarchy~\cite{softbody-lod-glui-cisse08}.
The simulation interface
allows adjustment of the LOD parameters,
including the whole algorithms at run-time as well as
co-existence of all dimensional objects in one simulation
scene~\cite{softbody-lod-glui-cisse08}.

To enable the LOD parameters and the interface to them
we extended the original Softbody Simulation Framework
with the notion of state as well as greatly enhanced
the interactivity of the simulation system by exposing the
state to the {\glui} interface at run-time.
The users and researchers
working in the field with the framework can easily observe
the effects of a vast variety of LOD parameters on the
physical-based softbody simulation visualization
at real-time without the need of altering the source code
and recompiling prior each simulation increasing the
usability of the system.

We identified a scale problem
with some of the LOD parameters when mapping to a GUI: for example, spreading
the mass over each particle or layer of particles is unmanageable
and has to be specified by other means when needed~\cite{softbody-lod-glui-cisse08}.

\item
\index{Contributions!Jellyfish Simulation!2D}
\index{Contributions!Jellyfish Simulation!3D}
Currently we have a complete 2D {\jellyfish}, to which we add haptic interaction
support and for 3D a single-layer Euler-integrated 7 slices
of 12 particles animated and interacted with in real-time
for the {\jellyfish}'s bell without tentacles~\cite{jellyfish-c3s2e-2012}.

\item
\index{Contributions!Softbody Simulation!GPU shading}
{\gpu}-based shading---we implemented the first draft version
of a general vendor-independent
API for shader use within the softbody system and provided
two implementations of that API: one that loads {\glsl}
vertex and fragment shaders and the other that loads
the cross-vendor assembly language for shaders~\cite{adv-rendering-animation-softbody-c3s2e09}.

\item
We implemented the Feynman algorithm alongside the
existing Euler, midpoint, and RK4, to further validate
the integration framework design and to compare~\cite{adv-rendering-animation-softbody-c3s2e09}.

\longpaper
{
XXXX:
A work in progress to publish the details about
the Feynman implementation, the center point layer
for attachment of softbody objects and the
non-uniform and otherwise sphere modeling under
these conditions~\cite{softbody-center-nonuniform-feynman}.
}

\item
\index{Contributions!Softbody Simulation!Bezier}
We implemented a rudimentary Bezier-curve %key-frame based
animation as a separate stand-alone application and we
produced the softbody system as a library to be
used later in this application and render the softbody objects
along the curve to demonstrate the ability to use the
softbody framework outside its own main simulation
application~\cite{adv-rendering-animation-softbody-c3s2e09}.

\longpaper
{
\item
We invested some amount of time in learning Fly3D
game engine with the purpose of trying out the
\file{libsoftbody} in it, but this experiment was not
completed~\cite{adv-rendering-animation-softbody-c3s2e09}.
}

\item
\index{Contributions!Softbody Simulation!Collision detection framework}
We abstracted the default penalty-based collision (see \xs{sect:softbody-jellyfish-cd})
detection implementation to allow for replaceable
other collision detection and response algorithms
integrated in the future for comparative studies~\cite{adv-rendering-animation-softbody-c3s2e09}.

\item
\index{Contributions!Softbody Simulation!OpenGL slides}
We completed the first proof-of-concept integration of
the {\softbodysys} and {\oglsf} frameworks.
We made a number of slides in a {\opengl}-based softbody
presentation typically found in lab/tutorial like
presentations, which are to be extended to a full
lecture-type set of slides~\cite{softbody-teaching-opengl-slides}.
Power-point slides in {\opengl} is a way to present
for demo and teaching how the softbody objects are
modeled and rendered~\cite{softbody-teaching-opengl-slides}.
\longpaper
{
The papers~\cite{opengl-slides-cisse08,opengl-slides-grapp09} present a base for teaching, learning, and
education of/for computer graphics and are 
integrated with the softbody framework~\cite{softbody-opengl-slides}.
This milestone significantly advances our contribution
to a good CG teaching module, suitable for use by
instructors to present the material in class as well as
for learning by providing its source code to the
students for study and extension to demonstrate their
CG projects at the end of a semester~\cite{softbody-teaching-opengl-slides}.
}

\item
\index{Contributions!Softbody Simulation!Stereo}
We have presented the initial iteration of the interface
for real-time stereoscopic\index{stereoscopic!modeling} modeling within {\maya} as 
a {\mel} and {\cpp} plug-in, with the {\maya}-independent
{\opengl} core that can be used in other 3D modeling,
animation, game engine, and medical VR tools~\cite{stereo-plugin-interface}.
The stereoscopic {\opengl}-based rendering\index{stereoscopic!rendering} plug-in~\cite{stereo3d}
originally for {\maya}, and then beyond, is another advanced rendering technique this work
was designed to include in order to render the softbody and {\jellyfish}
simulations in stereo in the future.
\longpaper
{
We have released
our code as open-source for the greater benefit
to the community and to solicit feedback from other
developers and stereo modelers and artists~\cite{stereo-plugin-interface}.
}

\end{itemize}

\longpaper
{
Statistics:

V vertices

S springs

L layers
{\todo}
}

\subsection{Limitations}
\index{Limitations}
\label{sect:softbody-jellyfish-limitations}

There are some assumptions and limitations to the
realization described here\longpaper{; thus, it is not an all-in-one
solution, but rather for a specific purpose presentations~\cite{softbody-teaching-opengl-slides}}.

\begin{itemize}
	\item 
	This work assumes the CG topics
	presented
	are renderable at real-time,
	like the {\softbodysys}~\cite{softbody-teaching-opengl-slides}.
		While it is of course possible to render the animation images offline first
		(possibly of photorealistic quality), and then replay them back, or
		use keyframe animation, but this is not what the challenge is at.
		However, combining real-time simulation and offline-made imagery
		can be used to enhance the overall perception of the scene and, e.g., to
	  playback video footage as dynamic texturing of polygons,
	  alongside the real-time animation that transforms the geometry,
		and this is what we do in our installation
		since one can playback AVI (standard and HD) movies in {\opengl}\longpaper{; however, this is not
		a primary learning source, though may be necessary at times
		(e.g. to teach how to play such things when/if needed)~\cite{softbody-teaching-opengl-slides}}.
	
	\longpaper
	{
	\item 
	Not suitable for presentation in online conferences, so have
	to make screenshots (not a big deal, but ideally, at least
	the screenshots should be taken automatically)~\cite{softbody-teaching-opengl-slides}.
	} % \longpaper
	
	\item 
	May be hardware dependent for real-time processing, though today's commodity
	hardware should generally be good enough.
	
	\longpaper
	{
		\item 
	Tedious to compose and debug---need to be a programmer~\cite{softbody-teaching-opengl-slides}.
	\begin{itemize}
		\item 
		A proposal to load import info from XML or text is made.
		\item 
		Has to be done well in advance before the teaching session.
		\item
		Assumption is made the instructors and the students are able
		to do {\cpp} or {\opengl} programming in a CG programming course.
	\end{itemize}
	} % \longpaper
	
	\longpaper
	{
	\item 
	Slides' rendering is resolution-sensitive---tidgets may have various
	spacing or stretch, not matching the softbody object being
	displayed in the scene~\cite{softbody-teaching-opengl-slides}.
	} % \longpaper

	\longpaper
	{
	\item 
	Rendering is an issue (not supported, or tedious/difficult) for now for
	math (formulas), source code, highlighting---can only be done as images~\cite{softbody-teaching-opengl-slides}.
	} % \longpaper

	\longpaper
	{
	\item
	Current limitations lie within the stereo
	window lacking some of the common
	menus that are typically found in Maya's view panels. 
	A color mask adjustment is required for the anaglyph
	rendering preview to prevent  the light green and yellowish appearance 
	\longpaper{as shown in \xf{fig:stereo-window},} caused
	by interaction with the user interface's background color settings.
	} % \longpaper

	\item
	We also do not have
	\linux{} support for the moment, only \macos{X}
	and \win{} platforms~\cite{stereo-plugin-interface}.

\end{itemize}

\subsection{Future Directions}
\label{sect:softbody-jellyfish-future-directions}
\index{Future Directions!Softbody}
\index{Future Directions!Softbody}
\index{Future Directions!Jellyfish!Kinect}
\index{Future Directions!Stereoscopy}
\index{Future Directions!Virtual Reality}
\index{Future Directions!Character Animation}
\index{Future Directions!Rendering Engines}
\index{Future Directions!Collision Detection}
\index{Future Directions!Disciplines}

For the most part the future work will
focus on the addressing and resolving some of the limitations, unfinished items
or experiments mentioned in the earlier
sections. Some specific items are
mentioned further in point form and details.

\longpaper
{
Working on the course lecture notes in OpenGL slides illustrating all techniques
to hand out as code samples.
} % \longpaper

\begin{itemize}

\item
\begin{itemize}
	\item
Our goal for the 2D {\jellyfish} is to render it automatically with an 
artistic appeal by texture mapping it accordingly %as exemplified in \xf{fig:p02-low-res}

	\item
Integrate with a real-time game and MoCap systems (see \xs{sect:softbody-jellyfish-future-vr}) %of Sha, Michael, TML, etc.

	\item
Harness the {\gpu} power to speed up the simulation and rendering processes

	\item
Public haptics~\cite{haptics-large-workspaces}
installation with a projector screen~\cite{jellyfish-c3s2e-2012}
\end{itemize}

\item
Allow softbody modeling and alteration as the Archimedean-based graphs
and different types of them than an octahedron~\cite{archimedean-graphs-wmsci08},
not just the number of subdivision iterations as a run-time LOD parameter~\cite{softbody-lod-glui-cisse08}.

\longpaper
{
\item
Interactivity through haptic devices~\cite{wiki:haptic-technology} with the softbody feedback
(e.g. for surgeon training or interactive cinema~\cite{haptics-cinema-future-grapp09}).
\item
Integration and interaction with the softbody
using an haptics device~\cite{haptics-large-workspaces,wiki:haptic-technology}.
} % \longpaper

\item
Provide animation state tracing, saving, and replay with the intermediate
values for analysis~\cite{softbody-lod-glui-cisse08}.
Specifically, allow the state dump and reload functionality in order to display each
particle and spring state (all the force contributions, velocity, and the position)
at any given point in time in a text or XML file for further import into a relational
database or an Excel spreadsheet for plotting and number analysis (perhaps by external
tools). Reloading of the state would enable to reproduce a simulation from some
point in time, a kind of a replay, if some interesting properties
or problems are found. This is useful for debugging as well~\cite{softbody-lod-glui-cisse08}.

\item
Continue with improvements to usability and functionality of the user interface
for scientific experiments~\cite{softbody-lod-glui-cisse08}.

\item
Port the source code fully to \linux{} and \macos{X}. Currently
it only compiles properly under \win{7} under Visual
Studio 2010 and basic softbody and curve-based animation also in \macos{X}.

\item
Release our code and documentation as open-source implementation (see \xs{sect:softbody-jellyfish-future-oss}) either
a part of the Concordia University Graphics Library~\cite{cugl} and/or
as part of a {\maya}~\cite{maya} plug-in and as a CGEMS~\cite{cgems}
teaching modules.

\item
Allow advanced interactive UI controls of the scenes and slides by
using haptics devices~\cite{haptics-cinema-future-grapp09} with
the force feedback, head-mounted displays and healthcare virtual reality
systems~\cite{stereo-plugin-interface} (see \xs{sect:softbody-jellyfish-future-vr}).

\longpaper
{
\item
Integrate stereoscopic effects into the presentation of softbody
objects (under way) with another open-source plug-in project under development
that implements OpenGL-based stereoscopic
effects~\cite{stereo3d,arloader-thesis-08,stereo-plugin-interface}.

\item
While working on this and other integration efforts, take
down and formalize software engineering requirements for systems
like ours to simplify the future development and integration
process of academic and open-source OpenGL and CG frameworks
and systems for physical based animation and beyond~\cite{soen-spec-cg-simulation-systems} (in progress).
} % \longpaper

\longpaper
{
\item
Provide automatic loading and display of the softbody simulation source code
on the slides with breakdown onto multi-page slides.
} % \longpaper

\item
Showcase various softbody shading techniques and shaders via the {\opengl} slides.
We already implemented the first draft version
of a vendor-independent API for shader use within the softbody system
and provided two implementations of that API---one that loads {\glsl}
vertex and fragment shaders and the other that loads
the cross-vendor assembly language for shaders.

\item
Demonstrate attachment of softbody objects to skeletons for
character animation (see \xs{sect:softbody-jellyfish-future-char-anim}). % via the slides.

\item
The application of softbody deformation in computer
graphics has significant value for medical research and media arts production. %achievement.
The resulting models from the simulated models, will help to create responsive/feedback
environments that will map the virtual feedback to physical through an haptic interface\longpaper{,
see \xs{sect:softbody-other-domains}}.

\item
Complete/make experiments with the game and rendering engines~\cite{fly3d,ogre}
(see \xs{sect:softbody-jellyfish-future-fly3d} and \xs{sect:softbody-jellyfish-future-ogre}).

\longpaper
{
\item
See the application and offline rendering if can produce
animation movies with high-quality images and their
possible integration into the film\longpaper{ or augmented reality
applications}.
} % \longpaper

\longpaper
{
\item
Apply stereoscopic framework~\cite{stereo3d,arloader-thesis-08,stereo-plugin-interface}
to the softbody to enable viewer a better experience and towards the
path of integrated augmented reality applications, and potentially
implemented it as a stereo GPU shader as an option.

\item
Test and perhaps implement own, softbody-related, shaders.

\item
Add {\glui} controls for the shaders and their parameters.
} % \longpaper

\item
Investigate the replacement of vertex normal computation as well as
other vertex properties in relation to animation and physics
on the {\gpu} to speed up performance on multiple iterations.

\longpaper
{
\item
Teaching some computer graphics techniques with advanced
rendering and physical-based real-time animation with
and without GPU support and publish the teaching
materials online similar to~\cite{gtt,cgems,teaching-graphics-with-osl}.
Integrate for that purpose with the OpenGL slide presentation
system~\cite{opengl-slides-cisse08,opengl-slides-grapp09}.
} % \longpaper

\item
In the longer term future, attempt to map the graphical simulated softbody {\jellyfish}
to electrical pulses to control physical synthetic {\jellyfish}
like the one made by Nawroth {\etal}~\cite{synthetic-engineered-jellyfish,synthetic-life-jellyfish-science20,artificial-jellyfish-discovery}.

\end{itemize}

\subsubsection{Fly3D Game Engine}
\label{sect:softbody-jellyfish-future-fly3d}
\index{Fly3D}

Supplying a game character and other attributes
with the the softbody objects was one of the
targets to test the softbody object properties
with the textbook's~\cite{AlanFabio03} Fly3D
game engine~\cite{fly3d}.
\longpaper
{
Due to some inherent architectural complexity of the project and
significantly reduced time availability this experiment
was not completed as of this writing, and left to be
completed as a future work.
}

\subsubsection{{\ogre}}
\label{sect:softbody-jellyfish-future-ogre}

The Object-oriented Graphics Rendering Engine, or simply {\ogre}~\cite{ogre},
is another popular open source engine that was considered for
experiments of integration and testing of the softbody objects
in. We did not reach the ability to experiment with this engine
yet, but its possibilities look very promising~\cite{adv-rendering-animation-softbody-c3s2e09}.
In particular it overlays its API over either {\opengl} or {\directx},
so an application has a chance to be deployed and run on Microsoft XBOX, for example.

\longpaper
{
\subsubsection{Stereo}
\label{sect:softbody-jellyfish-future-stereo}
\index{Stereoscopy}

\begin{itemize}
\item Extend our stereo interface support onto other modeling and game engine tools, specifically
{\blender}~\cite{blender}, {\ogre}~\cite{ogre}, 3D Max~\cite{3dsmax}.
\item Further improve the stereo window user interface support
to what normal {\maya} views have in terms menus and action buttons.
\item Provide \linux{} support.
\item Complete the documentation with the release at CGEMS~\cite{cgems} or SIGGRAPH.
\item Integrate with the physical based two-layer softbody simulation
system~\cite{softbody-framework-c3s2e08}, {\cugl}~\cite{cugl},
and OpenGL Slides framework~\cite{softbody-opengl-slides,opengl-slides-grapp09} for teaching
and educational activities in a computer graphics course.
\item Integrate with the Motek VR SDK~\cite{motek-vr-sdk}.
\end{itemize}
}

\subsubsection{Open Source Release}
\label{sect:softbody-jellyfish-future-oss}
\index{Softbody!open-source}
\index{open-source!Softbody}

We plan to refine our set of requirements and design decision
guidelines for interactive computer graphics physical-based
simulation systems, such as the {\softbodysys}, further
during our ongoing work. Specifically, our experiments and derived
metrics will be based on the system's integration with the {\cugl}~\cite{cugl},
the {\opengl} presentation slides framework~\cite{softbody-opengl-slides}
as well as the stereoscopic effects of another framework that
generalizes the handling of stereoscopy~\cite{stereo-plugin-interface}.
Furthermore, the haptic devices' sensory input as well as integration of
the softbody library \tool{libsoftbody}, which contains the complete
implementation of the current simulation framework into games and
other more realistic environments, such as virtual reality systems
and head-mounted displays for further studies in various domains
ranging from cinema production to medical research in pain management.
We plan to release the different builds and iterations mentioned above
as a part of the open-source initiative incrementally.

\subsubsection{Character Animation}
\label{sect:softbody-jellyfish-future-char-anim}
\index{Character Animation!Softbody}
\index{Softbody!Character Animation}

The functionality development of elastic 
simulation modeling for 3D software design and implementation has emerged as 
a new challenge in computer graphics. One of the existing software with the 
elastic modeling functionality is {\maya}~\cite{maya}, which provides shape deformation, 
especially facial animation, for a group of objects. It is more convenient 
than traditional frame animation. However, the elastic object movement is not 
attached to skeleton animation. Furthermore, this elastic simulation is not 
done in real time.
A possible future work that can be done based on the elastic simulation is to 
define a skeleton system and to map the body mesh onto it. The different 
parts of the body can be defined as with different degrees of deformability based 
on the elasticity. For example, the mesh is less elastic on the arms, legs; 
the mesh is more elastic on the fatty or softer areas, like belly, breasts, etc.
The weight of the elastic property of the muscles can be mapped and 
dynamically set according to the skeleton's joints. The system can then be
integrated into the advanced animation software as a plug-in. The skeleton
could a model or a motion-captured from the {\kinect} SDK~\cite{kinect-ms-sdk}
and its {\cpp} API.

\subsubsection{Collision Detection}
\label{sect:softbody-jellyfish-cd}

CD between soft objects is a complex phenomenon.
In our current 
system, we are using the penalty methods~\cite{MJ88}, which do not generate 
the contact surface between the interacting objects. This method uses the 
amount of inter-penetration for computing a force which pushes the objects 
apart instead. Even though the result is fair enough based on estimation, in 
reality, the contact surfaces should be generated rather than local inter-penetrations.
Especially, if we want to use computer animation to imitate 
organ surgery and help surgeon practice as if interact with real objects, the 
penalty method is no longer appropriate. There must be a more accurate 
algorithm to define the collision between rigid body and soft body, or soft 
body and soft body. Our software should be able to describe other soft body 
deformation, such as fractures, possibly benefiting from freshly released
Vega\index{Libraries!Vega}~\cite{vega-deformable-lib}.
An approximation of the softbody object collisions can be done by
forming a few temporary invisible springs between the nearest particles of the
would be colliding objects with the distance below certain
threshold and let the spring forces do work while the distance
is below the threshold.

On the other, artistic side, advanced CD is important for all kinds
of artistic animation.
Thus, the future work would be to ``import'' the tricks from traditional animation into
computer animation:

\begin{itemize}
\item
Give characters a pseudo personality

\item
Stretch and squeeze is used to highlight dynamic
action such as deceleration due to collisions

\item
Shape distortion
(the collision can be thought of as having two phases:
compression and restitution)

	\begin{itemize}
	\item
	In the compression phase, kinetic energy of motion changes
	into deformation energy in the solids
	\item
	If the collision is perfectly inelastic ($e=0$), then all of the
	energy is lost and there will be no relative motion along the
	collision normal after the collision
	\item
	If the collision is perfectly elastic ($e=1$), then all of the
	deformation energy will be turned back into kinetic energy in
	the restitution phase and the velocity along the normal will be
	the opposite of what it was before the collision
	\end{itemize}

\end{itemize}

While we have not managed to implement new collision
detection algorithms to compare with the existing one,
the existing algorithm was abstracted
to create a more flexible subframework to support multiple
collision detectors, i.e., one can switch between the
implementations of collision detection at run-time or
start time. The framework provides a general \api{CollisionDetector}
class and its concrete implementation for the penalty
method. Any new algorithm would have to subclass \api{CollisionDetector}
in order to participate in the framework~\cite{adv-rendering-animation-softbody-c3s2e09}.

\subsubsection{Virtual Reality}
\label{sect:softbody-jellyfish-future-vr}
\index{Virtual Reality}

The possible future work on the VR side would consist of expanding
the system and installation onto the advanced VR equipment and
various integration works to make the whole experience more immersive
and tangible and further turn it into a part of the interactive documentary
project. Some equipment I worked with was quite expensive and
I hope to be able to regain access to such equipment to produce
a great installation~\cite{vr-medical-research-docu}.

The major way of integration is via {\opengl} and the related
software technologies based on {\ogre} and {\vizard} with
{\cpp} and {\python} interfaces to the major CG parts of the
{\caren}/{\motek}/{\sensics} system and the HMD display respectively.
Thus, it would be nice to see
the HMD VR {\vizard}~\cite{vizard} integration with softbody
to completion.
In addition to that, I would like to complete the integration of the {\falcon} device
in the same VR context to enhance the feedback response to the audience.
Finally, relying on the previous and ongoing achievements, I would like to
provide a stereoscopic softbody interface to the system
\longpaper{(see \xf{fig:softbody-sequence-diagram})}
employing the
open-source stereoscopic plug-in\longpaper{ (see \xf{fig:stereo-modeling-views-maya}
and \xf{fig:stereo-rendered-multiview})}~\cite{vr-medical-research-docu}.

If I am lucky to have access to the equipment and will have completed 
the integration, I am considering creating or adapting an immersive
3D VR documentary installation based on this~\cite{vr-medical-research-docu}.

\longpaper
{
\begin{itemize}

\item
Design and build virtual rehabilitation center project.

\item
Continue working on VR {\caren} system to get kinematic data
from various groups of patients.

\end{itemize}
} % \longpaper

\subsubsection{{\libsdl}}

We plan to explore the use of the Simple Direct media Library ({\libsdl})~\cite{libsdl}
that provides an open cross-platform media API that works uniformly with {\opengl}
and many interaction peripherals and devices. It is a lot more comprehensive than {\glut},
and allows for better video playback.

\subsubsection{Softbody Jellyfish with {\kinect}}
\label{sect:softbody-jellyfish-kinect}
\index{Softbody!Jellyfish!Kinect}

Finally, it is very natural, interesting, and logical to produce
an interactive installation based on the recent technological advancements
to enable a human being skeleton to drive a 3D computer generated character with
the possibilities to proceed
to the game and cinema scenes to enrich them with the interactive
features. An initial PoC with our {\jellyfish}, {\kinect}, and
fluid simulation was partially achieved in a collaboration with
Fortin~\cite{fortin-interactive-fluid-flow-mcsthesis,jellyfish-grand-2011}
where the {\jellyfish} was swimming partially in motion-blur fluid with
limited {\kinect} connectivity~\cite{jellyfish-grand-2011}.

Another version of own production is in the works with more features
and interaction scenarios and integration with the works such as interactive
documentary and illimitable space installations discussed in the earlier
chapters.

We describe several details of modeling and implementation of an
interactive work of the earlier {\jellyfish} control HSC system
using {\opengl}, the Softbody and the Fluid Simulation System, {\kinect},
and {\jitter}~\cite{jellyfish-grand-2011}.

In this PoC prototype,
we integrated a number of technologies for the interactive
user-assisted control of a swimming {\jellyfish}. In the foundation
of this interactive installation are a number of hardware and software
platforms. The hardware included the same {\kinect} platform from Microsoft
as in \xc{chapt:interactive-docu} and \xc{chapt:illimitable-space}
and the {\jitter}/{\maxmsp}
setup. The real-time fluid flow and the described physical based softbody simulation
frameworks were designed and developed in-house~\cite{jellyfish-grand-2011}.
The PoC featured real-time physical based simulation of 3 or 4
{\jellyfish} swimming in a fluid-simulated environment. 
Jellyfish followed the fluid waves and could exert a small force back to fluid through its
inner compression. The audience not only could disturb the animation of fluid by their motion
blobs captured by {\kinect} device, but also could somewhat drive a {\jellyfish}.
The unattended jellyfish swim about randomly~\cite{jellyfish-grand-2011}.

\longpaper
{
The audience interacted with the {\jellyfish} via {\kinect}'s sensors that feed
the data to a {\jitter} controller ``patch'' (a {\jitter} data-flow program) that, in turn,
controls the {\jellyfish} rendered on the screen (e.g. computer screen or a
projector screen) in real-time by pushing or pulling on the {\jellyfish} geometry.
The resulting pushing and pulling forces cause the {\jellyfish} to change direction,
stretch, or squeeze. The {\jellyfish} are modeled based on the {\softbodysys}
where each {\jellyfish} is a 2-layer elastic spring-mass system of interconnected
vertices that respond and react to physical forces applied to them. The rendering
is performed using {\opengl}. The {\jitter} code also performs the fluid flow
simulation to enhance the realism and artistic appeal of the environment around
the jellyfish with liquid naval or deep-water colors~\cite{jellyfish-grand-2011}.

The user interaction with the jellyfish can also be performed with the traditional
input using a mouse (or planned haptic device)~\cite{jellyfish-grand-2011}.

The founding work for these comes from two master theses and the supported software
\cite{msong-mcthesis-2007,fortin-interactive-fluid-flow-mcsthesis,jitter,opengl}. 
} % \longpaper

The future work planned for {\jellyfish}
is to make it into an advanced real-time interactive game akin to
\gametitle{Blush}~\cite{blush-jelly-fish-game}.

\longpaper
{
\subsubsection{Compatibility with Other Applications}
\label{sect:softbody-jellyfish-compat}
}

\section{Interactive Documentary Summary}
\label{sect:interactive-docu-summary}
\index{Interactive Documentary}
\index{Documentary!Interactive}

We summarize our findings, achievements in \xs{sect:conclusion-interactive-docu-contributions}, as well as the
advantages (\xs{sect:interactive-docu-advantages}) and limitations (\xs{sect:interactive-docu-limitations})
of our approach and offer future directions (\xs{sect:interactive-docu-future-directions}).

\subsection{Interactive Documentary Contributions}
\label{sect:conclusion-interactive-docu-contributions}
\index{Interactive Documentary!Contributions}

When I described the interactive documentary system I have implemented,
I almost forgot the contents of those documentary films featured there.
From the international winning short film~\docutitle{I Still Remember}~\cite{song-still-remember-movie-combined}, which
I have made almost entirely by myself, including shooting, lighting, editing,
subtitling, with my daughter's involvement to be my main character who
truly shared her feelings and memories with the audience.
I think it might not be a coincidence that the film got several awards and got screened
at several film festivals, but the artistic creativity and skills accumulated
from my previous studies in arts and working experience in TV/film productions have helped here.
I am pleased that my artistic talents come back to me finally and
I have been successfully qualified by peers as an artist again after
many years of training in computer science and software engineering.
The art piece has a social value, which draws on the society's attendance
and increases the public awareness and garners sympathetic response to some children who
have similar experience as my daughter, the main character.
There are also more memory bubbles related to this original story,
such as the TV interview to the filmmaker about this film,
film festival screening scenes, the continuing life of the main character's life
after the film was made, the impact to the character and filmmaker after the film 
gain lots of successful experience (see \xs{sect:interactive-docu-media-content}).
Making a linear documentary film is a life-long personal project.

Now, look at the newly invented concept of interactive documentary design and system.
The tangible memory bubbles concept came from the content of the
original documentary itself of a little girl describing
her own memories as floating bubbles she could ``pick''
and ``see inside'' if she wanted to remember something in particular.
There is originality and a real need derived from a static documentary film by its character's
description to her memories instead of trying to make some fancy ideas
to show off the interactive media computer technologies.

At this moment, interactive documentary or cinema mainly is referred
to web-based works instead of installations~\cite{wiki:web-docu}.
Moreover, the interaction is limited to mouse drag-click.
The footage of cinematic materials is static, pre-filmed materials
instead of being able to dynamically displayed upon interaction or even record in place.
My interactive documentary system fills the insufficiently covered gaps between the static media.
At first,
I presented the initial ``poor woman'' way of approaching
an interactive documentary 
with a widely
available {\opengl} library and a set of online openly available
resources to play clips, texts, and images from a passive
storyline documentary into making it interactive. 
That working interactive prototype featured five bubbles
with three videos and two images (as illustrated in sample
screenshots in \xf{fig:bubbles-float-example} and
in \xf{fig:selected-bubble-zoom-in}).
Subsequently, in the {\xna}/{\kinect} version, the
audience could use their body movement, gesture, and speech to interact with the system.
Meanwhile, the system changes the method of existing interactive documentary,
by making audience live participation also a part of the documentary
project itself, audience become part of the interactive documentary project.
The interactive documentary non-linear story telling system is an eternal project
which will never become a PAST tense.

\subsection{Advantages}
\label{sect:interactive-docu-advantages}
\index{Interactive Documentary!Advantages}

We outline the advantages of both {\opengl} and {\xna}
installations first.

\subsubsection{{\opengl} Installation}
\index{Interactive Documentary!Advantages!OpenGL}

Advantages of the {\opengl} version include the fact that {\opengl} itself is an
open standard and runs on many OS platforms, desktop or mobile~\cite{opengl-redbook-ed8}.
This, and the author's familiarity with it, initially prompted the development of the first {\opengl}
\docutitle{Tangible Memories} version. Additional examples~\cite{falcon-manual,falcon-sdk} were
readily available of haptic-based interaction in {\opengl} that were planned for this installation.

\subsubsection{{\xna} Installation}
\index{Interactive Documentary!Advantages!XNA}

The advantages of the {\xna} based installation include the easier development
effort required as opposed to the {\opengl}- or {\directx}-level
to come up with runnable prototypes and as well as the recent
influx of many open- and shared source examples for it; including {\kinect}
examples~\cite{kinect-ms-sdk} and speech processing libraries. 
The author's growing familiarity
with the subject has also been influenced by preparing for and instructing
laboratory sections for the related computer graphics and animation courses.

\subsection{Limitations}
\label{sect:interactive-docu-limitations}
\index{Interactive Documentary!Limitations}

It should be noted that the interactive documentary approach described
here is more suited for home-use by a family or a few individuals rather than large audiences
in traditional cinema theatres as the interactivity aspect in movies does not scale
very well nor does it make much sense to have many people to interact with it unless
to create chaos (and document it). The home and small audience aspect, the technologies studied and developed
in this work should be affordable by regular home users, small education groups in
kindergartens, schools, colleges, universities, and other educational institutions.
Current limitations that plague interactivity in the
documentaries, specifically a profound difficulty
to for multiple people to interact with the same
documentary piece instance at the same time is
problematic (e.g., one would need to support multi-input
in a form of multiple mice, keyboards, haptic devices,
cameras or motion tracking sensors from more than one
individual from the same audience in the same space-time.)
We further summarize platform-specific limitation with some hints for
possible solutions.

\subsubsection{{\opengl} Installation}
\label{sect:interactive-docu-opengl-installation-limitations}
\index{Interactive Documentary!Limitations!OpenGL}

{\opengl}-based programming is generally more tedious as it is lower-level
than some other libraries or APIs like {\xna} and it therefore takes longer to prototype a piece.
However, open-source rendering engines, such as {\ogre}~\cite{ogre} could be of help here.

At this point we limit the
interaction by the audience to use
only mouse clicks or preset keys
in a {\glut} window to bring a bubble of choice or to move a synthetic
camera in the environment with a keyboard. Speech, haptic, and {\kinect} based
interaction did not make the cut at this time of this thesis but the work on those has
already started.
Moreover, the bubbles float in 3D space,
but the basic mouse interaction is 2D and clicks have
to be translated to the bubbles nearest in the $z$
dimension, which is confusing sometimes when $x$ and
$y$ are near for two or more bubbles, and the user
wants one bubble, but gets another one instead to
view.

In {\opengl}, we used only the AVI format playback (however more are offered
via {\libsdl}~\cite{libsdl}, {\jitter}/{\maxmsp}~\cite{jitter,maxmsp}, or {\puredata}~\cite{puredata}, but in this thesis
they are not explored).
{\kinect} with {\openkinect}~\cite{kinect-openkinect}, or {\jitter} or {\puredata} can also
be done as the latter have patches for its support~\cite{jit-freenect-grab}.

Despite the limitations mentioned, the author plans to continue exploring
the {\opengl} version with other technologies as time and resources permit.

\subsubsection{{\xna} Installation}
\label{sect:interactive-docu-xna-installation-limitations}
\index{Interactive Documentary!Limitations!XNA}

The main disadvantage of the {\xna}-based solution is that it is by
default Microsoft-only for the use primarily with Microsoft devices.
However, {\xna} could also at least partially run on {\mono}~\cite{wiki:mono-software} and {\monoxna}~\cite{monoxna}
(which uses {\opengl} calls to render things in \linux{} and \macos{X} for example) in {\csharp}.

Additionally, {\xna} video playback is only for \file{.wmv} files in a specific constrained
resolution settings, where the video has to be converted first before the {\xna} based
content resources can compile them into the internal \file{.xnb} representation like they
do for all the media content before it can be used by the application.

\subsection{Future Directions}
\label{sect:interactive-docu-future-directions}
\index{Interactive Documentary!Future Directions}

Some of the future directions have to do with expanding and
exhibiting the installations world-wide, not only as installations
or desktop applications, but also as web and mobile applications. In the meantime,
the other aspects would address some of the limitations
presented earlier (\xs{sect:interactive-docu-limitations}).

One aspect is the interview recording for the documentary can be done on the spot
and the content produced stored in one of the memory bubbles
while the installation is progressing at the same time
(with the care taken to temporarily stifle the speech recognition
pipeline to avoid the interview's spoken words interference).
Such submissions or users' interactions with the installation
documenting it can be the part of the ever growing footage available
at the installation and be different at each location where the
installation is deployed, including web.

\subsubsection{{\opengl} Installation}
\index{Interactive Documentary!Future Directions!OpenGL}

A more advanced bubble modeling is under consideration
in utilization of the softbody objects
(see \xc{chapt:softbody-objects})
instead of plain spheres to add realism and tangibility.
Bubbles can either modeled as softbody objects, or, e.g., as buckyballs
acting as animated containers~\cite{nasa-buckyballs-spitzer-2010,discovery-buckyballs-nebulae-2010}
with proper spring stiffness.

The amount of bubbles and their content is determined
presently via a preset set of media in the code
constants, but is planned to be more dynamic.

For the video (clip-based) bubbles
on zoom-in the sound is to be activated. The sound may also
optionally be activated on textual and photo bubbles if it is
present in a form of a narration.

Finally, more advanced interactivity is planned with haptics
and camera-marker motion capture, a {\wii} or {\cpp}-based
{\kinect} API implementation code with the techniques already
discussed.

\subsubsection{{\xna} Installation}
\index{Interactive Documentary!Future Directions!XNA}

An audience could query the contents of the bubbles
by some kinds of keywords, or create some brand new bubbles with 
realtime footage, such as dynamic recording of the additional footage of the users
interacting with the bubbles and retaining it in the 
bubbles beyond the eight currently active; perhaps in fancy
bubbles.
The audience will even be able to interact with the scene and
its objects in order to push forward the story scenarios via gestures.

\subsubsection{Rendering and Projection}
\index{Interactive Documentary!Future Directions!Rendering}
\index{Interactive Documentary!Future Directions!Projection}

An option is planned for the
projection to be stereoscopic. Another projection option is
the 180-degree screen VR system that we are considering
like {\caren}
as explained in the next section.

\subsubsection{VR and Documentary Production}
\label{sect:vr-and-cinema}
\index{Interactive Documentary!Future Directions!Virtual Reality}

Applying new media with VR equipment into documentary production definitely will be a
unique new model of representation in Cinema in the new digital era.
Again, typically, except for some rare occasions, documentary producers
and computer scientists and/or digital artists who work in computer graphics
are relatively far apart in their domains and rarely intercommunicate
to have a joint production; yet it happens, and perhaps more so
in the present and the future.
There were attempts to fuse interaction, drama, narration, and computers
in the past such as the works
\cite{interactive-rt-3d-drama,drama-edu-digital-2009,diamond-road-user-docu-2008,interactive-docu-golden-age-2009,interactive-docus-2003},
but not on such a scale as a complete VR system~\cite{vr-medical-research-docu}.

\subsubsection{Film Production and Studies}
\index{Interactive Documentary!Future Directions!Film Production}
\index{Interactive Documentary!Future Directions!Film Studies}

\longpaper{\paragraph{Documentary Film}}

Today, CG and animation are more advanced than traditional animation
according to the interactive, physical based (mimicking live-action camera), provide
more effective visual results, and impose less burden on the financial/budget
concerns~\cite{i-still-remember-opengl-remake-2011,role-cg-docu-film-prod-2009}.

In the past, the autobiographic documentary film type
was often constrained to for filmmakers themselves only as the
ones possessing the knowledge, resources, and equipment to do so.
With the wider accessibility of
digital content and high computing power these days, autobiographic
documentaries are now accessible to virtually any household~\cite{i-still-remember-opengl-remake-2011,role-cg-docu-film-prod-2009}.

Furthermore, the notion of script writing in terms of screenplay writing---i.e.,
the writing for the film screen or television screen for the new genre of documentary films
is an emerging, new research area requiring more work~\cite{role-cg-docu-film-prod-2009,i-still-remember-opengl-remake-2011}.

Moreover, teaching in the research and application
of ``CGI for documentary films'' is also an important
topic that is going to be inevitably introduced into
the core of Film Studies for scholarly research
and film production industry~\cite{role-cg-docu-film-prod-2009,i-still-remember-opengl-remake-2011}.

Finally, the
``rich woman'''s augmented and expanded interactive documentary installation~\cite{i-still-remember-opengl-remake-2011}
immersive and tangible experience will include the following 
items:
\begin{itemize}
	\item Research on and address the limitations mentioned earlier
	\item Feature haptic connectivity
	\item Softbody/bucky bubbles
	\item Black box and VR projection installation
	\item Stereoscopy
	\item Database-driven dynamic footage acquisition, selection, and display
\end{itemize}

\longpaper
{
\paragraph{Snowflake}

Short animation and feature fiction film.

{\todo}

} %\longpaper

\section{Illimitable Space Summary}
\label{sect:tangible-interaction-summary}
\label{sect:illimitable-space-summary}
\index{Illimitable Space!Conclusion}
\index{Conclusion!Illimitable Space}

We merge virtual and real performance augmenting the real performance 
with CG imagery driven in real-time by the actions of the performers. The CG 
images can be augmented in real-time with extra elements, clothing, 
extremities, etc., and driven by the real actors in the real-time and interact 
with the virtual beings. 
There are also a number of larger future directions to
extend this work expressed further in \xs{sect:tangible-interaction-future-directions}.
We begin, however, with the description of the contributions in this direction in
\xs{sect:tangible-interaction-contributions} and the correlation of the work
we presented earlier in \xc{chapt:illimitable-space} to the theatre elements in 
\xs{sect:tangible-interaction-theatre-elements}.

\subsection{Illimitable Space Contributions}
\label{sect:tangible-interaction-contributions}
\index{Illimitable Space!Contributions}
\index{Illimitable Space!Contributions!General Conceptual Design}
\index{Illimitable Space!Contributions!``Rich Theatre''}

Just as new media was shown to build on old media and include interaction
aspects into TV, cinema, and documentaries, a somewhat similar approach is
happening to theatre and performance arts.
The process begins with a historical record of Grotowski's experiments in ``poor theatre'',
both theory and practice~\cite{poor-theatre-1968}.
It is followed, among other works, by Giannachi in
her book~\cite{virtual-theatre-intro-2004} where she reviews interfacing
between digital arts and theatre production
with virtual reality.
And the most recent book by Salter about the influence of the
technology on the artistic performance, including historical overview of the
performance experimentation in theatre among several other fields, including
computational and responsive environments~\cite{entangled-2010}.

\begin{figure*}[htpb]%
	\centering
	\includegraphics[width=\textwidth]{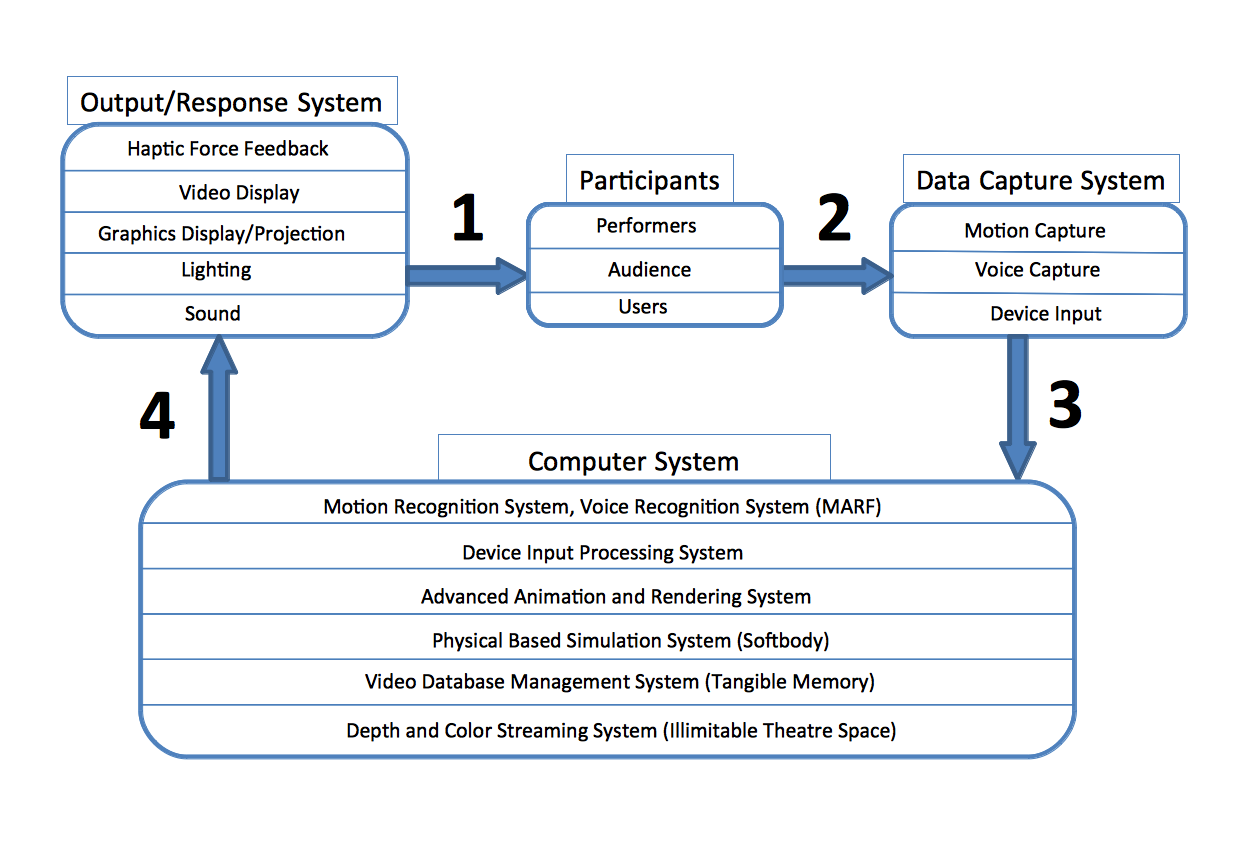}%
	\caption{Conceptual Design of a Generalized Interactive Documentary and Performance Arts Installation}%
	\label{fig:ConceptualDesign}%
\end{figure*}

My installation turns the ``Poor Theatre'' to an extremely ``Rich Theatre'',
in which one would create more dimensions than the traditional
theatre form allows.
It will allow the audience
to experience the freedom of the illimitable performance environment surrounding them,
rather than remain nearly fixed in their seats and stay at the passive observing level.
Moreover, the performance of actors also would be effectively interactive to 
computer generated virtual environment and improvised by the communication with the audience.

Additionally, this contribution reveals that the overall design can be generalized as shown in \xf{fig:ConceptualDesign}
to include and encompass all subsystems presented in this work so far, including
the interactive documentary and {\jellyfish} softbody simulation, and beyond.
As a result, as a more massive and general aspect of the interactive
concept design and technology have been transposed onto theatre production
advanced MoCap, tangible media, and audience participation in realtime computer
graphics effects.

This PoC installation has been featured in Concordia's Open House event in October 2012
(see \xf{fig:concordia-open-house}) and Exposcience 2012 at Stewart Hall, West Island,
Montreal.

\begin{figure*}[ht]
\hrule\vskip4pt
\begin{center}
	\subfigure[]%
	{\includegraphics[height=1.9in]{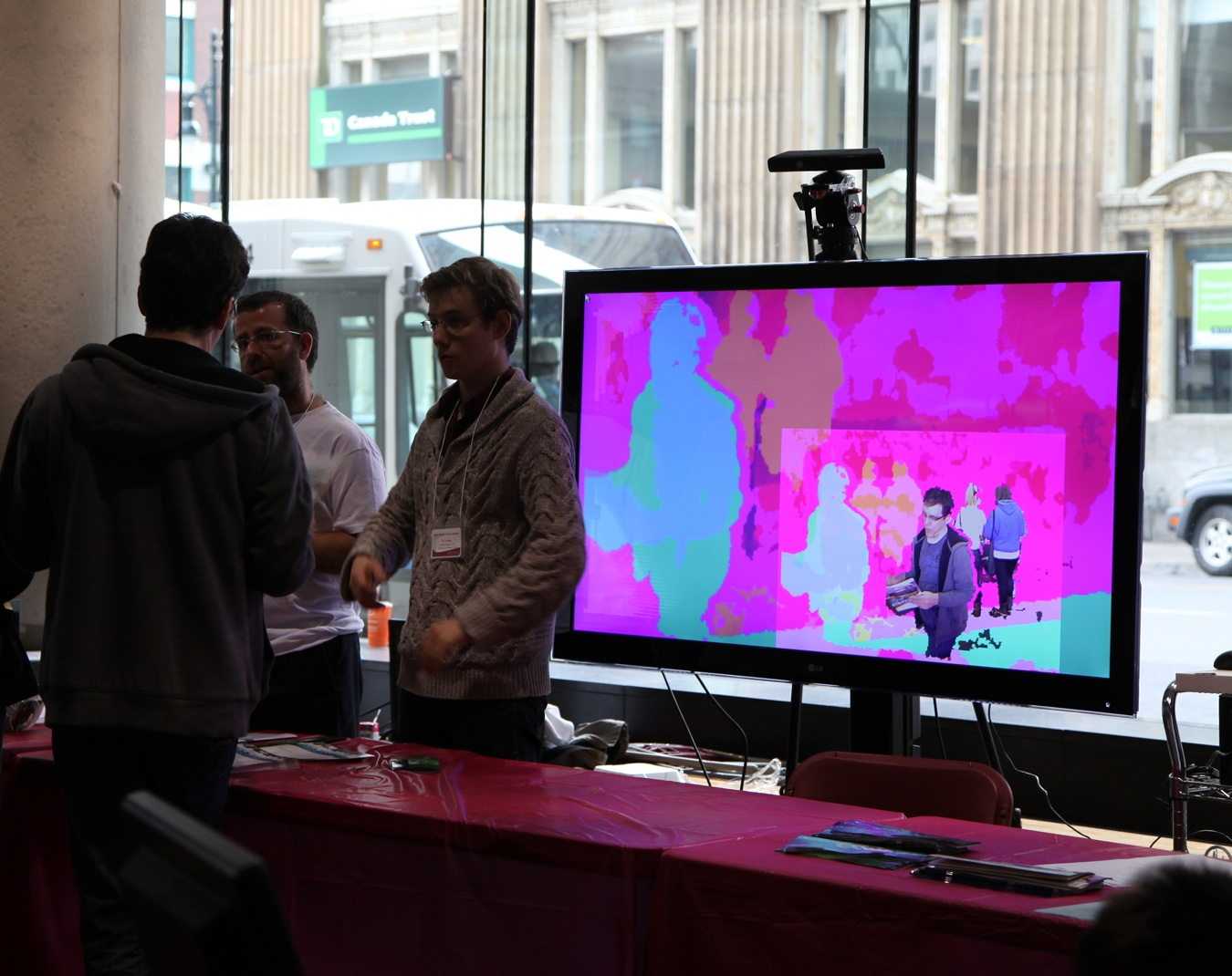}%
	 \label{fig:concordia-open-house-1}}%
	\hspace{.15in}%
	\subfigure[]%
	{\includegraphics[height=1.9in]{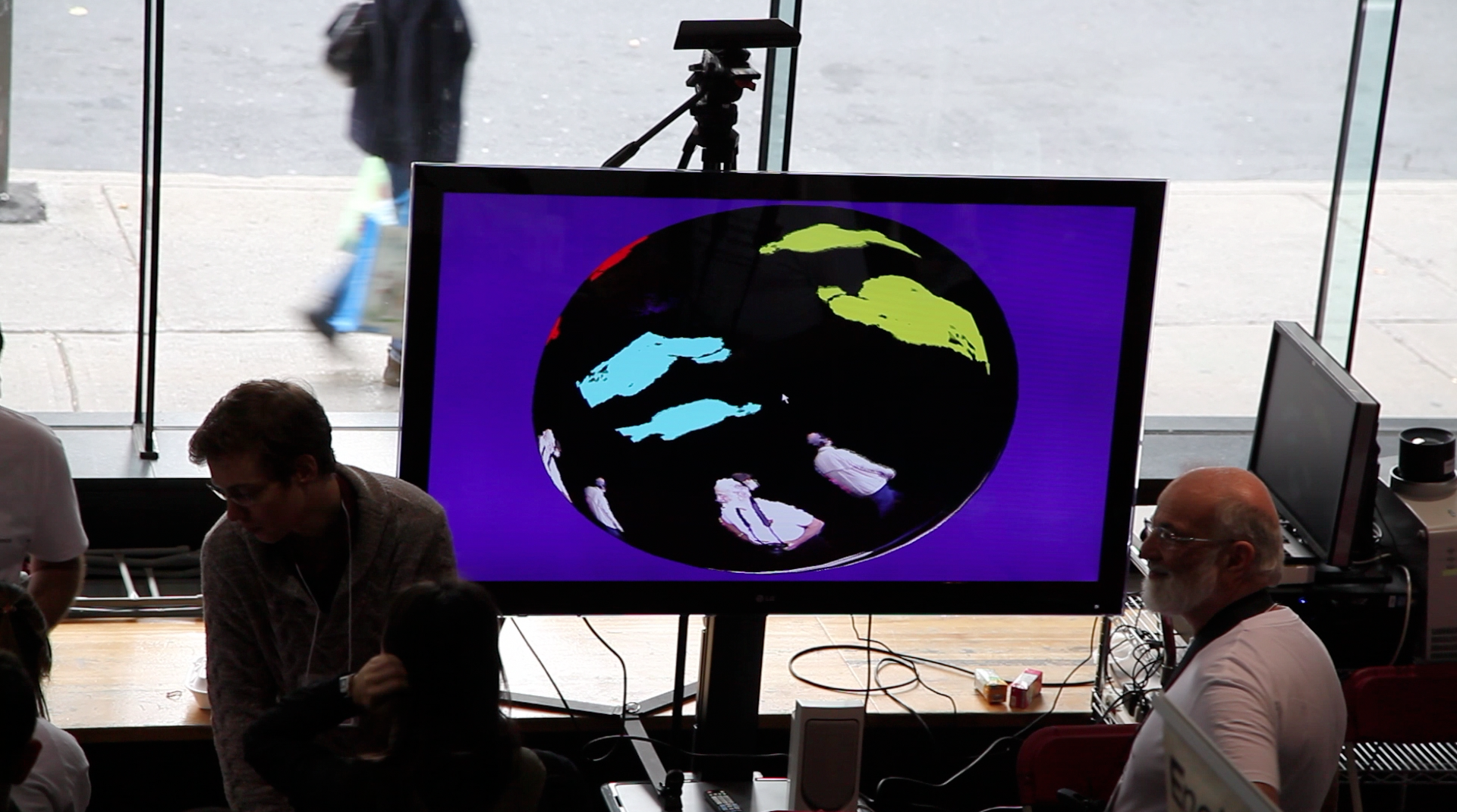}%
	 \label{fig:concordia-open-house-2}}%
\caption{Exhibition of the Installation and Concordia Open House, 2012, CSE}
\label{fig:concordia-open-house}
\end{center}
\hrule\vskip4pt
\end{figure*}

\subsection{Theatre Elements}
\label{sect:tangible-interaction-theatre-elements}
\index{Theatre elements}
\index{Illimitable Space!Theatre Elements}

In this section we describe our coverage of the theatre elements
in this interactive installation work. We decided to apply Aristotle's drama elements~\cite{poetics-aristotle}
to our contemporary theatre performance installation, such as \theatreelement{Plot},
\theatreelement{Character}, and \theatreelement{Spectacle} because they
have retained their importance through time,
are still the most widely used evaluative tools and general rules for artistic theatre performances. 

\subsubsection{\theatreelement{Plot}.}
\index{Illimitable Space!Plot}

In the six essential theatre elements, \theatreelement{Plot} is arguably
the most important one~\cite{poetics-aristotle,wiki:poetics-aristotle}.
If a performance does not carry a story, we would not
call it a theatre performance.
Therefore, a great story with dramatic storytelling approach
requires artistic creative talents.
Here, we won't give a concrete design of the story lines, but the storytelling approach.
All plots have a beginning, a middle, and an end. In conventional theatre,
actors could memorize their lines and have some flexibility in improvisation while they perform.
However, in our illimitable theatre performance system,
the story will not be told in a pre-planned linear way.
It requires more acting skills and talents from acting for an unplanned expedient and event
interrupted by audience.

One possible augmented example design could be that in the Cinderella's story, whether or not
Cinderella tries on the glass shoe when the messenger knocks on her door.
When she is hesitating via an interior monologue, the audience could apply gestures, such as wave or clapping to respond to her.
The suspension here is that the audience could help Cinderella to make a decision, either
she would succeed and so be with the prince forever, or she is doomed forever
as her step-family slave or the story somehow deviates and resolves differently or 
arrives to the same or similar ending via a different path.

\subsubsection{\theatreelement{Character}.}
\index{Illimitable Space!Character}

This element exists not only in a play, but also in TV, movies, and even video games.
Characters, who are agents of the \theatreelement{Plot}, provide the motivations for the events of the \theatreelement{Plot}.
However, the traditional theatre art has more limitations to portray characters than movies and games,
such as some fictionalized and deified figures with humanly impossible bodily or bodiless properties.
In traditional Chinese \emph{Nuo} opera, there are always ghosts and gods.
The performer plays a valiant god who dispels ghosts and devils.
The whole process how the god transfers into a human being's body,
is through the actor's body movement, facial expression, voice change, and different gestures.
With our illimitable system, we could project the computer generated ghost/god onto a media in front of actor,
or directed onto the actor, to show the mystery play.

\subsubsection{\theatreelement{Spectacle}.}
\index{Illimitable Space!Spectacle}

Everything that could be seen or heard on stage (such as actors,
sets, costumes, lights and sound) is a \theatreelement{Spectacle}.
In Japanese theatre, on the \emph{Noh} stage, the only ornamentation 
is the \emph{kagami-ita}, a painting of a pine tree at the back of the stage to
represent a distant view or prospect. 
It represents the tree through which \emph{noh} was passed down from heaven to mankind.
However, it is quite limited and boring in scenery expression.
Our illimitable system could expand the time and space on stage, which may improve the
\emph{Noh}'s actors' performance. When a \emph{Noh} actor walks far from \emph{kagami-ita},
the background CG scenery could even move based on his motion, speed and gesture
as if the space has been moving along with the actor from far to close.
Many new movie technologies, such as large scale projection with landscape and ocean floor, 
also could be experimented with here.

\subsection{Future Directions}
\label{sect:tangible-interaction-future-directions}
\index{Illimitable Space!Future Directions}

In the future plans we make a theatre production design,
a combination of all proposed components and elements below.

A logical extension and augmentation of the interactive documentary
\docutitle{Tangible Memories} would be the use of the illimitable space
concept applied to it and the advanced interaction specifically from \xc{chapt:illimitable-space}.
By being both the viewer and artist, in the \docutitle{Tangible Memories} project case,
for the dancer, the dance-based interaction with the memory bubbles is to playfully
scatter them around, meanwhile being the viewer to observe. To the audience, she is an
artist, and an actress.

The subsequent step is to give the illimitable space even more dimensions by
bringing it into the realm of the web-based installation, using, e.g.,
Unity (e.g., 14 locks---a \gametitle{Bart Bonte} game), HTML5, and Web~3
technologies combined with the database storage and retrieval and real-time
media streaming along with the webcams on the participant's computers
greatly scaling up the audience, similarly to the web-based interactive
documentaries~\cite{wiki:web-docu}.

\subsubsection{Augmenting Conceptual Design}
\index{Illimitable Space!Future Directions!Augmenting Conceptual Design}

At first, the audience and actors coexist in the closed stage space indicated
(see \xf{fig:stage}). It is not really a stage, but rather a space
divided into different zones.
The blue ring area is the four season zones: Spring, Summer, Autumn, and Winter.
When participants step into a zone, the projected graphics on the surrounding 
environment (walls) will be transformed.
The virtual seasons will be changed upon the participants' positions in the physical space.
The central circle area is where participants could explore
the dynamic lighting and sound.

\begin{figure*}[htpb]%
	\centering
	\includegraphics[width=.5\textwidth]{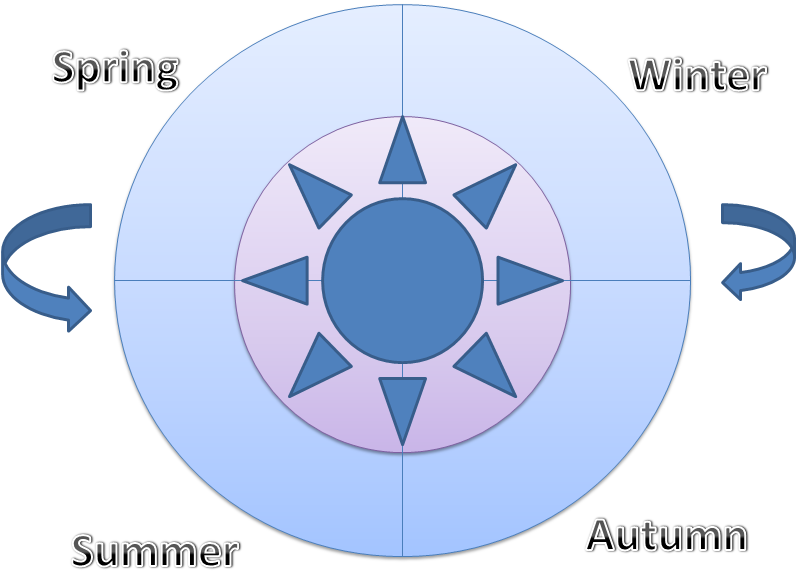}%
	\caption{Theatre Production Performance Space Design}%
	\label{fig:stage}%
\end{figure*}

The theatre performance will take place in a blackbox with a maze
to replace a standard stage.
There are different shapes of surfaces, such as hanging curtains,
altered floor, fog (used as projection screens for multimedia resources).
Actors perform in three different positions within the maze:
dance, sing, and monologue.
Audience walk freely in the environment and their joining
to the actors' performance, in gesture, voice, or texts
could directly contribute to the whole performance.  

\paragraph{Particles.}
\index{Illimitable Space!Future Directions!Particles}

Particles are a very efficient model for computer rendering, especially with animation.
Same groups of particles could be reused for 
Spring rain, Summer fireworks, Autumn leaves, and Winter snowflakes in the virtual environment,
accomplished with the corresponding physical based algorithms.
As mentioned in \xs{sect:tangible-interaction-animation} in the absence of
the interaction input from humans, the animation, lighting, and audio
revert to their default ``stable'' states and scripted scenarios, e.g.,
such as snowflakes falling down to the ground
if there is no wind.

\subsubsection{Remaining Theatre Elements}
\index{Illimitable Space!Future Directions!Theatre Elements}

For completeness, as a part of the future work, we need to finalize
the review of the remaining three Aristotle's drama elements~\cite{poetics-aristotle}:
\theatreelement{Diction}, \theatreelement{Melody}, and \theatreelement{Thought}.
The work on these has started already.

\subsubsection{Tools}
\label{sect:illimitable-tools}
\index{Illimitable Space!Future Directions!Tools}

\paragraph{{\maxmsp}/{\jitter}/{\puredata}.}

Delivering the installation and its implementation of interaction to the masses
and artists can be achieved via {\maxmsp} and {\jitter}~\cite{jitter,maxmsp} or their open-source
alternative of {\puredata} that create artistic media patches in their corresponding
data flow languages.
There is an \emph{external} (plug-in) available for {\maxmsp} in {\jitter} to connect
to {\kinect}---\api{jit.freenect.grab}~\cite{jit-freenect-grab}; as well as the Novint {\falcon}
haptics sensor was used with {\puredata}~\cite{puredata}~\cite{haptic-opengl-vr-env-pdcon07}
and {\opengl}. Both can be used to augment the interaction and be available via
{\jitter} or {\puredata} data-flow \emph{patches} (programs) for much easier 
perusal by the artists at large for more than the Microsoft's proprietary platforms
allowing us to re-enhance back again the {\opengl} \docutitle{Tangible Memories}
(see \xs{sect:interactive-docu-opengl-installation-limitations})
as well as the {\csharp} version can be used with {\mono}
and {\monoxna}~\cite{wiki:mono-software} (see \xs{sect:interactive-docu-xna-installation-limitations}).

\paragraph{{\maya}.}
\label{sect:illimitable-space-maya}

{\maya}~\cite{maya,maya-api-mpxtools,maya-mfncamera,maya-plugins-highend3d}
is a great tool to create virtual models of the 3D spaces and environments
and do their animation that can be the dynamic virtual 3D props in any designed
interactive theatrical projected installation (e.g. the maze).
Additionally, starting from a specific
version {\maya} began supporting stereo buffers for keyframed animation, which can be
very helpful for creation of the environment and space similar to that of
\animtitle{Spectacle}~\cite{spectacle-anim-film-2003},
but in stereoscopic 3D. {\maya} achieves that by providing implementation for graphics cards that
have hardware stereo buffers (as well as the plug-in we developed prior to that support for
the cards that do not have the built-in stereo capability~\cite{stereo3d,stereo-plugin-interface})
for greater perception of the virtual and augmented reality in the performative space.

\subsubsection{Augmenting Lighting and Audio}
\label{sect:illimitable-lighting-audio-future}
\index{Illimitable Space!Future Directions!Augmenting Lighting}
\index{Illimitable Space!Future Directions!Augmenting Audio}

To enhance the perception and participatory experience further,
I will use a few dynamic stage spot lights and 
some off-stage lights (in my production, they will be
positioned within props based on the story scenarios) to create
an artistic conceptual environment.
As proposed earlier,
the lighting and audio should be programmed to respond in real-time, 
dynamically stimulated by actors' or audience's actions instead of
statically configured each time.

The sound visualization and response are also planned to be changed
to be more representative of the mood, season, and tune of the audio,
voice, tune, and the like, especially during the transition moments
and dynamic music selection and playback made accordingly.

\subsubsection{Augmenting Projection}
\label{sect:illimitable-space-projection-future}
\index{Illimitable Space!Future Directions!Augmenting Projection}

The installation is planned to be projected in a blackbox
or an enclosed large room alike.
The space could be in any shape, which then will be divided into four season
zones with the actors movement influencing the digital
projections by their actions surrounding the audience seated in the
center of such a theatre with motion-enabled chairs and all-dimensional
$360^{\circ}$ projections including ceiling and the floor projector or
glass screens and even onto fog.
The projection requires multiple projectors with split video signals.
A more advanced technique I intend to use is 
stereoscopic rendering and projection (see \xs{sect:illimitable-space-maya}). 

\paragraph{Holographic Displays.}
\index{Illimitable Space!Future Directions!Holographic Display}

It is important to explore the newly emerged holographic display technology, which is
currently very expensive, but with time the costs should
come down and it will likely be an excellent performance tool among
other things; in the mean time I will seek various sources of funding.

\subsubsection{Nezha.}
\index{Illimitable Space!Future Directions!Nezha}

I rediscovered a Chinese animation film, \animtitle{Nezha Conquers the Dragon King} (1979),
which was made in 1970s by the Shanghai Animation Studio. 
It was incredibly beautiful and 
has accompanied me in my childhood in China.
This fabulous artwork piece attained international reputation 
in the past in various festivals and reviews.

The \theatreelement{Plot}
is based on a very ancient Chinese traditional myth story, which has only
been, and could have only been, produced in animation film because 
it is impossible for traditional theatre
to represent the mythology scenery of the story,
such as the \theatreelement{Character} of Nezha (a boy was born during the Shang Dynasty, 
always depicted as an incarnation of brave, protagonist, and superior power) 
has three heads and six arms and has the ability to spit fire.
There are many Chinese opera elements this film has borrowed, such as
costumes, characters' gestures and movements, music background,
and stage speech.

However, such a theatre performance could only be made with the \emph{Rich Theatre}
by using digital technology and the new media to create some special performance effects,
such as an under-sea environment, heaven, fire, etc.,
and also to include the interaction between a 3D dragon
and a Nezha-playing actor with the recent technological innovations 
such as {\kinect}, CG, projection, and others previously described.

\longpaper
{
\subsubsection{15 Years.}

Another proposed production may be a personal, additional project, inspired by Artaud's theory,
``We shall not act a written play, but we shall make attempts at direct staging, 
around themes, facts, or known works.''  
I would produce a theatre performance, only depending on facts and improvisation,
without story script and advanced rehearsal. Since I attended theatre training in Shandong
Fine Arts University in 1994 and after graduation most of our classmates haven't
seen each other for more than ten years. Luckily, the majority of my alumni are still devoting themselves
in theatre performance and direction career. The reunion will actually take place
on theatre stage. It will be interesting, unexpected, and a real-time improvising performance. 

``15 Year Later Reunion'' would distinguish theatre and film and prove
theatre is extraordinary art form compared to other art type, such as film.
} %\longpaper

\longpaper{\input{discussion}}

\longpaper
{
\section{My Next Journey in Postdoctoral Research}

This postdoctoral research joins the disciplines of Theatre Performance Arts,
Information Technology and Human Computer Interaction, documentary film-making,
and Computer Graphics with the aim of forming an alliance for both artistic
expression and technological development. The focus of this research is to
transform theatrical performance environments under the influence of the new
interactive digital technologies such as realtime physical based simulation
system, motion capture system, haptic responsive environment, interactive media
design, realtime theatrical sound and light technologies, human-computer interfaces,
stereoscopic virtual reality techniques, and drama therapy. The new
possibilities of theatre stage conception design, theatre performance and
direction, as well as strategies and ways of making for technical software
realization will be thoroughly explored, and elaborated and creatively
implemented.

My current proposed research is to extend original re search work of the SIP study
program not only by developing some new technical solutions to artistic
productions, but also, more importantly, making an in-depth the study and research
of the art and performance in theatre, documentary film, and human-computer
interaction. My goal is to contribute a new theory and approach bridging theatre,
film, and interactive media. Therefore, the intended research gradually and
incrementally will focus on the mentioned aspects and will include:
1.theatre arts and the performance; 2.theoretical and practical experience in
interactive documentary film; 3.extensive 3D computer animation knowledge and
programming skills.
b) The fact that so many disciplines converge in one and the whole of theatre, the
theatre itself gets a modern look and update without losing its existing precious
human elements is already original in itself.
c) The project work will concentrate primarily on three dimensions of research
that will be further investigated from my doctoral thesis with regard to the
concept and the realities of stage performance encompassing all the elements:
human-computer interaction as a field for the exploration of human self understanding
and exposition in performance environments impacted by digital and
computer graphics artifacts, mechanization and virtualization, motion tracking,
projection, real-time animation and physical based method for more immersive
sensory experience as well as education and theatre therapy.
The stage is a virtually illimitable space connecting virtual/physical locations
linked together in one performance pieces either via video and/or Internet links

and projections on various media and surfaces and stereoscopic displays.
d) The artistic process is both the further generalization and refinement of the
process detailed in my doctoral research with more pieces put together and more
ways of interaction and technologies used. Essentially, it amounts to the
following:
State 1: In a initial theatrical space, eg. blackbox, lighting, projection, sound
props, and related facilities are properly set up for actorsÕ performance,
audienceÕs participation, and other user groupÕs interaction devices.
State 2: Participants including actors, audience, and other users could freely
motion in the physical space. They could move their body with different
gestures, talk or sing, and could play musical instrument or touch haptic devices.
The state of human beingÕs performance, such as motion of actorsÕ body movements,
audienceÕs voices and the forces of users applying through a haptic device, could
be all captured through the integrated Data Capture System.
State 3: The captured data from the theatrical space is transferred to PACS
(Performance Assistant Computer System) to be analyzed. The computer system mainly
contains three big components: recognition, computing, and media output. The
motion recognition system, voice recognition system, and device processing system
are filtering, parsing, and sorting motion, audio, and device/sensor input
signals captured into system. The computing system then compute and simulate
physical based computer graphics and dynamic audio based on the mathematical and
physical based algorithms, and fetch proper video footage from video database
management system follow the querying algorithm.
State 4: The outputting system then reconfigures the theatrical technologies
after based on the computation results and state updates. The outputting system
then re-renders the newly computed computer graphics onto projectors, plays
generated audio signals through speakers, and even controls dynamic lighting
interacted with originally as a result of participantsÕ actions.
e) The dissemination of the research and creation will be in two major ways: the
peer-reviewed publication venues of conferences and journals primarily related to
the science technology aspects of it and public installations, expositions,
gameplay, web frontiers, festivals, and shows in world theatres, science centers,
cinemas, university spaces would be primarily for the artistic aspects of it.
f) The majority of the project will span over two years, where the majority of the
development and refinement of the doctoral thesis techniques will take place along
with experimental on-stage performances and setups. The 2nd year will focus more
on the public exhibitions, performances, and research publications throughout the
worldÕs events.

I will continue working on the multidisciplinary areas in my postdoctoral research,
which includes computer graphics, theatre performance, virtual reality, and video
arts. In my current doctoral research, I am finding the initial relationship
between science and arts, computer and human interaction (the subset of Information
and related technologies). I have implemented the three proof-of-concept
installations to demo the ideas:
- jellyfish: realtime physical based soft body simulation
- tangible memories: the new role of computer graphics in making interactive
documentary production
- illimitable theatre space: the realtime responsible environment for various
theatre performance
I have explored and achieved a lot in each discipline, all of them crossing
boundaries into each other emerging as one whole; however, many interesting future
work items listed in my dissertation could be done and need to be done to enrich
the works further... To summarize, how computer and human could interact
coherently, how computer technology could contribute to human being, how human
motion, gesture, could stimulate a dynamic physical based system and its
computation in real time... I want to answer these research questions not only
through theory, but also through practical experience, human performance, theatre
performance and applications.
Faculty of Information's iSchool at University of Toronto, which is committed to
understanding the role of information in nature and human endeavors, conducts
research into the fundamental aspects of information and related technologies. It
is the best suitable research place for me given all the options and constraints
because it has achieved not only international excellence in research and education
in Information, but also supports innovative and entrepreneurial individuals and
multidisciplinary research projects. I have contacted the Dean, Professor Seamus
Ross, who has tentatively accepted me, and is very interested in my research,
especially in the realtime soft body simulation, the role of CG in documentary
film, etc. Moreover, Professor Ross's research involves the preservation of
cultural heritage and scientific digital objects, humanities informatics, and the
application of information technology which are very relevant to my postdoc
research scientific and humanity components.
I also include the Central Academy of Drama in China as a co-host university for my
postdoctoral research because the theatre is the all-encompassing media and goes
back to my roots. The Central Academy of Drama is regarded as the most preeminent
artistic school in Asia and the most prestigious artistic schools in China. It has
artistic-driven laboratory including a black-box situated within the theater
school. Professor Ruru Ding, the Dean of Theatre Direction Department is very
interested in the realtime physical based simulation system and responsive theatre
environment I implemented over the doctoral studies. He will be very excited to
explore with me through the new Human Computer Interaction
approach, how the new digital technology could best serve the conventional theatre
performance.
Thus, it is certainly will be a great collaboration between the two
schools with two experts for my multidisciplinary postdoctoral
research.

\paragraph*{In Theatre}
Miao Song's research topic, Computer Graphics and Multimedia Applications in Theatre Production,
is a research project which combines computer technology and artistic creation in order to serve
artists to augment their creative capabilities to present their artistic and performance works. Such
an integration may portray the artistic imagination with real-time visual and auditory effects directly
on stage and expand the space of artistic expression to a larger extent so that directors and actors
could enrich and enhance the charm of theatre performance arts.
My research has been focused on spatial and narrative drama and non-linear storytelling in theatre
performance in the past few years. The goal of my research is to propose a new possible spatial
narrative and non-linear storytelling approach in theater performance in order to enrich this
artistic form, make it more expressive, and produce the unique artistic effects with the attractive
capabilities of the newly emerged computer technology.
Miao SongÕs proposed research in her postdoctoral studies concisely fits within my research
interests. I therefore hope she could work with me, not only to collaborate with me on my
research projects and theatre productions, but also for me to properly assist her to complete her
own research project.

Since the core of Miao SongÕs research project is how computer technology could enhance the
theatre performance and enrich artistic creation, it is very important to verify and validate her
research work, for example if her implemented innovative computer applications could actually
operate on stage and produce the desired effects in creative theatre practice. Therefore, in her
postdoctoral study period, I plan to combine her research results and the implemented application
to a theatrical creation and theatre performance which I am going to direct. She can verify her
research findings and improve her application through this theatre production opportunity as
practical validation experience.
The Central Academy of Drama is the highest institution of Chinese Theatre (Drama) Education
and Theatre Directing Department was one of established departments of when the university
was first founded. In its 60 years of teaching history, our Theatre Directing Department has
produced many well known national outstanding theatre performances.
Currently, the Theatre Directing Department produces two formal multi-act play theatre
productions and officially perform in theatre every year. We have tried to introduce the use of
multimedia computer technology in our artistic creation and performance at different levels in
previous theatre productions and itÕs time to take advantages of the newest interactive technology.
The accumulated drama directing and performance experience, regular first hand production
practice, and related research background to Miao SongÕs proposed research work will play a very
important role for her not only to complete her research work, but also her experimental
theatrical practices.

\paragraph*{Information Technology}

} %\longpaper

\phantomsection
\addcontentsline{toc}{chapter}{Bibliography}%
\small
\bibliography{msong-phd-thesis}
\normalsize

\phantomsection
\addcontentsline{toc}{chapter}{Index}%
\scriptsize%
\printindex%
\normalsize%

\phantomsection
\chapter*{Appendix}
\addcontentsline{toc}{chapter}{Appendix}
\appendix

\chapter{User Manuals}
\label{appdx:user-manuals}

\longpaper{
{\todo}
} % \longpaper

\section{Curve-Based Animation Control Keys}
\label{appdx:curve-anim-controls}

\longpaper{
{\todo}
} %\longpaper

The keys are as follows:

\begin{itemize}
\item \texttt{+} -- zoom in by decreasing \texttt{fovy} (not part of camera)
\item \texttt{-} -- zoom out by increasing \texttt{fovy} (not part of camera)
\item \texttt{W}, \texttt{8} -- tilt camera up
\item \texttt{X}, \texttt{2} -- tilt camera down
\item \texttt{D}, \texttt{6} -- turn right
\item \texttt{A}, \texttt{4} -- turn left
\item \texttt{ARROW KEY UP} -- move camera forward
\item \texttt{ARROW KEY DOWN} -- move camera backward
\item \texttt{ARROW KEY LEFT} -- move camera left
\item \texttt{ARROW KEY RIGHT} -- move camera right
\item \texttt{R} -- place camera into the initial position (not a full reset)
\end{itemize}

\section{Tangible Memories {\kinect} Interaction}
\label{appdx:kinect-interaction}

\longpaper{
{\todo}
} % \longpaper

\subsection{Keyboard Controls}

\begin{description}
\item [ESC] -- complete the viewing/interaction session
\item [V] -- toggle video playback
\item [K] -- toggle main Kinect video feed
\item [L] -- toggle Kinect skeleton capture
\item [D] -- toggle Kinect depth capture
\item [O] -- initialize and make visible skybox environment or disable it
\item [B] -- toggle the visibility of the skybox environment (if initialized)
\item [G] -- toggle the visibility of the floor
\item [P] -- toggle the camera projection of the floor
\item [C] -- toggle clean screen
\item [F] -- toggle full screen
\item [1] -- add simple LOD 2D memory bubbles to the scene
\item [2] -- add simple LOD 3D memory bubbles to the scene
\item [3] -- add fancy LOD 3D memory bubbles to the scene
\item [F1] -- exclusively add simple LOD 2D memory bubbles to the scene; exclude all others
\item [F2] -- exclusively add simple LOD 3D memory bubbles to the scene; exclude all others
\item [F3] -- exclusively add fancy LOD 3D memory bubbles to the scene; exclude all others
\item [S] -- toggle the speech-based updates. Only active when speech processing
             was actually initialized (with `T' or by default).
\item [T] -- initialize or unload speech processing. Implies turning off speech
             updates on unload and the reverse on load. Speech updates `S' can be
             selectively turned on and off with `S' when `T' is on.
\item [M] -- toggle music visualizer
\item [F11] -- toggle wireframe mode for the computed memory bubbles

\item [X] -- toggle collision detection among bubbles

\item [F4] -- simulate ``red'' command for testing
\item [F5] -- simulate ``orange'' command for testing
\item [F6] -- simulate ``yellow'' command for testing
\item [F7] -- simulate ``green'' command for testing
\item [F8] -- simulate ``cyan'' command for testing
\item [F9] -- simulate ``blue'' command for testing
\item [F10] -- simulate ``violet'' command for testing

\item [0] -- simulate ``play'' command for testing
\item [F12] -- simulate ``stop'' command for testing

\item [Left Control] -- toggle white/black background

\end{description}

\longpaper{
{\todo}
} % \longpaper

\subsection{Mouse Controls}

Rotation of the camera in the skybox world when it is enabled.

\longpaper{
{\todo}
} % \longpaper

\subsection{Voice Commands}
\label{sect:kinect-voice-commands}

\begin{description}
\item [red bubble, red bubble please] -- calls up red media bubble
\item [orange bubble, orange bubble please] -- calls up orange media bubble
\item [yellow bubble, yellow bubble please] -- calls up yellow media bubble
\item [green bubble, green bubble please] -- calls up green media bubble
\item [cyan bubble, cyan bubble please] -- calls up the light blue media bubble
\item [blue bubble, blue bubble please] -- calls up blue media bubble
\item [violet bubble, violet bubble please] -- calls up purple media bubble
\item [purple bubble, purple bubble please] -- calls up purple media bubble
\item [play] -- start/resume video playback
\item [stop] -- pause video playback
\end{description}

\longpaper{
{\todo}
} % \longpaper

\subsection{Gestures}

\begin{description}
\item [left hand over play button] -- start/resume video playback
\item [left hand over stop button] -- release the bubble held in the right hand from the viewer and stop the video playback in it
\end{description}

\longpaper{
{\todo}
} % \longpaper

\chapter{Glossary}
\label{appdx:abbrs}

\begin{description}
	\item[API]--- application programming interface\footnote{\url{http://en.wikipedia.org/wiki/API}} (see \xs{sect:bg-api})
	\item[ARB]--- OpenGL Architecture Review Board
	\item[AVI]--- Audio Video Interleave format
	\item[BJIFF]--- Beijing International Film Festival
	\item[BMP]--- bitmap image format
	\item[BSP]--- binary separating planes (see \xs{sect:bg-collision-detection})
	\item[BVHs]--- bounding volume hierarchies  (see \xs{sect:bg-collision-detection})
	\item[CD]--- collision detection (see \xs{sect:bg-collision-detection})
	\item[CG]--- computer graphics
	\item[CGEMS]--- computer graphics gems~\cite{cgems}
	\item[CGI]--- computer-generated imagery\footnote{\url{http://en.wikipedia.org/wiki/Computer-generated_imagery}}
	\item[CPU]--- central processing unit
	\item[{\cugl}]--- Concordia University Graphics Library~\cite{cugl} (see \xs{sect:softbody-jellyfish-api-cugl})
	\item[DLOD]--- discrete LOD (see \xs{sect:bg-subdiv-techniques})
	\item[FFT]--- fast Fourier transform~\cite{shaughnessy2000}
	\item[FPS]--- frames-per-second
	\item[{\glsl}]--- OpenGL Shading Language (see \xs{sect:bg-gpu})
	\item[{\glui}]--- {\glut}-Based User Interface Library~\cite{gluiManual}
	\item[{\glut}]--- {\opengl} Utility Toolkit
	\item[GPGPU]--- general purpose GPU computing (see \xs{sect:bg-gpu-shaders})
	\item[{\gpu}]--- graphics processing unit (see \xs{sect:bg-gpu})
	\item[GUI]--- graphical user interface
	\item[HCI]--- human-computer interaction
	\item[HD]--- high definition
	\item[HDL]--- haptics device library~\cite{falcon-sdk}
	\item[{\hlsl}]--- High-Level Shader Language (see \xs{sect:bg-xna-hlsl})
	\item[HMD]--- head-mounted display
	\item[ITV]--- interactive TV
	\item[LOD]--- level of detail (see \xs{sect:bg-lod-techniques})
	\item[MoCap]--- motion capture
	\item[MVC]--- Model-View-Controller architecture
	\item[{\oglsf}]--- OpenGL Slides Framework
	\item[{\ogre}]--- Object-oriented Graphics Rendering Engine~\cite{ogre}
	\item[{\opengl}]--- open graphics library~\cite{opengl} (see \xs{sect:bg-opengl})
	\item[OS]--- operating system
	\item[NPR]--- non-photorealistic rendering
	\item[PoC]--- proof-of-concept
	\item[RK4]--- Runge-Kutta 4th order differential equation integrator
	\item[SDK]--- software development kit
	\item[SDL]--- Simple Directmedia Layer library~\cite{libsdl}
	\item[SIGGRAPH]--- Special Interest Group on GRAPHics and Interactive Techniques\footnote{\url{http://en.wikipedia.org/wiki/SIGGRAPH}}
	\item[SIP]--- Special Individualized Program
	\item[STB]--- set-top box\footnote{\url{http://en.wikipedia.org/wiki/Set-top_box}}
	\item[VB]--- Visual Basic
	\item[VGA]--- video graphics array
	\item[VR]--- virtual reality
	\item[XML]--- Extensible Markup Language
	\item[{\xna}]--- recursive acronym for ``\textbf{X}NA's \textbf{N}ot \textbf{A}cronymed''~\cite{video-game-encyclopedia-2012}\footnote{\url{http://en.wikipedia.org/wiki/Microsoft_XNA}}
\end{description}

\end{document}